\documentclass[11pt,aps]{report}
\def\GeVc{\mathbin{\hbox{GeV}\mkern - 3mu/\mkern -1mu c}}
\def\GeV2c{\mathbin{\hbox{GeV}\mkern - 3mu/\mkern -1mu c^2}}
\def\lesim{\mathrel {\vcenter {\baselineskip 0pt \kern 0pt
\hbox{$<$} \kern 0pt \hbox{$\sim$} }}}
\def\gesim{\mathrel {\vcenter {\baselineskip 0pt \kern 0pt
\hbox{$>$} \kern 0pt \hbox{$\sim$} }}}
\def\geant4{{\sc geant4}}
\def\glg4sim{{\sc glg4sim}}
\def\inch{{\tt "}}

\usepackage{amsmath}
\usepackage{graphicx}
\usepackage{chapterbib}
\usepackage{color}
\usepackage{eso-pic}
\usepackage{fancyhdr}

\newif\ifpdf
 \ifx\pdfoutput\undefined
 \pdffalse                  
\else
 \pdfoutput=1               
 \pdftrue
\fi

\ifpdf
 \usepackage{url,boxedminipage}
 \usepackage{graphicx,thumbpdf}
 \definecolor{rltred}{rgb}{0.75,0,0}
 \definecolor{rltgreen}{rgb}{0,0.5,0}
 \definecolor{rltblue}{rgb}{0,0,0.75}
 \usepackage[colorlinks%
  ,bookmarks%
  ,urlcolor=rltblue%
  ,filecolor=rltgreen%
  ,linkcolor=rltred%
  ,pdftitle={pdftitLe}%
  ,pdfsubject={pdfsubject}%
  ,pdfkeywords={pdfkeywords}%
  ,pdfauthor={Joe User <joeuser@fnal.gov>}%
  ,pdfpagemode={UseOutlines}%
  ,bookmarksopen=true%
  ,bookmarksnumbered=true%
  ,pdfstartview={Fit}%
  ]{hyperref}
 \DeclareGraphicsExtensions{.pdf,.jpg,.jpeg}
\else
 \usepackage{graphicx}
 \DeclareGraphicsExtensions{.eps,.ps,.eps.gz,.ps.gz}
 \newcommand{\href}[2]{#2}                   
 \usepackage[dvips,pagebackref]{hyperref}    

\fi 

\begin{document}
\begin{figure}
\leftline{\includegraphics[scale=0.2]{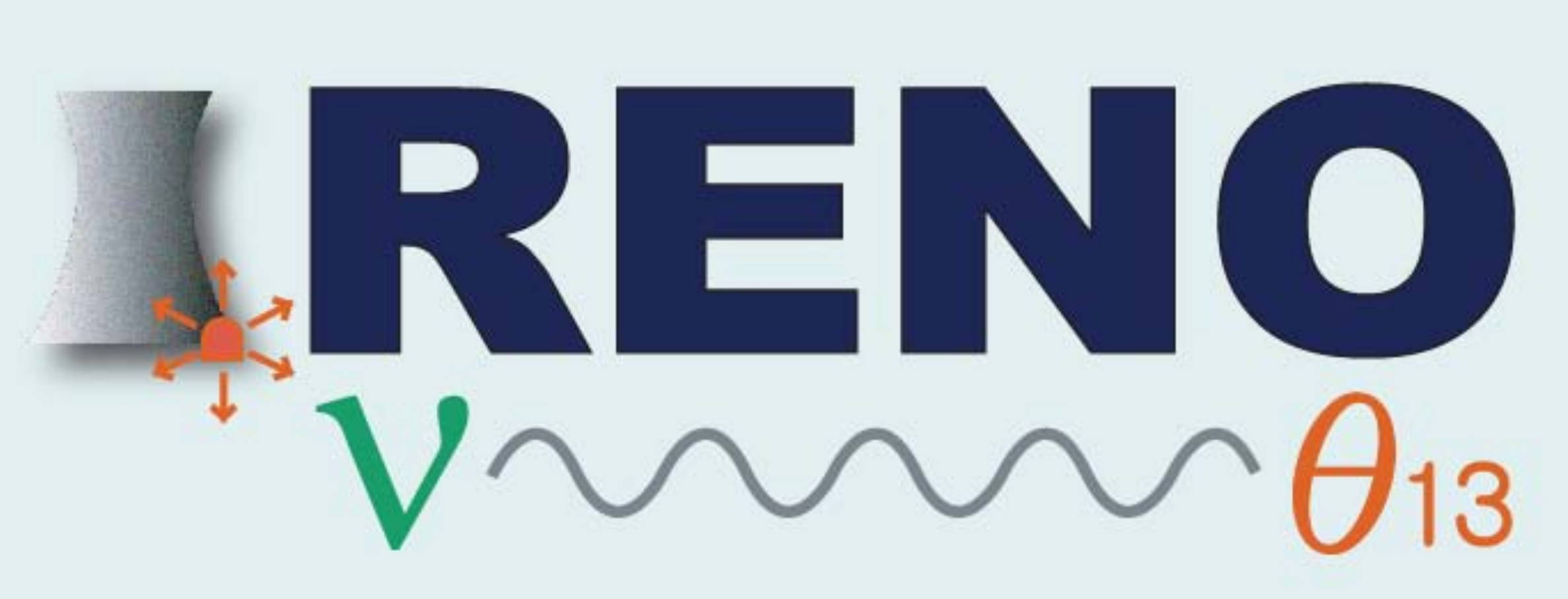}
}
\end{figure}
\title{
\vspace{0.2in}
RENO: An Experiment \\
for Neutrino Oscillation Parameter $\theta_{13}$ \\
Using Reactor Neutrinos at Yonggwang\\
\vspace{0.3in}
{\large (Proposal and Technical Design Report)
}
\vspace{1in}
\begin{figure}[!h]
\centerline{\includegraphics[scale=1.25]{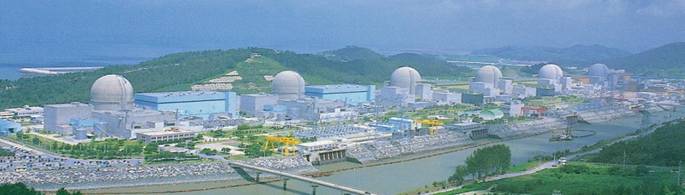}}
\end{figure}
}

\maketitle

\unitlength=1mm
\protect\begin{picture}(40,30)
\put(0,0){}
\end{picture}

\vspace{1.5in}
\font\eightit=cmti8
\def\r#1{\ignorespaces $^{#1}$}
\lefthyphenmin=2
\righthyphenmin=3
\author{
\noindent
\begin{center}
{
{\Large RENO Collaboration}\\
\vspace{10mm}
J.~K.~Ahn,\r{8}
S.~R.~Baek,\r{12}
S.~Choi,\r{11}  
Y.~Choi,\r{12}
I.~S.~Chung,\r{1}
N.~Danilov,\r{6}
J.~S.~Jang,\r{1}
E.~J.~Jeon,\r{9}
K.~K.~Joo,\r{1}
Y.~H.~Jung,\r{7}
B.~R.~Kang,\r{11}
W.~K.~Kang,\r{9}
A.~Kim,\r{7}
B.~C.~Kim,\r{1}
D.~H.~Kim,\r{12}
H.~S.~Kim,\r{2}
J.~Y.~Kim,\r{1}
S.~B.~Kim,\r{11}$^*$
S.~H.~Kim,\r{1}
W.~Y.~Kim,\r{7}
Y.~D.~Kim,\r{9}
Yu.~Krylov,\r{5}
V.~Kuznetsov,\r{7}
H.~S.~Lee,\r {8}
J.~Lee,\r {11}
I.~T.~Lim,\r {1}
K.~J.~Ma,\r {9}
G.~Novikova,\r{5}
Y.~D.~Oh,\r {7}
M.~Y.~Pac,\r{3}
C.~W.~Park,\r{12}
J.~S.~Park,\r{11}
K.~S.~Park,\r{10}
J.~W.~Shin,\r{11}
S.~Stepanyan,\r{7}
J.~S.~Suh,\r{7}
E.~Yanovich,\r{5}
I.~Yu,\r{12}
}
\end{center}
\begin{center}
\end{center}
\begin{center}
\r {1} {\it Chonnam National University, Kwangju, 500-757, Korea} \\
\r {2} {\it Chonbuk National University, Jeonju, 561-756, Korea} \\
\r {3} {\it Dongshin University, Naju, 520-714, Korea} \\
\r {4} {\it Gyeongsang National University, Jinju, 660-701, Korea} \\
\r {5} {\it Institute for Nuclear Research, Moscow, 117312, Russia} \\
\r {6} {\it Institute of Physical Chemistry and Electrochemistry, Moscow, 117071, Russia} \\
\r {7} {\it Kyungpook National University, Taegu, 702-701, Korea} \\
\r {8} {\it Pusan National University, Pusan, 609-753, Korea} \\
\r {9} {\it Sejong University, Seoul, 143-747, Korea} \\
\r {10} {\it Seokyeong University, Seoul, 136-704, Korea} \\
\r {11} {\it Seoul National University, Seoul, 151-742, Korea} \\
\r {12} {\it Sungkyunkwan University, Suwon, 440-746, Korea} \\
\end{center}
\vspace{30mm}
* Contact person: Soo-Bong Kim (sbkim@phya.snu.ac.kr)
}

\chapter*{Project Summary}
An experiment, RENO ({\it R}eactor {\it E}xperiment for {\it N}eutrino 
{\it O}scillation), is under construction to measure the smallest but
yet to be measured neutrino mixing angle ($\theta_{13}$) using 
anti-neutrinos emitted from the Yonggwang nuclear power plant in Korea
with world-second largest thermal power output of 16.4 GW.

A high precision measurement of reactor anti-neutrino oscillation can be
achieved by a multiple detector experiment because the experimental sensitivity
would be nearly unaffected by the uncertainties related to anti-neutrino
source and interaction~\footnote{Yu. Kozlov, L. Mikaelyan, and V. Sinev, 
``{\it Two Detector Reactor Neutrino Oscillation
    Experiment Kr2Det at Krasnoyarsk. Status Report},'' 
Phys. Atom. Nucl. 66 (2003) 469-471; Yad. Fiz. 66 (2003) 497-499.}.
Consideration of the RENO began in early 2004 based on ``White Paper for a
New Nuclear Reactor Neutrino 
Experiment''~\footnote{K. Anderson {\it et al.} (2004), hep-ex/0402041.}.
The Yonggwang site was chosen due to a large number of anti-neutrinos from 
the nuclear power plant at the site and a mountainous geography suitable for
constructing underground detectors.

The experimental setup consists of two identical 16-ton Gadolinium loaded
liquid scintillator detector located near and far from the reactor 
array to measure the deviations from the inverse square distance law.
The near and far detectors are to be placed roughly 290~m and 1.4~km
from the center of the reactor array, respectively. 
The near detector will be constructed at underground of a 70~m high
hill and the far detector at underground of a 260~m high mountain.

The basic feature of RENO experiment is to search for energy dependent
$\bar\nu_{e}$ disappearance using two identical detectors for comparison
of neutrino fluxes at two different locations. The detectors are 
necessary to be located underground in order to reduce backgrounds
from cosmic rays and cosmic ray induced spallation products. The detectors
need to be designed identically in order to reduce systematic uncertainties
to 1\% or less. Controlling of the relative detector efficiency, fiducial
volume, and good energy calibration are critical to the successful
measurement. 

The construction of experimental halls and access tunnels for both
near and far detector sites was completed in early 2009.
The experiment is planned to start data-taking in mid 2010. An expected
number of observed anti-neutrino is roughly 510 and 
80 per day in the near detector and far detector, respectively. 
An estimated systematic uncertainty
associated with the measurement is less than 0.6\%. Based on three years of
data, it would be sensitive to measure the neutrino mixing angle in the
range of $\sin^2(\theta_{13})>0.02$. This sensitivity is ten times 
better than the current limit obtained by CHOOZ.
The RENO collaboration is presently consists of 12 institutions from 
Korea and Russia. It anticipates more international institutions to
join the experiment.
The RENO experiment was approved by the Ministry of Science and Technology
in Korea in May 2005. 

A measurement of or stringent limit on $\theta_{13}$ would be crucial
as part of a long term program to measure the CP violation parameter of
$\delta$ using accelerators. A sufficient value of $\theta_{13}$ measured
in this reactor experiment would strongly motivate the investment required 
for a new round of accelerator based neutrino experiments.

\addcontentsline{}{chapter}{}
\tableofcontents
\newpage
\bibliographystyle{plain}
\chapter{Overview}
\section{Experimental Goals and Descriptions}
There has been great progress in understanding the neutrino
sector of elementary particle physics in the last decade. The
discovery of neutrino oscillations is a direct indication of 
physics beyond the Standard Model and it provides a unique new 
window to explore physics at high mass scale including unification,
flavor dynamics and extra dimensions. The smallness of neutrino
masses and the large lepton flavor violation associated with
neutrino mixing are both fundamental properties that give insights 
into modifications of current theories. Since neutrino oscillations
now have been established, the next step is to map out the parameters
associated with neutrino masses and mixings. The experimental programme
to accomplish this goal will require a wide range of experiments
using neutrinos from solar, atmospheric, reactor, and accelerator 
sources.

In the presently accepted paradigm to describe the neutrino oscillations,
there are three mixing angles ($\theta_{12}$, $\theta_{23}$, $\theta_{13}$)
and one phase angle ($\delta$). There is now a world-wide experimental 
programme underway to measure the parameters associated with neutrino
oscillations. One of the three mixing angles, $\theta_{12}$, is measured
by solar neutrinos and the KamLAND experiment, and another, $\theta_{23}$,
by atmospheric neutrinos and the long-baseline accelerator K2K experiment.
Both angles are large, unlike mixing angles among quarks. 
MiniBooNE is searching for $\nu_{\mu}\to\nu_{e}$ appearance signal in the LSND
$\Delta m^2$ region from 0.2 to 1.0 eV$^2$. Current longer-baseline
($\sim 700$~km) experiments are NuMI/MINOS at Fermilab and CNGS at CERN that
will study $\nu_\mu$ oscillations in the atmospheric $\Delta m^2$ 
region. Several new long-baseline experiments are planned which
will use off-axis beams including the approved J-PARC to Super-Kamiokande
experiment (T2K) and the developing NuMI off-axis experiment (NO$\nu$A).
Following these experiments, the next stage might be neutrino superbeam
experiments with even longer baseline that could possibly be combined with
large proton decay detectors.

The third angle, $\theta_{13}$, has not yet been been measured to be 
non--zero but is constrained to be small in comparison by the CHOOZ reactor
neutrino experiment. Future measurement of $\theta_{13}$ is possible
using reactor neutrinos and accelerator neutrino beams. 
Reactor neutrino experiment can provide $\theta_{13}$ measurement without
the ambiguities associated with matter effects and CP violation.
In addition, initially the reactor neutrino experiment does not necessarily 
have to have large detectors and the it does not need construction of a 
neutrino beam line. 
The previous measurement had a single 
detector which was placed about 1~km from the reactors. Future reactor 
experiments using two detectors of $10\sim 30$ tons at near ($100\sim 500$~m) 
and far ($1\sim2$~km) locations will have significantly improved sensitivity 
for $\theta_{13}$ down to the $\sin^2(2\theta_{13})\sim 0.01$ level. 
With $\theta_{13}$ determined, measurements of $\nu_\mu\to \nu_e$ and 
$\bar\nu_\mu\to \bar\nu_e$ oscillations using accelerator neutrino beams 
impinging on large detectors at long baselines will improve the knowledge 
of $\theta_{13}$ and also allow access to matter or CP violation effects.

During the past several years, there have been competitively proposed several 
reactor neutrino experiments to measure $\theta_{13}$. They include a new 
experiment at Chooz called Double Chooz, the Braidwood experiment in the US, 
the KASKA experiment in Japan, the Daya Bay experiment in China and 
the RENO experiment in Korea.
The experiments that have been approved for funding are summarized in 
Table~\ref{planned experiments}.

The basic feature of this reactor experiment is to search for distance 
dependent $\bar\nu_e$ disappearance using two or more detectors, for comparison
of neutrino fluxes at two different locations. The detectors need to be located 
underground in order to reduce backgrounds from cosmic rays and cosmic ray 
induced spallation products. The detectors need to be designed identically 
in order to reduce systematic uncertainties to 1\% or less. Controlling of 
the relative detector efficiency, fiducial volume, and good energy 
calibration are needed.  

The Yonggwang nuclear power plant with world-second largest thermal power 
output of 16.4~GW, is an intense source of low energy anti-neutrinos suitable 
for measuring neutrino oscillations due to $\theta_{13}$. The anti-neutrino 
fluxes from the nuclear reactors are measured nearby before their oscillations,
and measured again at a distance of about 1.4~km away from the reactor center. 
A neutrino mixing parameter $\theta_{13}$ is obtained by finding the reduction 
of neutrino fluxes by comparison of the two measured fluxes.  

The experimental setup consists of two 16 ton liquid scintillator detectors 
with one at a near site, roughly 290~m away from the reactor array center, 
and the other at a far site, roughly 1.4~km away from the reactor array 
center. The near detector will be located at underground of a 70~m high 
hill, and the far detector at underground of a 260~m high mountain. 

It is now widely recognized that the possibility exists for a rich programme of
measuring CP violation and matter effects in future accelerator based neutrino
experiments, which has led to intense efforts to consider new programmes at 
neutrino superbeam, off-axis detectors, neutrino factories, and beta beams. 
However, the possibility of measuring CP violation can be fulfilled only if 
the value of neutrino mixing parameter $\theta_{13}$ is such that 
$\sin^2(2\theta_{13})$ should be greater than the order of 0.01. It is 
believed that a timely new experiment using nuclear reactors, sensitive to 
the neutrino mixing parameter $\theta_{13}$ in this range has a great 
opportunity for an exciting discovery.

A measurement of or stringent limit on $\theta_{13}$ would be crucial as a 
part of a long term programme to measure CP violation parameters at 
accelerators, even though a reactor $\bar\nu_e$ disappearance experiment 
does not measure any CP violation parameter. A sufficiently large value of 
$\theta_{13}$ measured in this reactor experiment would strongly motivate 
the investment required for a new round of accelerator neutrino experiments. 
A reactor experiment's unambiguous measurement of $\theta_{13}$ would also 
strongly support accelerator based measurements by helping to resolve 
degeneracies and ambiguities. The combination of measurements from reactors 
and accelerator neutrino beams will allow early probe for CP violation 
without the necessity for long running at accelerators with anti-neutrino
beams.

\begin{table}
\begin{center}
\begin{tabular}{cccccc}\hline
        &       &Total Reactor &Detector &Overburden
&Target Mass\\
Experiment &Location &Thermal Output &Distance &Near/Far
&(Near/Far) \\
        &       &(GW$_{th}$)    &Near/Far (m)   &(mwe)     &(tons) \\\hline
Double Chooz    &France &8.7    &410/1067       &115/300 &10/10 \\
Daya Bay        &China  &11.6(17.4)   &360(500)/1985(1613)  &260/910        &$40\times 2$/10 \\
RENO    &Korea  &16.4   &292/1380       &110/450        &16.1/16.1 \\\hline
\end{tabular}
\end{center}
\caption{Planned reactor based neutrino oscillation experiments around the world. The detector
distance represents the distance of the detector from the center of the reactor group(s).}
\label{planned experiments}
\end{table}

\section{Experimental Setup}
\subsection{Yonggwang Nuclear Power Plant}
\begin{figure}
\begin{center}
\includegraphics[width=6in]{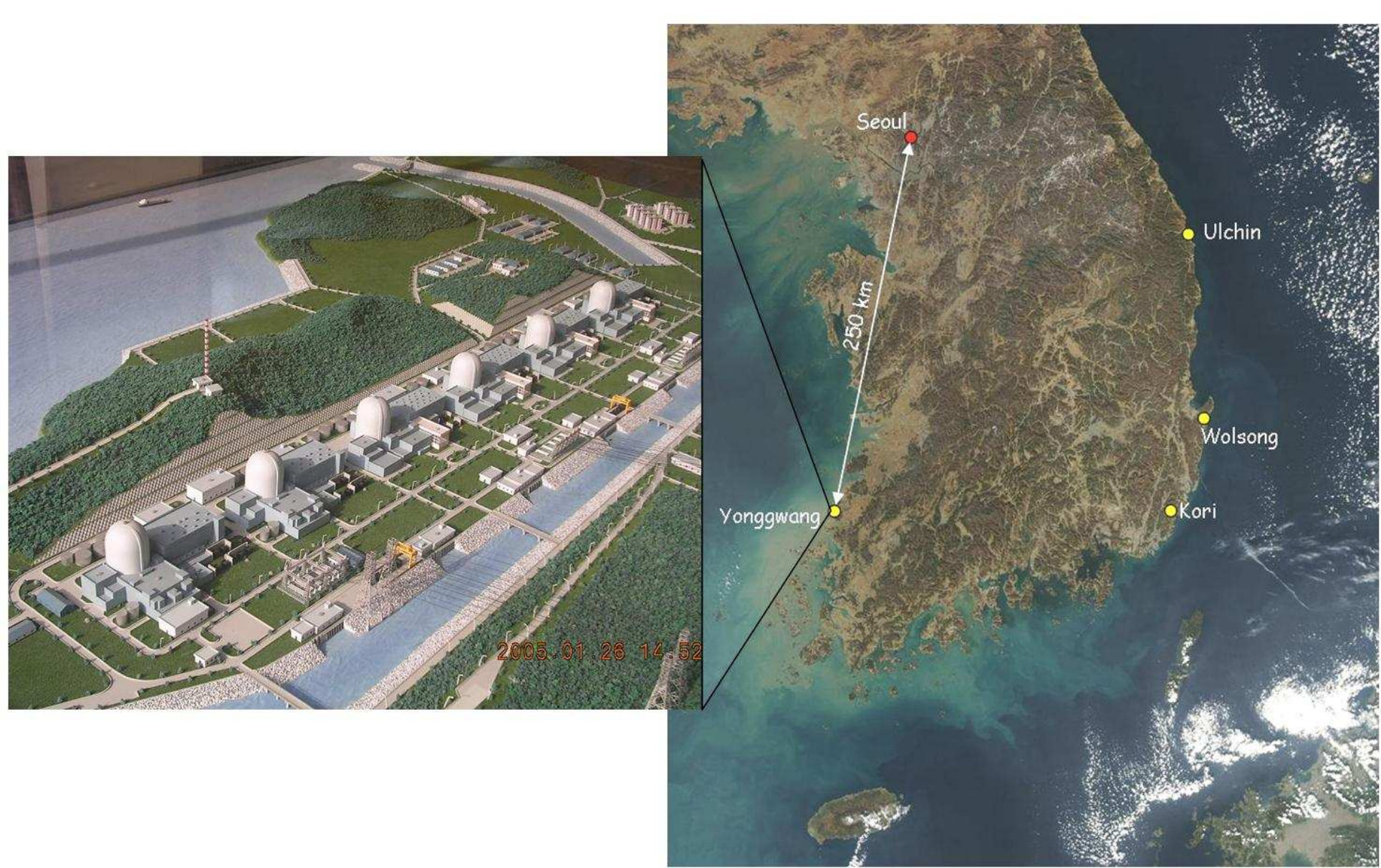}
\end{center}
\caption{Yonggwang nuclear power plant. The power plant is located about
250~km south of Seoul. Three other nuclear power plant sites in Korea are also
shown.}
\label{yonggwang overview}
\end{figure}

\begin{figure}
\begin{center}
\includegraphics[width=4in]{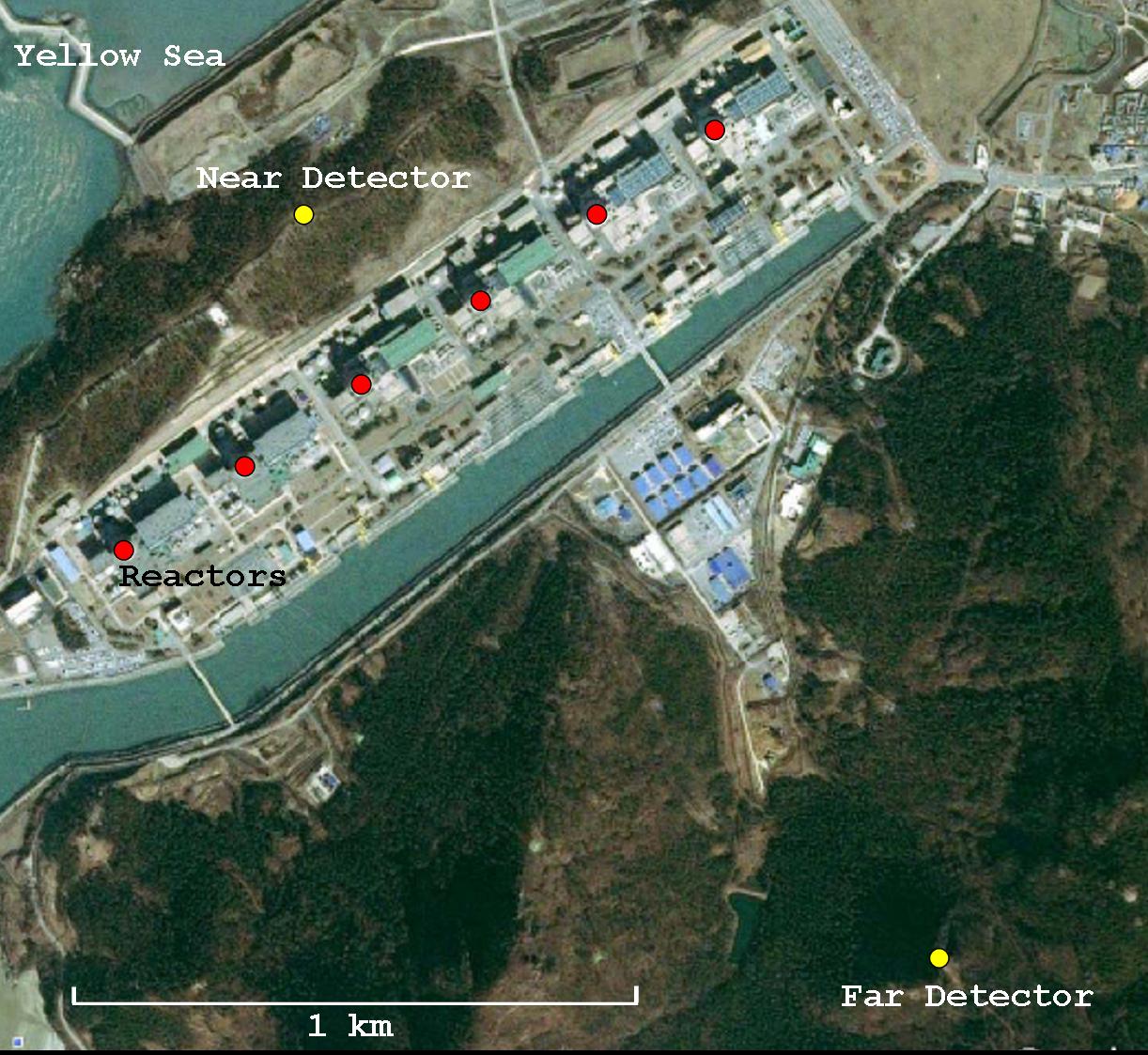}
\end{center}
\caption{The layout of the Yonggwang experiment site. Red dots and
yellow dots represent reactors and detectors, respectively. Six reactors 
are roughly equally spaced in a 1280~m span. The near and far detectors
are 290~m and 1380~m away from the reactor array, respectively. The image
taken from Google Earth$^{\rm TM}$ and copyrighted therein.}
\label{google}
\end{figure}

The Yonggwang nuclear power plant is located in the west coast of
southern part of Korea, about $\sim 250$~km from Seoul as
shown in Fig.~\ref{yonggwang overview}. 
The power plant has six reactors producing total thermal output
of 16.4~GW$_{th}$, the second largest in the world.
The reactors at the Yonggwang nuclear power plant are Pressurized
Water Reactors (PWR). There are six reactors and the reactor units 1 
and 2 are of the Combustion Engineering (CE, now Westinghouse) System 80 
design. Units 3 to 6 are of the Korean Standard Nuclear Power Plant (KSNP) 
design, which incorporates many improvements on the CE System 80. 
The first reactor, unit 1, became operational in 1986 and the last
one, unit 6, in 2002. 
These reactors are lined up in roughly equal distances and
spans $\sim 1.3$~km as shown in 
Fig.~\ref{google}.

A reactor core is comprised of 177 fuel assemblies and 73 control
element assemblies. The fuel assemblies are arranged to form a 
cylinder with an equivalent diameter of 3.12~m and an active length 
of 3.81~m. Reactor fuelling cycle varies from 12 months to 24 months
and refuellings are done with the plant shutdown.
The fuel is a low enrichment UO$_2$ type supplied by Korea Nuclear 
Fuel Co., Ltd. 
   
The average total thermal power output of the six reactor cores is 
$16.4~\hbox{GW}_{th}$ with each reactor core generating about equal power.
The average cumulative operating factors for all reactors are above 90\%. 
The power plant is operated by Korea Hydro \& Nuclear Power Co. Ltd. (KHNP).  

\subsection{Near and Far Detectors}
One of the main sources of systematic uncertainties is the
uncertainty in the number of antineutrinos coming from the
reactors. To minimize the effects of this problem,
two identical detectors, near and far detectors are needed.
Each detector will contain 18.7~m$^3$ of liquid scintillator
target doped with 0.1\% of Gadolinium by weight (see 
Chapter~\ref{chapter detector}).

Figure~\ref{google} shows the layout of six reactors and 
two detectors and Table~\ref{reactor-detector distances} shows
the distances between reactors and detectors. 
The near and far detectors are to be located 290~m and 1400~m 
from the center of the reactor array, respectively. 
The near detector is to be under an 70~m 
(AMSL) ridge with an overburden of $\sim 110$~mwe whereas
the far detector is to be located under a 260~m mountain 
with the overburden of $\sim 450$~mwe as shown in Figs.~\ref{google} and 
\ref{tunneling}.  

\begin{table}
\begin{center}
\begin{tabular}{ccc}\hline
Reactor No.  &Near Detector (m) &Far Detector (m) 
\\\hline
1	&667.9	&1556.5 \\
2	&451.8	&1456.2 \\
3	&304.8	&1395.9 \\
4	&336.1	&1381.3 \\
5	&513.9	&1413.8 \\
6	&739.1	&1490.1 \\\hline
\end{tabular}
\end{center}
\caption{Distances of the reactor cores from the near and far detectors.
}
\label{reactor-detector distances}
\end{table}

\subsection{Geographical Data}
An interesting feature of RENO is having a sufficient overburden for the near 
detector due to a 70~m hill of 2.8~g/cm$^3$ rock which is quite close (290~m) 
to the center of the reactor array. A detector close to the nuclear reactor is 
necessary for cancelling the systematic uncertainties related to the nuclear 
reactors such as ambiguities of the anti-neutrino flux and spectrum, as well 
as for reducing systematic uncertainties related to the detector and to the 
event selection. The near detector laboratory will be located inside the 
restricted area of the Yonggwang nuclear power plant.

\subsection{Tunneling and Experiment Halls}
\begin{figure}
\begin{center}
\includegraphics[scale=0.8]{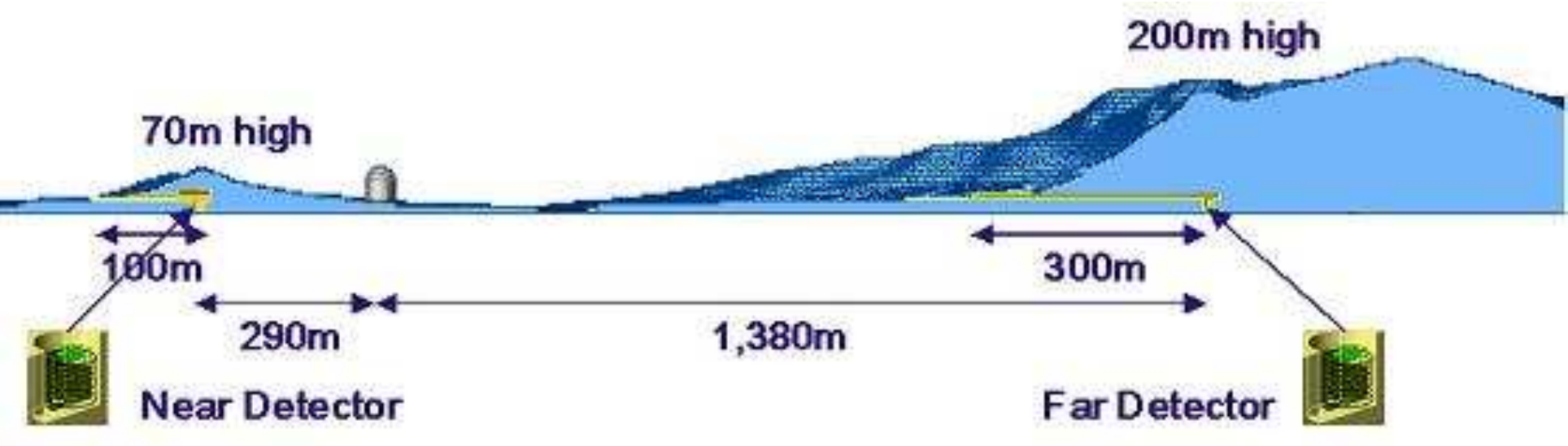}
\end{center}
\caption{Overall side view of RENO experiment. The near detector is under a 
70~m hill located within the perimeter of the power plant whereas the far 
detector is located under a 200~m mountain near the power plant.}
\label{tunneling}
\end{figure}

\begin{figure}
\begin{center}
\includegraphics[scale=0.8]{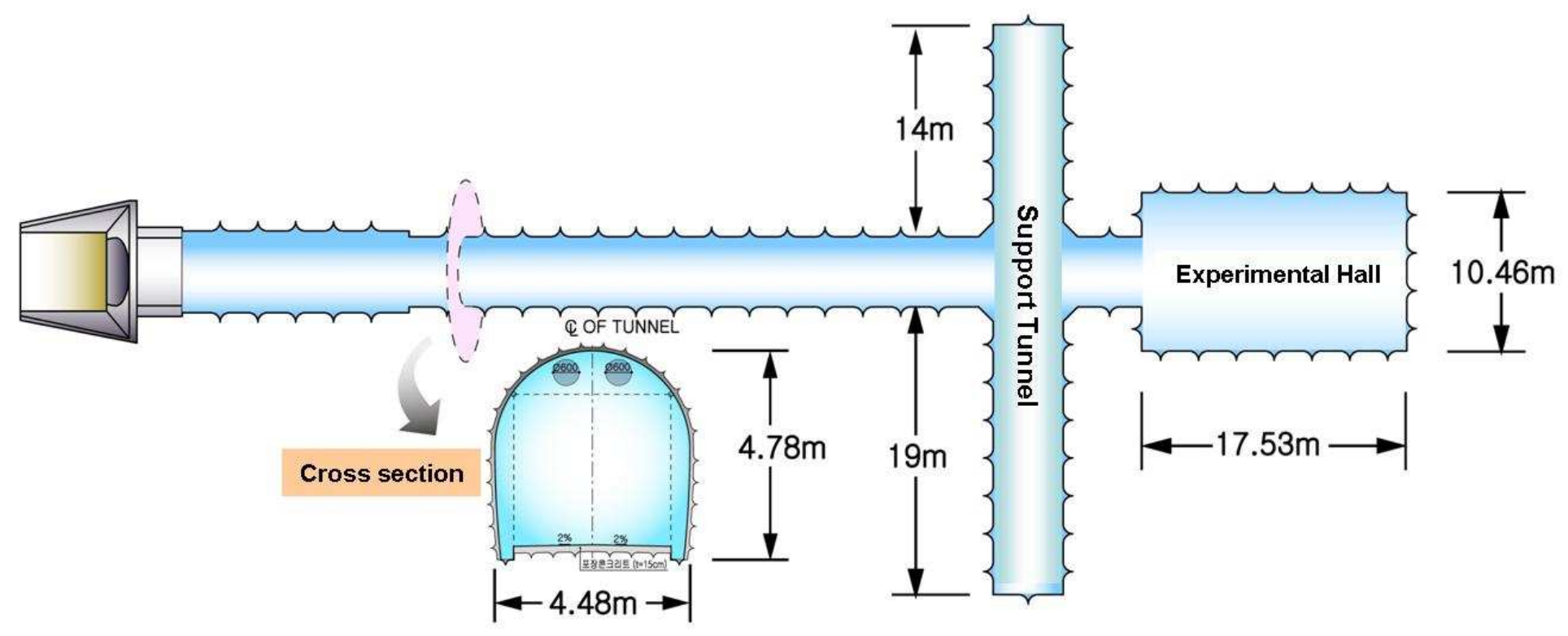}
\includegraphics[scale=0.5]{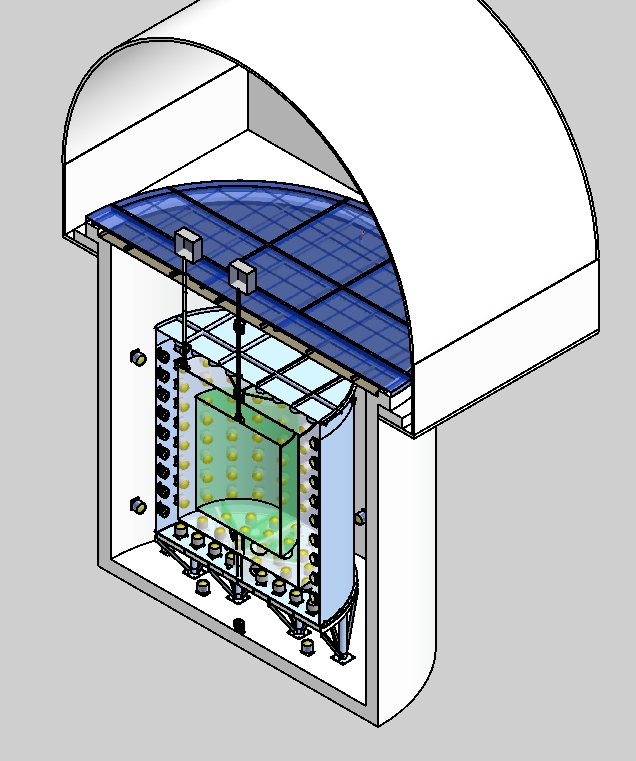}
\end{center}
\caption{Plan of the access tunnel (top) and 3D schematic for experimental hall 
(bottom). The tunnels are constructed using NATM.}
\label{tunnel cross section}
\end{figure}

The underground laboratories are constructed with two horizontal tunnels, 
100~m long for the near detector and 300 m long for the far detector, as 
shown in Fig.~\ref{tunneling}. 
The tunnels are constructed using NATM (New Austrian Tunneling Method).
The tunnel plan and schematic of experimental hall are shown in 
Fig.~\ref{tunnel cross section}

The access tunnels are 95~m and 272~m long for near and far detector
sites, respectively. The cross section of the access tunnel is shown in
Fig.~\ref{tunnel cross section}. The gradient toward the experimental
hall is $0.3\%$ for both tunnels for natural drainage.
It is designed to accommodate the passage of a 10 ton truck.

\section{Site Survey}
To check the suitability of constructing experimental 
halls and access tunnels at the experiment site, geological 
surveys were performed. The site surveys were 
conducted by Daewoo Engineering Co., Ltd. in May, 2007.

The rocks at the experimental site are precambrian granite gneiss 
covered with Cretaceous plutonic rocks forming unconformity between 
the layers. There are fault lines near the site running north-south.

\begin{figure}
\begin{center}
\includegraphics[width=7.5cm]{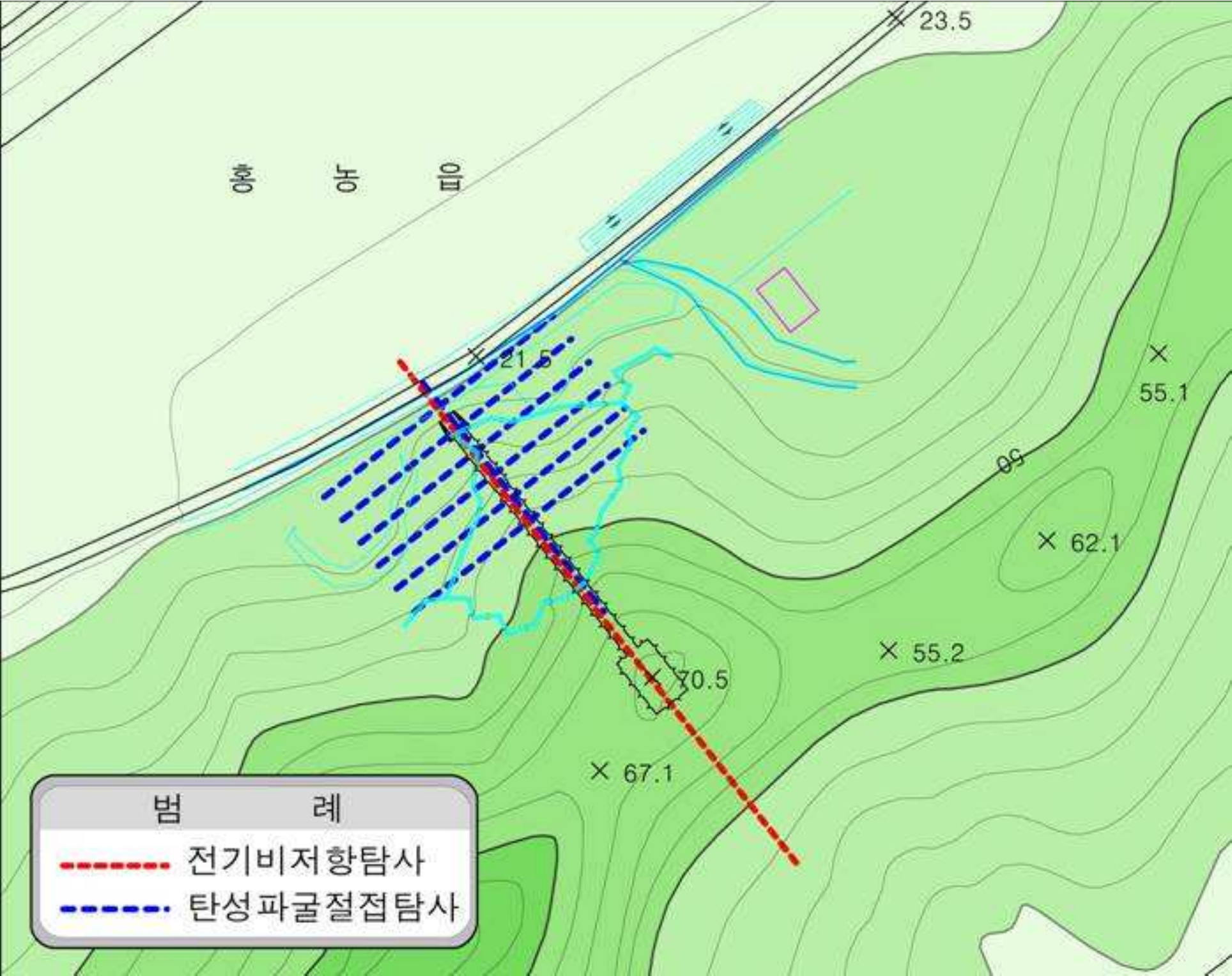}
\includegraphics[width=7.5cm]{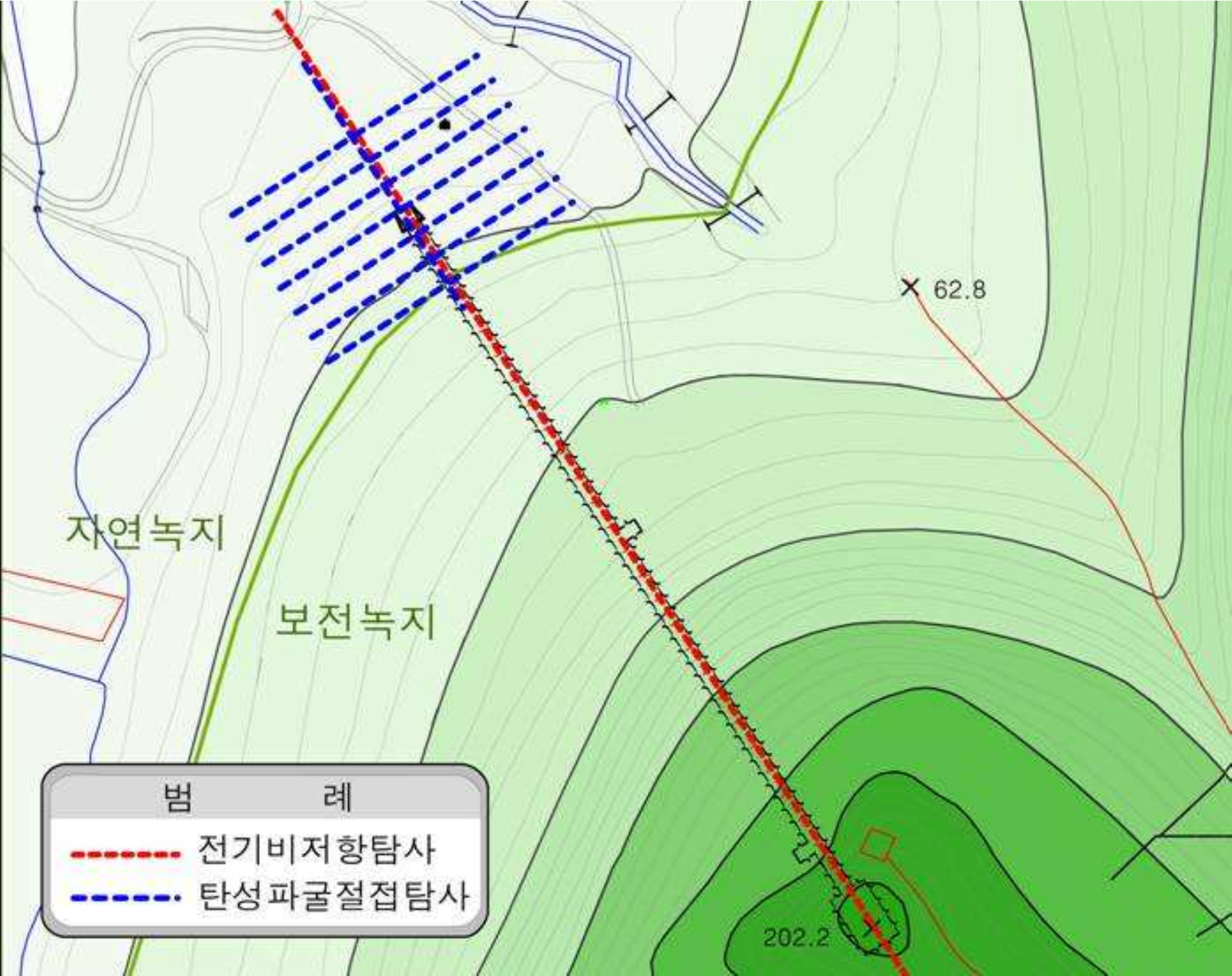}
\end{center}
\caption{Locations of the electrical resistivity survey (red) and 
seismic refraction test (blue) performed in near (left) and far (right) 
detector sites.}
\label{geophysical}
\end{figure}

\begin{figure}[ht]
\begin{center}
\includegraphics[width=9.5cm]{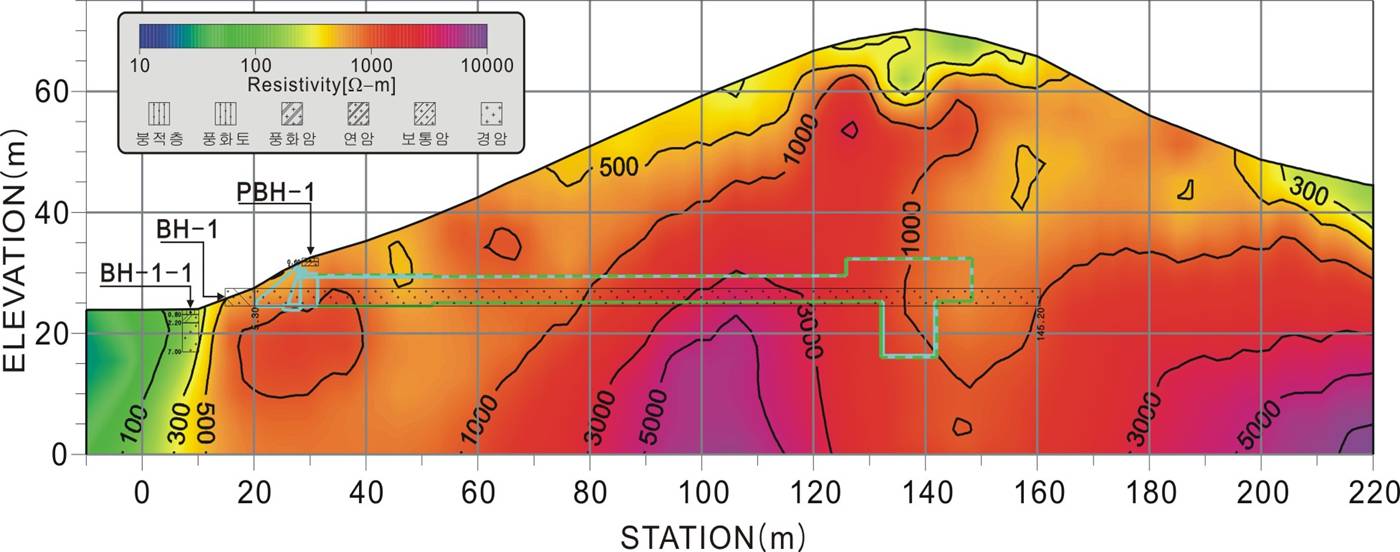}
\includegraphics[width=9.5cm]{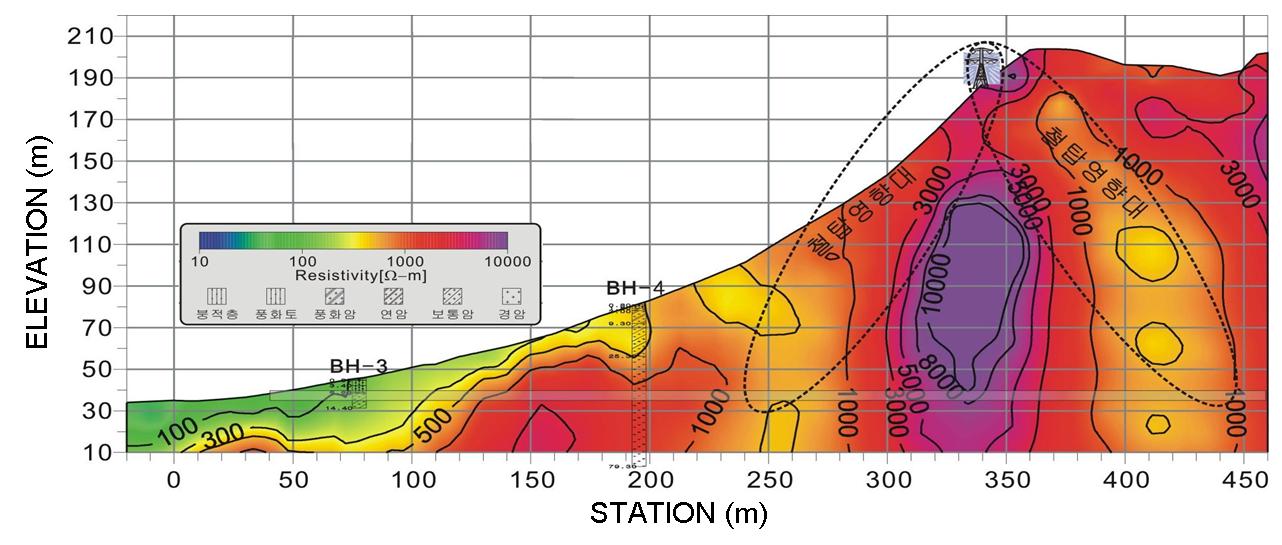}
\end{center}
\caption{Electrical resistivity survey results at the near (top) and 
far (bottom) detector sites along the access tunnel.}
\label{resistivity}
\end{figure}

\begin{figure}[h]
\begin{center}
\includegraphics[width=9.5cm]{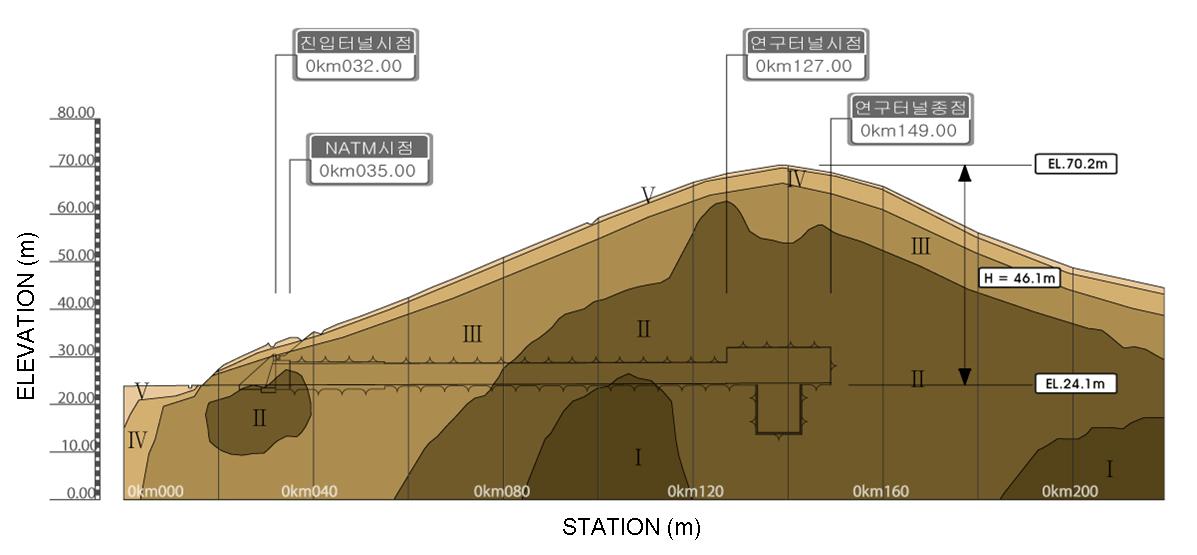}
\includegraphics[width=9.5cm]{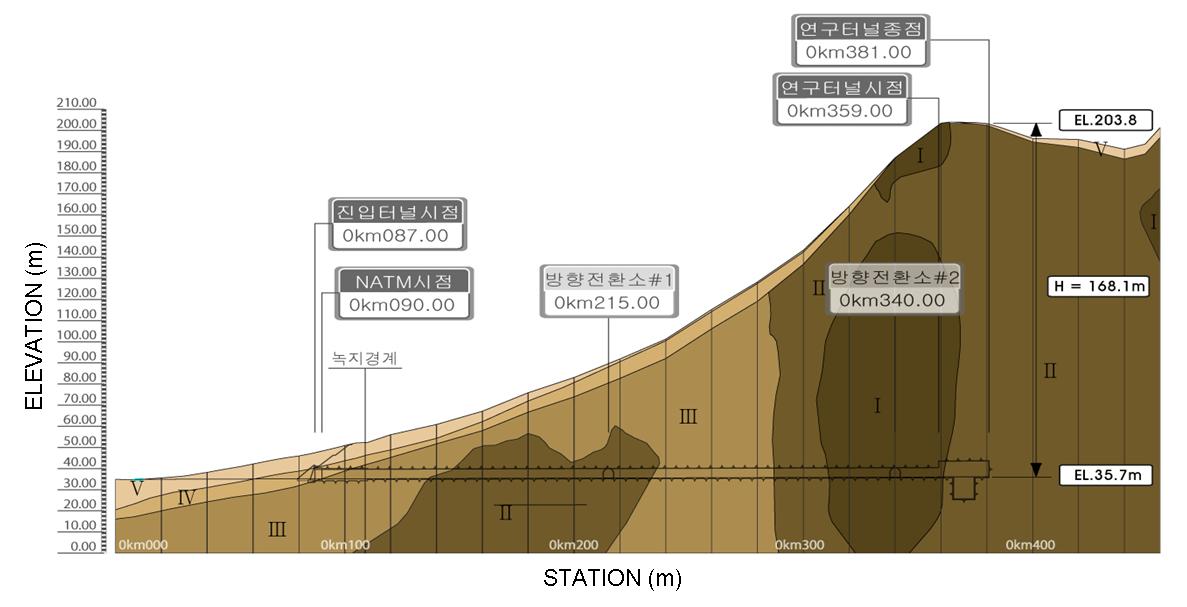}
\end{center}
\caption{Rock classification maps of the near (top) and far (bottom) 
detector sites along the access tunnel.}
\label{classification map}
\end{figure}

Two methods are used in the geophysical survey of the site;
electrical resistivity survey and seismic refraction imaging.
Figure~\ref{geophysical} shows the locations of the survey done
at the experimental site. 
The electrical resistivity survey results are shown in 
Fig.~\ref{resistivity}.
The resulting rock classification maps are shown in Fig.~\ref{classification map}.

Most of tunneling sections at both near and far detector sites
contains hard rocks of class II and III except near the entrance
of the tunnels as shown in Fig.~\ref{classification map}.
Also no significant faults are found along the tunnels.

\begin{figure}
\begin{center}
\includegraphics[scale=0.6]{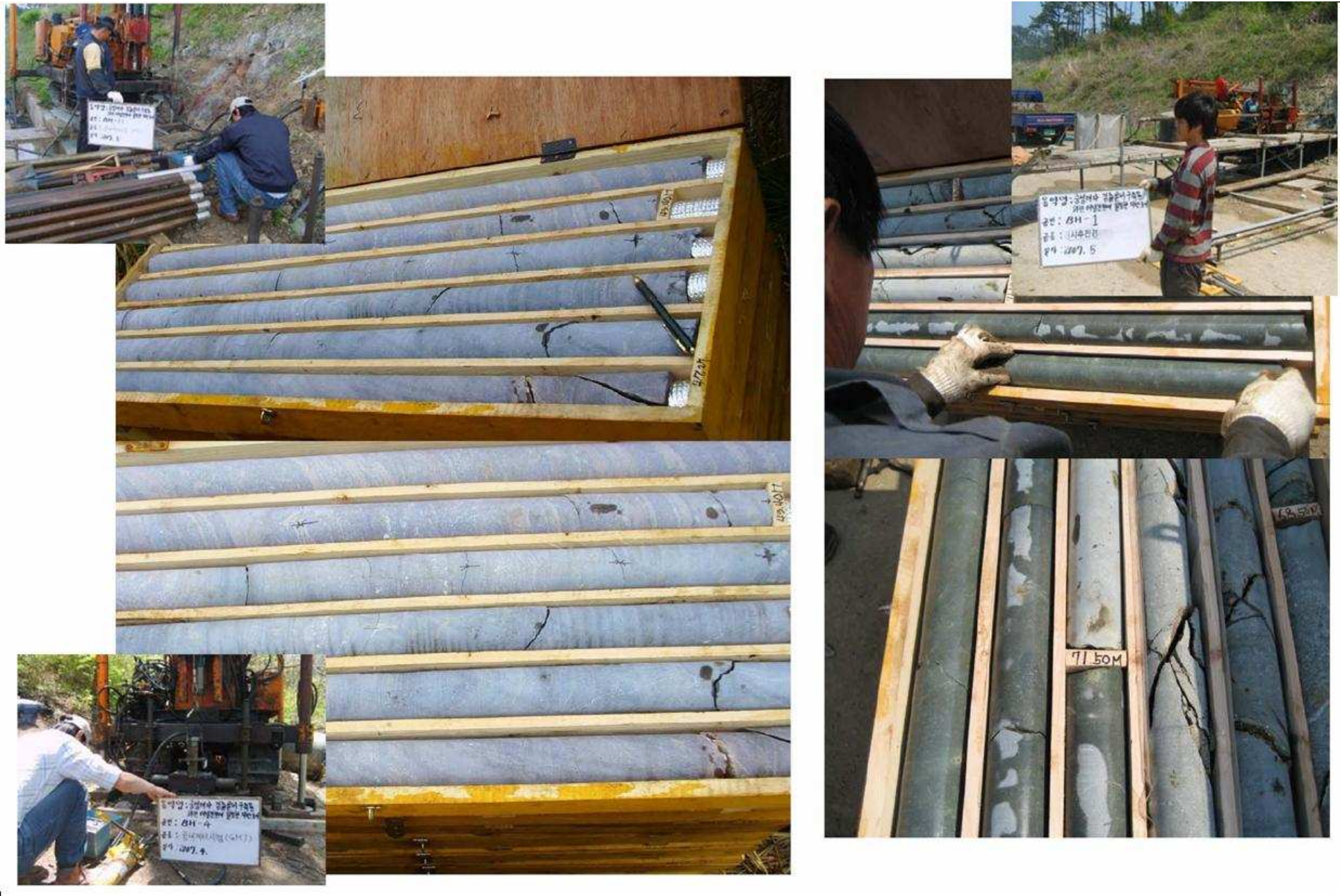}
\end{center}
\caption{Borehole rock samples from the near and far detector sites.}
\label{borehole samples}
\end{figure}

Based on the results from the electrical resistivity survey and 
the seismic refraction tests, four borehole positions and 
three borehole positions are selected for near and far
detector sites, respectively.
Figure~\ref{borehole samples} shows the rock samples from boreholes.
The samples are used to determine various properties of rocks, such as
chemical composition, compressive strength, density, and radioactivities.

Four on-site tests at the boreholes were also performed: televiewer, 
borehole loading, hydraulic fracture, and downhole.

\section{RENO Detectors}
Both RENO near and far detectors will consist of a cylindrical target of 
140~cm in radius and 320~cm in height, providing a volume of 18.7~m$^3$. 
Identical arrangement of the near and far detectors will significantly 
reduce the systematic errors of relative normalization to 0.6\%. However, 
they will have different cosmic ray background levels because of unequal 
overburdens ($\sim$110~mwe {\it vs.} $\sim$450~mwe). Although the near 
detector will suffer from higher cosmic ray background, it will observe 
much more signal events of reactor anti-neutrinos due to shorter distance 
from the nuclear cores and thus allows a high signal-to-background ratio. 
A detailed description of the backgrounds at both sites is given in 
Sect.~\ref{back sim}. 

The RENO detector consists of a neutrino target, a gamma catcher, a buffer 
and a veto. Target and gamma catcher vessels will be made from acrylic 
plastic material, having transparency to the light of wavelengths above 
400~nm. The acrylic vessels should hold aromatic liquids without leakage 
and its properties should not change for the duration of the experiment.
They should not develop any chemical reaction with the scintillating liquids 
of neutrino target, gamma catcher and buffer for a long time period. 
Overview of each detector component is given below.
\begin{figure}
\begin{center}
\includegraphics[scale=0.8]{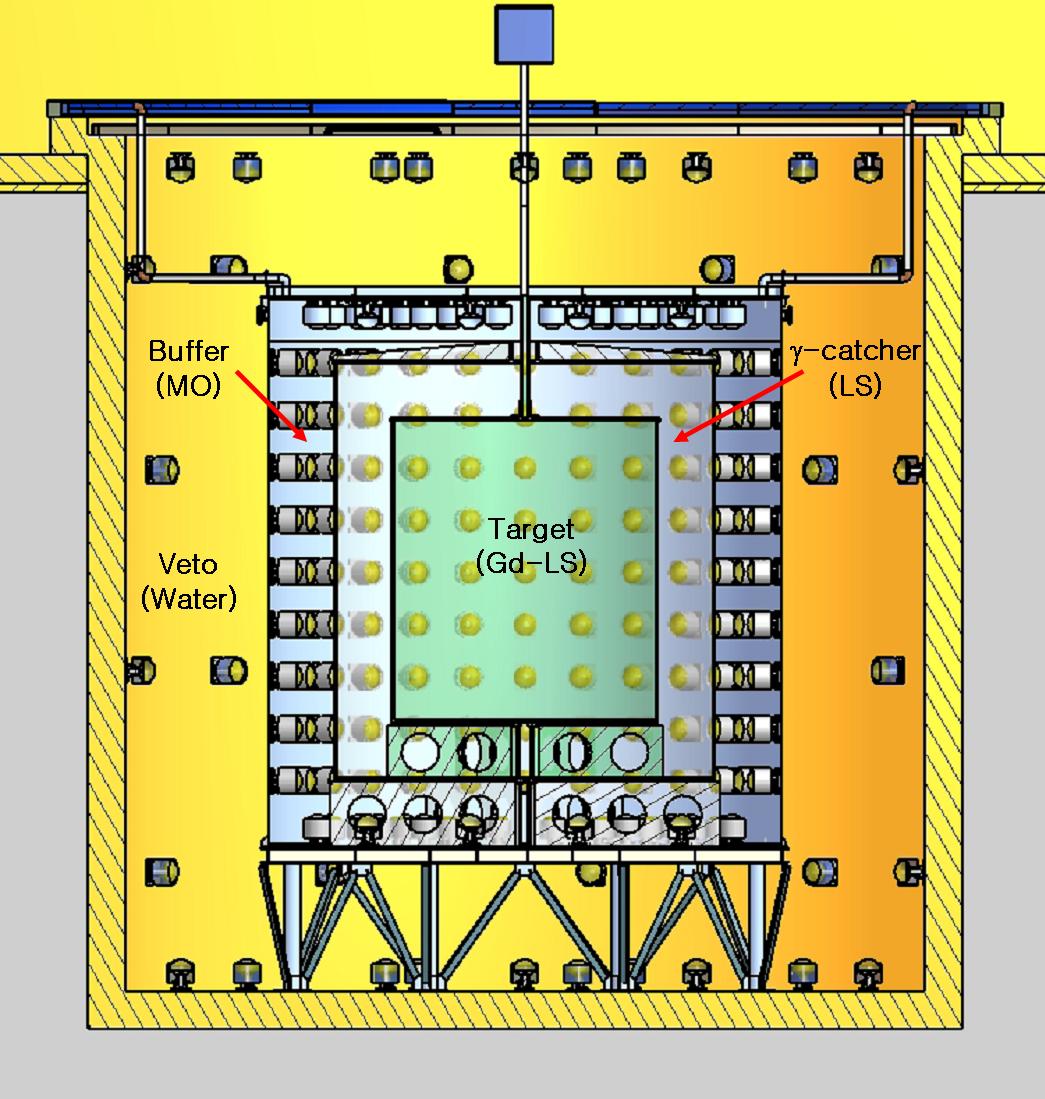}
\end{center}
\caption{A schematic view of RENO detector. A neutrino target
of 18.7~m$^3$ Linear Alkyl Benzene (LAB) based liquid scintillator
doped with Gd is contained in a transparent acrylic vessel, and
surrounded by 33.2~m$^3$ unloaded liquid scintillator of gamma catcher
and 76.5~m$^3$ non-scintillating buffer. 
There are 354 and 67 10-inch PMTs mounted on buffer and veto vessel walls,
respectively.
}
\label{tunnel and hall 2d}
\end{figure}

\subsection{Neutrino Target, Gamma Catcher, and Acrylic Vessels}
The neutrino target consists of 0.1\% Gd loaded liquid 
scintillator in a cylindrical acrylic container of 140~cm in radius, 
320~cm in height, and 25~mm in thickness. It has a total volume of 
18.7~m$^3$ and a target mass of 16.1~tons.  Gamma catcher surrounds 
the neutrino target with 60~cm thick {\it unloaded} liquid scintillator of 
33.2~m$^3$ in volume and 28.5~tons in mass. The gamma catcher is 
contained in a cylindrical acrylic vessel of 200~cm in radius, 440~cm 
in height, and 30~mm in thickness. The gamma catcher vessel should be 
chemically compatible with mineral oil of the buffer region as well as 
the scintillating liquids inside. This scintillating volume is necessary 
for efficient tagging of the gammas from neutron capture by Gd and 
from positron annihilation, and for rejecting the backgrounds from the 
fast neutrons. 

\subsection{Non-Scintillating Buffer and Stainless Steel Vessel}
A 70~cm thick non-scintillating liquid surrounds the gamma catcher 
to reduce the accidental backgrounds coming from outside (mainly 
from radioactivity in the photomultiplier tubes), by almost two 
orders of magnitude. A total of 76.5~m$^3$ (64.2~tons) mineral oil 
is contained in a stainless steel vessel of 270~cm in radius, 580~cm 
in height, and 6-12~mm in thickness. This buffer is necessary for keeping 
the background single rate below 10~Hz in the neutrino target and 
gamma catcher regions.

\subsection{PMT}
A total of 354 10-inch photomultipliers in a uniformly distributed array 
are mounted 
on the inner surface of the buffer vessel, providing a 14\% 
photo-sensitive surface area coverage. The cylindrical stainless steel vessel optically 
isolates the inner detector part from the outer veto system. 

\subsection{Veto System}
A 1.5~m thick water layer of 353~tons surrounds the whole inner detector.
A total of 67 10-inch PMTs 
are mounted on a cylindrical 
concrete tank painted with Titanium Oxide (TiO$_2$) reflector.
It is used for vetoing cosmic muons and reducing backgrounds coming 
from its surrounding rock.

Some of detector parameters or design may be changed afterwards 
according to the result of detection performance studies using 
a mock-up detector. A list of interesting parameters for the RENO 
experiment is given in Table~\ref{summary experiment parameters}. Some of 
RENO experimental parameters 
are compared with those of Double Chooz and Daya Bay experiments in 
Table~\ref{planned experiments}.

\begin{table}
\begin{center}
\begin{tabular}{lcc}\hline
Parameter &Value &Description \\\hline
Thermal Power (GW)   &16.4(average)/17.3(peak) &6 reactor \\
Target Size (ton) &16 (near/far) &Gd loaded Liquid Scintillator \\
PMT Coverage &14\% (near/far) &surface area\\
Baseline Distance(m)    &292 (near)/1380 (far) & \\
Overburden (mwe)    &110 (near)/450 (far) &Vertical\\
Number of Events per Year &$2.6\times 10^5$ (near)/$3.0\times 10^4$ (far)
&$\epsilon_{total}=$56\%(near)/72\%(far) \\
90\% CL Sensitivity (3 years)
&$\sin^2(2\theta_{13})\sim 0.02-0.03$ &$\Delta m_{13}^2 = (2-3)\times 10^{-5}~{\rm eV}^2$\\
\hline
\end{tabular}
\end{center}
\caption{Summary of RENO experimental parameters. }
\label{summary experiment parameters}
\end{table}

\section{Time Scale}
\begin{figure}
\begin{center}
\includegraphics[scale=0.47,angle=90]{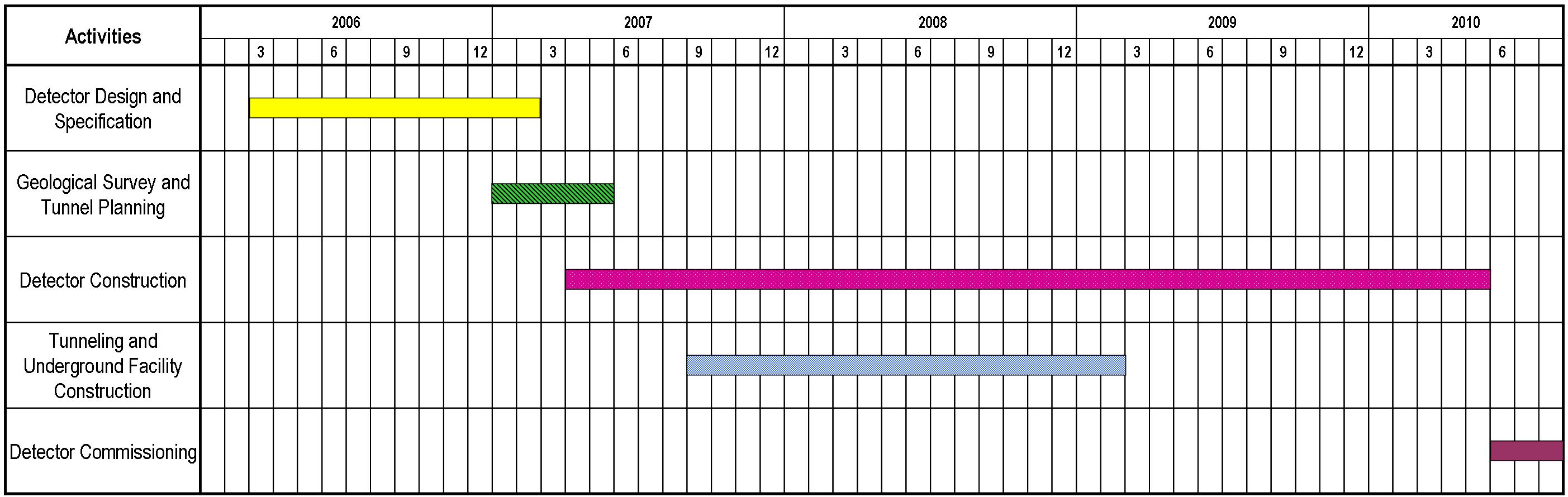}
\end{center}
\caption{RENO construction schedule.}
\label{figure schedule}
\end{figure}

An overall schedule of the RENO experiment is given in 
Fig.~\ref{figure schedule}. The KHNP, the company 
operating the Yonggwang nuclear power plant, has 
allowed us to carry out the experiment in the Yonggwang's  
restricted area. The local government and residents have 
also expressed their best cooperation for RENO. A company 
for geological survey and tunnel design was chosen through 
a bidding process in Feb, 2007. Civil construction for 
underground facility 
began in July 2008 and was completed in Feb. 2009.
RENO has gone through rapid development stages, to date, of project 
planning, fund approval, administrative negotiation, and detector 
design. This was possible only due to the great worldwide interest 
in finding $\theta_{13}$. Excavation of two tunnels and construction 
of underground facility was completed in late 2008. 
Construction of both near and far detectors will be completed by 
mid-2010. Data taking is expected to start shortly afterward. 
RENO will reach $\sin^2(2\theta_{13})$ sensitivity of 0.03 with a 
year of data in 2011, and 0.02 with three years of data, 
corresponding to the luminosity 
of ~400~ton$\cdot$GW$\cdot$yr, in 2013. The sensitivity will rely 
on the evaluated systematic errors and background levels. 

\newpage

\bibliographystyle{plain}
\chapter{Physics Goals}
\section{Motivation}
As stated earlier, $\theta_{13}$ is a key parameter in developing the 
future neutrino oscillation program. In addition to the general physics
arguments, there is a point that brings urgency to the reactor neutrino
experiments. Reactor experiments offer a straightforward and cost effective
method to measure or constrain the value of $\theta_{13}$. The sensitivity
of a two-detector experiment is comparable to that of the proposed
initial off-axis long-baseline experiment. Since a reactor experiment
would be relatively small and simple and use an existing reactor neutrino 
source with
a well understood neutrino rate, it should be able to be done fairly 
quickly and at reduced costs. It is likely that such an early measurement
of $\theta_{13}$ will be necessary before the community invests a large
amount of resources for a full off-axis measurement. For the longer
term, a reactor neutrino experiment would be complementary to the 
off-axis experiments in separating the measurement of $\theta_{13}$ from
other physics parameters associated with matter effects and CP violation.
A follow-up reactor experiment with much larger detectors at various 
baseline will continue to be an important component of the neutrino
oscillation program.

It is now widely recognized that the possibility exists for a rich program
of measuring CP violation and matter effects in future accelerator 
neutrino experiments, which has led to intense efforts to consider new 
programs at neutrino superbeams, off-axis detectors, neutrino factories,
and beta beams. However, the possibility of measuring CP violation can
be fulfilled only if the value of neutrino mixing parameter $\theta_{13}$
is such that $\sin^2(2\theta_{13})$ should be greater than or equal to the
order of 0.01. It is believed that a timely new experiment using nuclear
reactors sensitive to the neutrino mixing parameter $\theta_{13}$ in this 
range would provide a great opportunity for an exciting discovery. 

\section{Neutrino Oscillations}
\subsection{Neutrino Mixing}
Just as the CKM matrix parametrizes the mixing of the quark flavors,
the neutrino flavor states can be related to the mass states through 
the MNS (Maki-Nakagawa-Sakata) lepton flavor mixing matrix  
assuming three flavors,
\begin{equation}
U =
\left( \begin{array}{ccc}
1           &0           &0      \\
0           &c_{23}      &s_{23} \\
1           &-s_{23}     &c_{23} 
\end{array}\right)
\left( \begin{array}{ccc}
c_{13}      &0           &s_{13}e^{i\delta} \\
0           &1           &0 \\
-s_{13}e^{-i\delta}      &0            &c_{13} 
\end{array}\right)
\left( \begin{array}{ccc}
c_{12}      &s_{12}         &0 \\
-s_{12}           &c_{12}           &0 \\
0      &0            &1 
\end{array}\right)
\end{equation}
\begin{equation}
=
\left( \begin{array}{ccc}
c_{12}c_{13}                                
&s_{12}c_{13}                                
&s_{13}e^{-i\delta} \\ 
-s_{12}c_{23}-c_{12}s_{23}s_{13}e^{i\delta} 
&c_{12}c_{23}-s_{12}s_{23}s_{13}e^{i\delta} 
&s_{23}c_{13} \\
s_{12}s_{23}-c_{12}c_{23}s_{13}e^{i\delta} 
&-c_{12}s_{23}-s_{12}c_{23}s_{13}e^{i\delta} 
&c_{23}c_{13} \end{array}\right),
\nonumber
\end{equation}
where 
$c_{ij}=\cos\theta_{ij}$,
$s_{ij}=\sin\theta_{ij}$,
and $\delta$ is a Dirac $CP$ violating phase.
The mixing angle $\theta_{12}$ has been probed by solar and
reactor neutrino experiments and is often referred to as
$\theta_{sol}$, while the angle $\theta_{23}$ has been
investigated by atmospheric neutrino experiments and it is
often identified as $\theta_{atm}$.
Reactor neutrino experiments have probed the mixing
angle $\theta_{13}$.
If the mixing angle $\theta_{13}$ vanishes exactly, then
the $CP$ violating matrix elements vanish and
the $CP$ violation would not be observed in the lepton sector,
independently of the value of the phase $\delta$.

The probability of $\nu_\alpha$ with energy $E$ changing to $\nu_\beta$ 
($\nu_{\alpha,\beta}=\nu_{e, \mu, \tau}$) after travelling distance $L$ 
in vacuum is
\begin{equation}
P(\nu_\alpha\to\nu_\beta)
=\delta_{\alpha\beta}-2{\rm Re}\sum_{j>i}U_{\alpha i}U_{\alpha j}^{*}
U_{\beta i}^{*}U_{\beta j}\left(1-\exp{i\Delta m_{ji}^2 L\frac 2 E}\right),
\end{equation}
where $\Delta m_{ji}^2\equiv m_j^2-m_i^2$ and $m_i$ is the mass of the
$i$th eigenstate.

The results from solar neutrino experiments and a long baseline reactor 
experiment favor large mixing angle MSW solar solution
with $\Delta m_{21}^2=m_2^2-m_1^2=\Delta m^2_{sol}>0$, where $m_1$ is the mass eigenstate
with a larger electron neutrino component~\cite{Physics:Ashie05,Physics:Ahmad02,Physics:Eguchi03}. 
The large quadratic mass difference measured in the atmospheric neutrino
experiment is therefore the mass splitting between eigenstate 3 and more
closely spaced states 1 or 2. And it is unknown what the sign of the
splitting between the state 3 and states 1 or 2 is. This leads to 
an ambiguity in the sign of $\Delta m_{32}^2= m_3^2-m_2^2 = \Delta m^2_{atm}$.  
A further theoretical description of neutrino mixing can be found
in Ref.~\cite{Physics:Moh05a}.

\subsection{Experimental Results of Neutrino Oscillation}
Various experiments using solar, atmospheric, reactor and accelerator 
neutrinos have observed oscillations among different flavors of neutrinos, 
providing rich information on the flavor structure of the lepton 
sector~\cite{Physics:ref. 1, Physics:ref. 2, Physics:ref. 3, Physics:Apollonio03, Physics:Boehm00}. 
Based on a global 
analysis with $\pm 2\sigma$ ($\sim 95$\% C.L.)
ranges~\cite{Physics:ref. 5}, neutrino oscillation data have determined; (1) 
$\sin^2\theta_{12}$ to 18\% and $\Delta m^2_{21}$ to 9\%, (2) 
$\sin^2\theta_{23}$ to 41\% and $\Delta m^2_{32}$ to 26\%, and
(3) upper bounds on $\theta_{13}$ due to null oscillation results.
The CP phase angle $\delta$ will be likely hard to measure
with current and near future oscillation experiments. The results from 
the global fits are as follows:   
\begin{eqnarray}
\sin^2\theta_{13} = 0.009^{+0.023}_{-0.009} & &  \\
\nonumber
\sin^2\theta_{12} = 0.314(1^{+0.18}_{-0.15}) & & 
\Delta m_{21}^2 = 7.92(1\pm 0.09)\times 10^{-5}~\hbox{eV}^2\\
\sin^2\theta_{23} = 0.44(1^{+0.41}_{-0.22}) & &  
\Delta m_{23}^2 = 2.4(1^{+0.21}_{-0.26})\times 10^{-3}~\hbox{eV}^2.
\nonumber
\end{eqnarray}

Another global analysis~\cite{Physics:ref. 6} including the first MINOS results has found 
similar oscillations parameters that overlap significantly with the above 
results. There are three unmeasured neutrino oscillation parameters of 
$\theta_{13}$, the Dirac CP phase $\delta_{CP}$, and the sign of 
$\Delta m^2_{32}$ that determines the hierarchy of neutrino masses. 
Based on the measured mixing angles, the MNS matrix can be approximately 
written by 
\begin{equation}
U\simeq  
\left( \begin{array}{ccc}
0.8 &0.5 &<0.2  \\
0.4 &0.6 &0.7 \\ 
0.4 &0.6 &0.7
\end{array} \right)
\end{equation}

The elements of the MNS matrix are quite different from those of the CKM 
matrix which is nearly diagonal. The $U_{e3}$ element including 
$\sin^2\theta_{13}$ is small compared to other elements. This peculiar 
feature needs to be explained by the unified theory of elementary 
particles and thus the smallness of $\theta_{13}$ may play an important role when 
constructing the unified theory.

\section{Mixing Angle $\theta_{13}$}
\subsection{Current Knowledge of $\theta_{13}$}
\begin{figure}
\begin{center}
\includegraphics[width=4.0in]{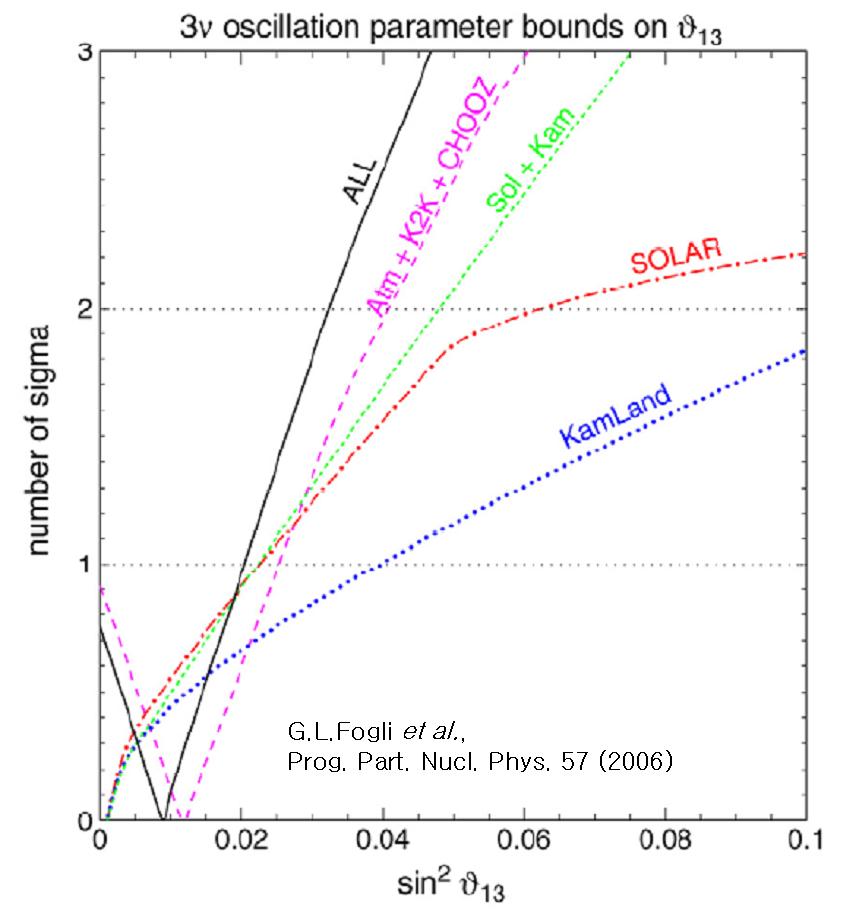}
\end{center}
\caption{Bounds on $\sin^2\theta_{13}$ for different data sets.
This figure is taken from Ref.~\cite{Physics:ref. 5}.}
\label{bounds 1}
\end{figure}

\begin{figure}
\begin{center}
\includegraphics[width=4.0in]{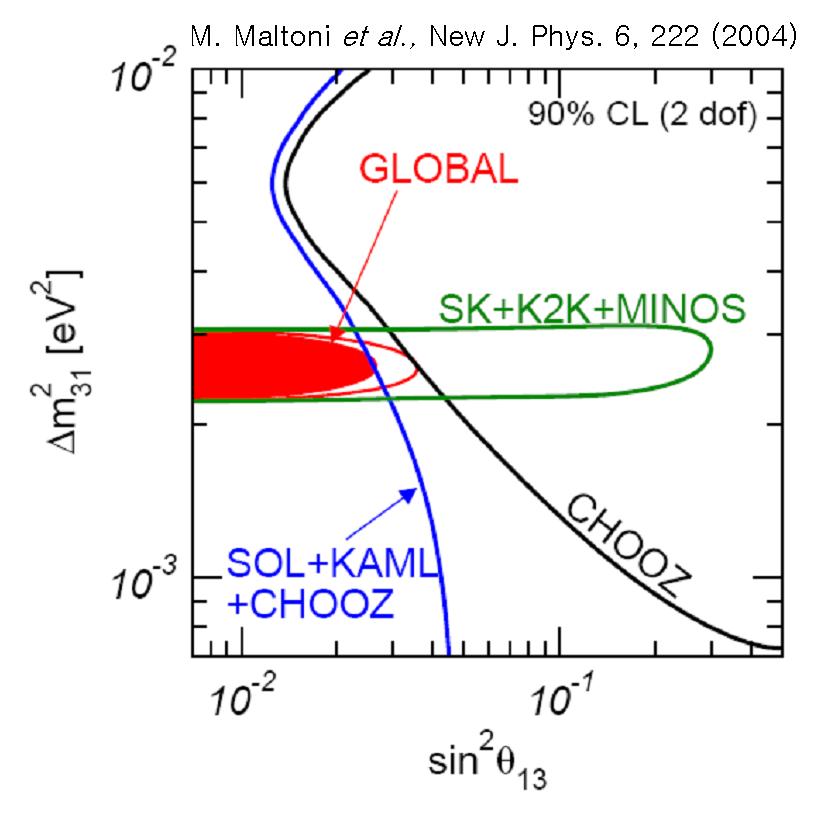}
\end{center}
\caption{Bounds on $\sin^2\theta_{13}$ from the interplay of the
global data. This figure is taken from Ref.~\cite{Physics:ref. 6}.}
\label{bounds 2}
\end{figure}

Figure~\ref{bounds 1} in Ref.~\cite{Physics:ref. 5} shows several fit results 
of $\theta_{13}$
based on different inputs. 
Those fits with solar and atmospheric experiments separately favor values 
near $\sin^2\theta_{13}=0$, but their global analyses together find minimum  
$\chi^2$ at non-vanishing values of $\theta_{13}$  as listed above. 
The other global fit also has obtained a similar bound on 
$\theta_{13}$ as shown in Fig.~\ref{bounds 2}~\cite{Physics:ref. 6}.

At 95\% C.L., the upper bound of $\sin^22\theta_{13}$ from Ref.~\cite{Physics:ref. 5}
is approximately 0.12, corresponding to the $\theta_{13}$ value of 
$10^\circ$. This could be compared to the upper limit of 0.17 at
90\% C.L. obtained by CHOOZ~\cite{Physics:Apollonio03}. 
The mixing angle $\theta_{13}$
is very small compared to $\theta_{12}$ and $\theta_{23}$.

\subsection{Significance of $\theta_{13}$}
Genuine three flavor oscillation effects occur only for a finite value of 
$\theta_{13}$. Therefore, it is necessary to measure a finite value of 
$\theta_{13}$ as one of the next milestones in further studies of neutrino 
oscillations. In addition, $\theta_{13}$ is important in theoretical 
model building of neutrino mass matrix, which can serve as a guide to 
the theoretical understanding of physics beyond the Standard Model.  

Leptonic CP violation is also a three flavor effect, but it can only be 
tested if $\theta_{13}$ is finite. The CP phase angle $\delta_{CP}$
appears always in the combination $U_{e3}=\sin\theta_{13}e^{-i\delta_{CP}}$.
If $\theta_{13}$ is zero then it is not possible to probe leptonic CP 
violation in neutrino oscillation experiments. If $\sin^2(2\theta_{13})>0.01$, 
then the design of experiments to measure the sign of $\Delta m^2_{32}$
and the CP phase $\delta_{CP}$ becomes straightforward extensions
of current experiments~\cite{Physics:Anderson04, Physics:APS04}. For this reason, there 
is a strong 
motivation that a $\theta_{13}$ measurement should be the prime goal 
of the next round of experiments~\cite{Physics:Freund01}. 

On the theoretical side it is quite interesting to know if the 
value of $\theta_{13}$ happens to be small without any reason or 
has some theoretical backgrounds such as some symmetry argument being 
required to explain a tiny value. 
A reason for expecting a 
particular value of $\theta_{13}$ does not clearly exist as long as one
extends the SM only minimally to accommodate neutrinos masses. $\theta_{13}$ 
is then simply some unknown parameter which could take an arbitrary
small value, including zero. The situation changes in models of 
neutrino masses. Even then one should acknowledge that in principle 
any value of $\theta_{13}$ can be accommodated. Therefore, today there 
is no particular reason to expect the third angle, $\theta_{13}$, to
be extremely small or even zero.  

\begin{table}
\begin{center}
\includegraphics[width=4.0in]{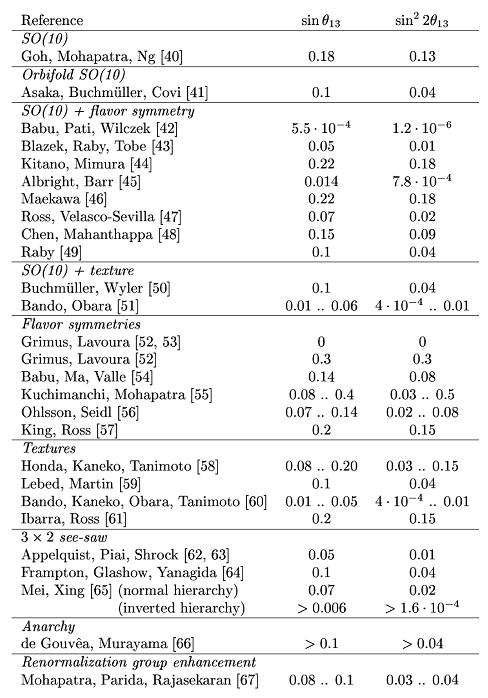}
\caption{Various predictions of $\theta_{13}$ based on different 
theoretical models taken from Ref.~\protect\cite{Physics:Anderson04}. 
The reference numbers in the table are the reference numbers in 
Ref.~\protect\cite{Physics:Anderson04}.}
\label{summary of theta13}
\end{center}
\end{table}

Some neutrino mass models often have a tendency to predict also a 
sizable value of $\theta_{13}$. This is the case for models in 
the framework of Grand Unified Theories and for models using flavor 
symmetries. There exist also many different texture models of 
neutrino masses and mixings, which accommodate existing data and try 
to predict the missing information by assuming certain elements of 
the mass matrix to be either zero or identical. In general, 
if $\theta_{13}$
is not too small e.g., close to the current upper limit of 
$\sin^2(2\theta_{13})\simeq 0.01$ and $\theta_{23}\neq \pi/4$, 
the neutrino mass matrix does not have any special symmetry features, 
sometimes referred as anarchy models, and the specific values of   
$\theta_{13}$ can be understood as a numerical accident. However, if 
$\theta_{13}$ is much smaller than the current limit, special 
symmetry of neutrino mass matrix will be required. For a large value
of $\theta_{13}$, it leaves open questions on quark-lepton unification.
Summary of various predictions is shown in Table~\ref{summary of theta13}.

Altogether there exist very good reasons to push the sensitivity limit 
from the current CHOOZ value by an order of magnitude and to hope that 
a finite value of $\theta_{13}$ will be found. But as already mentioned, 
at this precision even a negative result would be very interesting, 
since it would test or rule out many neutrino mass models and restrict 
parameters relevant for quantum corrections to masses and mixings. 

\section{Reactor Neutrino Experiment}
\subsection{Reactor Neutrinos}
Nuclear reactors have played crucial roles in experimental neutrino 
physics. The discovery of the neutrino was made at the Savannah River 
Reactor in 1956 by Reines and Cowan~\cite{Physics:Cowan56}. KamLAND observed disappearance 
of reactor antineutrinos and distortion in the energy spectrum because 
of neutrino oscillations due to mixing angle $\theta_{12}$. Furthermore, reactor 
neutrino experiments have the potential of uniquely determining $\theta_{13}$ at a 
low cost and in a timely manner.

The Yonggwang nuclear power plant has six Pressurized Water Reactors (PWR),
with average total thermal power output of 16.4~$\mbox{GW}_{th}$.
The fissile material in the reactors mainly consists of $^{235}$U and
$^{239}$Pu, which undergoes thermal neutron fission. 

The dominant $^{238}$U is fissile only for fast neutrons but also undergoes 
fission process by thermal neutron capture and
produces $^{239}$Pu,
\begin{equation}
n+ ^{238}\rm{U} \to ^{239}\rm{U}\to ^{239}\rm{Np} \to ^{239}\rm{Pu}.
\end{equation}
Similarly, $^{241}$Pu is generated from $^{239}$Pu, 
\begin{equation}
n+ ^{239}\rm{Pu} \to ^{240}\rm{Pu}\to ^{241}\rm{Pu}.
\end{equation}

Four fissile isotopes, $^{235}$U, $^{239}$Pu, $^{238}$U, and $^{241}$Pu,
are important and others contribute only at the 0.1\% level.
Fission fragments from these isotopes sequentially $\beta$ decay and emit
electron antineutrinos.
The purity of the antineutrinos is very high and electron-neutrino 
contamination is only at a $10^{-5}$ level above an inverse $\beta$ decay
threshold of 1.8~MeV.

\begin{figure}
\begin{center}
\includegraphics[width=3.6in]{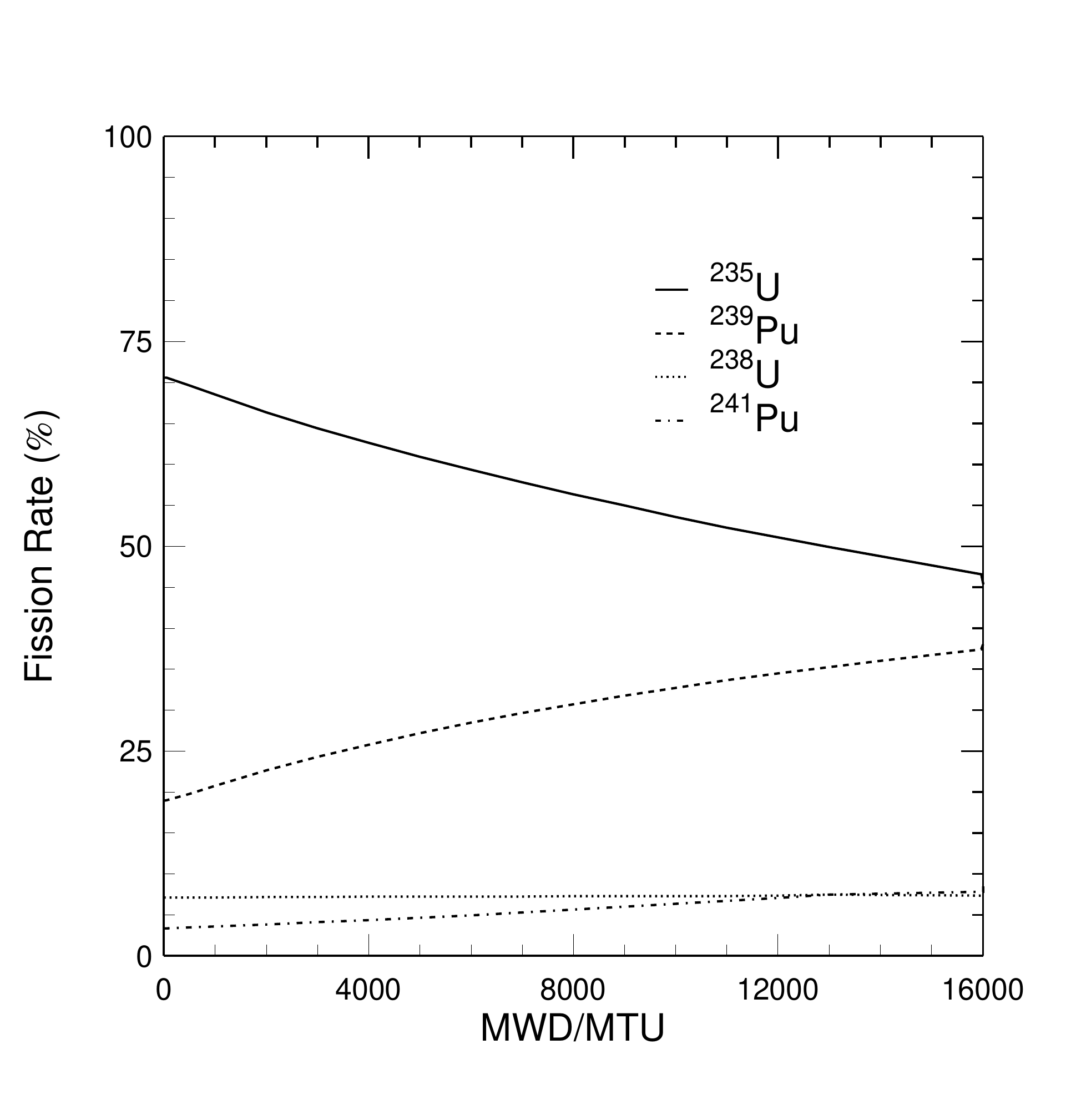}
\end{center}
\caption{The evolution of fission rate of the four dominant fissile 
isotopes of a typical refuelling cycle. Courtesy of Ulchin Nuclear 
Power Plant Co.} 
\label{fission rates}
\end{figure}

The fission rates of the fissile isotopes are shown in 
Fig.~\ref{fission rates}.
These four isotopes release similar energy, as shown in 
Table~\ref{mean energy}, 
when they undergo fission~\cite{Physics:Kop88a}.
Therefore, even though the makeup of the fissile material in the reactor 
changes over
a refuelling cycle, the average mean energy per fission does not change
significantly.
Assuming $\sim 200$~MeV per fission, there are $3.1\times 10^{19}$ 
fissions per GW$_{th}$.
Since one fission causes about six neutrino emissions above $\sim 2$~MeV on 
average~\cite{Physics:Sch85a,Physics:Hah89a,Physics:Kla82a,Physics:Vog81a}, the neutrino intensity can be 
estimated to be
$\sim 2\times 10^{20}~/({\rm GW}_{th}\cdot {\rm s})$.

\begin{table}
\begin{center}
\begin{tabular}{cc} \hline
Isotope &Mean Energy Per Fission (MeV)  \\\hline
$^{235}$U &$201.8\pm 0.5$ \\
$^{238}$U &$205.0\pm 0.7$ \\
$^{239}$Pu &$210.3\pm 0.6$ \\
$^{241}$Pu &$212.6\pm 0.7$ \\\hline
\end{tabular}
\end{center}
\caption{Mean energy emitted per fission for four main 
isotopes in nuclear fuel~\protect\cite{Physics:Kop88a}.}
\label{mean energy}
\end{table}

The neutrinos are radiated isotropically from the reactor core and, 
therefore, the
inverse square law applied on the neutrino intensity at a distance.
The neutrino energy spectrum from a reactor is shown in Fig.~\ref{neu flux}.

\subsection{Inverse Beta Decay}\label{inverse beta decay}
When an electron antineutrino enters matter, it can be captured
by a free proton via inverse neutron decay 
\begin{equation}
\bar{\nu}_{e} + p \to e^{+}+n,
\end{equation}
which has an 1.8 MeV antineutrino energy threshold.
The resulting neutron is subsequently captured by a 
proton in the following process:
\begin{equation}
n+p\to D+\gamma,
\end{equation}
where $D$ is a deuterium.
The mean time for neutron capture is $\sim 200~\mu$s.
The incident antineutrino energy is directly related to the energy
of the positron by 
\begin{equation}
E_{\bar{\nu}_e}=E_{e^+}+(m_n-m_p)+{\cal O}(E_{\bar{\nu}_e}/m_n),
\end{equation}
where $E_{e^+}$ is the energy of the positron coming out from 
the inverse neutron decay and $m_{n}(m_{p}$) is the neutron
(proton) mass.
The positron deposits its energy and then annihilates, yielding
two photons each with $0.511~\mbox{MeV}$, thus experimentally visible
energy is $(E_{e^+}+0.511~\mbox{MeV})$ with the minimum energy
of 1.022~MeV.
An electron antineutrino event then can be identified by a distinctive
signature of a prompt positron signal followed by a photon from the 
delayed neutron capture. 

\begin{figure}
\begin{center}
\includegraphics[width=5.0in]{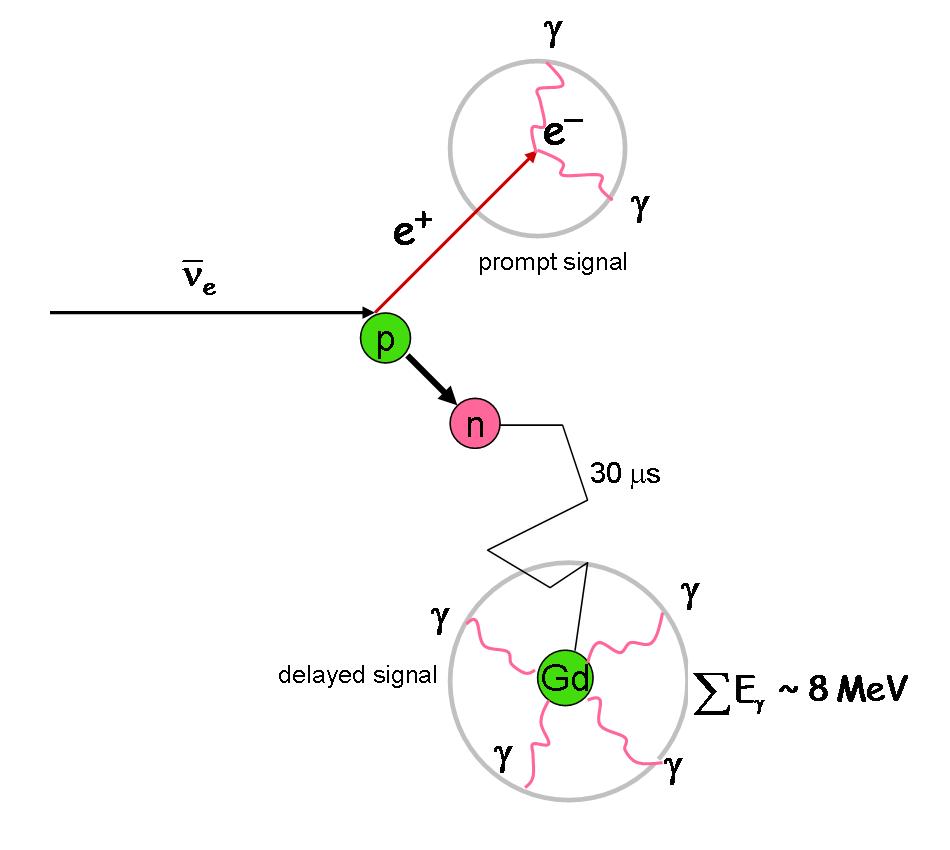}
\end{center}
\caption{An electron antineutrino would be detected by a coincidence
signal of a prompt positron and a delayed captured neutron. The 
neutrino energy is directly related to the measured energy of the
outgoing positron.}
\label{inverse beta decay cartoon}
\end{figure}

However, if a neutron is captured by Gd in which a proton is bound, then 
the capture cross section becomes larger and additional gammas are 
produced to have total energy of about 8~MeV. The experimental signature 
for reactor neutrinos is a prompt energy deposit of 1-8~MeV, due to 
the positron kinetic energy and the annihilated $e^+e^-$ masses, 
followed an average 30~$\mu$s later by 8~MeV energy deposit of gammas 
from neutron capture on Gd. Exploiting the delayed coincidence is key 
to controlling backgrounds. Figure~\ref{inverse beta decay cartoon} 
shows both prompt and delayed 
signals produced by a reactor neutrino. 

The inverse neutron decay process has the cross section of the form,
\begin{equation}
\sigma(E_{e^{+}}) \simeq {2\pi^2 \hbar^3\over m_e^5 f \tau_n}
p_{e^+}E_{e^+},
\end{equation}
where $p_{e^+}$ and $m_e$ are the momentum and the mass of the positron,
respectively, 
$\tau_n$ is the lifetime of a free neutron, and $f = 1.7152$ is the free 
neutron decay phase space factor~\cite{Physics:Wil82}. 

\begin{figure}
\begin{center}
\includegraphics[width=3.3in]{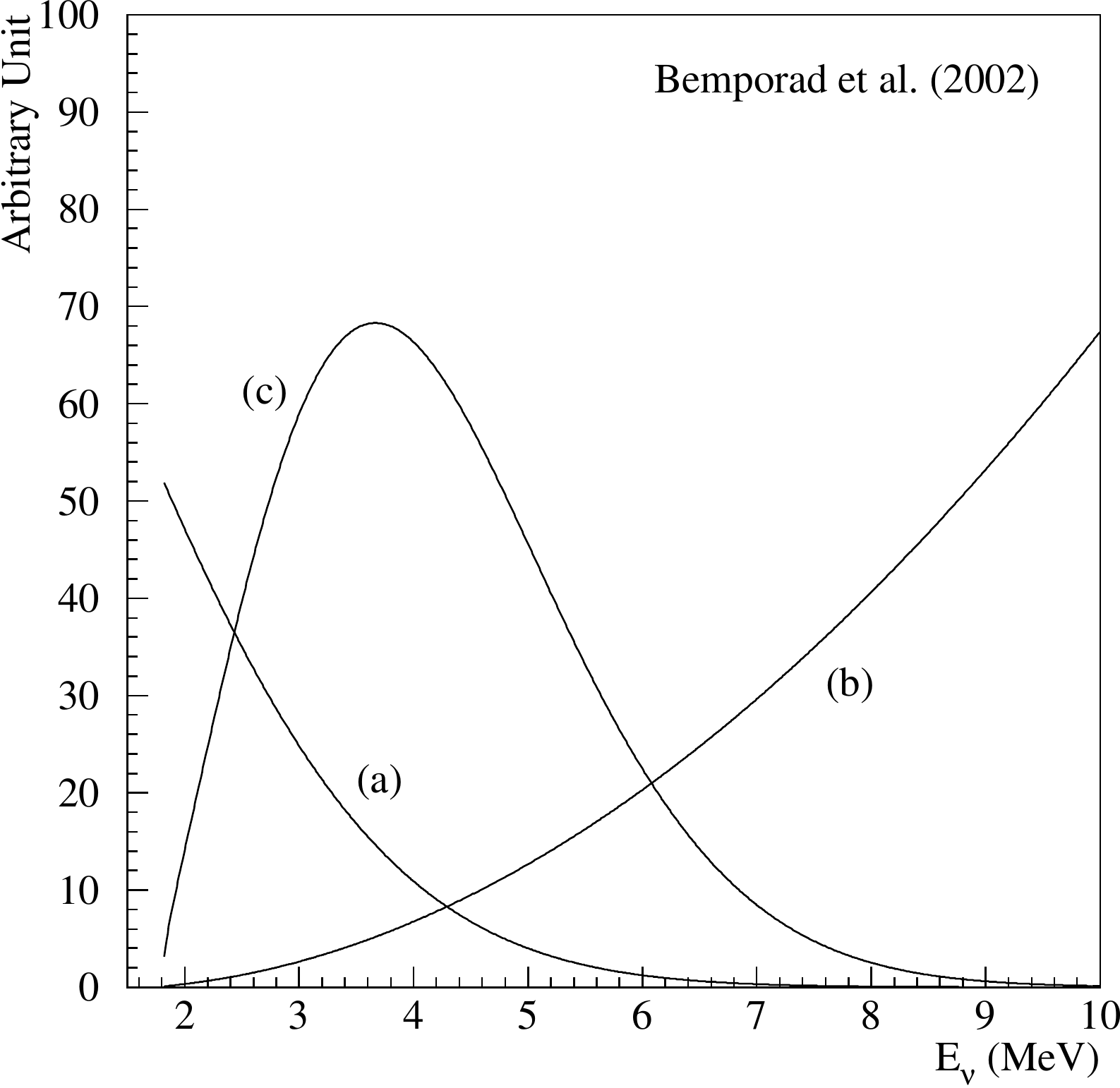}
\end{center}
\caption{Reactor $\bar\nu_{e}$ flux (a), inverse beta decay cross section (b),
and interaction spectrum at a detector based on such reaction (c) in 
Ref.~\protect\cite{Physics:Bem02a}. The cut-off at 1.8~MeV is due to the minimum 
neutrino energy required for inverse beta decay process.}
\label{neu flux}
\end{figure}

Figure~\ref{neu flux} shows the neutrino flux, inverse beta decay cross 
section, and interaction spectrum at a detector in arbitrary units 
calculated in Ref.~\cite{Physics:Bem02a}.
The most probable neutrino energy interacting at a detector is
$\sim 3.8$~MeV.

\subsection{Neutrino Oscillations in Reactor Experiments}
\label{oscillation subsection}
Because neutrinos from reactors have low energy of the order of
a few MeV, they do not have enough energy to produce muons
or taus through charged current interaction.
Therefore, any reactor experiments can only be disappearance experiments,
which measure the survival probability $P(\bar\nu_e\to \bar\nu_e)$.

\begin{figure}
\begin{center}
\includegraphics[width=3.6in]{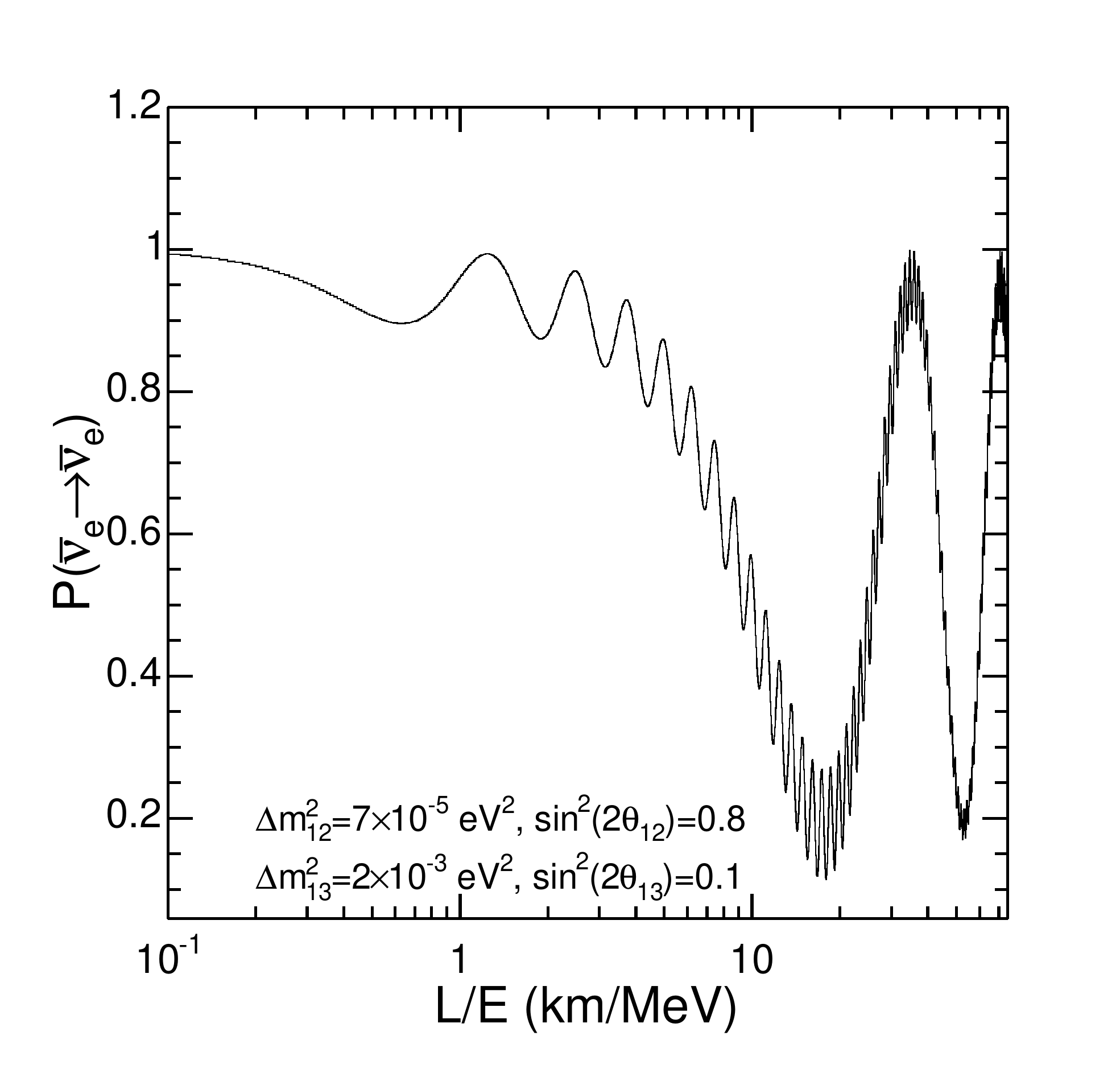}
\end{center}
\caption{The survival probability of $\bar{\nu}_{e}$ vs the ratio of
the distance to the neutrino energy ($L/E$) with $\sin^2(2\theta_{13})$
set at the maximum value allowed in the current limit~\protect\cite{Physics:Apollonio03}.}
\label{oscillation}
\end{figure}

It was shown in Ref.~\cite{Physics:Min99a} that the survival probability does not
depend on the $CP$ phase $\delta$.
And because of the low energy neutrinos 
and short baseline, matter effects are negligible in reactor 
experiments~\cite{Physics:Min03a}. Thus one can use the neutrino survival 
probability in vacuum to model the neutrino oscillations in the reactor 
experiments. Assuming a mass hierarchy of $m_1<m_2<m_3$, the expression 
for the $\bar\nu_e$ disappearance probability is written 
as~\cite{Physics:Min03a}
\begin{eqnarray}
\label{survival eqn1}
1-P(\bar\nu_{e}\to\bar\nu_{e}) 
&=& 4\sum_{j>k}|U_{ej}|^2|U_{ek}|^2\sin\left({\Delta m^2_{jk}L\over 4E}\right) \\
\nonumber
&=& \sin^2(2\theta_{13})
              \sin^2\left({\Delta m_{31}^2L\over 4E}\right) \\
\nonumber
          &+& \cos^4\theta_{13}\sin^2(2\theta_{12})
              \sin^2\left({\Delta m_{21}^2 L\over 4E}\right)\\
\nonumber
          &-& {1\over 2}\sin^2\theta_{12}\sin^2(2\theta_{13})
              \sin\left({\Delta m_{31}^2 L\over 2E}\right)
              \sin\left({\Delta m_{21}^2 L \over 2E}\right)\\
\nonumber
          &+& \sin^2\theta_{12}\sin^2(2\theta_{13})
              \cos\left({\Delta m_{31}^2 L\over 2E}\right)
              \sin^2\left({\Delta m_{21}^2 L\over 4E}\right).
\label{disap eq}
\end{eqnarray}
One can see that the oscillations are governed by two quadratic mass 
splittings, $\Delta m_{21}^2$ and $\Delta m_{31}^2$. 
Equation~\ref{survival eqn1} is plotted as a function of $L/E$ in
Fig.~\ref{oscillation} with the current best values of $\Delta m^2$s
and $\sin^2(2\theta_{12})$, and $\sin^2(2\theta_{13})$
at the upper bound~\cite{Physics:Apollonio03}.

The KamLAND experiment has observed a 40\% disappearance of 
$\nu_e$ at the baseline of 180~km~\cite{Physics:Eguchi03}. The detected deficit 
is presumably associated with the second term of Eq.~\ref{survival eqn1}.
Since the uncertainties in $\Delta m^2$s are large, it is  
clear that the measuring the first dip in the survival probability
is the key to the analysis.

\begin{figure}
\begin{center}
\includegraphics[width=3.6in]{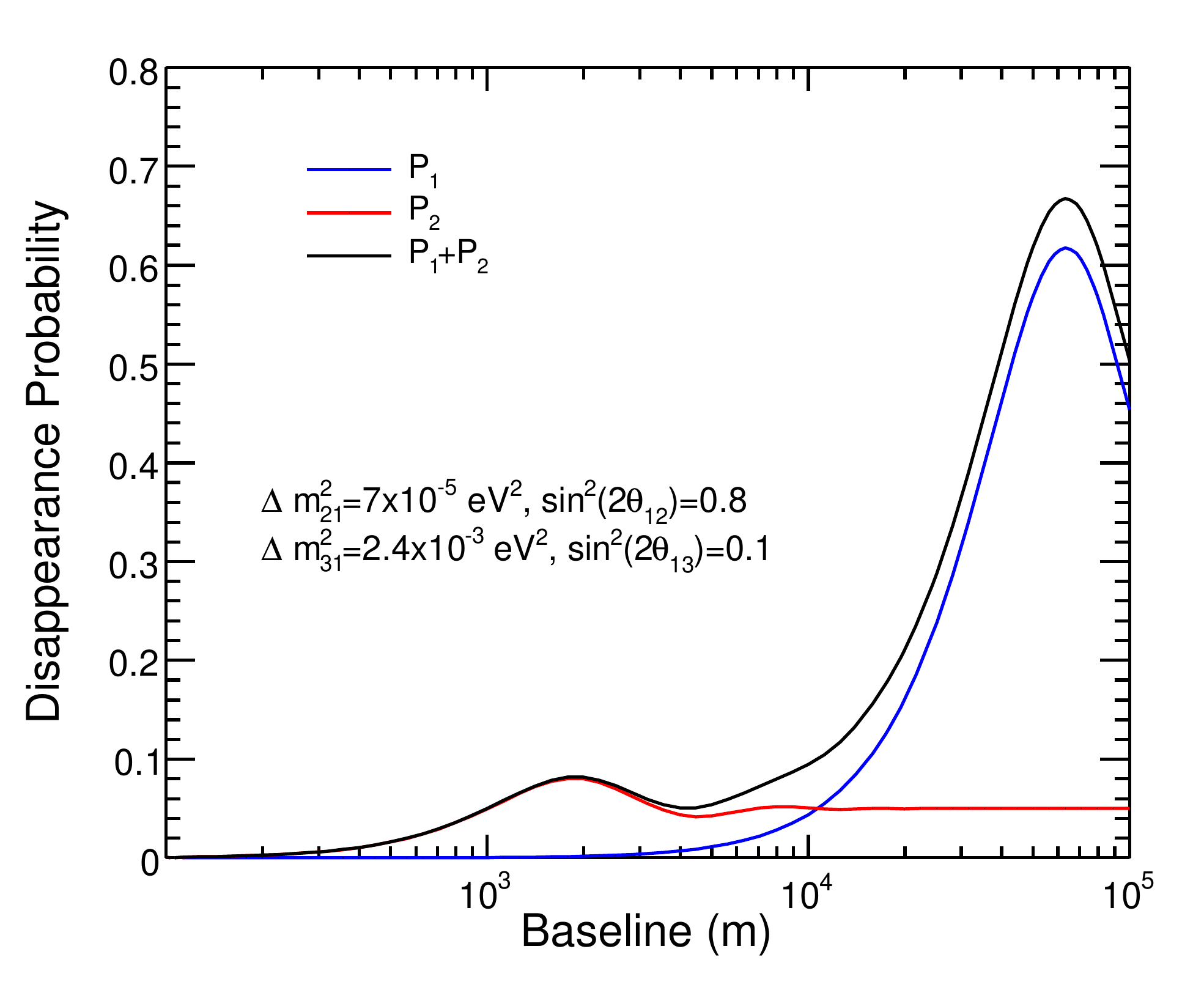}
\end{center}
\caption{Reactor neutrino disappearance probability integrated over 
neutrino energy from 1.806~MeV to 10~MeV as a function of distance 
from the source. $P_{1}$ and $P_{2}$ are defined in the text.} 
\label{disappearance}
\end{figure}

Figure~\ref{disappearance} shows the disappearance probability 
($=1-P(\bar\nu_e\to\bar\nu_e$)) of observable neutrinos as a function 
of distance from a reactor neutrino source. Here the term observable 
neutrinos refers to neutrinos undergoing inverse beta decay interaction, 
had there been no oscillation. The probability is integrated over energy 
from 1.806~MeV to 10~MeV.
The probability term $P_1$ and $P_2$ are the first and the rest terms 
on the right side of Eq.~\ref{survival eqn1}, respectively.  
The first maximum in the disappearance probability occurs $\sim 2$~km
almost soley due to $P_1$, which is independent of $\theta_{12}$ or
$\Delta m_{21}^2$.
The second peak is dominantly due to $P_2$, especially the second term 
in Eq.~\ref{survival eqn1}.
The contributions from the third and fourth terms on the right side of 
Eq.~\ref{survival eqn1} are inherently small compared to the first two 
terms due to the cross terms 
in $\Delta m_{21}^2$ and $\Delta m_{31}^2$ for the currently known values. 

Since the first and the second terms in Eq.~\ref{disap eq} can be 
combined due to $\Delta m^2_{31} \simeq \Delta m^2_{32}$, 
the survival probability can be approximated as
\begin{equation}
P(\bar\nu_{e}\to\bar\nu_{e}) \simeq 1-\sin^2(2\theta_{13})
              \sin^2\left({\Delta m_{31}^2L\over 4E}\right) 
          - \cos^4\theta_{13}\sin^2(2\theta_{12})
              \sin^2\left({\Delta m_{21}^2 L\over 4E}\right).
\label{survival eqn2}
\end{equation}

\section{Determination of $\theta_{13}$}
\subsection{Measurement of Reactor Neutrino Flux}
The reactor $\bar\nu_e$ flux is measured by a detector placed
at a short distance from the reactor before the oscillation
occurs. 
The RENO target consists of 0.1\% Gd loaded liquid scintillator, 18.7~m$^3$
in volume and 16.1 tons in mass. 
The liquid scintillator is made of Linear Alkylbenzene (LAB) with a
small amount of fluors and the Gd compound. This corresponds to 
$1.21\times 10^{30}$ free protons available for the inverse beta decay 
reaction. 

The reactor $\bar\nu_e$'s are detected by the inverse beta decay 
$\bar\nu_e+p\to n+e^+$ reaction. The ionization energy loss and subsequent 
annihilation of the positron from the inverse beta decay reaction result 
in a prompt signal. After an average of 30~$\mu$s, 
a delayed signal will follow due to the neutron capture by Gd and 
subsequently emitting several gammas with a total energy of about 8~MeV. 
Exploiting the delayed coincidence is a key to controlling backgrounds.

The detector should be located underground in order to reduce background 
from cosmic rays and cosmic ray induced spallation products, such as 
neutrons and the radioactive isotope $^9Li$. One also needs to remove 
backgrounds due to gamma rays from natural radioactivity in the detector 
material and the surrounding rocks.

\subsection{Effects on Reactor Neutrino Flux Due to $\theta_{13}$}
RENO will measure the survival probability for  at a baseline of 
about 1.4~km with neutrino energy ranging from 1.8 MeV to about 10 MeV. 
The disappearance probability $P_{dis}$ of reactor 
antineutrinos is obtained from the survival probability $P_{sur}$ of 
Eq.~\ref{survival eqn2}
\begin{equation}
P_{dis} = 1 - P_{sur} = 
\sin^2(2\theta_{13})\sin^2\left({\Delta m^2_{31}L\over 4E}\right) 
+\cos^4\theta_{13}\sin^2(2\theta_{12})
\sin^2\left(\Delta m^2_{21}L\over 4E\right).
\end{equation}
The disappearance probability can be expressed as sum of a 
$\theta_{13}$-dominant term $P_{13}$ and a $\theta_{12}$-dominant 
term $P_{12}$, 
\begin{equation}
P_{dis} = P_{13}+P_{12}
\end{equation}
where $P_{13}= \sin^2(2\theta_{13})
\sin^2\left({\Delta m^2_{31}L\over 4E}\right) $ and 
$P_{12} = \cos^4\theta_{13}\sin^2(2\theta_{12}
\sin^2\left(\Delta m^2_{21}L\over 4E\right)$.
Here, $P_{13}$ is $P_1$ and $P_{12}$ is an approximation of $P_2$ in
Sect.~\ref{oscillation subsection}.

The value of $\theta_{13}$ will be obtained from $P_{13}$ as a result 
of subtracting $P_{12}$ from the experimental measurement of $\bar\nu_e$ 
disappearance probability of $P_{dis}$. In Fig.~\ref{disappearance}, 
the disappearance 
probabilities of $P_{13}$, $P_{12}$, and $P_{dis}$ are shown as a function 
of the antineutrino flight distance from 100 m to 250 km. We chose 
$\sin^2(2\theta_{13})=0.01$ for a convenient discussion. The rest of 
parameters are taken as follows
\begin{equation}
\Delta m^2_{21}=7.9\times 10^{-5}~\hbox{eV}^2,
~~\Delta m^2_{31}=2.5\times 10^{-3}~\hbox{eV}^2.
\end{equation}

Figure~\ref{disappearance} shows that the $P_{12}$ contribution is 
negligible and thus the 
disappearance probability ($P_{dis}$) is almost equal to $P_{13}$, within 
a few kilometers of the baseline. Since $P_{13}$ is proportional to  
$\theta_{13}$
exclusively, the disappearance measurement of reactor antineutrinos 
will directly probe the mixing angle of $\theta_{13}$. The first 
oscillation maximum of $P^{max}_{13}(=\sin^2(2\theta_{13})$ occurs 
near the baseline of $\sim2$~km. 
The best measurement of $\theta_{13}$ could be possible at the first 
oscillation maximum. Beyond the first oscillation minimum the $P_{12}$ 
contribution grows rapidly, and $P_{13}$ and $P_{dis}$ deviate from 
each other. As the baseline goes longer than ~50 km, the  $P_{12}$ 
contribution becomes dominant in $P_{dis}$. 

\begin{figure}
\begin{center}
\includegraphics[width=3.00in]{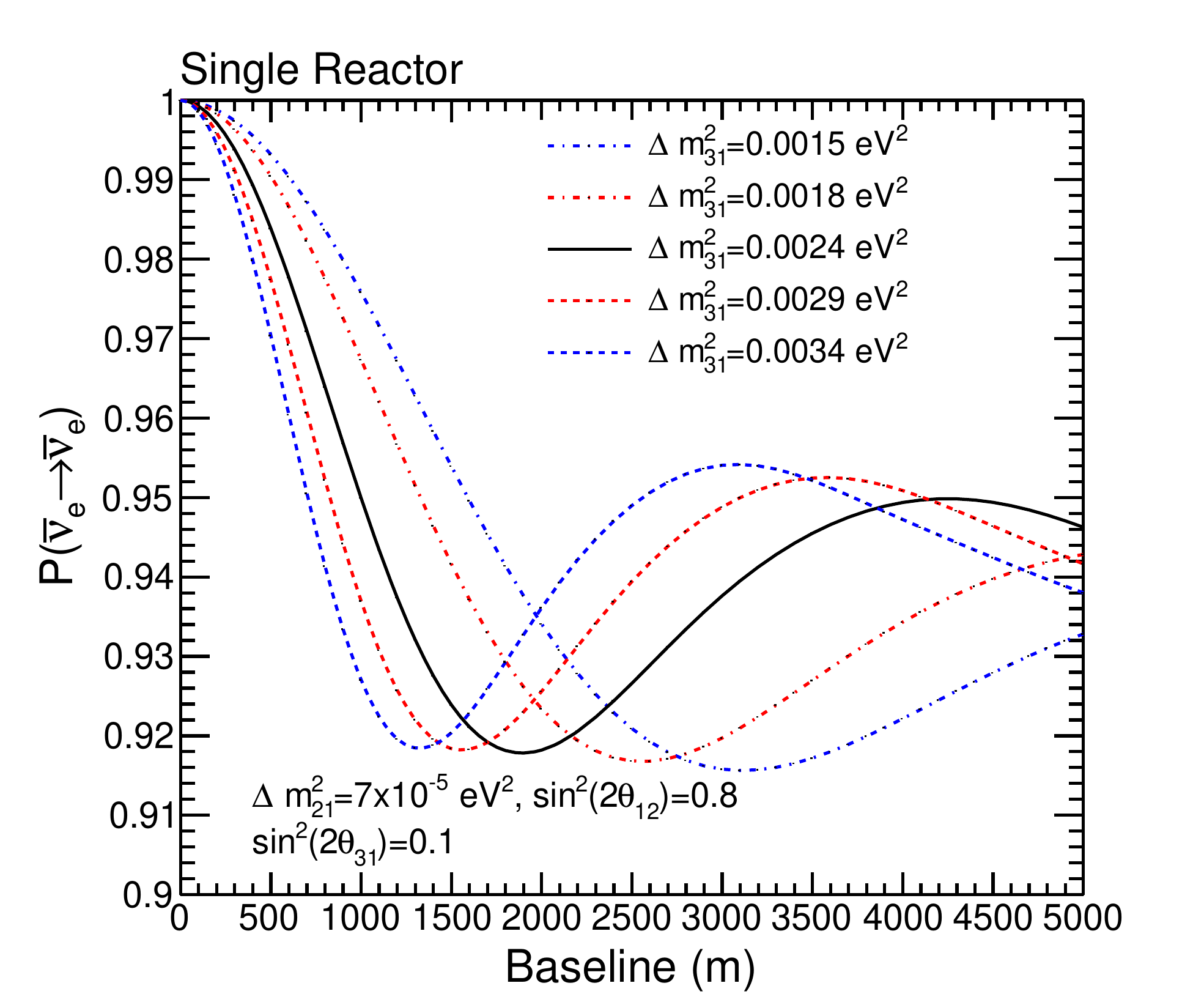}
\includegraphics[width=3.00in]{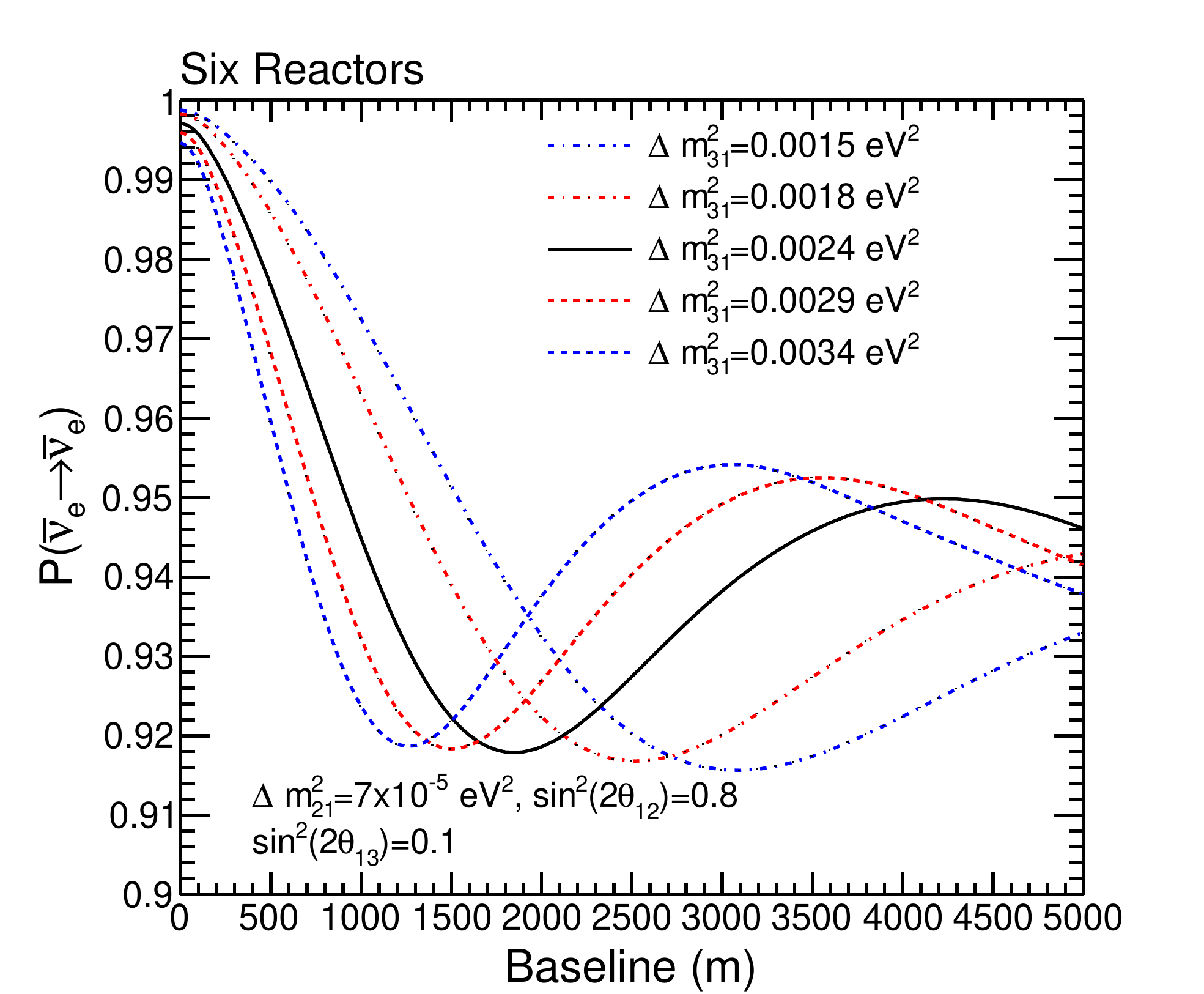}
\end{center}
\caption{
Survival probabilities of the reactor 
neutrinos with $\sin^2(2\theta_{13})=0.1$ and various values of $\Delta m_{31}^2$ 
as a function of the distance from a single reactor (left)
and six reactors
arranged as shown in Fig.~\protect\ref{google} (right).
$\Delta m_{31}^2 =0.0024$~eV$^2$ represents the most probable 
value and the ranges $0.0018\sim 0.0029$~eV$^2$ and $0.0015 \sim 0.0034$~eV$^2$ 
represent 63\% and 90\% CL, respectively. There is little difference
between these two cases except at very small distances.}
\label{survival probability}
\end{figure}

The maximum location differs within 
the measured error of $\Delta m^2_{23}$.
Figure~\ref{survival probability} shows $P_{13}$ integrated over neutrino energy 
from 1.8 to 8~MeV, 
as a function of the baseline $L$ for three values of $\Delta m^2_{32}$ 
in its 95\% C.L. allowed range. The curves show that $P_{13}$ is 
sensitive to $\Delta m^2_{32}$, and its oscillation maxima occur at 
baselines of 1.5 to 2.5 km. 

Since the measurement of disappearance probability includes the $P_{12}$ 
contribution, determination of $\theta_{13}$ from $P_{13}(=P_{dis}-P_{12})$ 
will suffer from the uncertainties of $\theta_{12}$ and $\Delta m^2_{21}$. 
At the first maximum the fraction of $P_{12}$ relative to $P_{13}$ is  
about 2.6\% to 25\% when $\sin^2(2\theta_{13})$ varies from 0.01 to 0.10. 
The uncertainty in determining $\sin^2(2\theta_{13})$ due to the uncertainty 
of $P_{12}$ is less than 0.005. For $\sin^2(2\theta_{13})>0.001$, 
the $P_{12}$ contribution to $P_{dis}$ can be ignored. Therefore, the 
survival probability of reactor antineutrinos can be written without  
$\theta_{12}$ and $\Delta m^2_{21}$ and  if the detector is located near 
the first oscillation maximum in
\begin{eqnarray}
P_{sur} &=& 1-\sin^2(2\theta_{13})\sin^2\left({\Delta m^2_{31}L\over 4E}\right) 
\nonumber
\\
&=&1-\sin^2(2\theta_{13})\sin^2\left(1.27
{\Delta m^2_{31}[10^{-3}~\hbox{eV}^2]L[km]
\over E_{\bar\nu_e} [\hbox{MeV}]}\right).
\label{survival prob}
\end{eqnarray}

\subsection{Experimental Extraction of $\theta_{13}$}
\subsubsection{Single Detector Measurement}
The current best limit on $\theta_{13}$ comes from null results of 
neutrino oscillations by CHOOZ  and Palo 
Verde~\cite{Physics:Apollonio03,Physics:Boehm00}. These 
experiments were at a baseline distance of about 1 km and thus 
were more sensitive to the second term of Eq.~\ref{survival eqn2}. 
Both experiments looked for a deficit in the $\bar\nu_{e}$ flux at 
the location of the detector by comparing the observed rate with 
the expected rate from the reactors based on no oscillation. 
Those experiments with only one detector at a fixed baseline 
could not have greatly improved sensitivity to $\theta_{13}$ 
because of uncertainties related to knowledge of neutrino flux 
from the reactors and to the detector acceptance. A single detector 
measurement had to calculate the expected  rate relying on the 
reactor operation data such as the generated thermal power as 
a function of time and the nuclear fuel composition. 

CHOOZ and Palo Verde detected the reactor $\bar\nu_e$ events by 
the inverse beta decay reaction utilizing the 0.1\% Gd loaded 
liquid scintillator. The value of $\sin^2(2\theta_{13})$ was 
determined by comparing the observed antineutrino rate and 
energy spectrum with expected ones based on no oscillation. 
If the observed event rate is smaller than the expected one, 
the number of detected antineutrinos ($N_{\bar\nu_e}$) finds 
the value of $\sin^2(2\theta_{13})$ by the following equation
\begin{equation}
N_{\bar\nu_e} = {N_p\over 4\pi L^2}\int\int
\epsilon_{det} P(\bar\nu_e\to \bar\nu_e){d\sigma\over dE_{e^+}}
{d\phi_{\bar\nu_e}\over dE_{\bar \nu}}dE_{e^+}dE_{\nu},
\end{equation}
where $N_p$ is the number of free protons in the detector target, $L$ 
is the distance of the detector from the reactor, $\epsilon_{det}$ is 
the efficiency of detecting an antineutrino, $P(\bar\nu_e\to\bar\nu_e)$
is the survival probability from oscillation as a function of 
$\sin^2(2\theta_{13})$, ${d\sigma\over dE_{e^+}}$ is the differential 
cross sections of the inverse beta decay reaction, and 
${d\phi_{\bar\nu_e}\over dE_{\bar nu}}$ is the differential 
energy distribution at the reactor. 

For CHOOZ, the background rate was $1.41\pm0.24$ events per day 
in the 1997 run, and $2.22\pm 0.14$ events per day after its 
trigger was modified in 1998. The background events were subtracted 
from the observed event rated before extracting the value of 
$\sin^2(2\theta_{13})$. The systematic uncertainties and efficiencies 
of CHOOZ are summarized in Tables~\ref{chooz antineutrino uncertainty table} 
and \ref{chooz antineutrino detection uncertainty table}, respectively.
\begin{table}
\begin{center}
\begin{tabular}{lc}\hline
Systematic Sources	&Relative Uncertainties (\%) \\\hline
Reaction Cross Section	&1.9	\\
Number of Protons	&0.8	\\
Detection Efficiency	&1.5	\\
Reactor Power		&0.7	\\
Energy Released per Fission	&0.6\\\hline
Combined		&2.7	\\\hline
\end{tabular}
\end{center}
\caption{Systematic uncertainties in the absolute antineutrino normalization
of CHOOZ.}
\label{chooz antineutrino uncertainty table}
\end{table}

\begin{table}
\begin{center}
\begin{tabular}{lcc}\hline
Selection Criteria      &$\epsilon$~(\%)  &Relative Uncertainties (\%) \\\hline
Positron Energy		&97.8	&0.8 \\
Positron Geode Distance		&99.9	&0.1 \\
Neutron Capture		&84.6	&1.0 \\
Capture Energy Containment &94.6	&0.4 \\
Neutron Geode Distance		&99.5	&0.1 \\
Neutron Delay		&93.7	&0.4 \\
Positron-neutron Distance		&98.4	&0.3 \\
Neutron Multiplicity		&97.4	&0.5 \\\hline
Combined                &69.8   &1.5 \\\hline
\end{tabular}
\end{center}
\caption{Antineutrino detection efficiency of CHOOZ.}
\label{chooz antineutrino detection uncertainty table}
\end{table}

Neither CHOOZ nor Palo Verde observed any deficit in the observed 
antineutrino rate. The negative result gave rise to set a limit on 
the neutrino mixing angle  as shown in Fig.~\ref{bounds 1}. CHOOZ obtained 
the best limit of 0.17 for $\sin^2(2\theta_{13})$ for  at the 
90\% C.L.

\subsubsection{Multi Detector Measurement}
Since the effective disappearance will be extremely small, new 
experiment for $\theta_{13}$ would need to improve the previous 
systematic limitations. This could be achieved by two or more 
identical detectors. 

Mikaelyan and Sinev pointed out that the systematic uncertainties 
can be greatly suppressed or totally eliminated when two detectors 
positioned at two different baselines are utilized~\cite{Physics:Mika00}. 
The near detector close to the reactor is used to measure the flux 
and energy spectrum of the antineutrinos before oscillation effects 
take place, and thus relaxes the requirement of knowing the details 
of the fission process and operational conditions of the reactor. 
The value of $\sin^2(2\theta_{13})$ can be measured by comparing 
the flux and energy distribution of antineutrinos observed with the 
far detector to those of the near detector after considering a 
reduction factor due to distance squared. 

With multiple detector setup, one will obtain the ratio in the
number of observed antineutrinos with energy between $E$ and
$E+dE$, at a far distance $L_{far}$ to that at a near distance
$L_{near}$ as follows:
\begin{equation}
{N_{far}^{\nu}\over N_{near}^{\nu}} = \left({L_{near}\over L_{far}}\right)^2
\left({N_{far}^{p}\over N_{near}^{p}}\right)
\left({\epsilon_{far}\over \epsilon_{near}}\right)
\left[{P(\bar\nu_e\to\bar\nu_e;E,L_{far})\over P(\bar\nu_e\to\bar\nu_e;E,L_{near})}\right],
\end{equation}
where $N_{near(far)}^{p}$ and $\epsilon_{near(far)}$ are the number of 
target protons and the detection efficiency at near (far) 
detector.
If the two detectors are identical and thus have the same efficiency
and the same number of target protons, the ratio is given only by the ratio of 
detector distances and the ratio of survival probabilities. Suppose the near
detector is located fairly close to a single reactor core, the value of 
$\sin^2(2\theta_{13})$ is approximately given by
\begin{equation}
\sin^2(\theta_{13}) \simeq {1\over \sin^2\left[1.267\Delta m_{31}^2 ({\rm eV}^2)
\times 10^3 {L_{far} ({\rm km})\over E({\rm MeV})}\right]}
\left[1-\left({N_{far}^{\nu}\over N_{near}^{\nu}}\right)
\left({L_{near}\over L_{far}}\right)\right],
\end{equation}
where $L_{far}$ and $E$ are given in the units of km and MeV, 
respectively.
From this simplified discussion, it is clear that the two
detector scheme is an excellent approach in the further sensitive 
measurement of $\sin^2(2\theta_{13})$.

The projected statistical uncertainty of RENO is 0.3\% with
three year data-taking. The goal of RENO is to reduce the total 
systematic uncertainty to less than 0.6\%. Due to the multiple
($>2$) reactor with a single near detector configuration, the 
reactor related systematic uncertainties are expected to cancel
out because of two identical detectors and isotropic reactor
neutrino fluxes, and will be less than 0.1\%. The RENO detector
design differs slightly from CHOOZ in a sense that a non-scintillating
buffer region shields the active region (target and $\gamma$ catcher)
from the intrinsic PMT radioactivity. This allows us to remove a few
selection cuts and expected systematic uncertainty will be $\sim 0.3$\%.
The systematic uncertainty of H/C ratio and the target mass will be
significantly reduced to $\sim 0.2$\% due to the two identical 
detectors and accurate measurement of detector volumes. In overall,
based on an order of magnitude smaller than CHOOZ in both
statistical and systematic uncertainties as shown in 
Table~\ref{summary of systematic uncertainty},
RENO is expected to measure the value of $\sin^2(2\theta_{13})$ above
0.02.  
\begin{table}
\begin{center}
\begin{tabular}{llcc}\hline
\hline
\multicolumn{2}{c}{Uncertainty Source} &CHOOZ &RENO (Goal) \\\hline
\hline
Reactor Related &Neutrino Flux and Cross Section&1.9    &$<$0.1 \\
                &Reactor Power                  &0.7    &0.4    \\
                &Energy Released per Fission    &0.6    &$<$0.1
\\\hline
Detector Related
        &H/C Ratio                      &0.8    &0.2 \\
        &Target Mass                    &0.3    &0.2 \\
        &H/Gd Ratio                     &1.0    &$<$0.4 \\
        &Positron Energy                &0.8 &0.1 \\
        &Positron Geode Distance        &0.1 &-- \\
        &Capture Energy Containment     &0.4    &0.2 \\
        &Neutron Geode Distance         &0.1 &-- \\
        &Neutron Delay                  &0.4    &$<$0.1\\
        &Positron--Neutron Distance     &0.3    &-- \\
        &Neutron Multiplicity           &0.5    &-- \\\hline\hline
Combined        &               &2.7    &$<$0.6 \\\hline\hline
\end{tabular}
\end{center}
\caption{Systematic uncertainties of CHOOZ~\cite{Physics:Apollonio03} and RENO.
}
\label{summary of systematic uncertainty}
\end{table}

\section{Additional Physics }
The main goal of the RENO experiment is to measure the value of 
neutrino mixing angle. It is worthwhile to explore other physics 
that can be done with this experiment. 

\subsection{Supernova Neutrinos}
Liquid scintillator detectors will be sensitive to a burst of 
neutrinos of all flavors from a Galactic supernova in the energy 
of a few to tens of MeV range. The time scale of the burst is 
tens of seconds. The RENO detector background in a 10 second 
period is enough low for a successful observation of the supernova 
signal. Identical near and far RENO detectors have roughly 100 tons
of liquid scintillators in total, sensitive to the supernova 
neutrinos. The RENO detector contains $6.0\times 10^{30}$ free protons, 
$4.5\times 10^{31}$ carbons, and $3.3\times 10^{31}$  electrons, 
and thus would observe 35 events from a supernova at 
10 kpc~\cite{Physics:Schol07,Physics:Cado02}. Twenty-six events are expected via 
the inverse beta decay, $\bar\nu_e+p\to e^++n$. 
Neutral current interactions of 
$\nu(\bar\nu)+^{12}$C$\to \nu(\bar\nu)+^{12}$C$^*$ 
would produce 7.5 events with a 
15.5 MeV de-excitation $\gamma$-ray from $^{12}$C$^*$~\cite{Physics:Kolb02}. 
Charged current interactions of $\bar\nu_e+^{12}$C$\to ^{12}$B$+e^+$
and $\nu_e+^{12}$C$\to ^{12}$N$+e^-$ would produce 1.5 events. Elastic 
neutrino-electron scattering $\nu+e^-\to\nu+e^-$ would produce 1.6 events. 
The observation will probably require an accurate clock and an effective 
trigger of minimal dead time.

\subsection{Sterile Neutrinos}
The discovery of sterile neutrinos would have a revolutionary impact on 
neutrino and particle physics. The idea of sterile neutrinos was initially 
introduced by Pontecorvo in 1967~\cite{Physics:Ponte68} and has been considered 
later by many physicists~\cite{Physics:Cald93,Physics:Pelt93,Physics:Bile98,Physics:Bena97,Physics:Kays98}. 
Further information and references on 
sterile neutrinos can be found in the paper by Berezinsky, Narayan, 
and Vissani~\cite{Physics:Berez03}.

While recent neutrino oscillation results are understood in the framework 
of 3 active neutrino mixing, they do not completely exclude admixture of 
sterile neutrinos. An experimental hint in favor of sterile neutrinos 
comes from the unconfirmed observation by the LSND collaboration on 
$\nu_\mu\to \nu_s$~\cite{Physics:LSND96}. The mixing with sterile neutrinos based 
on the LSND signal predicts disappearance of reactor neutrinos with 
$\Delta m^2\sim{\rm eV}^2$ very close to the current upper bound 
from the Burgey experiment~\cite{Physics:Malt02}.  

In case of only one sterile neutrino one obtains for the survival 
probability at nuclear reactors
\begin{equation}
P_{\bar\nu_e\to\bar\nu_e}=1-\sin^2(2\theta_{13})\sin^2\left({\Delta m^2_{atm}L
\over 4E}\right)
-\sin^2(2\theta_s)\sin^2\left(\frac{\Delta m_s^2 L}{4E}\right),
\end{equation}
where $\theta_s$ and $\Delta m^2_s = m^2_4-m^2_1$ are the mixing parameters 
of the sterile neutrinos. It is evident from this equation that a sterile 
neutrino would have modification in the measured neutrino flux of the 
RENO experiment if the associated mixing parameter $\sin^2(2\theta_{s})$ 
is not too small and the mass difference is the relevant range.

If $\Delta m^2_s\sim \Delta m^2_{atm}$ and the total rate measurement only 
is available, it will be rather difficult at the RENO experiment to 
disentangle $\bar\nu_e\to\bar\nu_{\mu/\tau}$ oscillations due to 
$\theta_{13}$ from  oscillations due to $\theta_s$. The RENO experiment 
might be able to separate the two oscillation effects if $\Delta m^2_s$ 
differs sufficiently from $\Delta m_{atm}^2$ and/or enough spectral information is available.

If $\Delta m^2_s\gg \Delta m^2_{atm}$ and the oscillations due to $\theta_s$ 
are already averaged out at the near detector, no information on the 
sterile neutrino mixing can be obtained from the comparison of the far 
and near detectors, and the transformation into the sterile neutrinos 
will not affect the $\theta_{13}$ measurement. In this case, the 
information on the sterile neutrino mixing can be obtained from the 
near detector if relatively precise measurement of the initial reactor 
neutrino flux is available, or if a very-near detector at $\sim10$~m 
could be installed.

\subsection{Mass Varying Neutrinos}
The idea of mass varying neutrinos came from a scalar field of acceleron 
associated with the dark energy of the 
universe~\cite{Physics:Hung00,Physics:Gu03,Physics:Fardon04,Physics:Recc05}. Possible 
couplings of acceleron to matter fields could introduce a very different 
feature of neutrino oscillation parameters~\cite{Physics:Kap04}. A possible effect 
due to 
the mass varying neutrinos may be possibly tested in the RENO experiment 
because of different path lengths in air and matter. A different 
parametrization of $\theta_{13}$ and $\Delta m^2_{31}$ for air and matter 
introduces arbitrary oscillation effects different in air and 
matter~\cite{Physics:Schwetz06}. 
Combination of reactor and accelerator neutrino experiments with different 
path lengths in air and matter will give meaningful information on the 
mass-varying neutrinos.

\subsection{Geo-neutrinos}
\begin{figure}
\begin{center}
\includegraphics[angle=0,width=3.5in]
{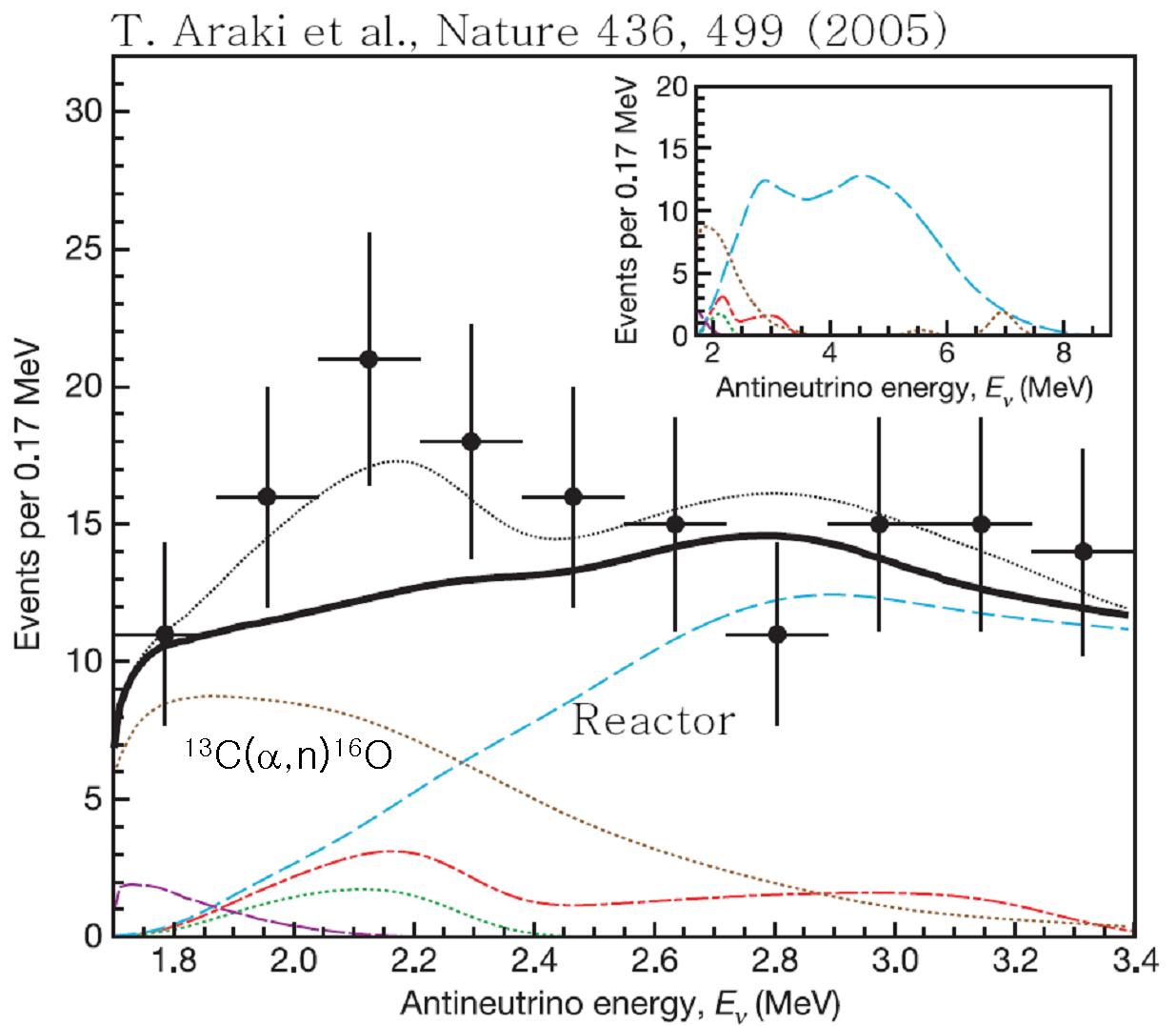}
\end{center}
\caption{KamLAND observed 152 
events in the geo-neutrino energy range from the 749 live days of data. 
The extracted signal events of geo-neutrinos are $25^{+19}_{-18}$ 
while backgrounds are 127 events (82 reactor antineutrinos (dashed light blue line), 
42 fake events from $^{13}$C$(\alpha,n)^{16}$O (dotted brown line), and 3 
random coincidences (dashed purple line)). Figure is taken from
Ref.~\protect\cite{Physics:Araki05}. 
}
\label{geoneutrinos}
\end{figure}

Geo-neutrinos, the antineutrinos from the progenies of U, Th and K decays 
in the Earth, provide the surface information on the content of 
radioactive elements in the whole planet. Their detection can shed 
light on the sources of the terrestrial heat flow, on the present 
composition, and on the origins of the Earth. A recent review on 
geo-neutrinos is found in Ref.~\cite{Physics:Fior07}. The first measurement of 
geo-neutrinos was made by the KamLAND detector in 2005 (see
Fig.~\ref{geoneutrinos})~\cite{Physics:Araki05}. 
KamLAND and Borexino 
detectors are collecting geo-neutrino data, while several planned 
experiments ({\it e.g.} SNO+, LENA, HANOHANO, and EARTH) have 
geo-neutrino measurements among their primary goals.

All experiments, either running or in preparation, use the inverse beta 
decay on free protons as the reaction for geo-neutrino detection. The 
measured shape of neutrino spectrum will be essential for determining 
observation of geo-neutrinos and their radioactive progenitors. The 
RENO detector is not large enough for the sensitive geo-neutrino 
measurement, but may observe some number of geo-neutrinos.

\subsection{Precise Measurement of $\theta_{12}$}
After completing $\theta_{13}$ measurement effort, the RENO experimental scope 
could be extended to the precise measurement of $\theta_{12}$ by adding 
one more detector of a few hundred tons for the target mass at a distance 
$50\sim70$~km from the reactor center. The near and far detectors of 
RENO could be used as near detectors, and thus would reduce relevant 
systematic uncertainties significantly for the new precise measurement
of $\theta_{13}$. 
For baselines longer than 50~km, the reactor antineutrino oscillations 
due to $\Delta m^2_{31}$ average out and the survival probability becomes
\begin{equation}
P=\cos^4\theta_{13}\left[1-\sin^2(2\theta_{12})
\sin\left({\Delta m^2_{21}L\over 4E}\right)\right].
\end{equation}
The oscillations due to $\theta_{12}$ and $\Delta m_{21}^2$ were observed in the KamLAND experiment. 

The optimal baseline for measuring $\theta_{12}$ is about 70~km because 
the antineutrino survival probability becomes minimal for 
$\sin^2\left({\Delta m^2_{21}L\over 4E}\right)\approx 1$, {\it i.e.}
$P\approx 1-\sin^2(2\theta_{12})$  very sensitive to the value 
of $\theta_{12}$~\cite{Physics:Band03}. The extension version of RENO detector is 
expected to improve the error of the $\theta_{12}$ value.

On the other hand, an intermediate baseline detector with 
$L\sim 20-30$~km from the reactor will see the maximum of survival 
probability for $\sin^2\left({\Delta m^2_{21}L\over 4E}\right)$ and 
becomes highly sensitive to the value of 
$\Delta m^2_{21}\ll1$~\cite{Physics:Petcov02,Physics:Chou03}.

\subsection{Reactor Physics}
The RENO near detector will detect an order of one-million neutrino 
events per year, and measure the flux and energy distribution of the 
reactor neutrinos with a greater accuracy than ever. This information 
would lead to meaningful comparison of thermal power and reactor fuel 
loading between measurements and calculations. Such a successful 
comparison will allow us to do a real-time and direct measurement of 
reactor thermal power with a neutrino detector. In addition, a precise 
determination of the reactor neutrino spectrum might be useful for 
reducing the flux uncertainty. There might be an interesting spin-off 
of such a precise measurement of reactor neutrino spectrum and flux, 
to reactor design and operation although it is not quite sure at the 
moment. Another application could be the direct check of nuclear 
non-proliferation treaties.

\subsection{Study of the Directionality}
The near detector of RENO will detect enough statistics of antineutrino 
events, and will allow us to do a detailed exploration of the 
directionality effect of incident reactor neutrinos. The incident 
neutrino direction could be determined from the forward scattering of 
the outgoing neutron in the inverse beta decay reaction. Successive 
multiple scatterings of the neutron introduce a broad distribution of 
their capture location. Therefore, the incoming neutrino direction 
cannot be determined on an event by event basis. However, the neutron 
would have a slightly larger probability to be captured in forward 
direction. This directional correlation between the reactor neutrino 
direction and neutron captured direction may be measured with sufficient 
high statistics and resolution. This directional effect was first seen 
in the Burgey experiment~\cite{Physics:Declais95} and even better in the 
CHOOZ experiment~\cite{Physics:Apollonio03}.

The near detector will record data when some of six reactor cores are 
off, or all on. Comparison of these different data sets, in combination 
with a modeling of the expected event distributions, will allow us to 
understand and test the directionality much better. The directionality 
information could be used for astrophysics, reactor physics, or 
geo-neutrino detection.


\newpage

\chapter{Detector}\label{chapter detector}
\section{Overview}
The RENO experiment uses two identical detectors, 
one located at $\sim 300$~m from the reactor array baseline 
and another at $\sim 1400$~m. 
They are called near detector and far detector, respectively.
By using identical design for both detectors, a number of 
systematic uncertainties cancel out due to normalization
of the neutrino flux at the far detector using that of the 
near detector. 

\begin{figure}
\begin{center}
\includegraphics[width=7in]{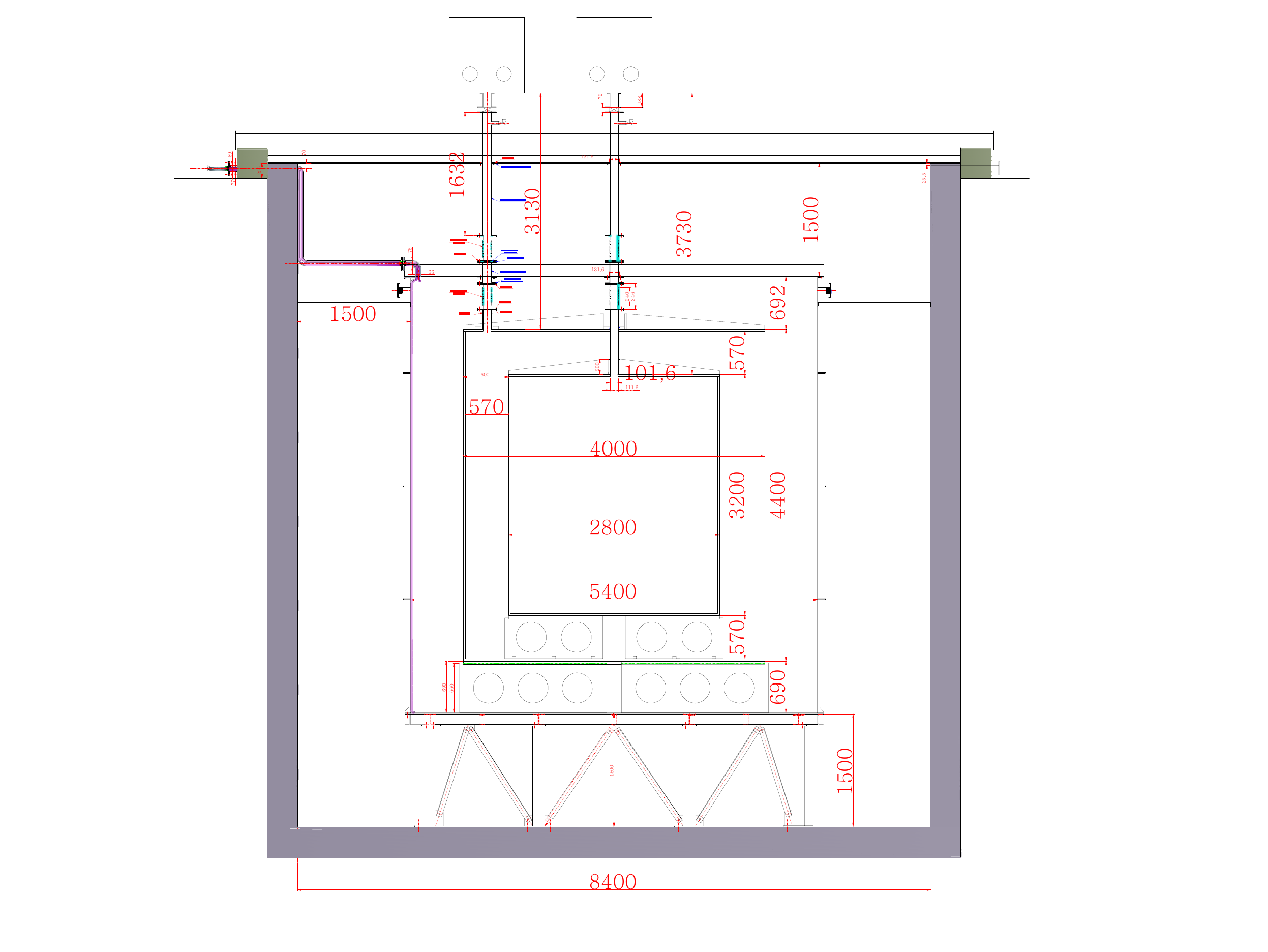}
\end{center}
\caption{RENO detector. From the center, there are liquid scintillator filled target
and gamma catcher with transparent acrylic vessel, mineral oil filled buffer
with stainless steel vessel, and water filled veto layers.
The PMTs for the inner and outer detectors are inwardly mounted on buffer
and veto vessels, respectively. 
} 
\label{detector 2d}
\end{figure}

The detectors have a layered structure similar to other reactor
neutrino experiments,
{\it i.e.} Daya Bay and Double Chooz experiments.
The RENO detectors consist of four cylindrical shape layers. They are, 
from the center, target, $\gamma$-catcher, buffer, and veto, where an 
outer layer almost enclosing an inner layer.  
The PMTs for detecting neutrino interaction will be in the buffer layer. 
The cutaway view of a RENO detector is shown in Fig.~\ref{detector 2d}.

The 
``target'' is Gadolinium (Gd) doped liquid scintillator contained in a 
transparent cylindrical vessel made of acrylic plastic. An inverse
beta decay event produces a positron and neutron pair.
The positron loses energy through scintillating process before 
being
converted into two gammas via a pair annihilation. The neutron
thermalizes, and then is captured by Gd nucleus producing several gammas. 
The gammas produced close to the boundary of target can escape
target without completely depositing its energy in scintillator.
To contain the energy carried by gammas escaping from the target,
the ``$\gamma$-catcher,'' another liquid scintillator layer, surrounds target. 
Unlike the target, the liquid scintillator in the $\gamma$-catcher is not
loaded with Gd since this layer is intended to augment the target
in energy measurement of gammas emitted in target. 
As with target, a transparent cylindrical acrylic vessel contains
$\gamma$-catcher liquid.  

Surrounding the $\gamma$-catcher is a non-scintillating liquid layer called
the ``buffer.'' Mineral oil is used as the buffer and is contained in a cylindrical
vessel made of stainless steel.   
The photomultiplier tubes (PMTs) are mounted on the inner surface of the
buffer vessel immersed in buffer.
The buffer acts as a shield against gammas mainly coming from radioactive
isotope contained in PMTs entering the scintillating volume.  

The outermost layer of the RENO detector is the ``veto,'' a 
water Cerenkov detector layer.
Its purpose is to reduce background gammas and neutrons 
from the surrounding environment (such as rocks) 
as well as cosmic muon induced background events. 
The veto container is constructed with 40~cm thick concrete 
and the top lid is made of stainless steel.
PMTs are mounted on the inner surface of veto container
for detecting Cerenkov light from cosmic muons.

The various design parameters have been determined for optimal
performance using detailed simulation. The simulation includes
background $\gamma$s from PMTs and surrounding rocks, cosmic
muons reaching the detector site as well as inverse beta decay from
the reactor anti-neutrinos.
The detector layers and vessels are summarized in Table~\ref{full detector dimensions}.

\begin{table*}
\begin{tabular}{cccrcrr} \hline\hline
Detector &Outer &Outer &Thickness&Material&Volume&Mass\\
Component   &Diameter(mm) &Height(mm) &(mm)& &(m$^3$)&(tons)\\ \hline
Target                  &2750 &3150 &--   &Gd-loaded LS &18.70  &16.08\\
Target Vessel           &2800 &3200 &25   &Acrylic     &0.99   &1.18\\ 
$\gamma$-catcher        &3940 &4340 &570  &LS          &33.19  &28.55\\
$\gamma$-catcher Vessel &4000 &4400 &30   &Acrylic     &2.38   &2.83\\ 
Buffer                  &5388 &5788 &694  &Oil         &76.46  &64.22\\
Buffer Vessel           &5400 &5800 &6/12*    &SUS         &1.05   &8.39\\ 
Veto                    &8400 &8890 &1500 &Water       &354.84 &354.84\\
\hline\hline
\end{tabular}
\caption[0]{Dimensions of the mechanical structure of
the detector. (*)The buffer vessel thickness is 6~mm for the
top and barrel sections and 12~mm for the bottom section.
} 
\label{full detector dimensions}
\end{table*}

\section{Target and Gamma Catcher}
\subsection{Structure}
The two innermost layers, target and $\gamma$-catcher, will be contained 
in vessels made of acrylic plastic. This acrylic plastic is transparent to 
photons with wavelengths above 400~nm. Two important issues for these layers 
are considered; chemical compatibility between the contents and the vessel, 
and mechanical stability.

As for the chemical compatibility, the liquid scintillating material
for both the target and $\gamma$-catcher should not chemically interact with the
vessel for the duration of the experiment. At the same time, the $\gamma$-catcher
vessel should be chemically inert to the mineral oil in buffer layer.
There have already been extensive studies on chemical compatibility 
of these materials for CHOOZ experiment and others. The RENO
collaboration has conducted various R\&D on the chemical 
interaction of acrylic plastic and other materials used in the 
experiment. The compatibility test results are in 
Chapter~\ref{Scintillator Chapter}.

Mechanically, these vessels are required to withstand the mechanical
stresses that they are subjected to during the all phases
of the experiment and maintain structural integrity. 
When loaded with liquids, the volume of the vessels 
can change slightly from the nominal volume. This change
should be within specified tolerances.

The target vessel an acrylic cylinder with a height of 3.2~m, a diameter of 2.8~m,
and a thickness of 25~mm.
The mass of the target vessel is 1.2~tons. 
The target vessel has an inner volume of 18.7~m$^3$ and the combined mass of
the target liquid and vessel is $\sim 17$~tons.
Inside the $\gamma$-catcher vessel, the target vessel is mounted on the 
supporting structure made of the same acrylic plastic.
When both target and $\gamma$-catcher are filled, the net load on the target
supporting structure will be $\sim 300$~kg including buoyancy. 
At the center of the top of the vessel, a pipe connects the
target volume to the outside of the detector for filling target liquid 
and inserting calibration sources.

The design of the $\gamma$-catcher is similar to that of the target but
about three times larger in volume.
The $\gamma$-catcher vessel is an acrylic cylinder with a height 
of 4.4~m and a diameter of 4.0~m. Its wall is 3~cm thick. 
The $\gamma$-catcher vessel is mounted on the supporting
structure made of acrylic plastic and placed inside the buffer vessel.
It has a pipe connecting the top of the $\gamma$-catcher vessel and the
outside of the detector for liquid filling and calibration source insertion.
The $\gamma$-catcher vessel has a mass of $\sim 2.8$~tons. The combined mass
of the $\gamma$-catcher vessel and $\gamma$-catcher liquid scintillator is 
31.4~tons. When the $\gamma$-catcher is immersed in buffer liquid, the total 
load on the $\gamma$-catcher supporting structure would be 2.2~tons.

\begin{figure}
\begin{center}
\includegraphics[width=4.5in]{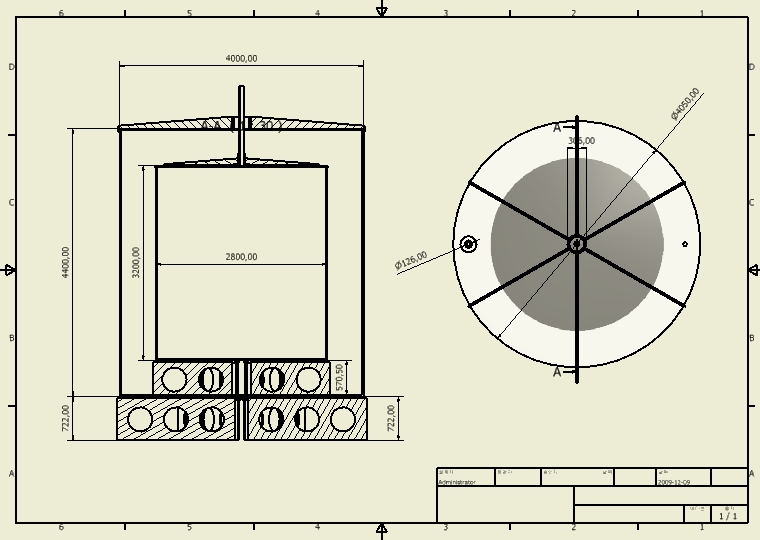}
\end{center}
\caption{Target and gamma catcher vessels with supporting structures. 
The supporting structures are 
made with the same transparent acrylic plastic as the vessels.}
\label{target catcher 3d transparent}
\end{figure}

\subsection{Acrylic Vessels}
The target and $\gamma$-catcher vessels are made of transparent acrylic 
plastic called Polymethyl-methacrylate (PMMA).
The molecular formula of PMMA is $(\hbox{C}_{5}\hbox{O}_{2}\hbox{H}_{8})_n$.
PMMA can be found under trade names like Plexiglas, R-Cast, and Lucite.
The properties of PMMA are shown in Table~\ref{acrylicproperties}. 
With additional ingredients in PMMA the UV below 400~nm is absorbed.

The target and $\gamma$-catcher vessels are made from cast acrylic 
sheets (Plexiglass, GS-233) supplied by Degussa GmbH, Germany. 
The cast acrylic sheet has better mechanical and chemical properties 
than the extruded acrylic sheet. The production of the vessels is 
done by KOA Tech in Korea. 
These vessels are manufactured in several pieces for ease of 
manufacture and will be assembled mostly at the manufacturing site. 
The vessel parts will be bonded by polymerization and the joined
sections will be treated with annealing process.

The manufacturing precision of the vessels will be 0.1\% in volume (2~mm in
1 dimension), therefore, 0.14\% difference in target vessel volume between 
near and far detector could incur. 
This difference could be measured and corrected by mass flow meter
and weight measurement.

\begin{table*}
\begin{center}
\begin{tabular}{cc} \hline
Properties       &  Value \\ \hline
Density          & 1.19 g/cm$^3$ \\
Melting point    & 130-140 $^{o}$C \\
Refractive index &  1.491  \\
Transmittance       & 92\%   \\\hline
\end{tabular}
\end{center}
\caption[0]{The mechanical and optical properties of cast acrylic, such as 
Plexiglas GS-233 from Degussa GmbH, Germany and R-Cast from Reynolds Co., USA.}
\label{acrylicproperties}
\end{table*}

\subsection{Chimney}
Each target and $\gamma$-catcher has a chimney for filling liquids
and transporting calibration sources from the top lid of the veto
vessel into and out of either target or $\gamma$-catcher. 
A chimney is made of about 4-inch transparent acrylic tubing with a 
flexible convoluted PTFE tube connecting the buffer vessel and the
acrylic tubing to ease the stress. From the top lid of the buffer
the chimney is made of stainless steel pipes extending to the top lid of the
veto vessel. 

\section{Buffer}
The buffer vessel is a stainless steel cylinder of 5.8~m height and 5.4~m diameter
containing target, $\gamma$-catcher, and buffer liquid. 
The buffer contains non-scintillating oil to shield the scintillating
volume within from background sources outside, including radioactivity
in PMTs. 
The buffer vessel also acts as the PMT mounting surface where 
354 PMTs are mounted pointing inward and optically isolates 
these PMTs from the veto volume.
The size of the buffer vessel has been determined from the 
MC simulations. 

\begin{figure}
\begin{center}
\includegraphics[width=4.5in]{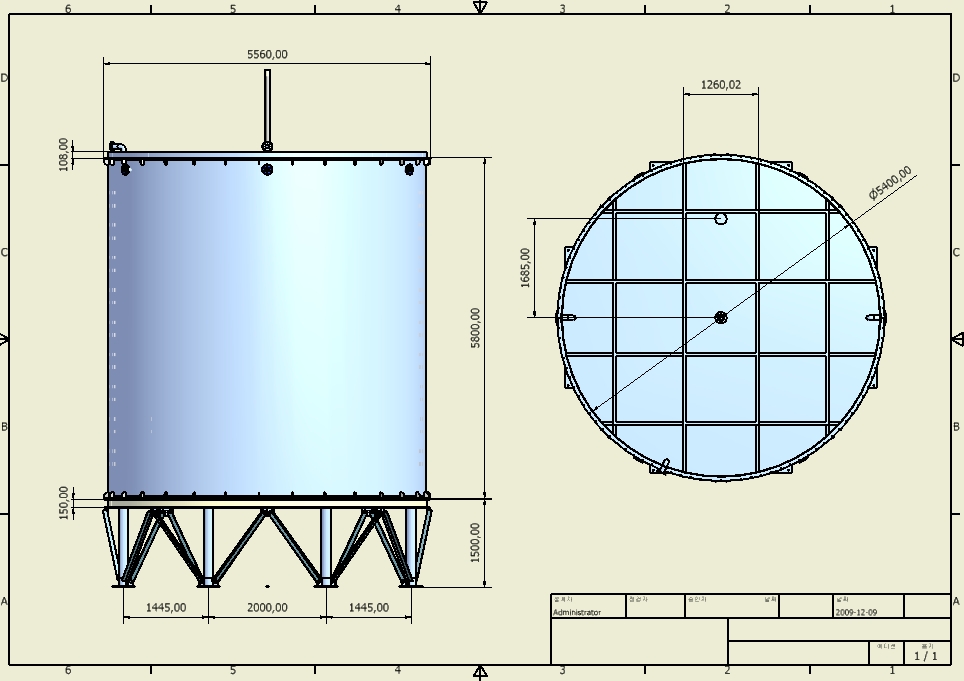}
\end{center}
\caption{External view of the buffer vessel. The vessel is made of 
stainless steel (304L) and the supporting truss structure is 
made of nickel--plated steel pipes and rods.}
\label{buffer 3d}
\end{figure}

The buffer vessel should be chemically inert against mineral oil inside 
and water outside. Also, it should withstand the stress from the loads 
coming from liquids and structures contained within. 
The buffer vessel is constructed with 6~mm thick 304L stainless steel for
the top lid and barrel section, 12~mm thick for the bottom plate for
extra mechanical support. The external view is shown in Fig.~\ref{buffer 3d}.
The surface of the vessel is not polished.
When the detector is filled with required liquids, the buffer vessel will
experience buoyant force due to the density difference between the organic
liquids inside the buffer vessel and water in veto layer. The buoyant force
is estimated to be 11.5~tons and the buffer vessel supporting structure 
is designed to counter this force.

The buffer vessels are manufactured by Nivak co., Korea.
They are transported as segmented pieces to the experiment site and assembled
in the experimental halls. The barrel section consists of eight 
segments and top and bottom plates each consist of three 
segments. The bottom plate is welded to the barrel section 
and the top plate is bolted to the barrel section.
 
\begin{figure}
\begin{center}
\includegraphics[width=4.5in]{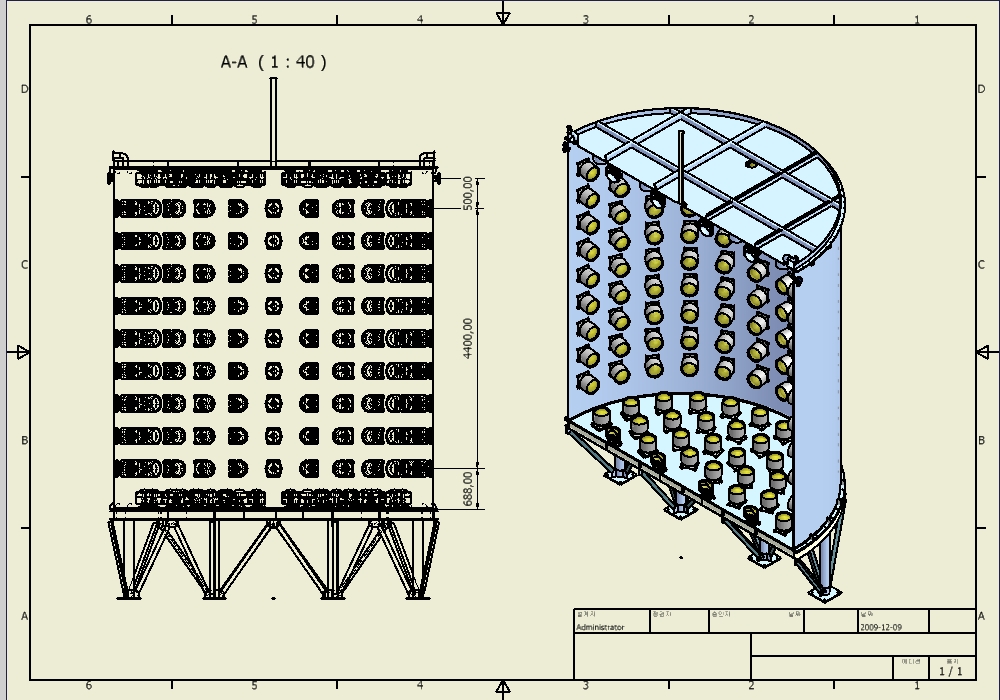}
\end{center}
\caption{Inner detector PMT array on the buffer vessel. 
The PMTs are mounted on the vessel wall using the PMT holders.
There are total 354 10-inch PMTs.
}
\label{pmt-holder scheme inner}
\end{figure}

There are total 354 10-inch PMTs mounted on the inner side walls 
of the buffer vessel, 234 PMTs mounted on the barrel section, 60 
each on the top and bottom plates,
as shown in Fig.~\ref{pmt-holder scheme inner}. 
PMTs will be mounted upright on the walls using the PMT 
holding structure described in Sect.~\ref{PMT Holder}.

\section{Veto System}
\subsection{Design Criteria}
The veto system is located just outside of the buffer tank. The main background 
of the experiment is caused by cosmic muons, and it is very important to identify 
the entering muons since they can produce neutrons by muon-nucleus interaction 
in the detector.
There are also correlated backgrounds by $^{9}$Li and $^{8}$He in the target and $\gamma$-catcher
produced by muons. 
Although the veto system will not be 
included in the trigger, the muon signals in the veto system will be used to identify
muon related background events for each candidate event from neutrino interaction.
The veto vessel should be chemically compatible with water and strong enough to
support all the three inner chambers before filling the liquids. 

The rate of inverse beta neutrino events is about 
$1~\rm{events}\cdot\rm{km}^{2}/(\rm{GW}\cdot\rm{ton}\cdot\rm{day})$ for
oil based liquid scintillator. One of the main backgrounds of RENO experiment 
will be fast neutrons which are generated in the surrounding rock and enter 
into the detector volume. We require that the fast neutron rate and
the accidental coincidence rate from single rates of $e^+$ signal 
$(1~{\rm MeV} < E < 8~{\rm MeV})$ and neutron signal 
$(6~{\rm MeV}<E< 10~{\rm MeV})$ are less than 1\% of the inverse beta
neutrino rate. The accidental rate from two uncorrelated single rates of 
$R_{1}$ and $R_{2}$ with a coincidence time window of $\Delta T$ is 
$R_{accidental}$ = $R_{1}R_{2}\Delta T$.  The veto system should shield 
effectively the ambient gammas and also reduce the muon related background 
events by imposing an offline veto timing cut after each muon passing
through the detector system. The results of Monte-Carlo simulations are 
described in Chapter~\ref{Simulation Chapter}. As a result, we set 
the thickness of the veto layer to be 1.5~m.

\subsection{Structure}
\begin{figure}
\begin{center}
\includegraphics[width=9cm]{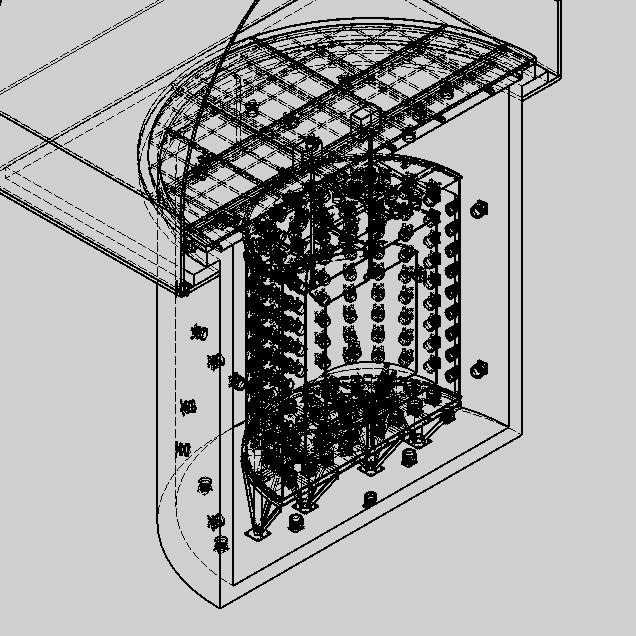}
\end{center}
\caption{Transparent view of PMT arrays showing both the inner and outer PMTs.}
\label{pmt-holder scheme outer}
\end{figure}

The inner diameter and height of the veto vessel are 8.4~m and 8.8~m, 
respectively. The vessel is constructed with a 40~cm thick concrete vessel. 
Inner surface of the concrete vessel was water-proofed with epoxy resin.
The water will be purified at filling and continuously be circulated through 
a water purification system.
There are 67 10-inch water-proof PMTs (R7081 Hamamatsu) attached on the inner 
surface of the veto vessel. The outer surface of buffer vessel and the inner 
surface of veto vessel will be coated with TiO$_2$ paint to increase the light 
collection of Cerenkov photons in the water. The whole PMT arrangements of 
both buffer and veto vessels are shown in Fig.~\ref{pmt-holder scheme outer}.

\section{Detector Assembly}
The construction of the acrylic vessels, target and $\gamma$-catcher vessels, 
will be done at the manufacturer's facility but integrated into the detector 
system in the detector pit at the experiment site.  
The assembly procedure is as follows.
\begin{enumerate}
\item The target vessel with complete supporting structures and a chimney 
is assembled at the manufacturer's site.  The $\gamma$-catcher vessel is
also assembled at the manufacturer's facility. The lid and chimneys are 
for the $\gamma$-catcher vessel are not attached here. 
\item The buffer vessel is constructed at the experiment hall pit. 
Shaping of the parts is done on site.
\item The $\gamma$-catcher and target vessels are transported to the
experimental hall and integrated into the detector. 
The $\gamma$-catcher vessel is closed with a lid and chimneys are 
installed.
\item PMTs for the bottom and barrel sections are installed on both buffer
and veto vessels.
\item PMTs are mounted on the buffer lid and the lid will be installed.
\item PMTs are mounted on the veto lid and the lid will be installed.
\end{enumerate}

The detector construction will be done at the detector pit and 
in the service tunnel adjacent to it. 
The detector assembly area will be made a clean room of class 10\,000.
Environmental control units will keep the constant temperature and 
humidity. 
Near and far detectors will be constructed concurrently
to reduce potential differences in the assembly procedure.
After completing the assembly, all vessels will be filled simultaneously 
maintaining the differences in liquid levels within a few centimeter.

\section{PMT and HV System}
\subsection{PMT Requirements and Specification}
The scintillation lights from target and $\gamma$-catcher 
will be detected with PMTs attached on the inner surface of the 
buffer vessel.
The number of detected photoelectrons 
is estimated to be 150 photoelectrons per MeV for an event
occurring at the center of the target.
Since the minimum energy deposited in the detector by a positron 
emitted in the inverse beta decay is 1.022~MeV, the average 
number of photoelectron per PMT in the buffer layer will be about 0.5.
Therefore, the PMTs should be able to measure the single photoelectrons 
with high efficiency. The peak-to-valley ratio and the single 
photoelectron resolution of the PMTs are important parameters.

The main reason 
for having a non-scintillating buffer region is 
to shield the $\gamma$-catcher and the target from PMT's radioactivity. 
The radioactivity of PMTs needs to be studied to understand the 
rate of background originating from PMTs.
The PMT background events are mainly in the low energy region of
less than 2~MeV and could be misidentified as signals by accidental 
coincidence with neutron-like background events.

Since the PMTs will be immersed in a layer of mineral oil, 
it is also important that the whole PMT assembly is chemically 
inert to mineral oil. The oil proofing should be stable for 
the duration of the experiment.

The quantum efficiency of each PMT will be measured
We will measure the quantum efficiencies of all PMTs with a 
relative accuracy less than 5\%.
The outlying PMTs will be excluded from installation
in the detectors.

\begin{table}
\begin{center}
\begin{tabular}{ccccc}\hline

	& R9512	& R7081	 & XP1806	& XP1804 \\ \hline
Gain($\times 10^{7}$)	& 1.0 @1500~V	& 1.0 @ 1500~V	& 1.0 @1600~V	& 1.0 @1600~V \\
QE @ peak (nm)	& 22\% @390	& 25\% @390	& 24\% @420	& 24\% @420 \\
DC (nA)	        & 50	& 50	& 30	& 30 \\
Size (inch)	& 8 	& 10	& 8	& 10.6 \\
Weight (g)	& 720	& 1150	& 880	& 1744 \\
Rise Time (ns)	& 3.8	& 4.3	& 5	& 5 \\
TTS (ns)	& 2.4	& 2.9	& 2.4	& 2.4 \\
Afterpulse	& 2\%	& 2\%	& 12\%	& 12\%  \\
Peak-to-valley ratio	& ~4	& ~3.5	& ~3.5	& ~3.5 \\
\hline
\end{tabular}
\caption{The specifications and measurements of the candidate PMTs.
R5912 and R7081 are from Hamamatsu and XP1806 and XP1804 are from 
Photonis.
}
\label{pmtspec}
\end{center}
\end{table}
 
\subsection{Tests on PMT performance}
We have tested four candidate PMTs for the RENO experiment;
R5912 (8\inch) and R7081 (10\inch) from Hamamatsu, and XP1806 
(8\inch) and XP1804 (10\inch) from Photonis. 
Table~\ref{pmtspec} shows the company specifications and our 
measurements for the PMTs considered. 

\subsubsection{Single photoelectron measurement}
\begin{figure}
\begin{center}
\includegraphics[width=7cm]{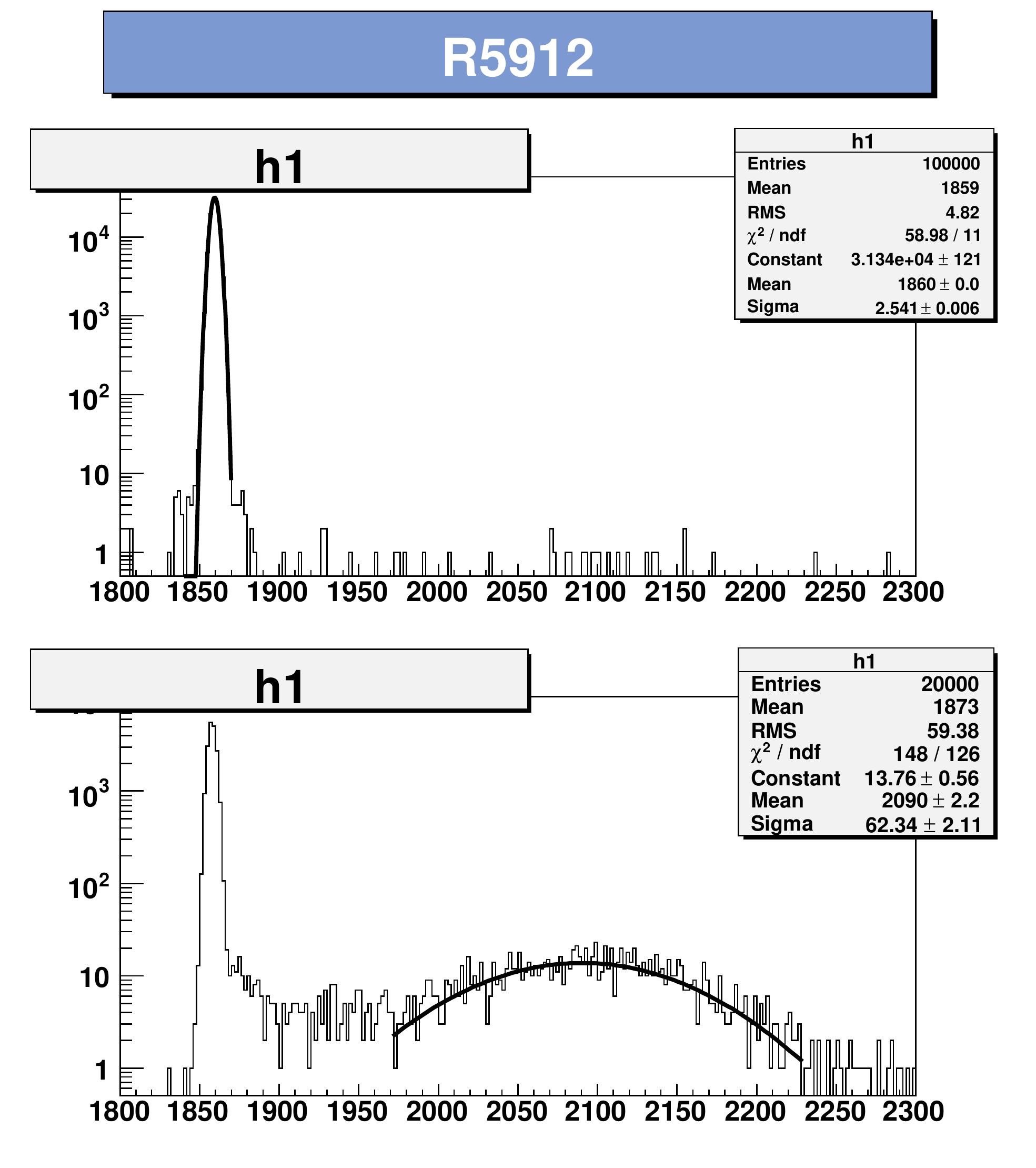}
\includegraphics[width=7cm]{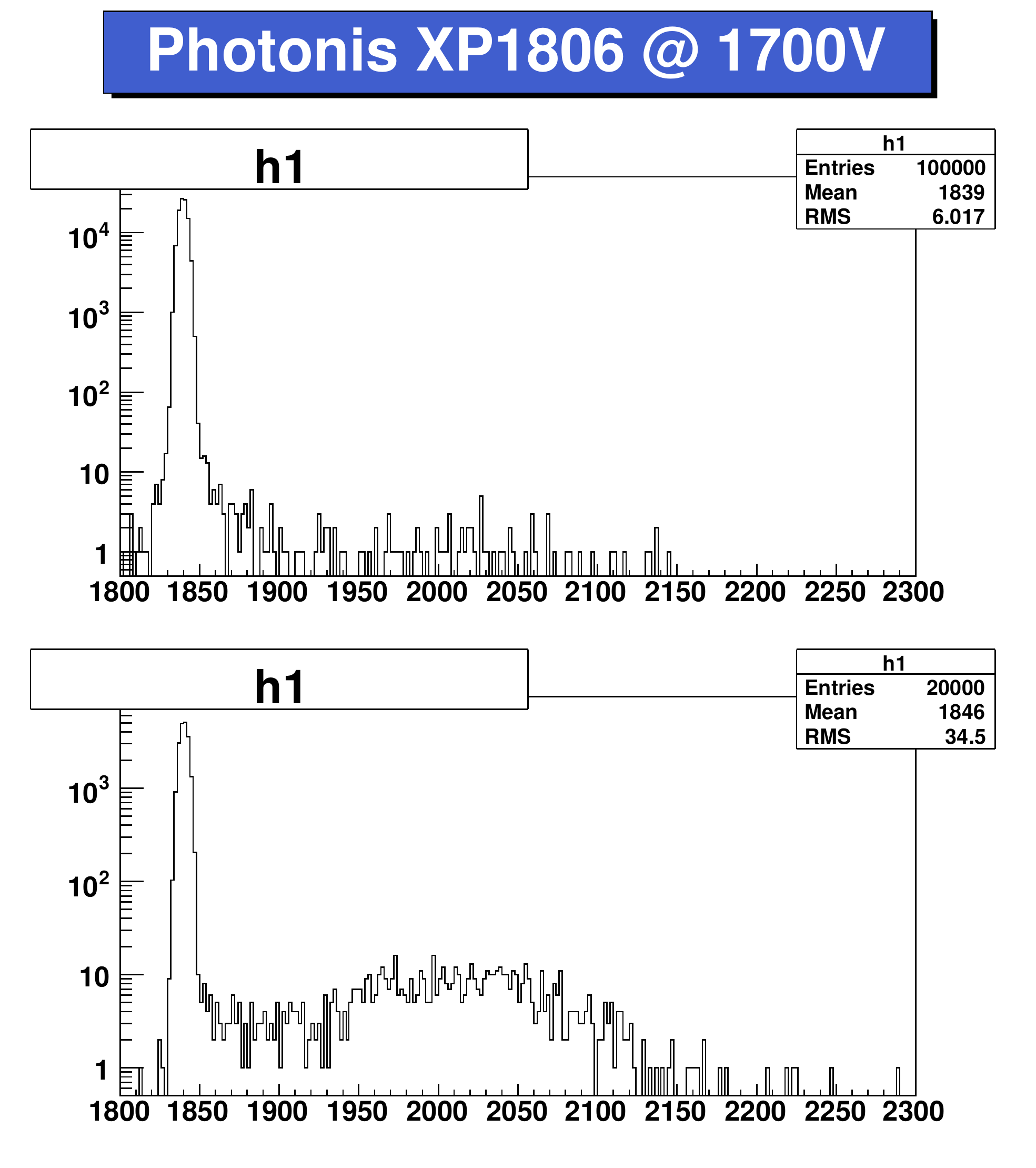}
\end{center}
\caption{Single photoelectron spectra of 8-inch PMTs from Hamamatsu 
Co.
and Photonis Co. Upper (lower) figure is obtained with LED off (on) for 
each PMT. X-axis is in ADC channel number.}
\label{singlephotoelectron}
\end{figure}

The single photoelectron spectra were measured 
with an LED flash system. 
The single photoelectron spectra of R5912 and XP1806 are shown in 
Fig.~\ref{singlephotoelectron}. 
The width-to-peak ratio
of the single photoelectron measured with R5912 was 3.7. 
The operating voltages of R5912 and XP1806 were set to have
the same gain of $1 \times 10^{7}$ for the measurement.

\subsubsection{Background Measurements}
The radioactive isotopes U, Th, and K in PMTs are 
most serious sources of the backgrounds. 
PMTs are located in the buffer filled with non-scintillating oil.
The buffer layer thickness should be enough 
to reduce the single rates of gammas with energy over 1~MeV in the 
target and $\gamma$-catcher region to less than 20~Hz.
The radioactivity of PMT is directly measured at the underground 
experimental facility of DMRC (Dark Matter Research Center) at  
Yangyang, Korea. A sample PMT was located on top of a low-background 
HPGe detector and the gamma rates from the whole PMT were measured. 
Even though the loci of radioactivity on PMTs are not measured, 
the rate of backgrounds coming from PMT can be reasonably estimated
since the distance between PMTs and the outer boundary of $\gamma$-catcher
is comparable to the size of PMTs (see 
Sect.~\ref{sect: radioactive backgrounds}).
{\sc geant4} simulation is used to get the
efficiency of the HPGe detector and a secular equilibrium is assumed 
for U and Th activities. The measured activities for the PMTs tested are 
summarized in the Table~\ref{pmtbackground}.
\begin{table}
\begin{center}
\begin{tabular}{cccc}\hline
PMT               &   $^{238}U$    &   $^{232}Th$    &  $^{40}K$      \\ \hline
R5912(Hamamatsu)  & 1.2 $\pm$ 0.04  & 0.74 $\pm$ 0.02   &  3.0 $\pm$ 0.2  \\
R7801(Hamamatsu)  & 4.8 $\pm$ 0.07  & 2.2 $\pm$ 0.03   &  13.1 $\pm$ 0.4  \\
R7801(Hamamatsu)-Low  & 0.72 $\pm$ 0.1  & 0.59 $\pm$ 0.07   &  3.3 $\pm$ 0.3  \\
XP1806(Photonis)  & 2.5 $\pm$ 0.05  & 0.35 $\pm$ 0.01   &  5.2 $\pm$ 0.1  \\
XP1804(Photonis)  & 5.9 $\pm$ 0.2  & 0.63 $\pm$ 0.06   &  9.4 $\pm$ 0.2  \\ \hline 

\end{tabular}
\caption{The radioactivity of candidate 8- and 10-inch PMTs from Hamamatsu and
Photonis. The units are Bq/PMT.}
\label{pmtbackground}
\end{center}
\end{table}

The activity of $^{238}$U in XP1806 PMT is about a factor of two 
higher than the Hamamatsu R5912 PMT. However, $^{232}$Th 
activity in XP1806 is lower than that in R5912. 
If the PMTs tested are used in RENO detector,
the rate of single events with energy over 1~MeV is expected to be 
20--50~Hz.
PMTs with lower radioactivity are available, and the expected 
single event rate will be about 5--10~Hz.

To reduce the background from PMTs, Hamamatsu makes 10-inch PMTs with 
low background glass (R7081-Low). We measured radioactivities on two 
R7081-Low PMT samples.
The background level of R7081-Low is found to be about 25\% of that 
of the normal glass R7081 PMTs. 
This value is higher than the values given by Hamamatsu, but
the measurements were performed over the whole PMT including 
base and cable. Also it was assumed that the spatial distribution
of radioactivity was uniform in the glass part, so the true value 
of radioactivity is questionable. However, the ratio between normal 
glass and low background glass would be still valid.
The measured level of radioactivity of low background glass PMTs 
would be acceptable for RENO experiment.

\subsubsection{Gain Drift and Afterpulse}
The nominal gain of PMTs tested is $\sim 10^7$. However, the gain
is temperature dependent and is measured to increase by about
0.2\% per degree for the samples tested. 
The temperature will be monitored with a precision better than 
0.5$^\circ$C, so the gain error will be an order of 0.1\%. This will 
give a negligible systematic uncertainty in the energy resolution and 
neutrino detection efficiency.

\begin{figure}
\begin{center}
\includegraphics[width=9cm]{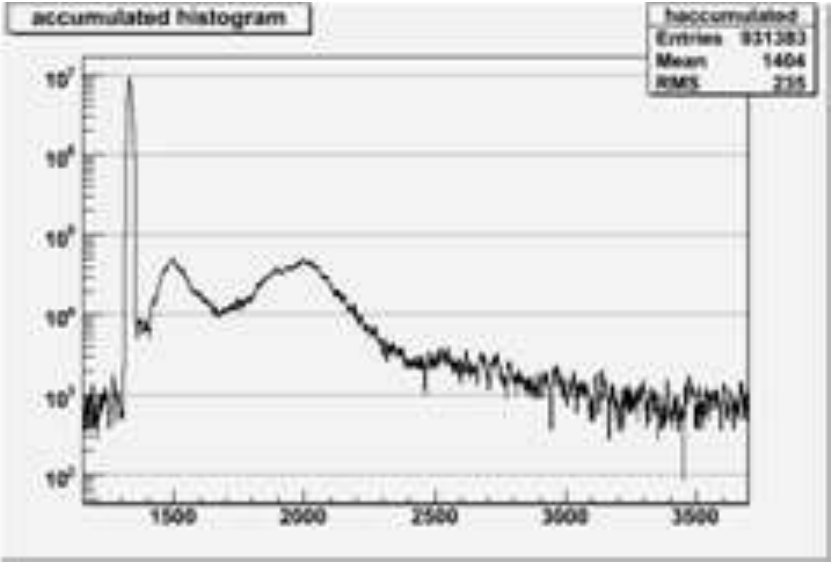}
\end{center}
\caption{Afterpulse measurements of R5912 PMT. Two peaks occur at 
2 and 7~$\mu$s after the primary pulse. The unit of x axis is 10 ns.
}
\label{afterpulse}
\end{figure}

Afterpulse of 8- and 10-inch PMTs occurs at several $\mu$s 
after the primary pulse and its size is 1--10\% of the primary pulse
depending on the quality of individual PMTs. All the tested PMTs show 
two afterpulse peaks at 2 and 7 $\mu$s after the primary pulse
as shown in Fig.~\ref{afterpulse}.
The amount of afterpulse is 4--5\% for Hamamatsu 8- and 10-inch PMTs, 
which is a factor of 2 higher than the company specification. 

Even though the afterpulse has timing characteristics similar to the 
inverse beta decay events, the energy distributions between the primary and 
secondary pulses are quite different. For high energy muon events 
the size of the afterpulse could amount to a few 
MeV, while its primary pulse should have a huge signal in the same PMT. 
However, the occurrence of such an event is expected to be a few percent
of the total high energy muon events, thus
will not cause any significant impact on the experiment.

After considering various performance parameters, such as single 
photoelectron resolution, afterpulse rate, radioactivity in PMT, and 
overall detector performance to cost ratio, we 
decided to use 10-inch low background PMTs by Hamamatsu (R7081-Low) 
for the RENO experiment. 

\subsubsection{PMT base}
\begin{figure}[h]
\centering
\includegraphics[width=17.0cm]{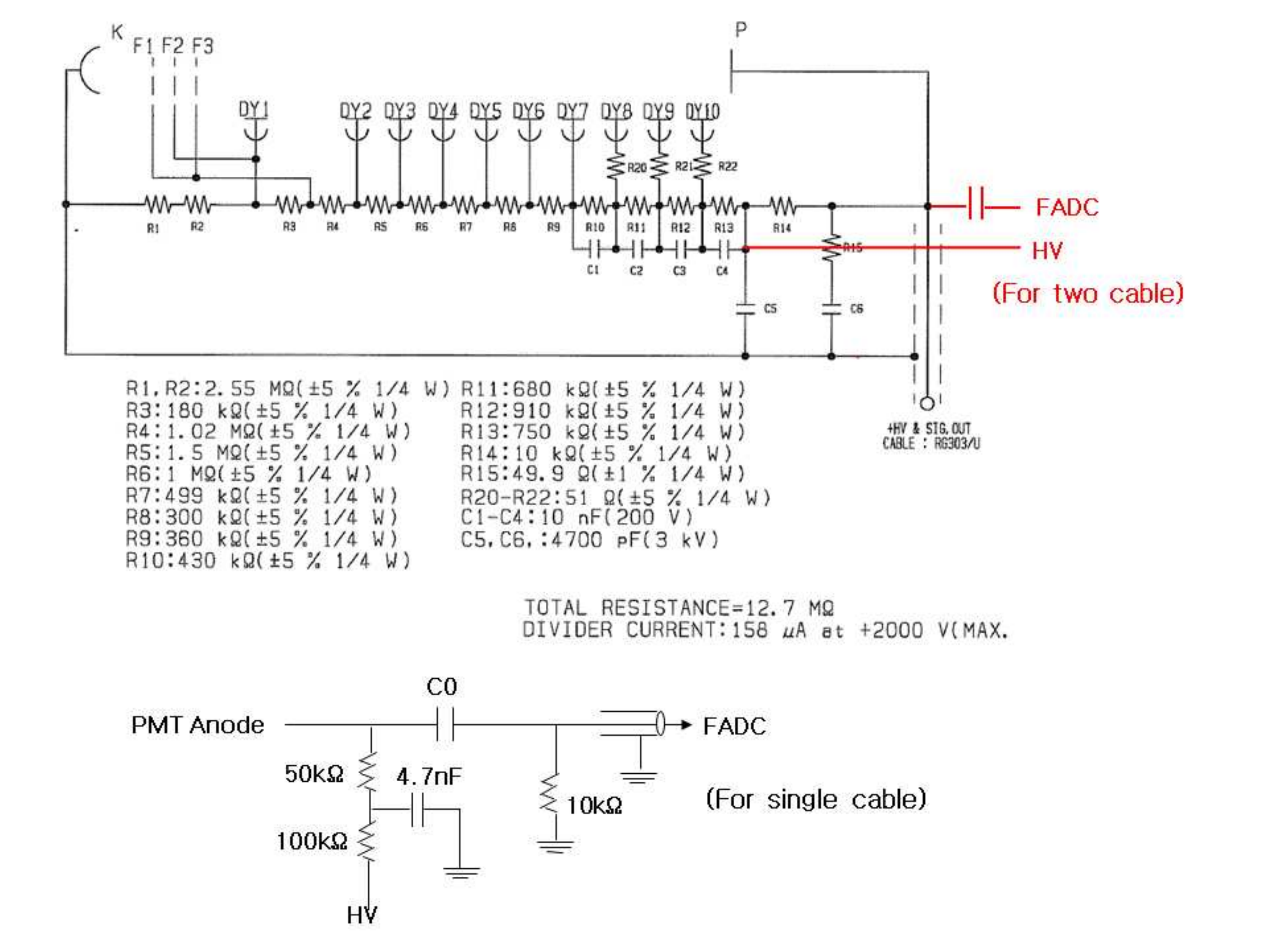}
\caption{Single-cable (HV and signal via a single cable) and two-cable 
(HV and signal in separate cables) base configurations 
for the 12.7~M$\Omega$ base. 
Top figure is the decoupling circuit for two cable configuration with
cables shown in red lines.}
\label{basedrawing}
\end{figure}

We have tested different base configurations with 
a sample R7081 PMT. The bases were home-made with Hamamatsu sockets 
following the voltage division given Hamamatsu. The bases also 
included back termination.
First, we compared single-cable (HV and signal in a single cable) 
and two-cable (HV and signal in separate cables) base configurations.
Second, two voltage divider configurations, $9.4~{\rm M}\Omega$ 
and 
$12.7~{\rm M}\Omega$ 
were tested.
For all four base-cable configurations, the PMT signal gain was set 
at $1.5 \times 10^{7}$. The decoupling schemes for both single and two 
cable cases are shown in Fig.~\ref{basedrawing}.

\subsubsection{SPE spectra}
\begin{figure}[h]
\centering
\includegraphics[width=12.0cm]{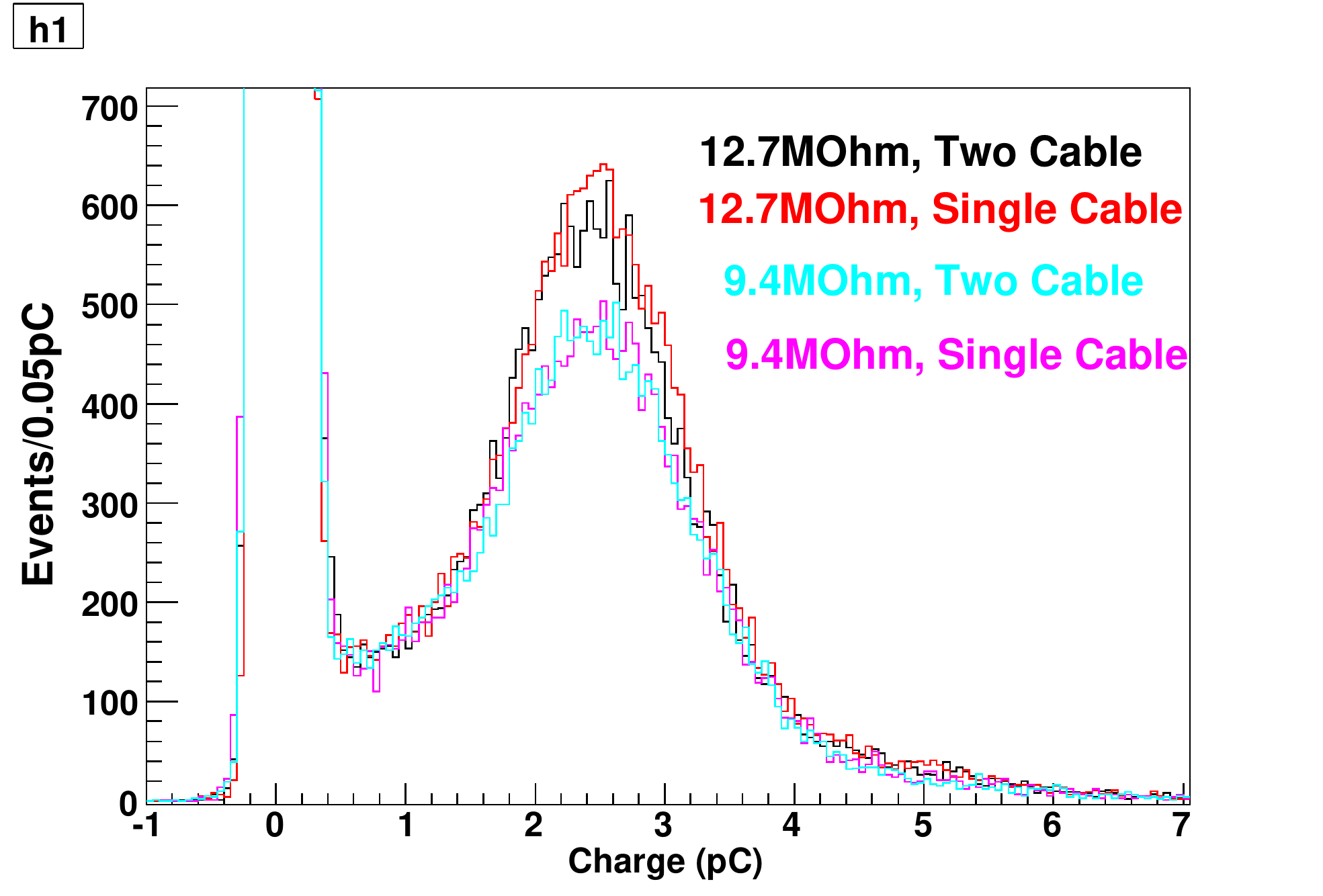}
\caption{Single photoelectron spectra for four base-cable configurations.}
\label{spe}
\end{figure}

Single photoelectron spectra for all four base-cable configurations are 
shown in Fig.~\ref{spe}. There is little difference among four 
configurations. The 12.7~M$\Omega$ base appears to have
a little higher quantum efficiency, maybe due to the efficient electron 
collection from photocathode. 
The 9.4~M$\Omega$ base has a higher gain than the 12.7~M$\Omega$ base,
consistent with data provided with Hamamatsu.

\subsection{PMT Holder}\label{PMT Holder}
\begin{figure}
\begin{center}
\includegraphics[width=10cm]{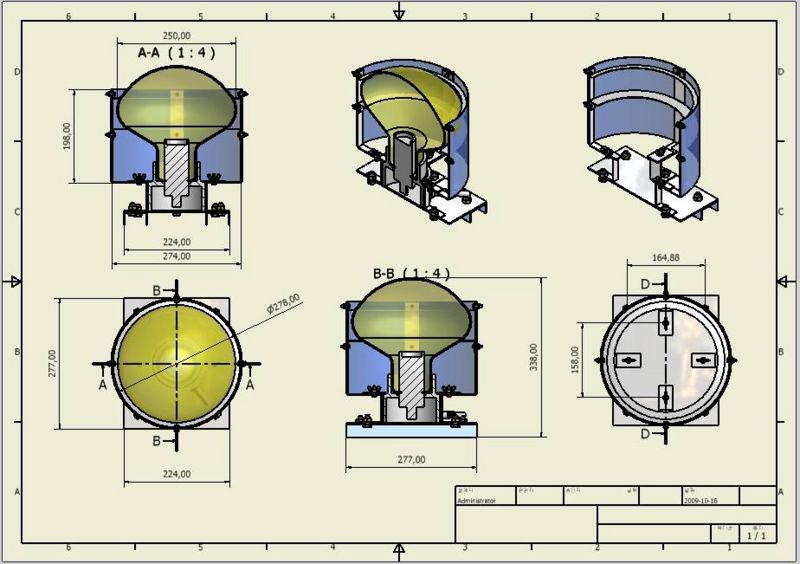}
\end{center}
\caption{A design of PMT holder. Two stainless steel rims hold the glass 
of 10-inch PMT. A cylinder shape mu metal sheet will be outside of the 
rings surrounding the individual PMT to reduce the effect of the magnetic 
field.}
\label{pmt-holder}
\end{figure}

PMTs will be mounted on the inner wall of stainless steel buffer vessel. 
We want to minimize the amount of material while 
keeping the holding structure as stable as possible. 
Also the distance between PMT photocathode surface and 
buffer vessel should be minimized.
The PMT holder is made of 1.5$\sim$2~mm thick stainless steel. The
schematic is shown in Fig.~\ref{pmt-holder}.
Two rings hold the glass bulb section of the PMT and the front ring
defines the photosensitive area. The inner diameter of the rings is 
12.3~cm. A mu-metal sheet surrounds the side of the 
structure to reduce the effects of external magnetic fields. 
The height of the mu-metal shielding will be determined based on the
magnetic field survey at the experiment halls.

\subsection{HV system}
\begin{figure}
\centering
\includegraphics[angle=0,width=13.0cm]{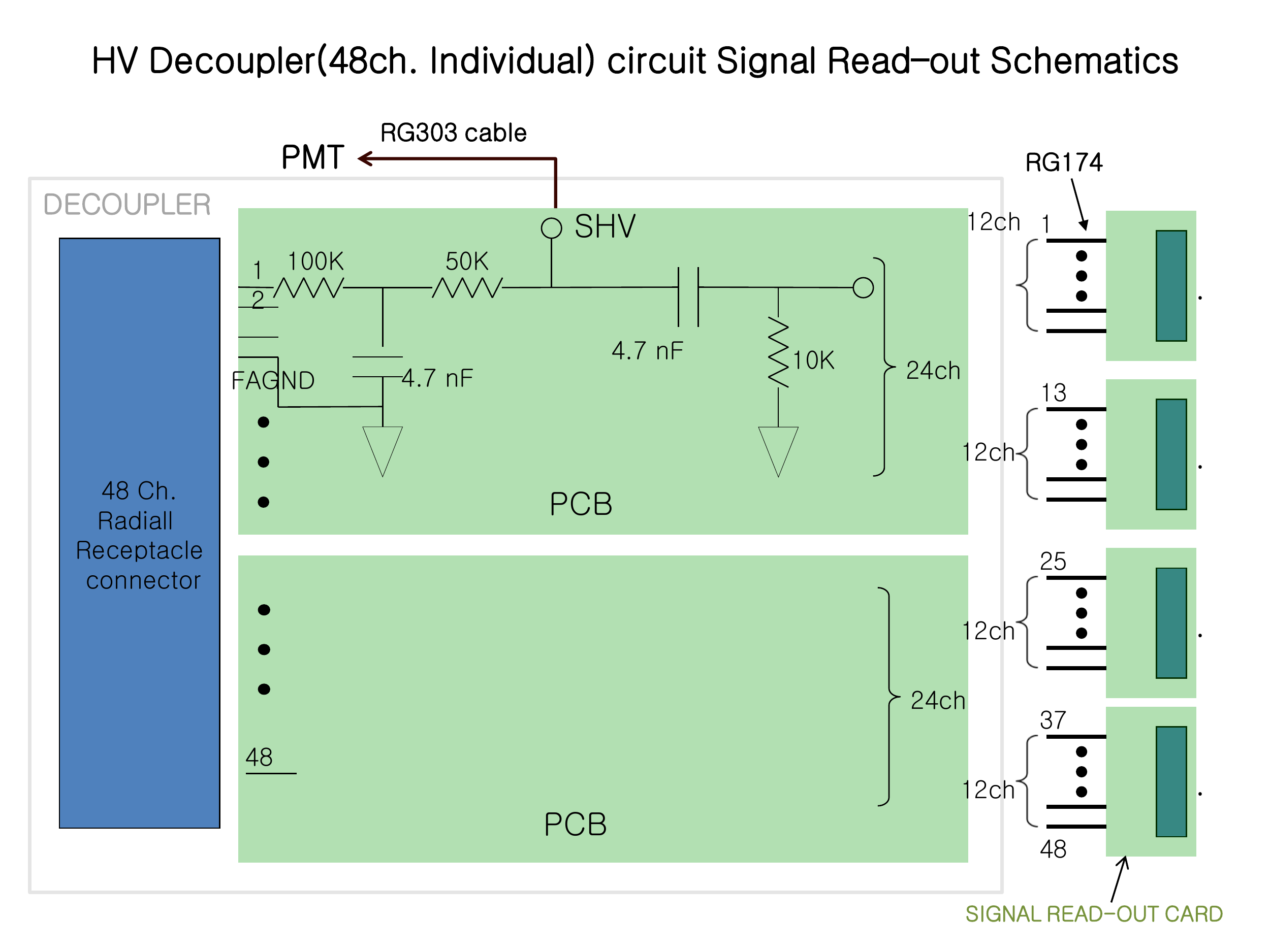}
\caption{A schematic of 48 channel decoupler with a 48 channel HV power 
supplier module A1932P.}
\label{decoupler}
\end{figure}

The PMTs are biased in such a way that the anode is connected to a 
positive high voltage and the cathode to ground. This prevents current 
leakage caused by discharges at the PMT glass.
A schematic of the voltage divider of the PMT is shown in Fig.~\ref{decoupler}.
A single cable (RG-303) 
will be used for carrying both high voltage and signal. 
A capacitor will decouple the high voltage and pulse signal from 
anode as shown in the schematic. 
The voltage difference between the photocathode and the first dynode is 
622~V for an applied voltage of 1550~V.
A 50~$\Omega$ termination will
be in the PMT base side to match the impedance and the signal will be 
attenuated by a factor of two compared to the two cable case where
signal is carried by a separate cable. Therefore, a higher voltage 
will be applied to compensate for the signal amplitude reduction. 
The cable length is 25 meter, identical for all the PMTs mounted in 
the buffer vessel.

For the high voltage module, two models of high density modules, A1932A 
(48 channel, 0.5~mA/channel, 8 channel/group) and A1535 (24 channel, 
1~mA/channel), are being considered. For both modules, the schematic shown 
in the Fig.~\ref{decoupler} can be applied. For A1535 modules, one 
HV channel will be used to bias four PMTs.
This channel splitting scheme has been tested on the mock-up detector 
and no noticeable cross talk or increased noise level were
observed.

\begin{figure}
\centering
\includegraphics[angle=0,width=13.0cm]{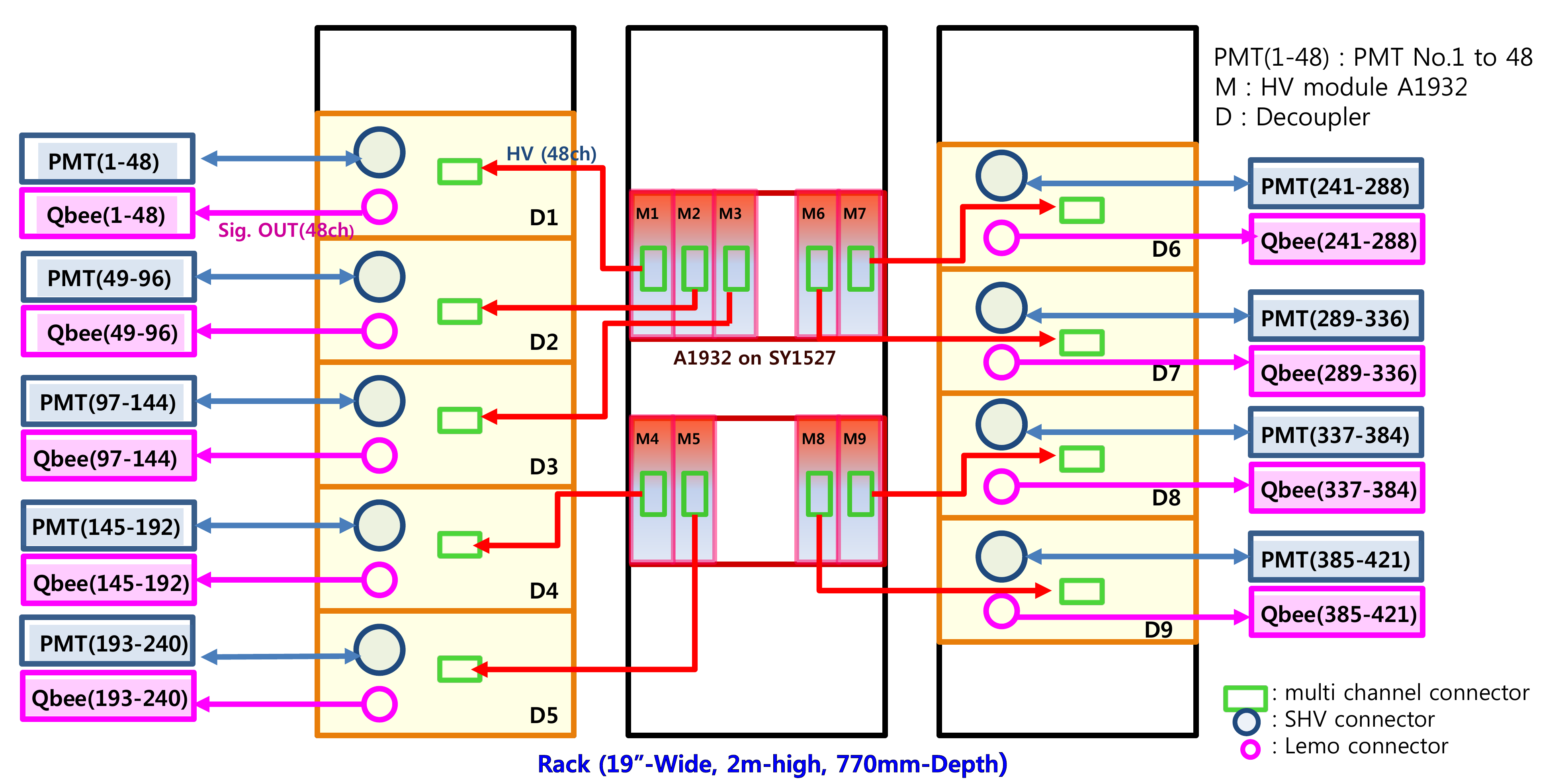}
\caption{HV supplying and decoupling system for a detector of 421 PMTs.
Each group consists of one A1932 48 channel HV supplying module and 48 channel decoupler box.}
\label{hv_rack}
\end{figure}

For a detector, we have 421 PMTs. These PMTs will be in nine groups in HV supplying
system as shown in Fig.~\ref{hv_rack}. Each group consists of one A1932 48 channel
HV supplying module and 48 channel decoupler box as shown in Fig.~\ref{decoupler}.
Three 19'' racks will house the whole system.

\section{Prototype Detector}
The small prototype detector is built to test 
properties of liquid scintillators and validate the Monte 
Carlo simulation model based on {\sc geant4}.

\subsection{Detector Design}
\begin{figure}
\begin{center}
\includegraphics[height=2.3in]{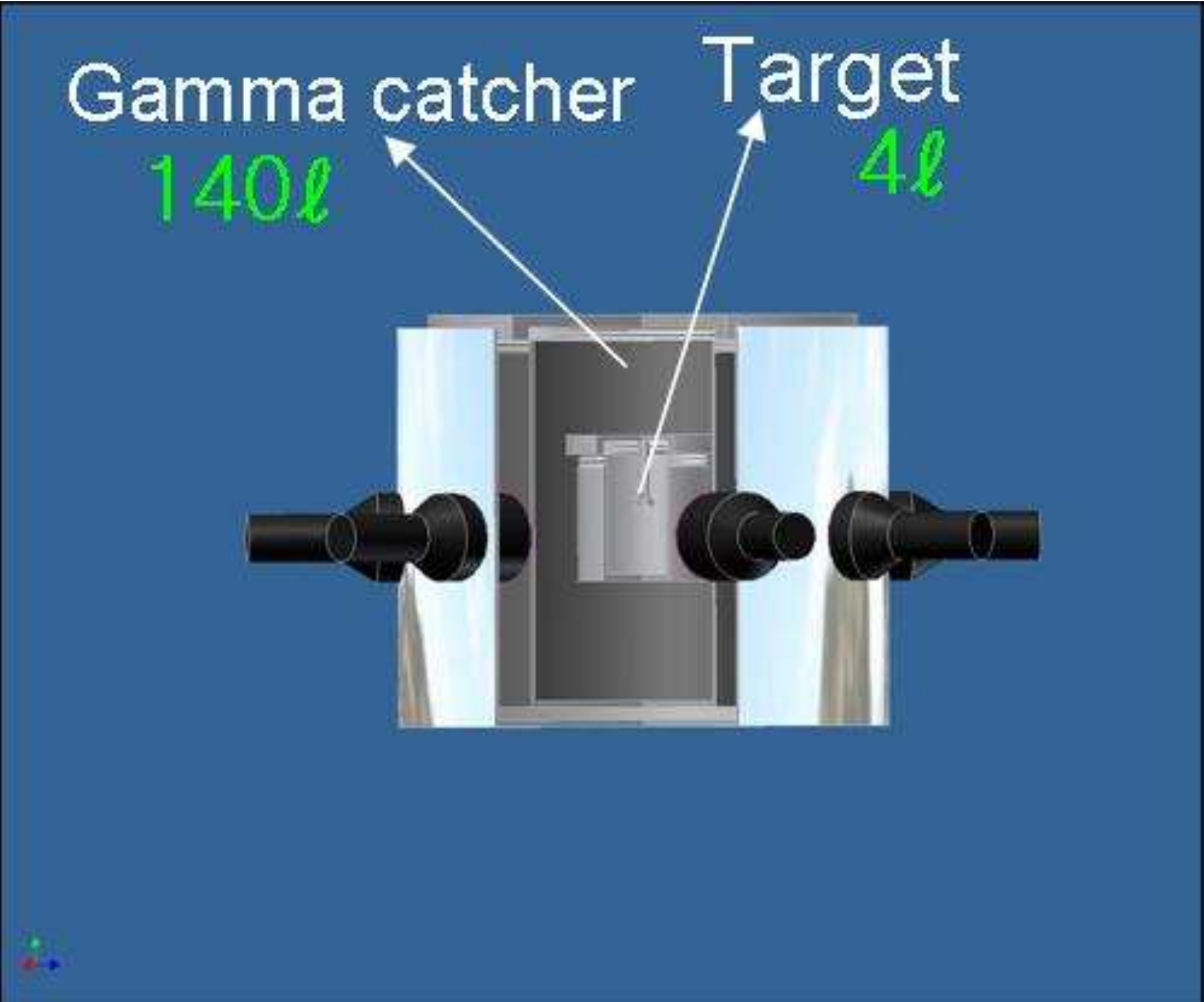}
\includegraphics[height=2.3in]{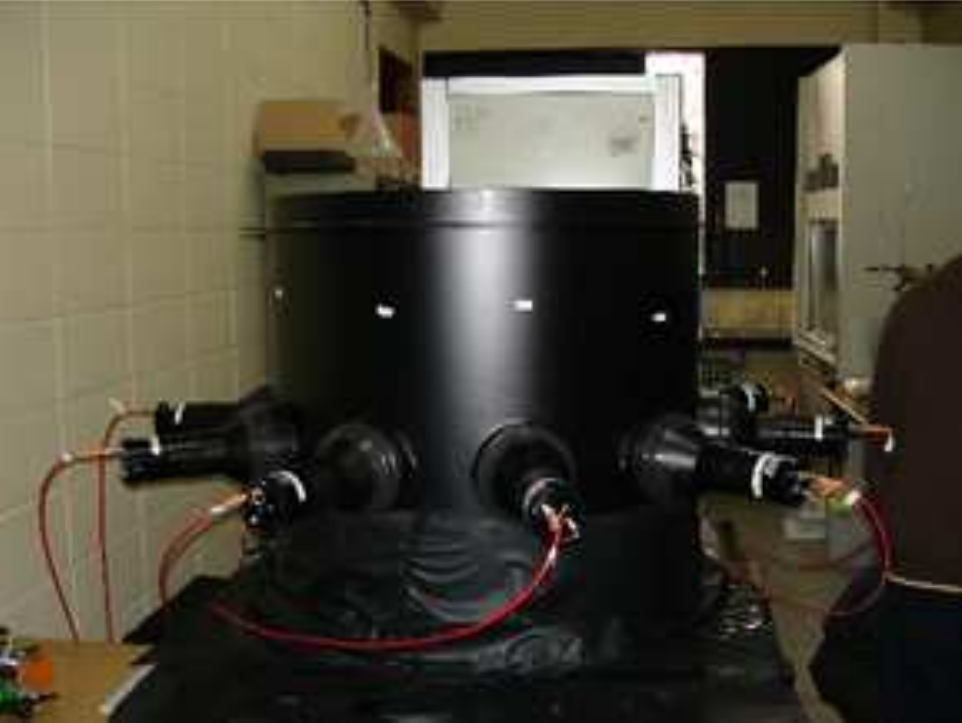}
\end{center}
\caption{The prototype detector. Cut-away view of target and gamma 
catcher (left) and a picture of the exterior of the detector (right).}
\label{proto_geo_1}
\end{figure}

The prototype detector consists of two concentric cylindrical vessels
filled with liquid scintillators with a cylindrical dark box surrounding 
the vessels. The vessels are 100~mm and 210~mm in diameter and 300~mm and 610~mm
in height, respectively, and made of 10~mm thick transparent acrylic
plastic. The smaller vessel is filled with Gd-loaded liquid scintillator 
(target) and the larger one with liquid scintillator ($\gamma$-catcher). 
There are ten equally spaced 5-inch PMTs mounted on the barrel section 
of the dark box through holes. 
The schematics and pictures of the assembly are shown in Fig.~\ref{proto_geo_1}. 

\subsection{Liquid Scintillator Mixture}
The liquid scintillator used in the prototype detector consists of solvent,
admixture of 40\% pseudocumene (PC) and 60\% mineral oil (MO) by volume, 
added with 3~g/l PPO and 0.05~g/l bis-MSB as fluor and wavelength shifter, 
respectively. This solvent mixture gives a light output of $\sim85$\% with 
respect to that of pure PC. 
The target liquid scintillator is loaded with 1~g/l Gd.
Table~\ref{prototype_elements} summarizes the liquid scintillator mixture. 

\begin{figure}
\begin{center}
\includegraphics[width=3in]{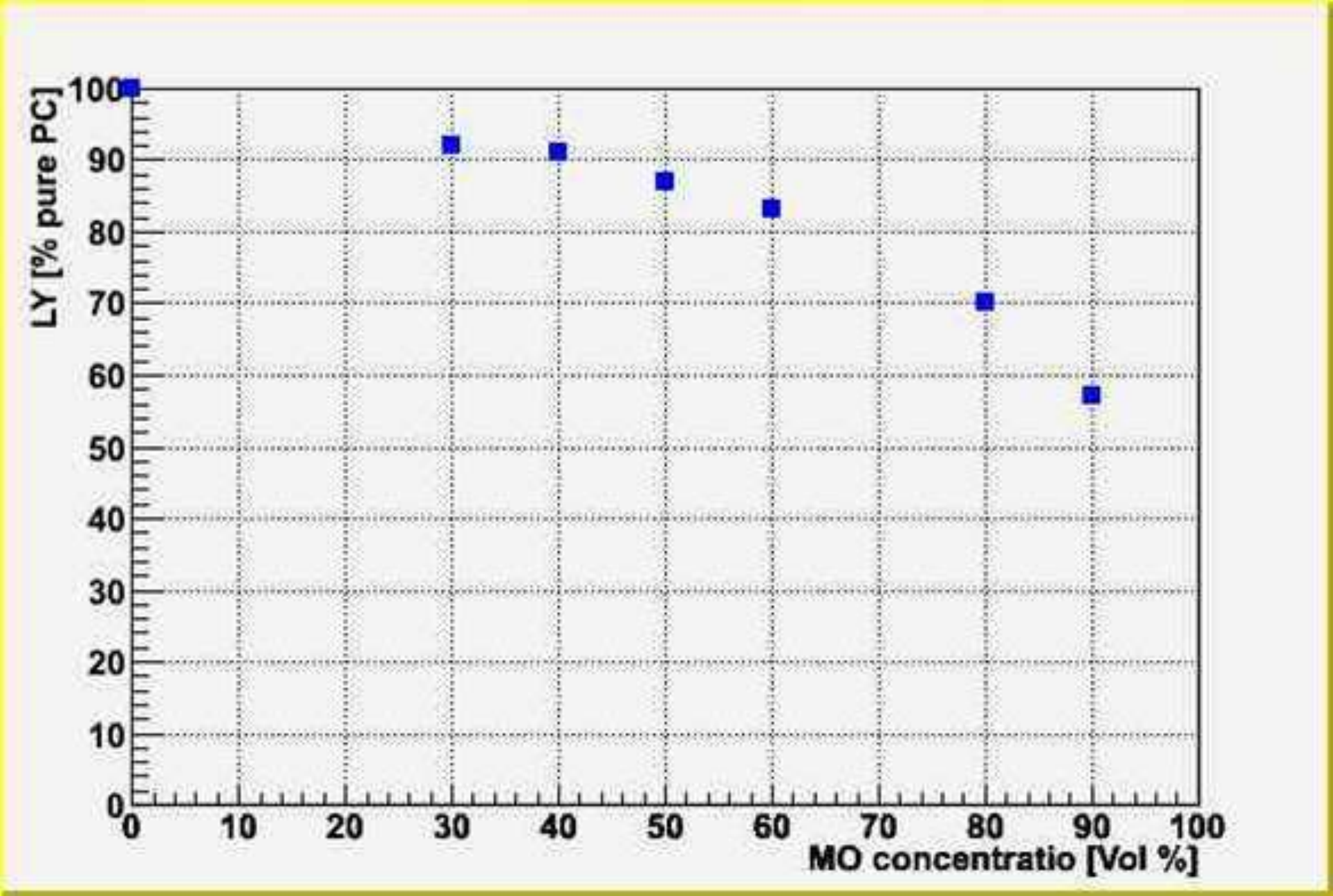}
\includegraphics[width=3in]{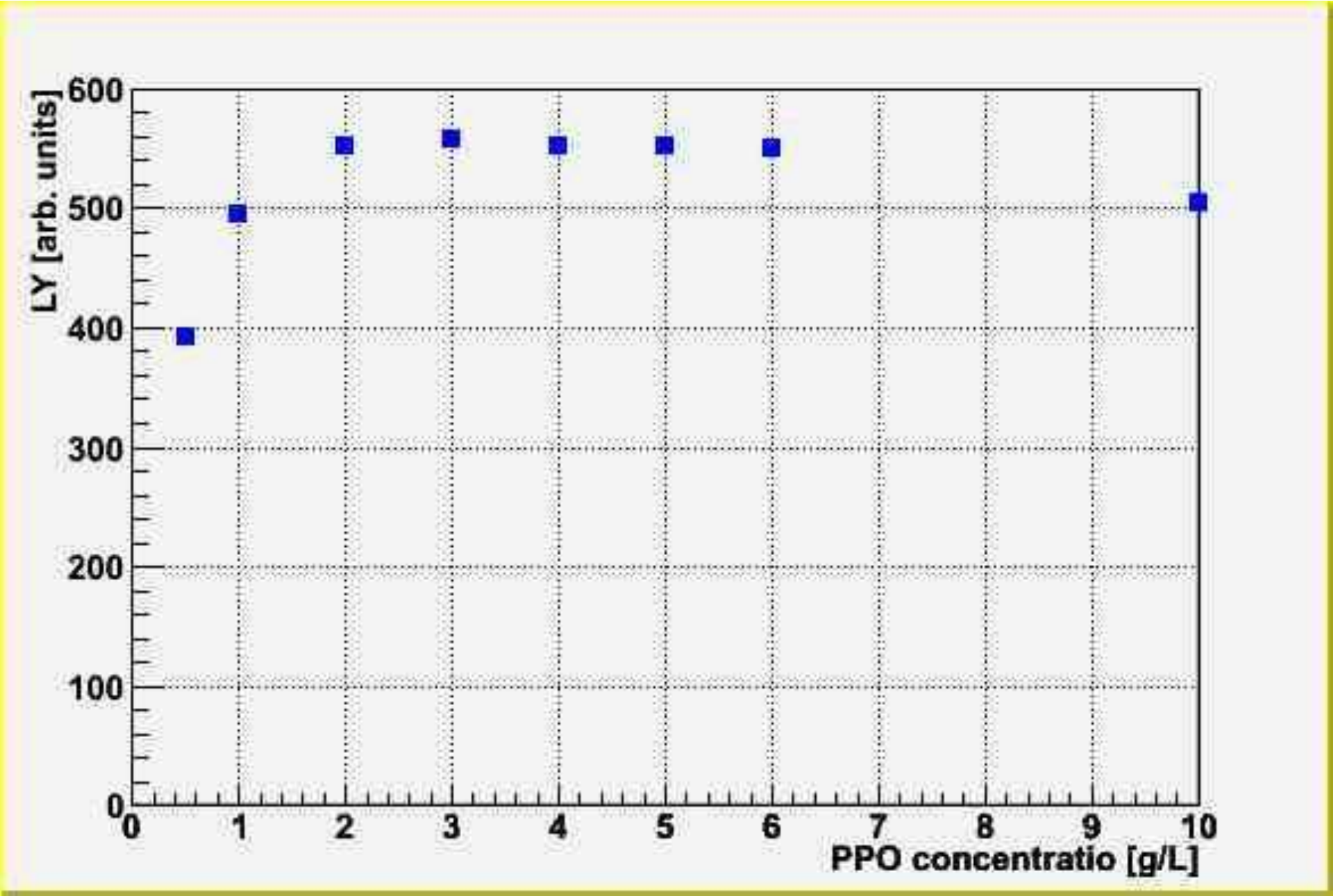}
\end{center}
\caption{The light yield of PC and MO mixture as a function of MO 
concentration (left) and PPO concentration (right). About 85\% light 
output is shown at the 60\% MO with respect to pure PC. The highest 
light yield is shown at the 3~g/l PPO concentration.}
\label{proto_mix_1}
\end{figure}

To prevent the deterioration in light yield by the oxidation of the solvent
through the contact with air, Argon gas was bubbled through the liquid scintillator 
whenever the liquid scintillator was exposed to air. 
The material compatibility of the liquid scintillator with acrylic vessel 
was checked in a long term test.
The performance of the liquid scintillator with various mixing ratios
were tested. The results are shown in Fig.~\ref{proto_mix_1}. 
For a detailed description of liquid scintillator, see Sect.~\ref{Scintillator Chapter}.

\begin{table}
\begin{center}
\begin{tabular}{cccccc}\hline\hline
               &PC  &MO  &PPO &bis-MSB  &Gd  \\ \hline
Target         &1.6~l    &2.4~l   &12~g       &0.2~g             &4.0~g \\
Gamma catcher  &56.0~l     &84.0~l    &620~g      &7.0~g               &0.0~g \\\hline\hline
\end{tabular}
\caption{Summary of liquid scintillator admixtures used in the prototype 
detector.} 
\label{prototype_elements}
\end{center}
\end{table}

\subsection{Electronics, Trigger, and DAQ Systems}
\subsubsection{Electronics}
For the front-end electronics of the prototype detector, we utilized 
off-the-shelf commercial products. We used an 8-channel flash 
analog-to-digital converter (FADC) VME module with an 100~MHz 
sampling rate.
The FADC has a 10-bit pulse height digitization.
Since a typical signal pulse has a width of about 20~ns, FADC with
an 100~MHz sampling rate (10~ns bin) has an inadequate time resolution 
to see the signal pulse structure.
Therefore, the PMT signal was stretched five times in time 
before digitization.

\subsubsection{Trigger and DAQ}
A simple trigger and DAQ system was 
used for the prototype.
We required a two-channel coincidence to collect events
from $\gamma$ sources.
If any of two PMT channels fired, then the event was accepted
and sent to FADC for digitization (see Sect.~\ref{proto-performance}). 

\subsection{Performance}\label{proto-performance}
We analyzed the single photoelectron spectra using an LED light source. 
The energy resolution for the SPE events was measured to be $\sim 50$\%. 
The trigger threshold for each channel was set to fire for signal above 
0.5 photoelectrons. 

To measure the detector performance, a $\gamma$ source, either 
$^{137}$Cs or $^{60}$Co, was placed at the center of the target.
The results were compared to the simulation based on {\sc geant4}.
The threshold behaviour observed in data was modelled and applied
to the Monte Carlo samples and the same trigger conditions were
used for event selection.  
  
\begin{figure}
\begin{center}
\includegraphics[width=3.1in]{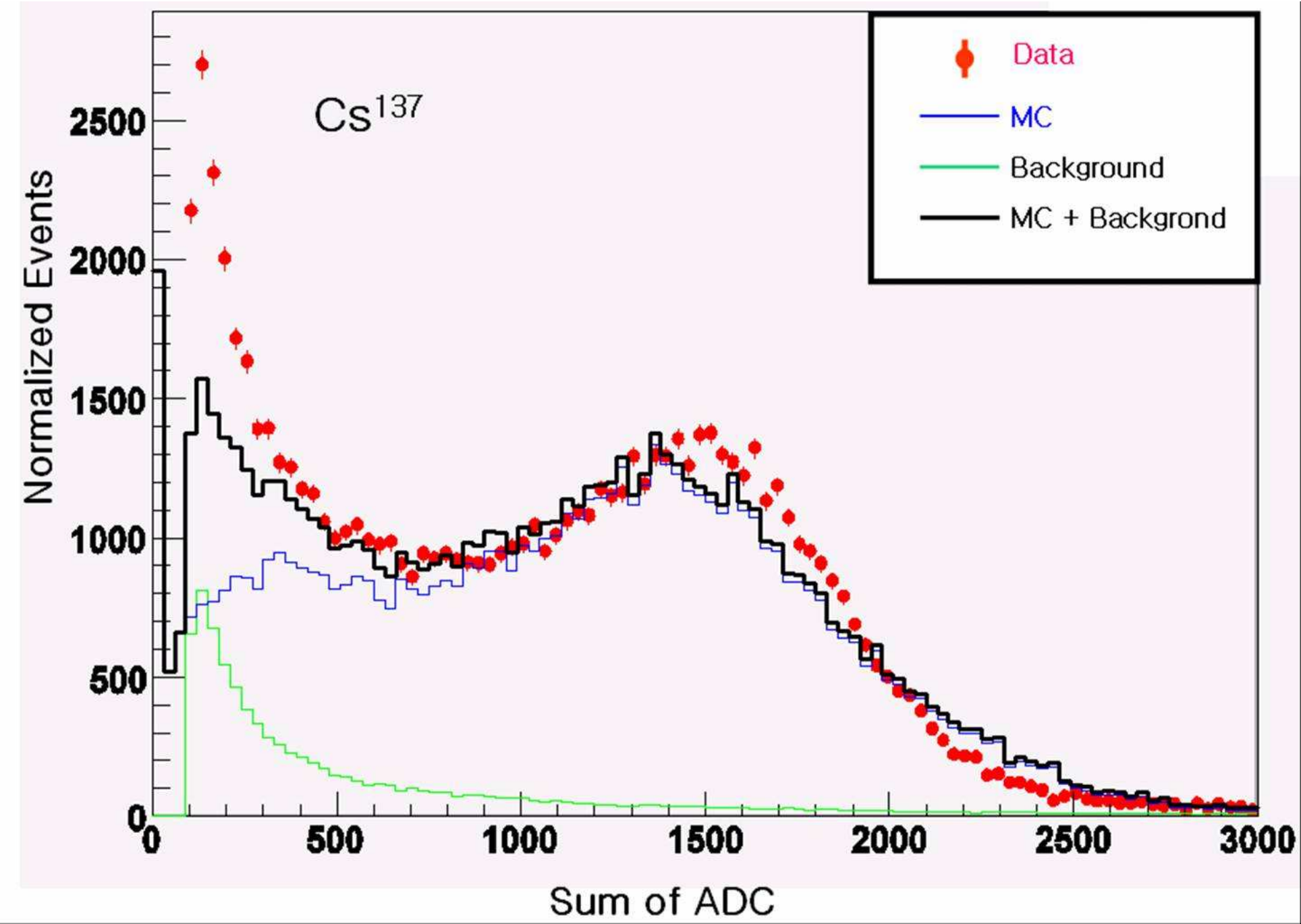}
\includegraphics[width=3in]{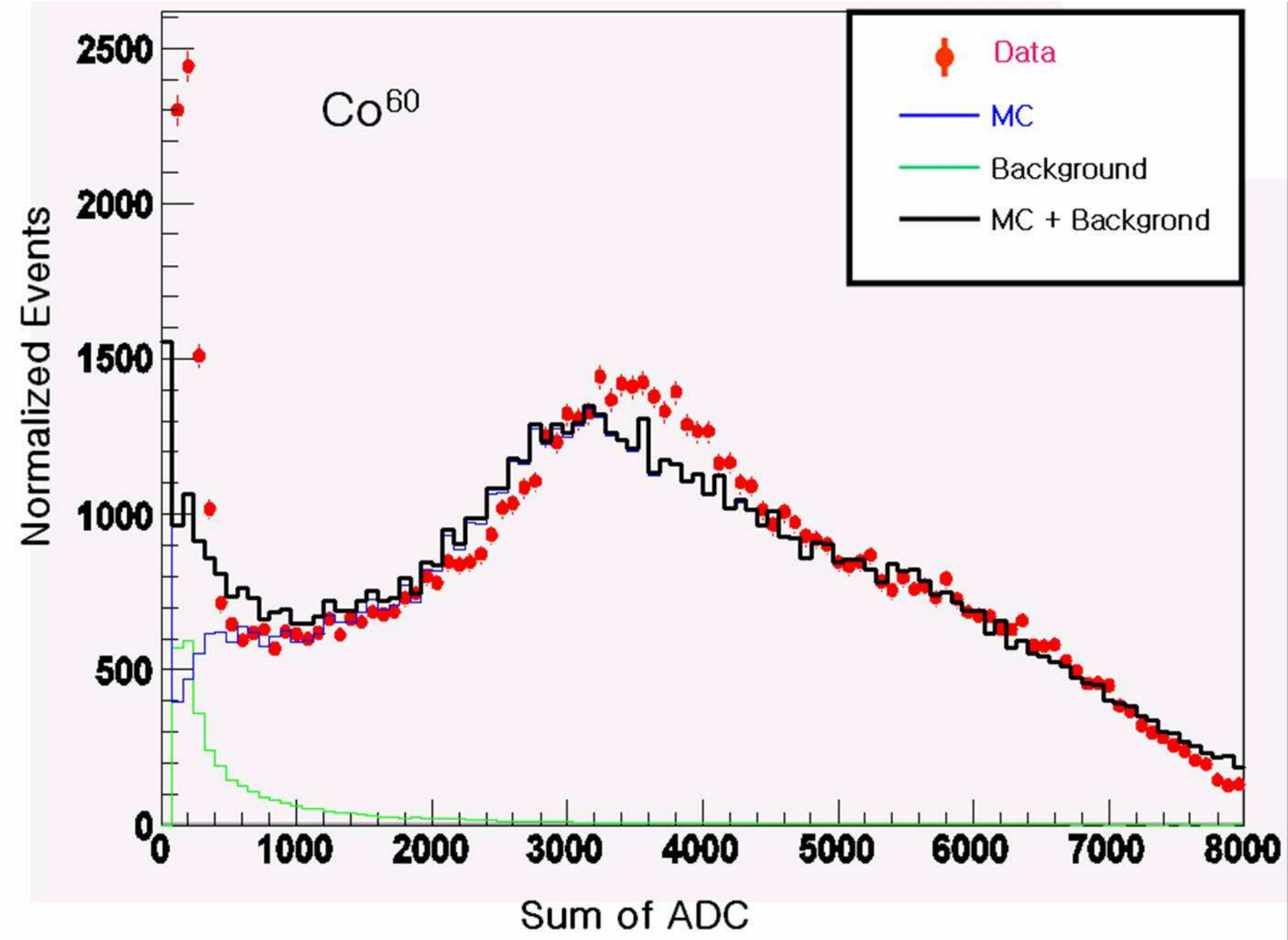}
\end{center}
\caption{Energy spectra 
for $^{137}$Cs (left) 
and $^{60}$Co (right)
measured with the prototype detector.
The horizontal axis is the sum of ADC values over all 
PMT channels. 
The red dots indicate the on-source data and bold black histogram 
indicate the Monte Carlo 
sample with background estimation included.
The Monte Carlo sample distributions are normalized to the data.
}
\label{proto_calib_1}
\end{figure}

Figure~\ref{proto_calib_1} shows the energy spectra obtained for $^{137}$Cs 
and $^{60}$Co. 
The background distribution was obtained from data collected without
any sources in the detector.
The energy response is estimated as 102 photoelectrons per MeV. 
We see discrepancies between data and Monte Carlo samples 
near the peak in both cases. It is under investigation.

\section{Mock-up Detector}\label{Mock-up Detector}
The mock-up detector is built to test the detailed design features as
well as various subsystems to be incorporated into the full detector.
It is also used to do a long term test on the performance of
liquid scintillator and its compatibility with acrylic.

\subsection{Design of the Mock-up Detector}
\begin{figure}
\begin{center}
\includegraphics[width=7cm]{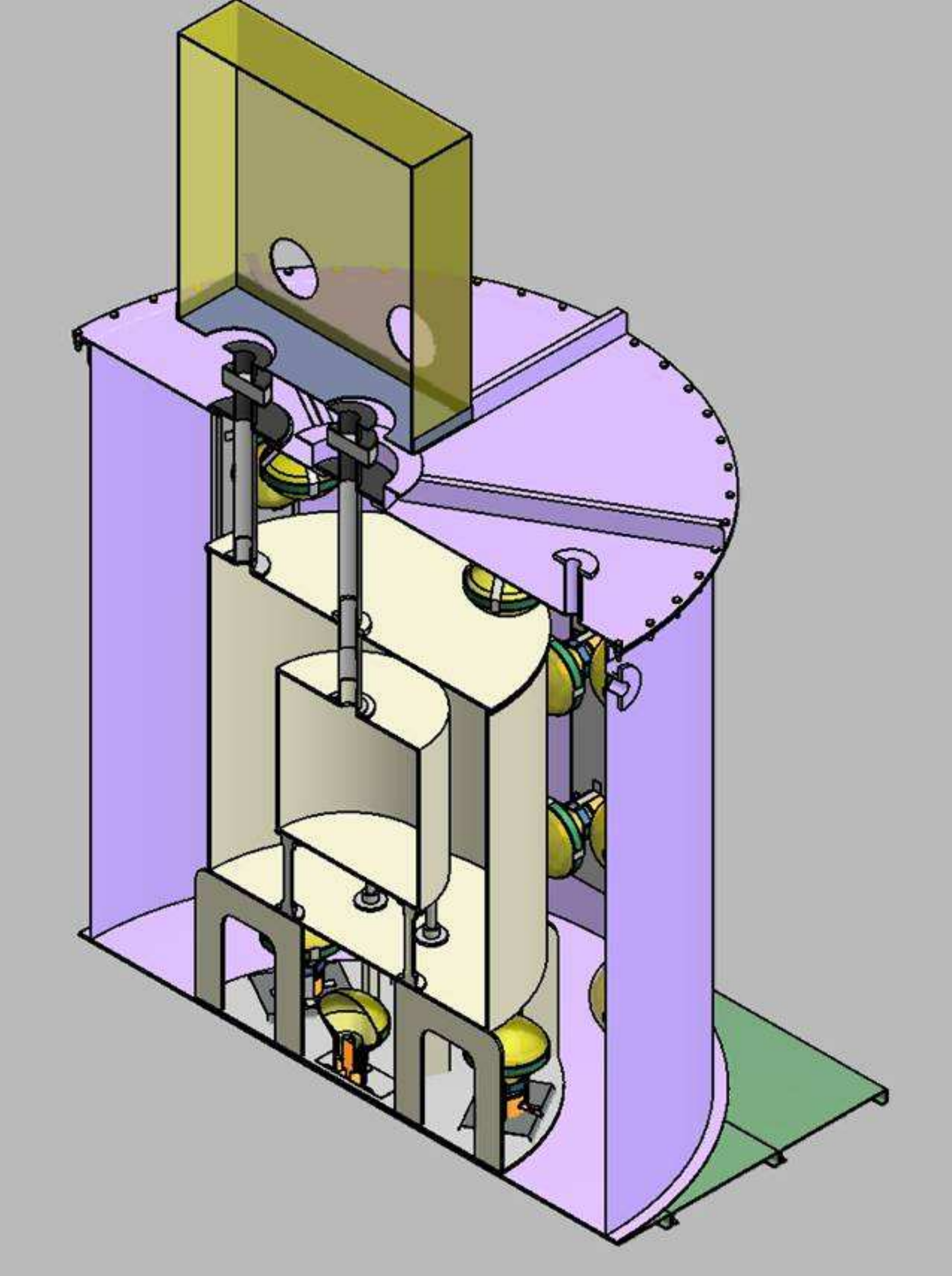}
\end{center}
\caption{Cut-away view of mock-up detector. It has two layers of liquid 
scintillator (target and $\gamma$-catcher) contained in acrylic vessels 
surrounded by a non-scintillating mineral oil (buffer) in a 
stainless steel vessel. There are 31 10-inch R7081 PMTs mounted on the 
buffer vessel walls. The mock-up is about 1/15 of the full size 
detector in volume. A glove box housing calibration source driving system 
is installed on the top of the detector. Sources can be inserted in the 
target and $\gamma$ catcher.}
\label{mockupdesign}
\end{figure}

The mock-up detector is about 1/15 of the full detector in volume. 
However, it has most features of the full detector to check the 
validity of the design features in the full detector as well as the 
performance of various detector components.
The mock-up detector has structural similarities to the full detector;
cylindrical target, $\gamma$-catcher, and buffer layers. However, it
lacks a veto layer. The vessels are made of the same materials as those 
of the full detector. 
The same model of 10-inch PMTs as in the full detector are used.
Figure~\ref{mockupdesign} shows the overall design of the 
mock-up detector  and Table~\ref{mockupdimension} shows the
dimensions of the detector. 

The sizes of the target and $\gamma$-catcher were determined 
such that the detector can measure most of neutron capture signal.
The buffer layer is 50~cm thick to accommodate 10-inch PMTs and is 
filled with mineral oil. The Linear Alkyl-benzene (LAB) based 
scintillator is filled through chimneys connecting to target and 
$\gamma$-catcher vessels.
The buffer vessel, made of stainless steel, is 4~mm thick for the 
side and 5~mm for top and bottom. 
The target vessel is constructed with a casted acrylic tube from 
Reynolds Co., USA.
The $\gamma$-catcher vessel is made with GS233 Plexiglass casted 
acrylic supplied by 
Degussa GmbH, Germany and shaped as a cylindrical tube.

\begin{table}
\begin{center}
\begin{tabular}{ccccc}\hline
Chamber	& Diameter	& Height	 & Thickness of vessel \\ \hline
Target  & 60~cm & 60~cm & 10~mm \\
Gamma Catcher & 120~cm & 120~cm & 10~mm \\
Buffer & 220~cm & 220~cm & 4~mm/5~mm$^{*}$\\ \hline
\end{tabular}
\caption{Dimensions of the mock-up detector. ((*)The thickness of SUS 
at side is 4~mm and the top and bottom are 5~mm.)}
\label{mockupdimension}
\end{center}
\end{table}

There are 31 10-inch PMTs(R7081) installed on the inner surface of 
the buffer vessel; 7 on the bottom, 6 on the top, and 18 on the side. 
The surface coverage of the PMT photocathode is 8\%. 
PMTs are held in place by the PMT holders made with a 2~mm thick 
stainless steel strips. In mineral oil the side mounted PMTs shifted 
upward about 2~mm from the nominal position by buoyant force but 
stayed stable.
All PMT cables are bundled together and potted with epoxy for sealing 
and are extracted to the outside of the buffer chamber. 

A glove box is on top of the buffer vessel housing the calibration 
source driving system. The calibration sources are inserted in target 
or $\gamma$-catcher through chimneys. 
\begin{figure}
\begin{center}
\includegraphics[width=6cm]{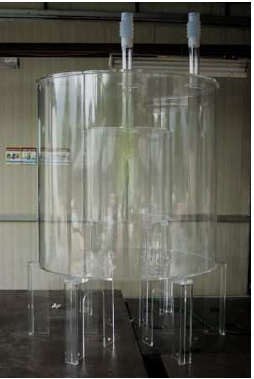}
\includegraphics[width=6cm]{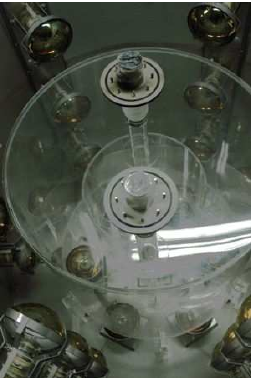}
\end{center}
\caption{Target and $\gamma$-catcher vessel assembly of mock-up detector 
(left) and with PMTs attached on the inner surface of the buffer vessel 
(right). The thickness of acrylic vessel is 1~cm.
}
\label{mockup-photo-assembly}
\end{figure}
Two chimneys, one for target and the other for $\gamma$-catcher, are
installed for filling liquids and transporting calibration sources. 
A part of each chimney is made with a 3-inch PTFE flexible 
tubes to relieve the stress between the acrylic vessels and the buffer 
vessel. The calibration source driving system is described in the 
Sect.~\ref{Source Driving System}. Each calibration source is enclosed 
in a 3~cm acrylic container and can be placed at various depths in the 
target and $\gamma$-catcher.
Figure~\ref{mockup-photo-assembly} shows the pictures of the constructed 
acrylic vessel assembly. 

\subsection{Performance of Mock-up Detector}
The background rate us high because the mockup detector is installed
at the ground level without much shielding.
The 
50~cm-thick mineral oil layer in the buffer reduces the ambient gamma 
background by more than a factor of ten. The background rate was 
reduced to $\sim 10$~kHz over 100 keV energy threshold. 

The gains of the PMTs were adjusted to be $1.5 \times 10^{7}$. 
Both target and $\gamma$-catcher were filled with LAB based liquid 
scintillator. %
The DAQ system is based on a 400~MHz FADC described in 
Sect.~\ref{DAQ system for mockup detector}. The time window 
for an event was set at 320~ns and the signal starts at around 80~ns. 
The threshold of PMT signal was set to be 3~mV. At least five PMTs are 
required to have signal above the threshold to accept the event. 
This condition makes the total energy threshold at about 100~keV. 

\begin{figure}
\begin{center}
\includegraphics[width=7.5cm]{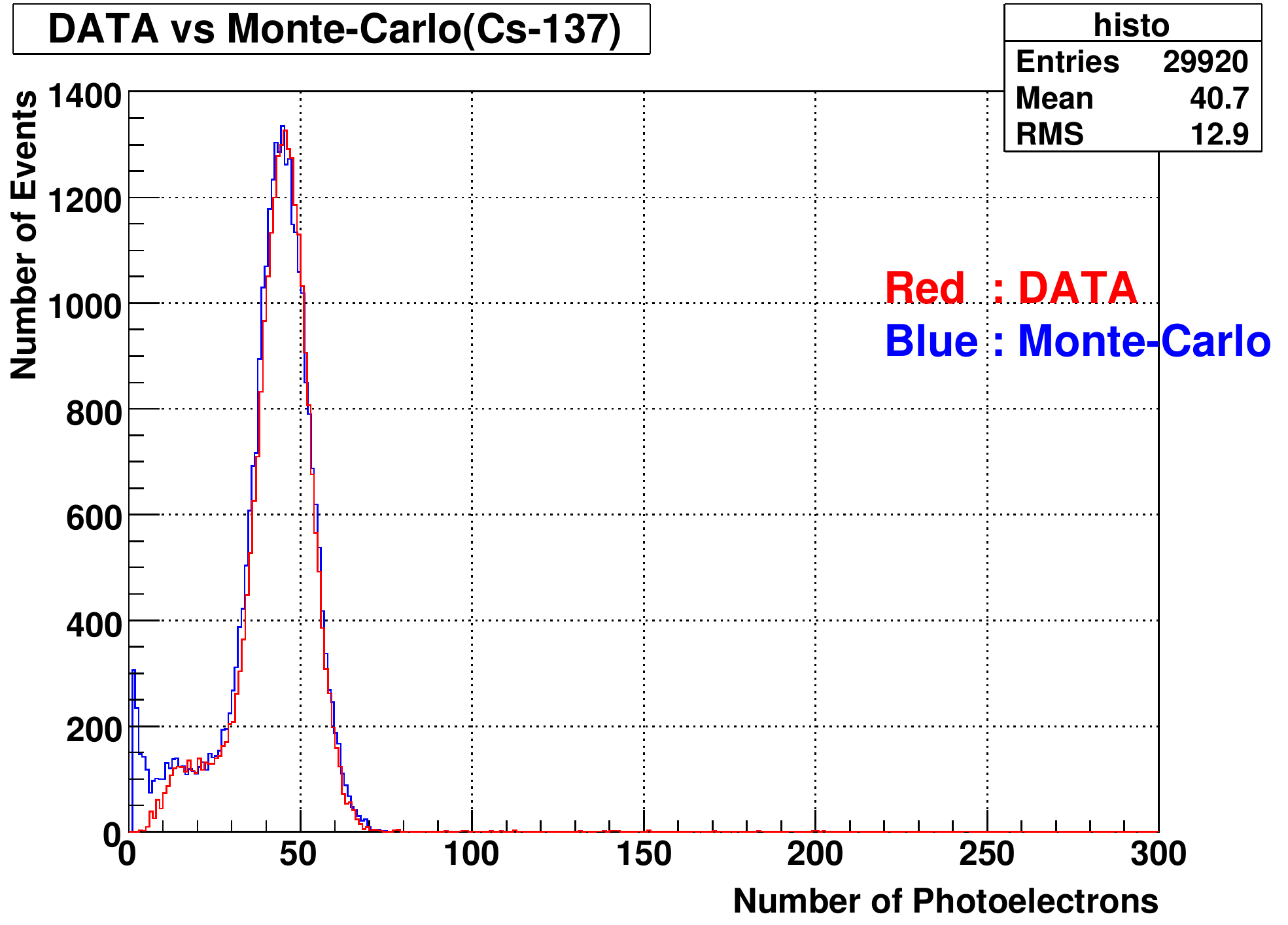}
\includegraphics[width=7.5cm]{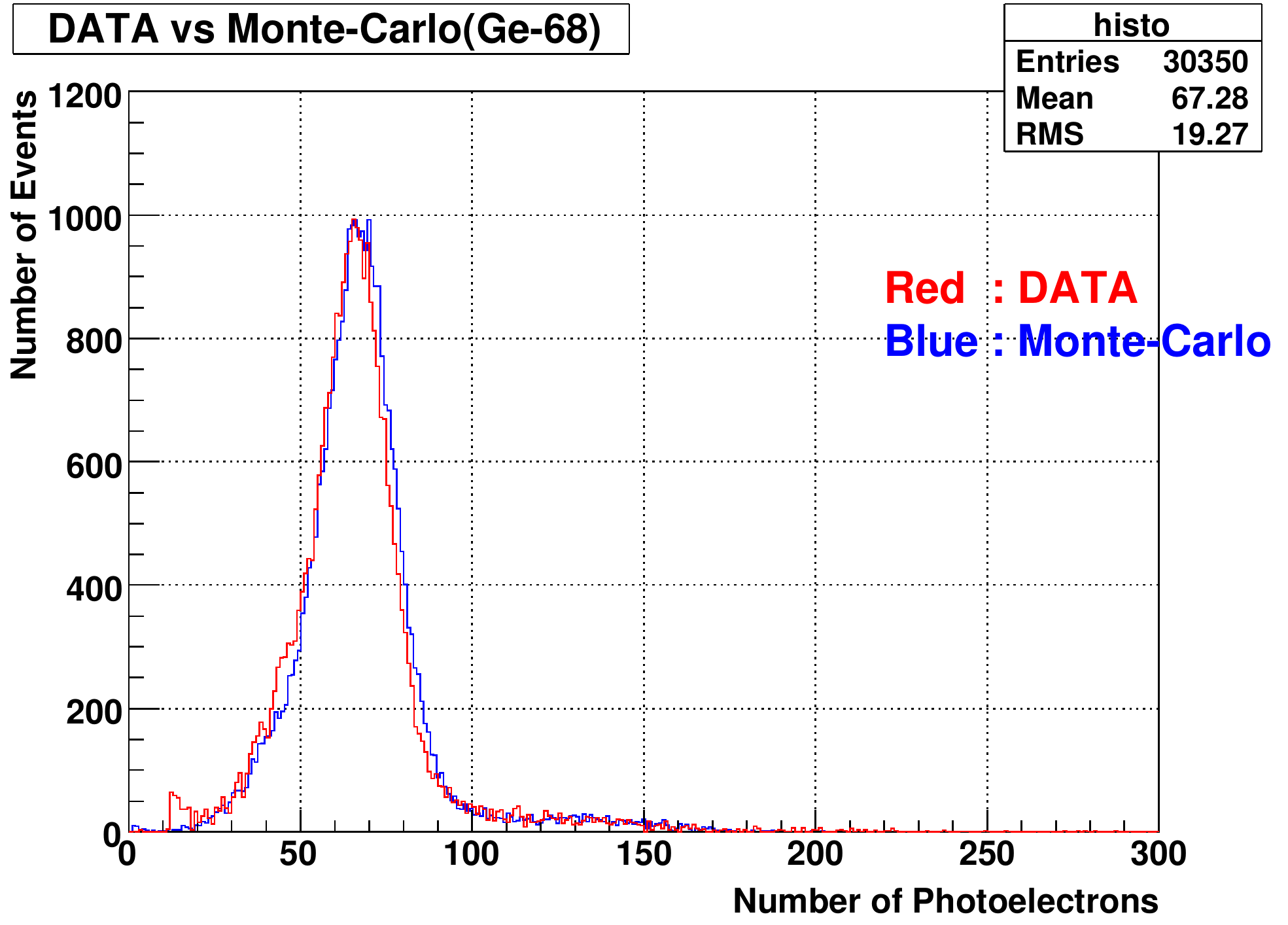}
\includegraphics[width=7.5cm]{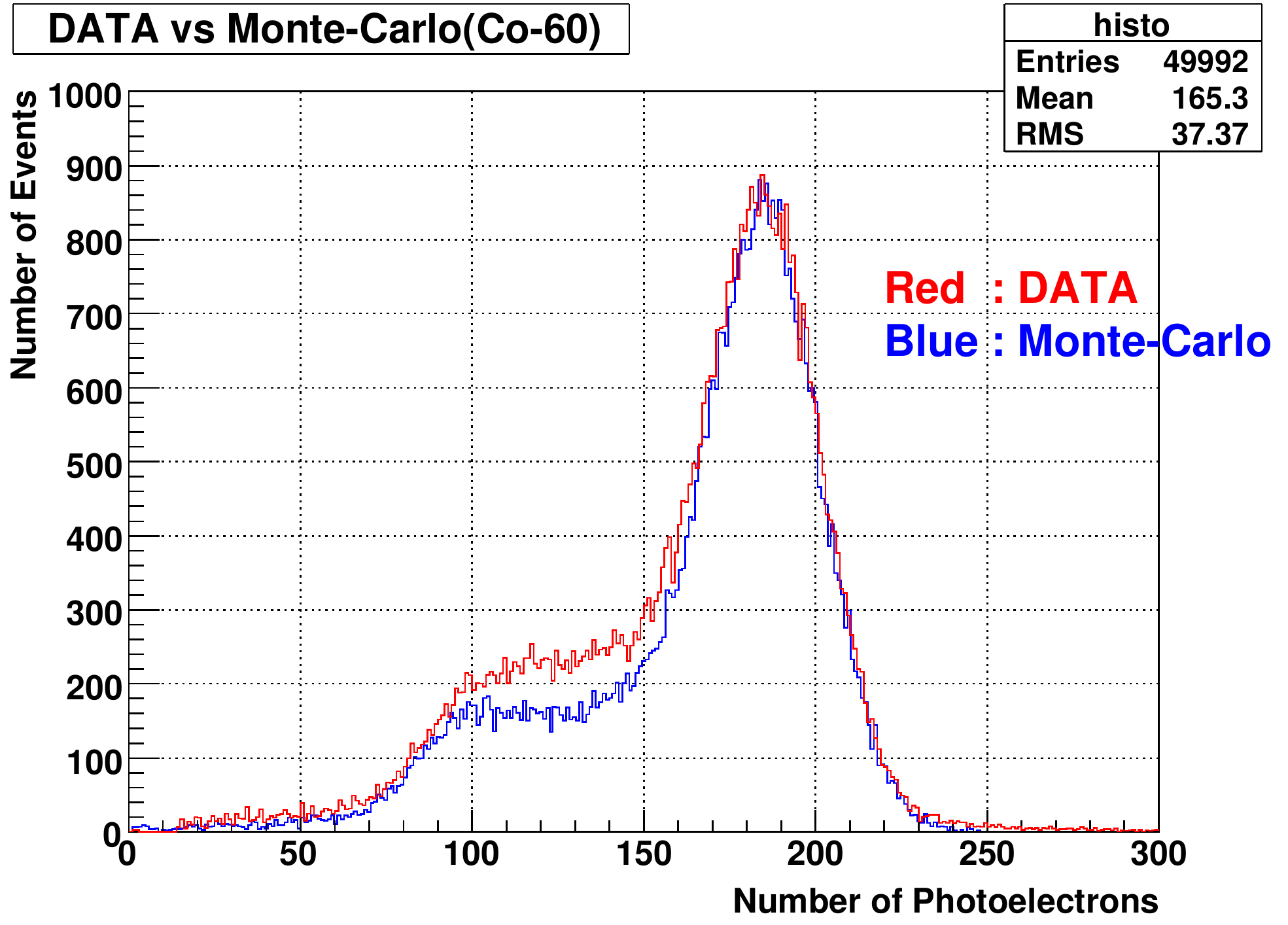}
\includegraphics[width=7.5cm]{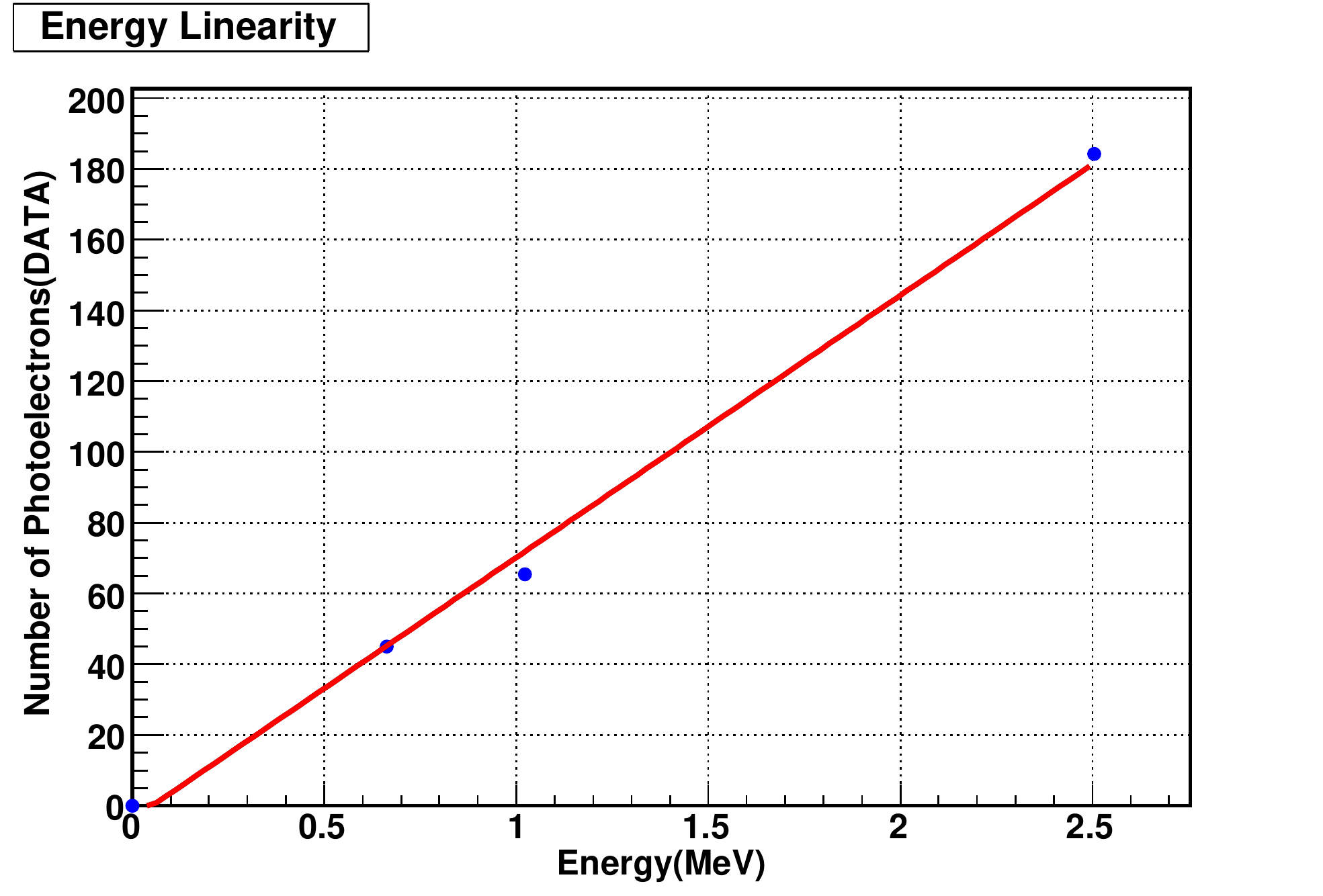}
\end{center}
\caption{The number of photoelectron distributions of $^{137}$Cs (top left), 
$^{68}$Ge (top right), and $^{60}$Co (bottom left) sources placed at the center of 
target. The red histograms are the data and the blue histograms 
are the results of {\sc geant4} simulations. The bottom right figure 
shows the linear correlation between the measured and simulated number of 
photoelectrons.}
\label{mockup-analysis}
\end{figure}

Figure~\ref{mockup-analysis} shows the number of photoelectron distributions
obtained with $^{137}$Cs, $^{68}$Ge, and $^{60}$Co sources. 
The expected background contribution obtained from data taken without 
the source was subtracted.
The measured number of photoelectrons per MeV is 70. The overall spectral 
shapes of the source data were reproduced well with the {\sc geant4} based 
simulation.
An acceptable linearity between energy values of the 
radioactive sources and the number of measured photoelectrons.

\section{DAQ System for Mock-Up Detector}\label{DAQ system for mockup detector}
\begin{figure}
\begin{center}
\includegraphics[width=5in]{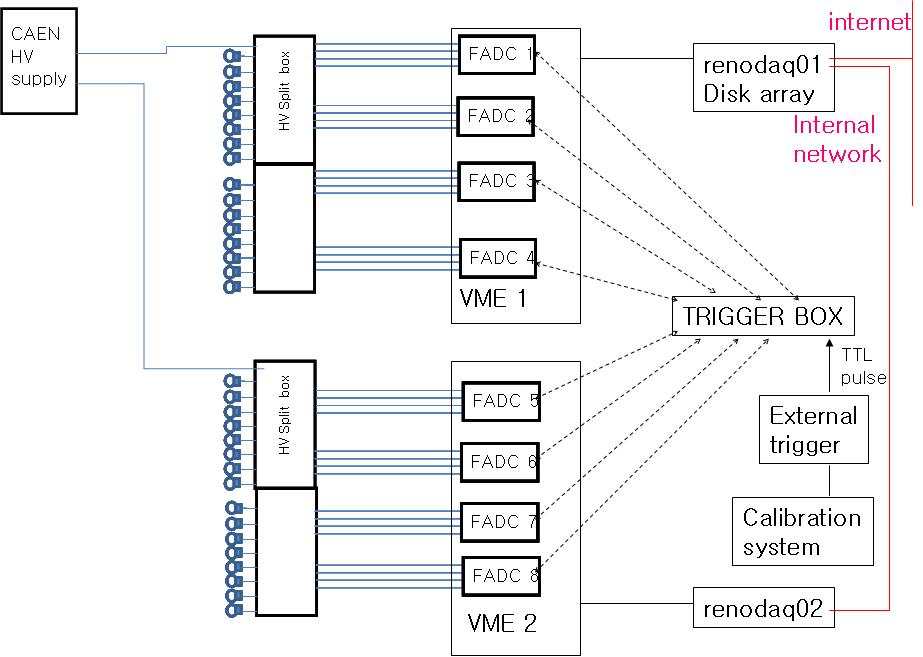}
\end{center}
\caption{A schematic diagram of DAQ system for the mock-up
detector. Each 400~MHz FADC board reads and digitizes the signals
from four PMTs. An event is accepted if there are hits above
3~mV on more than five PMTs.}
\label{mockup_electronics}
\end{figure}

\begin{figure}
\begin{center}
\includegraphics[width=10cm]{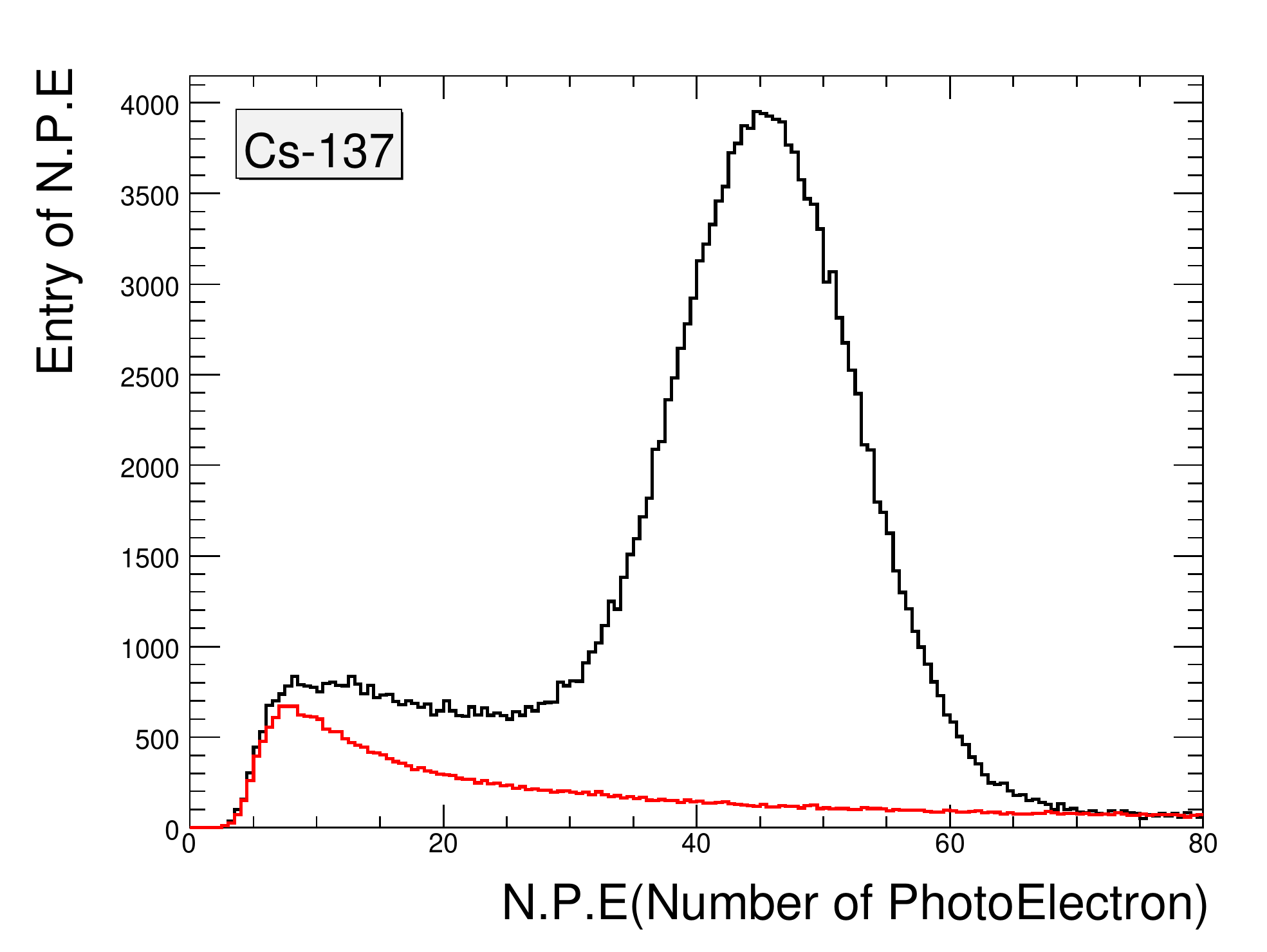}
\end{center}
\caption{Distribution of photoelectron of $^{137}$Cs source.
         Data is taken with the threshold at 3~mV and multiplicity of 5.
        }
\label{mockupCs137}
\end{figure}

The data acquisition system is based on 400~MHz flash analog-to-digital
converter (FADC) as shown in Fig.~\ref{mockup_electronics}. Eight FADC boards are 
installed in the two VME crates and two
personal computers are used for data taking and online monitoring. After timing
synchronization between the hit data from the VME crates, events are built. 
Figure~\ref{mockupCs137} shows the photoelectron distribution produced by a $^{137}$Cs source
inside of mockup detector.
The reconstructed events are used 
to understand the performance of the mockup detector.

\newpage

\bibliographystyle{plain}
\chapter{Liquid Scintillator}\label{Scintillator Chapter}
\section{Introduction} 
The reactor anti-neutrinos are detected through the inverse 
beta decay process followed by a neutron capture. The prompt 
positron yields $1\sim 8$~MeV of visible energy. 
When a neutron is captured by a hydrogen, which forms a deuteron
giving off photons with total energy of $\sim 2.2$~MeV. 
However, when the liquid scintillator (LS) is loaded with Gadolinium 
(Gd), which has a very large thermal neutron capture cross section
than a free proton, the delayed neutron capture signal is enhanced 
significantly over the radioactive background by producing photons 
with total energy of $\sim 8$~MeV.

Liquid scintillators are contained in the target and 
$\gamma$-catcher layers of the RENO detector. Detection of 
a small energy deposit will be possible only if the 
scintillator has excellent light output and optical clarity. 
It is also required that the liquid scintillator should be relatively
easy and safe to handle, cost effective to produce, and have
desirable physical and chemical properties.   

\section{Liquid Scintillator and Buffer Fluids}
The RENO detector uses organic liquids in target, $\gamma$-catcher, and buffer
as summarized in Table~\ref{P}. 
The acrylic vessel holding the target liquid is surrounded by the $\gamma$-catcher. 
The densities of the liquids should be similar in all volumes to minimize 
stress exerted on detector structures due to buoyancy force.

Important considerations for the liquid scintillator are light yield, 
stability, and radiopurity. Light yield requirement could be satisfied 
by the selection of adequate solvent and optimization of fluor 
concentration. 

\begin{table}
\begin{center}
\begin{tabular}{ccccc}\hline
Region & Radius(mm)      & Height(mm)  & Volume (m$^3$) & Type \\ \hline
Target                   & 1388 &3176  & 19.21    &0.1$\%$ Gd loaded LS \\
Target vessel                  & 1400 &3200  & 0.48     &Acrylic \\
$\gamma$-catcher         & 1985 &4370  & 34.37    &Unloaded scintillator \\
$\gamma$-catcher vessel                  & 2000 &4400  & 1.20     &Acrylic \\
Buffer                   & 2694 &5788  & 76.64    &Non-scintillating oil     \\ 
\hline
\end{tabular}
\end{center}
\caption{Organic liquids used in various parts of the RENO detector.}
\label{P}
\end{table}

\subsection{Specification}
The organic liquids should have good transparency, large attenuation length,  
high radiopurity, and chemical stability.
Also, each liquid should have the following additional requirements:
a) Gd-loaded scintillator (target) should have good light yield and
high H/C ratio, b) scintillator ($\gamma$-catcher) should have a good
light yield, and c) mineral oil (buffer) should be non-scintillating and 
has a similar density to liquids scintillator. 

\subsection{Organic Solvent}
Liquid scintillator consists of aromatic organic solvent, fluor,
and wavelength shifter as shown in Table~\ref{Q}. A benzene (C$_6$H$_6$) or 
benzene compounds have been used as aromatic solvent 
because of its excellent light transmission properties.
Several aromatic scintillation liquids have been studied for 
RENO experiment. 

Pseudocumene (PC or TMB, C$_{9}$H$_{12}$, 1,2,4-trimethylbenzene) is
the most commonly used solvent for Gd-loaded liquid scintillator (Gd-LS).
PC gives the highest light output among the widely used 
liquid scintillators. 
However, it attacks acrylic materials and has a low H/C ratio of 1.66. 
It is also flammable with low flash point (48$^o$C) and develops harmful fume. 
Therefore, PC is usually used in an admixture with diluent solvents.
The concentration of PC is determined by optimizing for the flash point, 
light output, and transparency. 
It is nearly insoluble in water, but well soluble in ethanol 
and benzene. 

Another aromatic solvent, 1,2-dimethyl-4-(1-phenylethyl)-benzene 
(phenyl-o-xylythane, PXE, C$_{16}$H$_{18}$), can be a candidate.  
Its flash point is 145$^o$C and density is 0.980 $\sim$ 1.000 g/cm$^3$ 
at 15$^o$C, but it has an even lower H/C ratio ($\sim$ 1.37) than PC. 
Borexino experiment~\cite{LS1} used PXE as the organic liquid scintillator 
due to its high density and high flash-point. 
Double Chooz experiment has proposed to use PXE~\cite{LS2}.

Organic solvent should have good compatibility with acrylic vessel. 
Although PC has good light yield and optical properties, 
the compatibility with acrylic vessel is not good.
Dilution component is added to PC to increase compatibility with acrylic 
vessel. Mineral oil (MO, C$_n$H$_{2n+2}$, where n = 11 $\sim$ 44) is used 
for a diluent solvent. 
Its density varies 0.7 $\sim$ 0.9 g/cm$^3$ depending on the products and
manufacturers.
A widely used liquid scintillator is a mixture of 40\% PC and 60\% MO  
and has a H/C ratio of $\sim$ 1.87.

Instead of MO as a diluent solvent, dodecane (C$_{12}$H$_{26}$) can be used. 
Normal dodecane does not have double chemical bonding nor a circular 
structure. 
It means that it is chemically stable and immune to oxidization. 
Normal dodecane has a higher flash point (83$^o$C) than PC and, therefore, 
adding 
normal dodecane to aromatic solvent with a low flash point significantly 
improves 
the safety of liquid scintillator. Dodecane has a high H/C ratio of 2.17. 
Normal dodecane is produced by distillation normal paraffins within a very
narrow temperature window. Therefore, its purity level is very high. 

Because the liquid scintillator serves as a neutrino target in reactor 
neutrino experiments, it should have a high H/C ratio and its proton 
density should be known precisely.
The solvents used in Palo Verde and CHOOZ experiment are admixtures 
of PC and MO.
Mineral oil is liquid paraffin which has a high H/C ratio and 
a good light yield. The light yield of liquid scintillator is 
measured with Compton edge of gamma rays emitted from a 
radioactive source.
For the light yield measurement, {$^{60}$Co and $^{137}$Cs}
are used.\footnote{ $^{60}$Co ($E_\gamma$ = 1173.2, 1332.5 KeV) has two 
Compton edges and $^{137}$Cs ($E_\gamma$ = 661.6 KeV) has one Compton 
edge (447 KeV).}$^,$\footnote{For 
light yield measurement, Dodecane, Decane, and several MO samples with
different kinetic viscosity values were prepared;
Aldrich, KF-50, KF-70, KF-250, and KF-400.
A larger KF number refer to higher kinetic viscosity. 
KF-series MO and decane are produced by SeoJin Chemical Co. and
``Aldrich'' MO and dodecane are purchased from Aldrich.}
The light yield of MO increases as kinetic viscosity decreases.
Dodecane based liquid scintillator has small light yield relative to MO based
one because it has single bond between neighboring atoms whereas MO has some 
double bonds between neighboring atoms.

Initially we took the admixture of 40\% PC and 60\% MO, by volume, 
which is used by Palo Verde experiment~\cite{LS3}, as a starting point 
of our organic solvent R\&D. The addition of MO to PC reduces the light 
yield, but it improves the chemical compatibility with the acrylic. 
However, LAB has shown many advantages over the PC-diluent admixture and it 
will be used as the solvent for liquid scintillator for our experiment. 
This is described later in this chapter.

\subsection{Various Fluors}
2,5-diphenyloxazole (PPO, C$_{15}$H$_{11}$NO) is used as the primary 
scintillation solute. Its emission spectrum peaks at $\sim$360 nm. PPO is 
a fluor which has been widely used in liquid scintillators for high energy 
physics. In addition, BPO (2-(4-bipheny)-5-phenyloxazole) and p-terphenyl 
(PTF) can be used. Unlike PPO, BPO does not need secondary wavelength shifter.
However, BPO is more expensive than PPO. 

\begin{figure}
\begin{center}
\leavevmode
\includegraphics[width=5in]{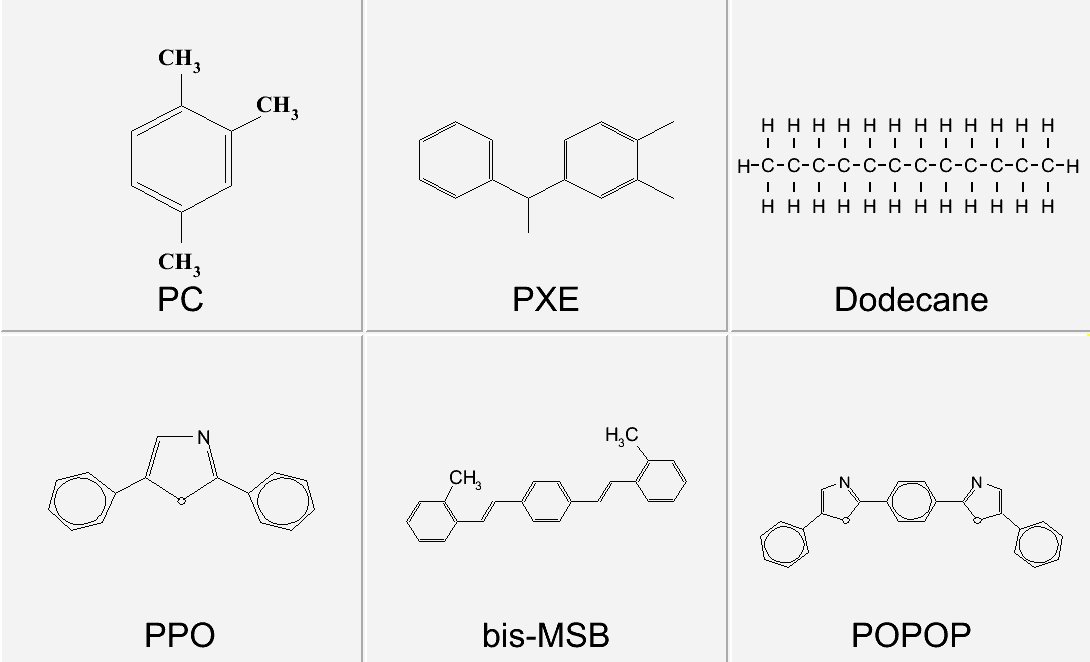}
\end{center}
\caption{Chemicals used in liquid scintillators for the reactor neutrino 
experiments.}
\label{formular}
\end{figure}

1,4-bis(phenyl-2-oxazolyl)-benzene (POPOP, C$_{12}$H$_{16}$N$_{2}$O$_2$) is 
used as a wavelength shifter, whose absorption spectrum peaks at 385 nm and 
emission spectrum at 418~nm. 
For a secondary wavelength shifter, 1,4,-bis(2-methylstyryl)-benzene (bis-MSB) 
could be used. 
Figure~\ref{formular} shows molecular structure of various chemicals used as 
solvents, fluors, and wavelength shifters. 

\begin{table}
\begin{center}
\begin{tabular}{|c|c|c|c|}\hline
Aromatic Solvent& Diluent Oil & Fluor   & WLS  \\ \hline
PC, PXE  & Mineral Oil, Dodecane &PPO, BPO, PTF        &bis-MSB, POPOP \\
DIN, PCH & Decane, Tetradecan    &p-pTp, PMP, PDB      &  \\\hline
\end{tabular}
\end{center}
\caption{Chemicals generally used for liquid scintillator admixtures.}
\label{Q}
\end{table}

\subsection{Target}
The hydrogen atoms (``free proton'') in the liquid scintillator serve as the 
antineutrino target in the inverse beta decay reaction.
When a neutron is captured by a free proton, gamma rays with a total
energy of $\sim$2.2~MeV are emitted.  
On the other hand, a neutron capture on a Gd atom leads to an emission of gamma rays 
with a total energy of $\sim$8~MeV, much higher than the energies of the 
gamma rays from natural radioactivities which are normally below 3.5~MeV.
The mean thermal neutron capture cross section of Gd isotopes is 
four orders of magnitude larger than that of proton.
Hence the liquid scintillator doped with a small amount of Gd is 
ideal for detecting inverse beta decay events.

\begin{figure}
\begin{center}
\leavevmode
\includegraphics[width=4.5in]{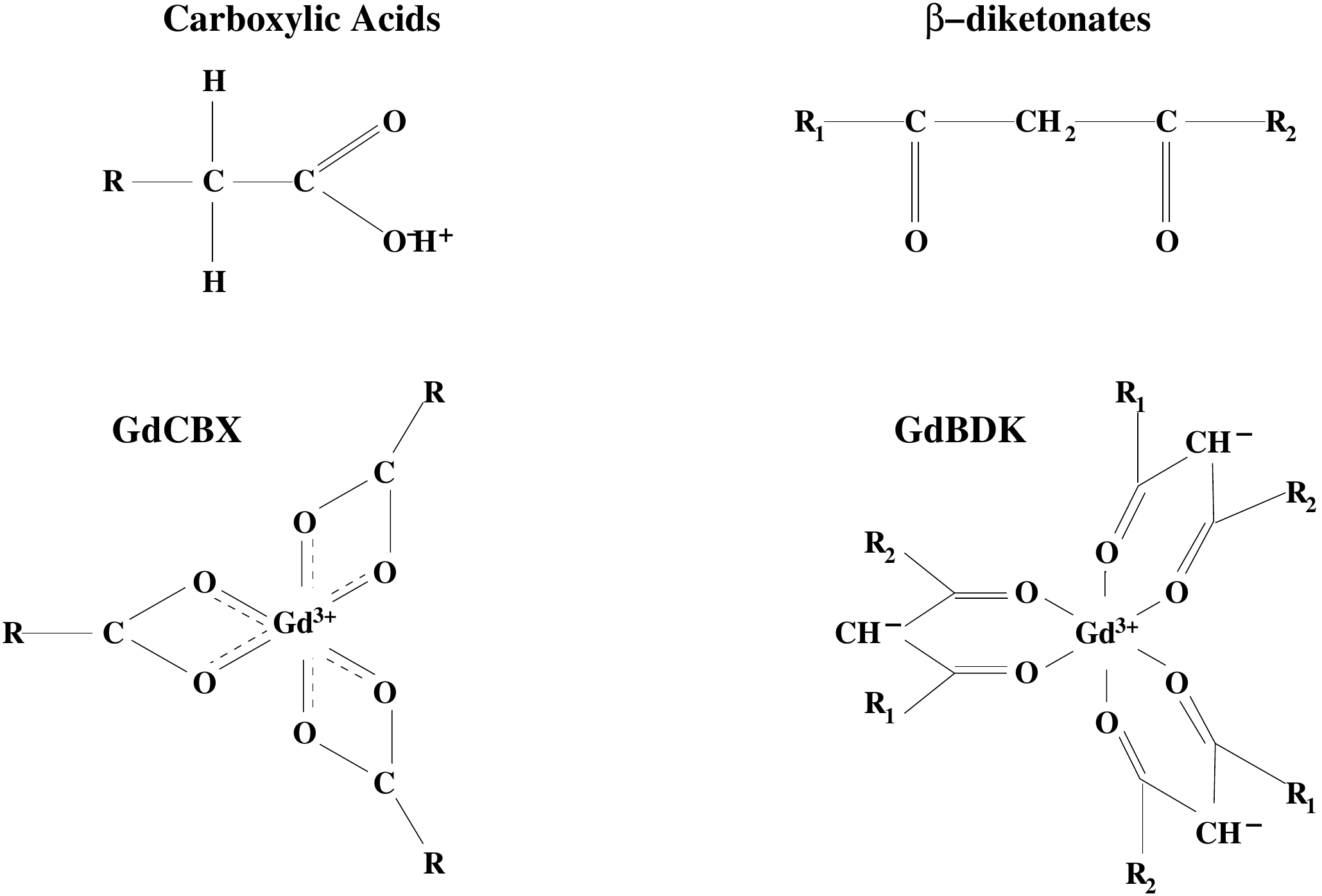}
\end{center}
\caption{Gd compound structures of carboxylic acid and $\beta$-diketonate 
ligands. A range of liquid carboxylic acid radicals with different alkyl 
chains exist: C2 (acetic acid), C3 (propionic acid), C4 (isobutyl acid), 
C5 (isovaleric acid), C6 (2-methylvaleric acid, C5H11COOH, HMVA), 
C8 (ethyl-hexanoic), and C9 (trimethyl-hexanoic).}
\label{acids}
\end{figure}

Gadolinium is a silvery white soft ductile metal belonging to the 
lanthanide group. It is one of the most abundant rare-earth elements.
It is never found as free element in nature, but is contained in many rare 
minerals. The metal does not tarnish in dry air, but oxide film forms in
moist air. Gadolinium reacts slowly with water and dissolves in acids. 
It can form stable organometalic complexes with ligands such as carboxylic 
acids (R-COOH) and $\beta$-diketones. Figure~\ref{acids} shows molecular 
structures of Gd compounds with ligands.

It is difficult to add inorganic Gd salt to organic liquid scintillator
to make a stable Gd loaded liquid scintillator. 
However, two formulations for Gd loaded liquid scintillator have shown
promising results; liquid scintillators with Gd binding with carboxylate (CBX) 
ligands and with $\beta$-diketonate (BDK) ligands.
Double Chooz and Daya Bay experiments report that both BDK and CBX Gd
loaded liquid scintillators have excellent performances.
Among these, TMHA is reported to be most promising~\cite{ligand}.

\subsection{Synthesis of Gd-Complex}
\begin{figure}
\begin{center}
\leavevmode
\includegraphics[width=3in]{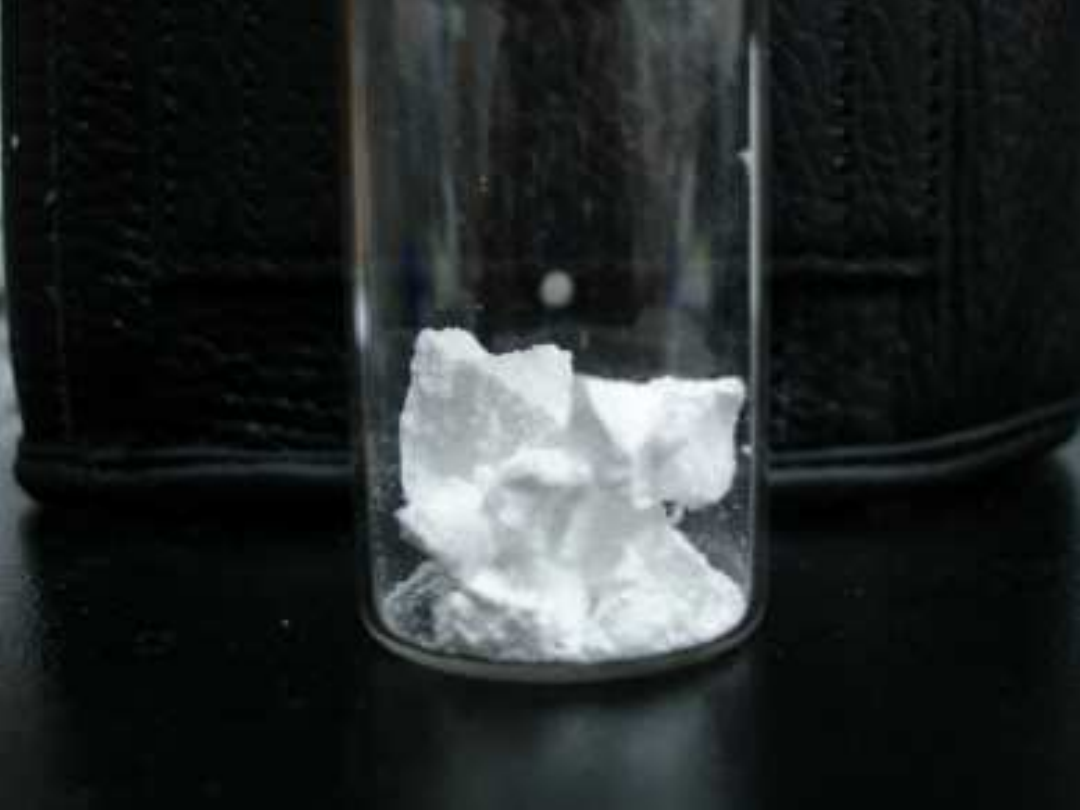}
\end{center}
\caption{White Gd-TMHA salt after filtration with 0.2~$\mu$m pore size 
Teflon membrane filter.} 
\label{Gd-TMHA}
\end{figure}

Since metal Gd by itself cannot be dissolved in the organic solvent, Gd 
salt with ligands is used. 
We chose to use CBX as our basis for ligands after consideration.
There are three 
steps in synthesizing the Gd-carboxylate compound:
\begin{enumerate}
\item\label{step1} $\rm{Gd}_{2}\rm{O}_{3} + 6\rm{HCl}
                   \rightarrow 2\rm{GdCl}_{3} + 3\rm{H}_{2}\rm{O}$          
\item\label{step2} $\rm{RCOOH}+ \rm{NH}_{3}* \rm{H}_{2}\rm{O} 
                   \rightarrow \rm{RCOONH}_{4} + \rm{H}_{2}\rm{O} $ 
\item\label{step3} $3\rm{RCOONH}_{4} (aqueous)+ \rm{GdCl}_{3} (aqueous) 
                   \rightarrow \rm{Gd(RCOO)}_{3} + 3\rm{NH}_{4}\rm{Cl} $ 
\end{enumerate}
First, we need to make $\rm{GdCl}_{3}$ solution from $\rm{Gd}_{2}\rm{O}_{3}$ 
based on step \ref{step1}. In step \ref{step2}, 3,5,5-trimethylhexanoic acid 
(TMHA) is neutralized with ammonium hydroxide. In step \ref{step3}, two aqueous 
solutions from steps \ref{step1} and \ref{step2} are mixed to produce Gd salt. 
When two solutions are mixed, white Gd-carboxylate compound (Gd-TMHA) 
precipitates immediately. These reactions are very sensitive to pH. Precipitated 
Gd-TMHA is thoroughly rinsed with 18~M$\Omega$ ultra pure water several times 
and then dried in vacuum desiccator. The final Gd-TMHA product is shown in 
Fig.~\ref{Gd-TMHA}. The yield of the synthesis is about 83\%.
On the other hand, if we purchase $\rm{GdCl}_{3}$ directly from vendor, 
we do not need step \ref{step1}. But $\rm{Gd}_{2}\rm{O}_{3}$ is much 
cheaper and we start from step \ref{step1}.

\begin{figure}
\begin{center}
\leavevmode
\includegraphics[width=4in]{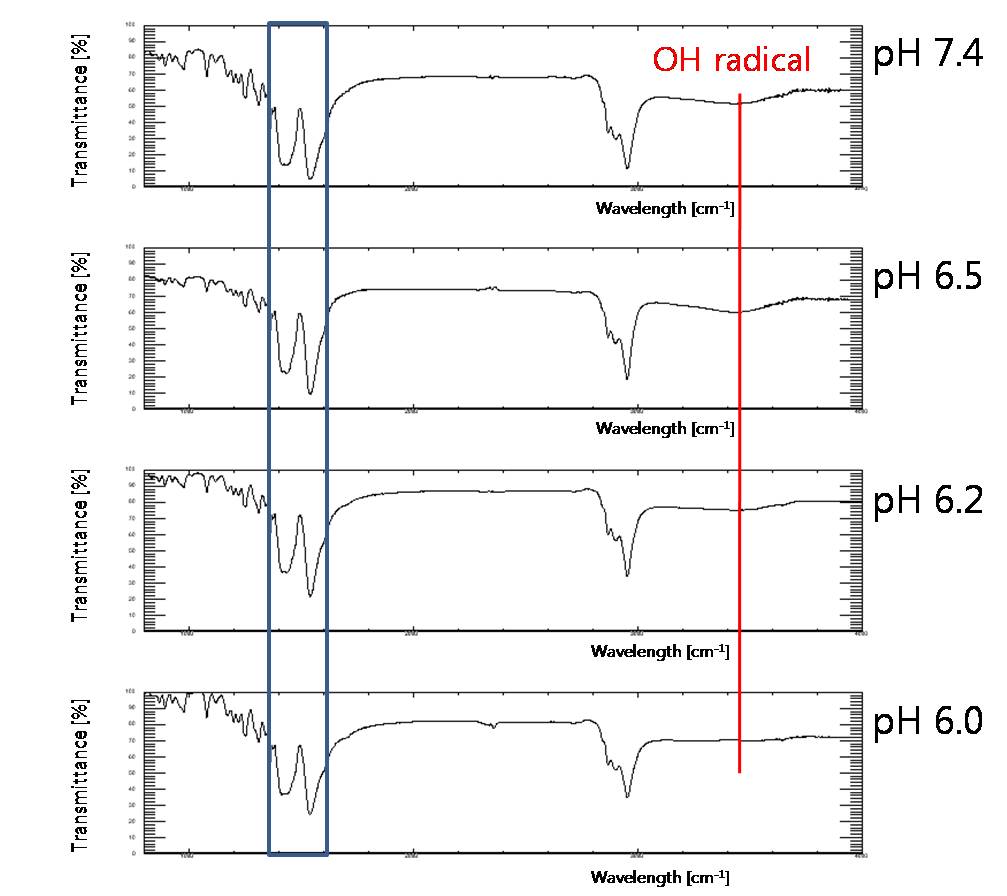}
\end{center}
\caption{Various FT-IR spectra for Gd-TMHA salt made at different pH conditions
in step (3). From top to bottom, values of pH in step \protect\ref{step3} of the
Gd-THMA synthesis are 7.4, 6.5, 6.2, and 6.0. The box shows where the carboxylic 
peaks are. The location of OH$^-$ radical peak is also shown.} 
\label{GdR3-IR}
\end{figure}

To study molecular structure and chemical bonds in organic compounds, we 
use Fourier Transform Infrared (FT-IR) spectrum technique. 
We require three conditions for our final sample to make pure Gd-TMHA. 
\begin{itemize}
\item No OH$^-$ radical in FT-IR ($3200 \sim 3500~\rm{cm}^{-1}$).
\item No free acid peak ($\sim 1700~\rm{cm}^{-1}$) in FT-IR. 
\item Presence of carboxylic peak ($\sim 1420$ and 1580~cm$^{-1}$).  
\end{itemize}
Figure~\ref{GdR3-IR} shows various FT-IR spectra for the Gd-TMHA 
with different pH conditions in step \ref{step3}. We can see that 
there are no OH$^-$ radicals or free acid group left
in pH = 6 case.    
    
\subsection{Gamma Catcher}
The purpose of $\gamma$-catcher is to contain gamma rays escaping from 
target region thereby providing correct energy measurement. 
The $\gamma$-catcher is a liquid scintillator, without Gd, enclosed in 
an acrylic vessel.

\begin{figure}
\begin{center}
\leavevmode
\includegraphics[width=3in]{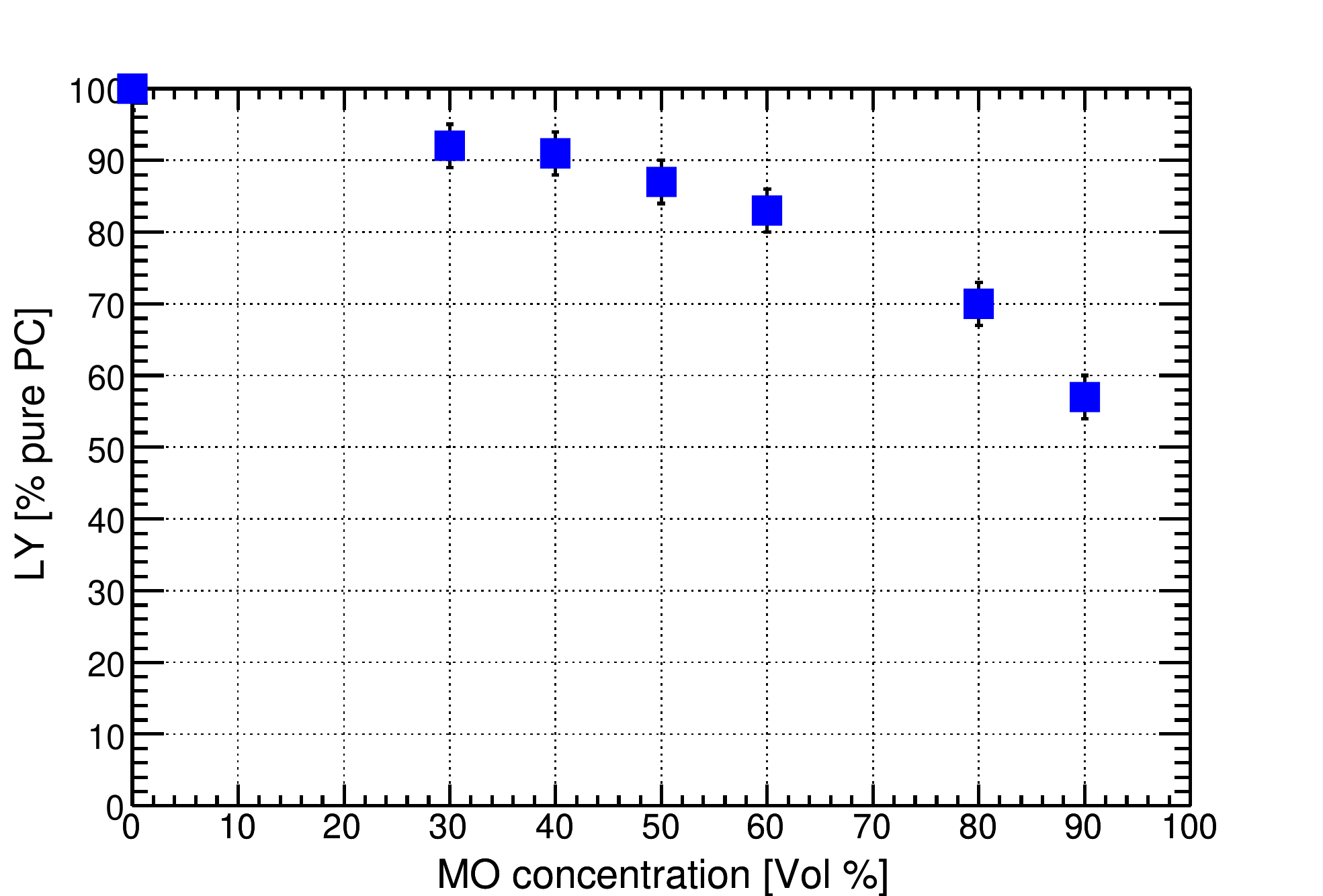}
\end{center}
\caption{Scintillation light yield of PC and MO mixture based scintillator with
varying mixture ratio with respect to that of 100\% PC based scintillator. The 
PPO concentration is kept constant at 3~g/l.}
\label{LY-MO}
\end{figure}

Figure~\ref{LY-MO} shows the scintillation yield of the non Gd-loaded, PC based 
scintillator as a function of MO concentration. The light yield of a sample 
is determined by measuring Compton edge from a $^{137}$Cs source. 
A scintillation yield of 80$\%$ with respect to pure PC is observed at a 
volume fraction of 60$\%$ MO.
This figure shows how the light yield decreases for increasing dilution of the 
primary solvent by MO.  
The light transmittance of a scintillator with a solvent mixture of 60$\%$ MO and 
40$\%$ PC with varying PPO concentration was measured with spectrophotometer. 

\subsection{Buffer}
To decrease the level of the accidental radioactive backgrounds, mainly 
coming from the PMT glass and surrounding rocks, non-scintillating oil 
is used to shield the scintillating layers from radioactive sources.
Since the PMTs are immersed in a non-scintillating buffer, the buffer
oil should have a good transmittance to the light coming from the
scintillating layers. 
Also, the buffer oil should have a density similar to those of the target and the
$\gamma$-catcher. MO and dodecane are possible candidates for the buffer oil.

\begin{figure}[htbp]
  \centering
  \includegraphics[width=7.7cm]{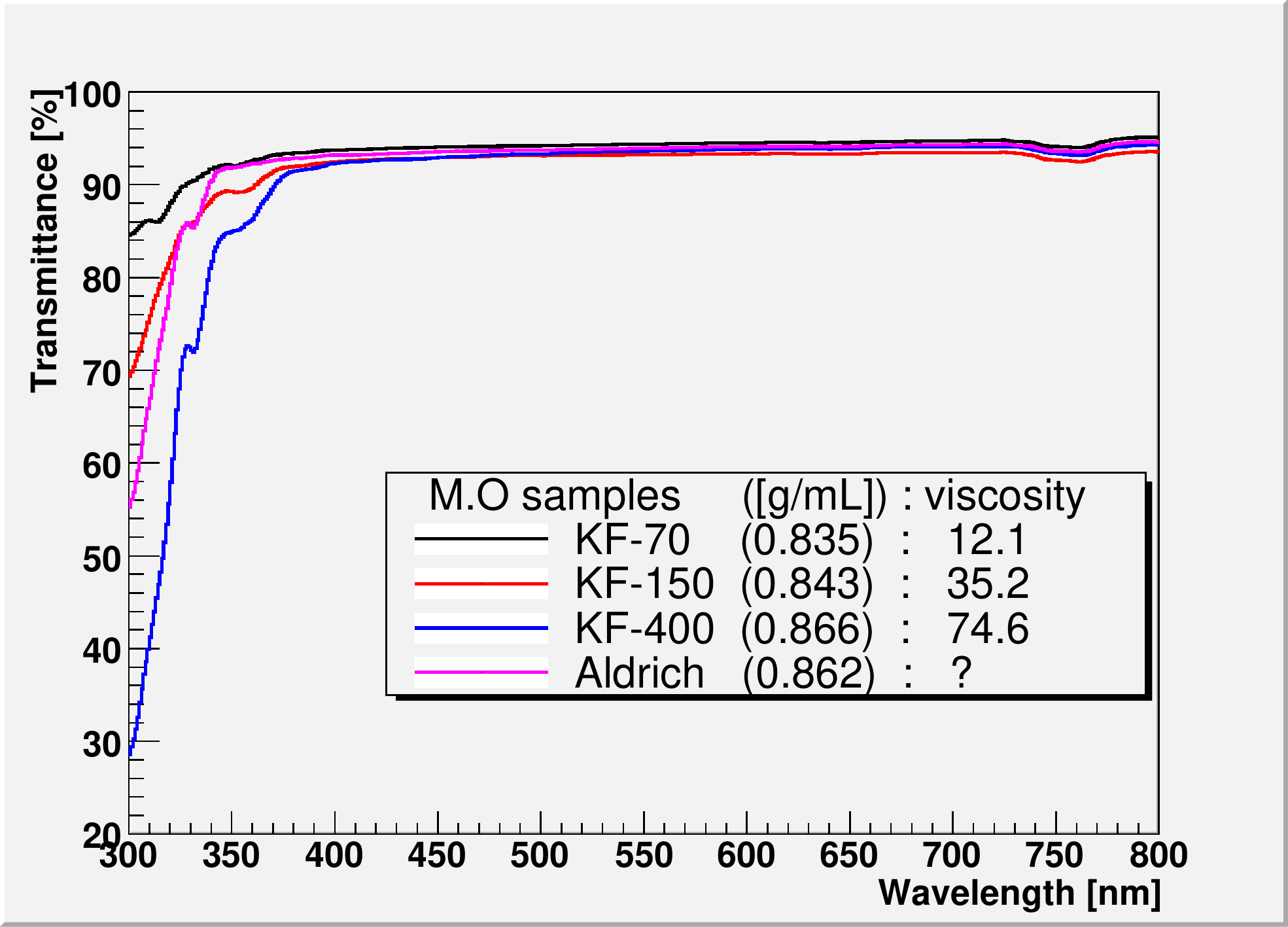}
  \caption[solution]{Transmittance of MO with different kinetic viscosity values.
MO with low kinetic viscosity has better transmittance below $\sim 400$~nm.}
\label{fig_transmittance1}
\end{figure}

Spectrophotometry has been used to measure optical transparency of MO. 
The transmittance of a collimated light beam through a sample in a 10-cm 
quartz cell is measured with a spectrometer.
Figure~\ref{fig_transmittance1} shows transmittance of MO with various 
viscosity values.
At wavelength above 400~nm the transmittance is flat without much differences
among MOs with different kinetic viscosity values,
whereas below 400~nm transmittance drops quickly with varying degrees 
depending on kinetic viscosity of solvent.

\section{Fluor and Wavelength Shifter}
\begin{figure}[htbp]
  \centering
  \includegraphics[width=7.7cm]{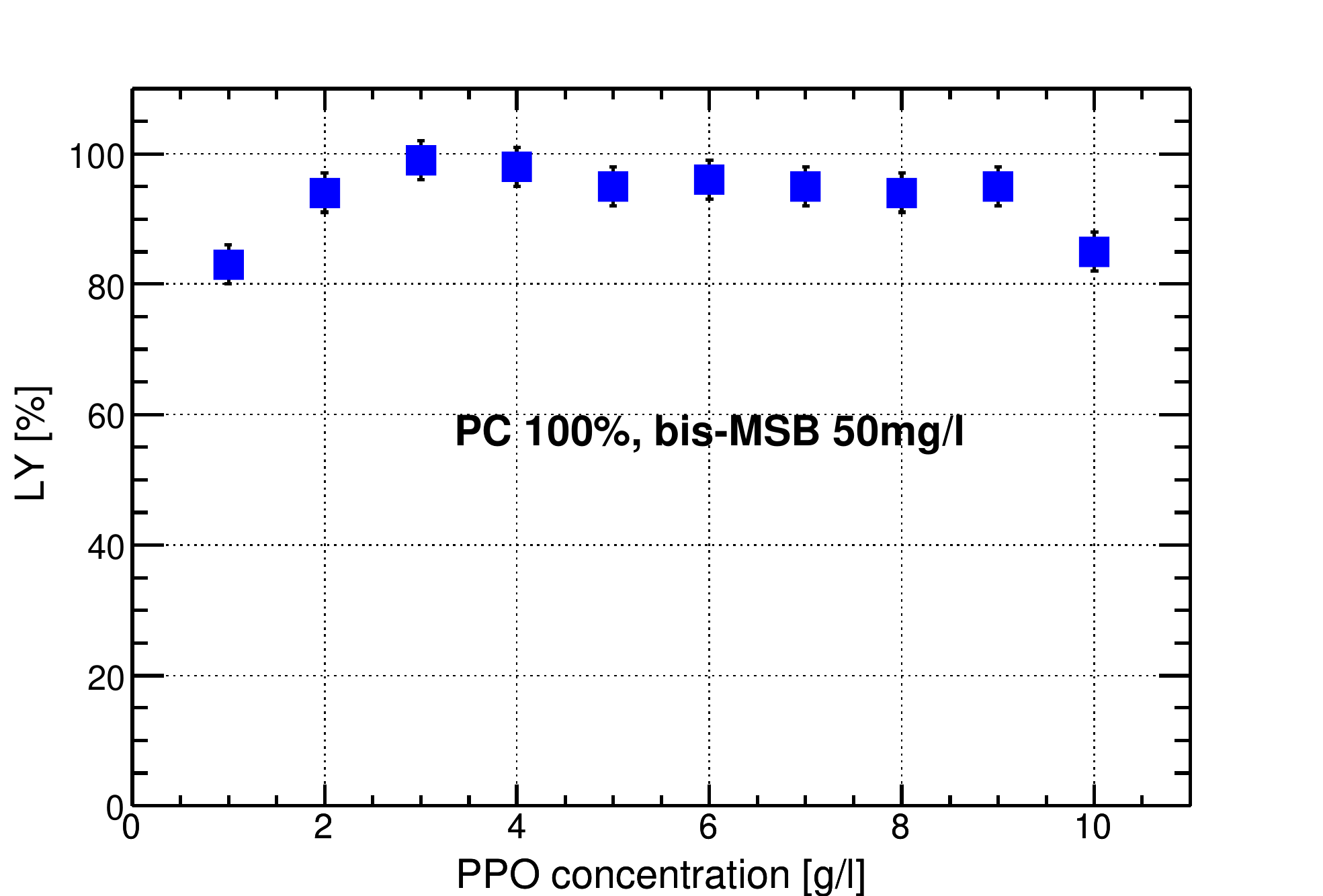}\hfill
  \includegraphics[width=7.7cm]{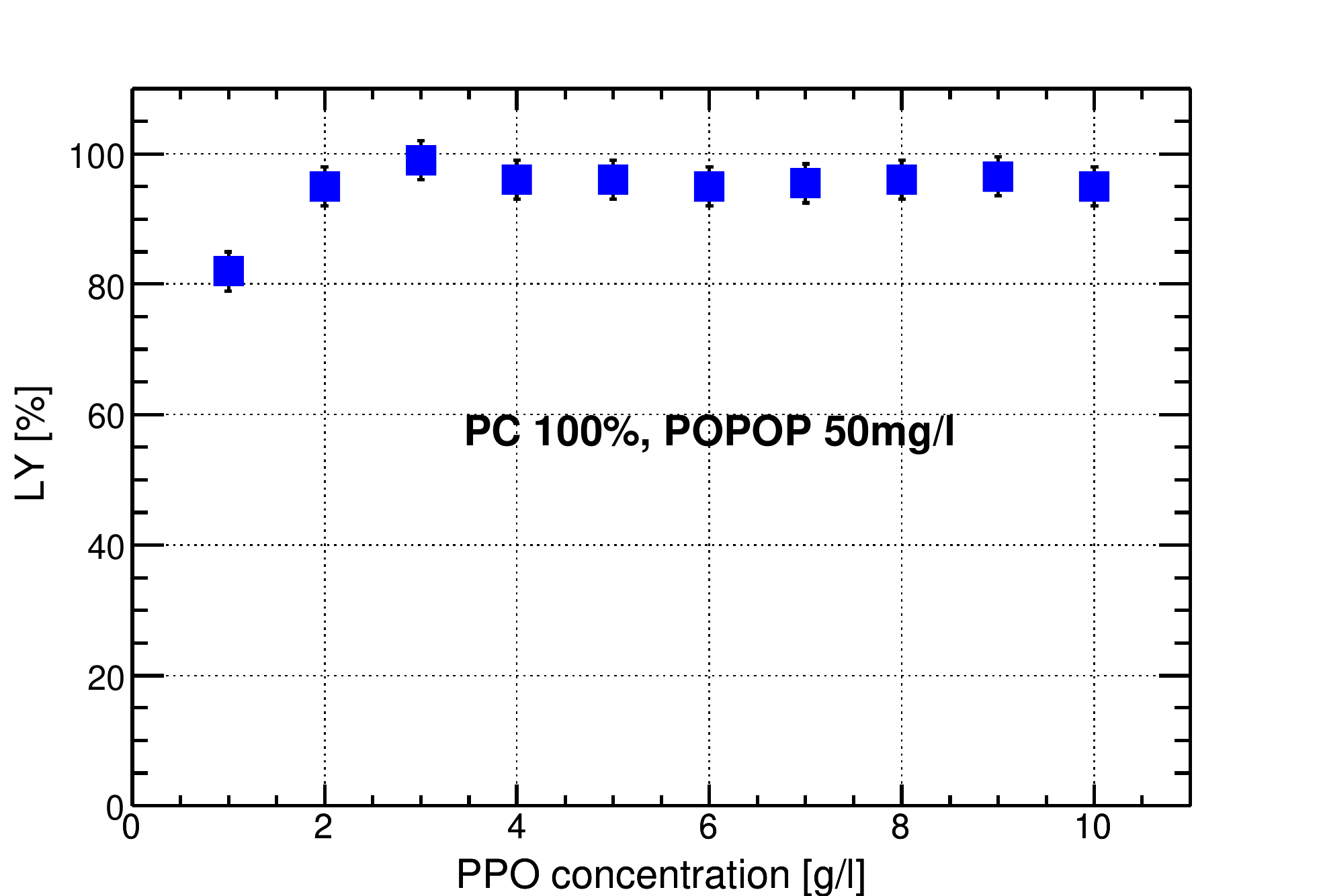}
  \caption[solution]{Relative light yield with different PPO concentrations 
in PC with respect to the light yield at 3~g/l PPO concentration. 
Bis-MSB (left) and POPOP (right) concentrations are both kept constant 
at 50~mg/l.}
\label{fig_fluors}
\end{figure}

The light yield was measured at different fluor concentrations from 1
to 10~g/l.
Figure~\ref{fig_fluors} shows the maximum light yield at PPO concentration 
about 3~g/l for both scintillators either with bis-MSB or POPOP as a wavelength 
shifter. 
At low PPO concentrations, the light output of the liquid scintillator is 
small. 
At concentrations above 3~g/l, the change in light output is very small. 
POPOP and bis-MSB are widely used as second wavelength shifter.
Although the light yield is similar for the liquid scintillator with POPOP 
and the one with bis-MSB, it takes longer to dissolve POPOP in the solvent 
than bis-MSB.
A second wavelength shifter is needed to change the wavelength of light 
from a fluor to about 410~nm, which is a sensitive region of bi-alkali 
type photocathode.

\section{Long-term Stability}
Besides the light yield, stability of the Gd loaded scintillator is 
another crucial matter. Gd-LS should be chemically stable for the duration 
of the experiment, {\it i.e.} several years. From Palo Verde and CHOOZ reactor 
experiments, unexpected problems with Gd-LS had been reported. Palo Verde had 
problems with precipitation, condensation, and slow deterioration of Gd-LS 
developing in time. In CHOOZ experiment, Gd-LS turned yellow a few months 
after deployment. A very rapid decay of attenuation length in Gd-LS had 
been measured~\cite{LS2}.

Organic solvents can be oxidized in the presence of oxygen or water and
develop coloration.
This oxidization is accelerated by UV light and heat. 
Therefore, care should 
be taken to assure
that moisture or humid air do not enter the sample. 
The liquid scintillators are flushed with nitrogen gas to purge oxygen and 
stored in air tight containers.

The long-term stability of the Gd-LS and liquid scintillator is 
investigated by means of spectrophotometric techniques. 
The transmittance is routinely measured in the wavelength range 
of $300~\hbox{nm} \leq \lambda \leq 800$~nm.   
The absorbance, $A$, is defined as  
\begin{equation}
A = -\log_{10}\left(\frac{I}{I_o}\right),
\end{equation}
where $I_o$ and $I$ are the intensities 
of the incident and the emerging lights, respectively. It is the 
absorbance that displays a simple dependence on the density and 
sample path length. 

For extracting the attenuation length, Beer-Lambert-Bouguer law is used. 
It shows absorption of a beam of light as it travels through 
liquid for a distance $L$.
The attenuation length, $\lambda$, at 420~nm is expressed as 
\begin{equation}
\lambda = 0.4343 \left(\frac{L}{A_{420}}\right), 
\end{equation}
where $L$ is path length in the sample and $A_{420}$ is the absorbance 
at 420 nm. 
For a long term stability test, closed type cuvettes have been used to reduce
oxygen contamination.

\section{Material Compatibility}
The target, $\gamma$-catcher, and buffer are in contact with acrylic vessels, 
so material compatibility of organic liquids with acrylic is crucial. Material 
compatibility tests for Gd-LS were performed. 
The admixture of 40\% PC and 60\% MO by volume provides
sufficient material compatibility and scintillation light yield. 
Material tests with other solvents also had been carried out. 
It was reported by Double Chooz collaboration that the Gd-CBX 
scintillator could react with steel~\cite{LS1}. 
Therefore, during the production of Gd-LS, any contact with steel surface 
will be avoided for the stability of Gd-LS. 


\section{R$\&$D on LAB based Liquid Scintillator}
\subsection{Introduction}
Recently, it has been brought attention of researchers to Linear Alkyl 
Benzene (LAB) as a possible solvent replacing PC and MO.
LAB has been reported as having good light yield as well as desirable
optical properties, {\it i.e.} high transmittance and large attenuation
length. Unlike PC, it is non-toxic and bio-degradable. Also, it
has a high flash point and, thus, can be handled safely.
And it does not chemically interact with acrylic and stainless
steel.
It is commercially mass produced, cheap, and readily available.

\begin{figure}
\begin{center}
\leavevmode
\includegraphics[width=3.0in]{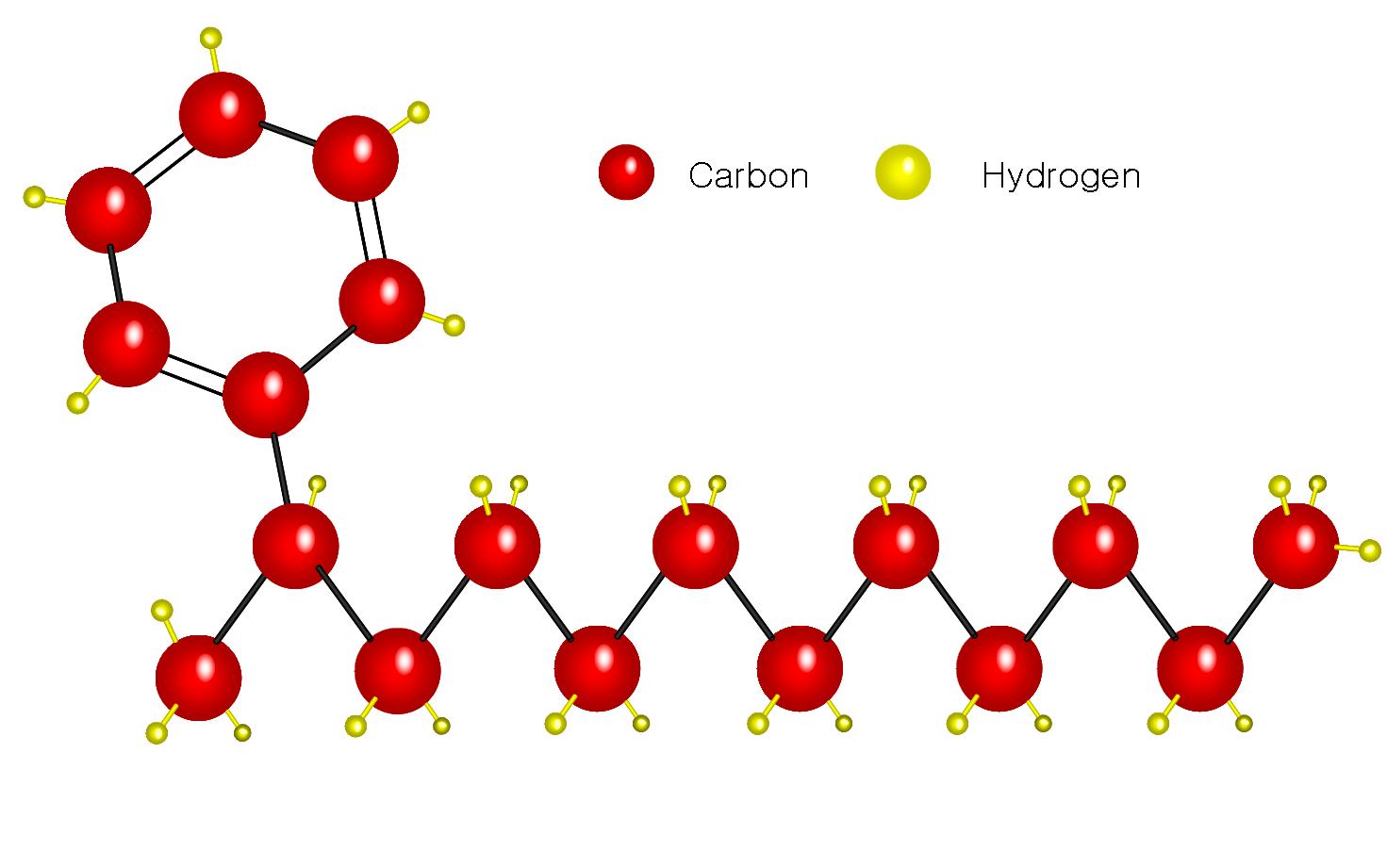}
\includegraphics[width=2.0in]{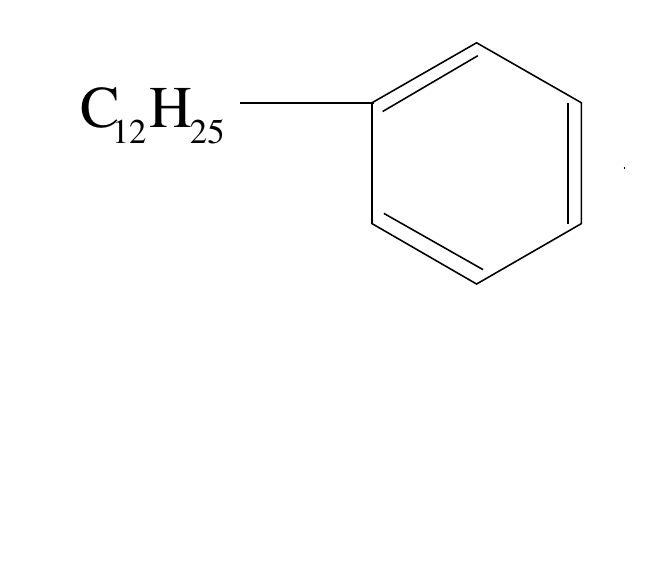}
\end{center}
\caption{Molecular structure of LAB with a linear alkyl chain C$_{12}$H$_{25}$.}
\label{MOL}
\end{figure}

As shown in Fig.~\ref{MOL}, LAB (C$_n$H$_{2n+1}$-C$_{6}$H$_{5}$, 
$n = 10 \sim 13$) is composed of a linear alkyl chain of $10 \sim 13$ 
carbon atoms attached to a benzene ring. 
Its density is 0.86~g/ml and is compatible to other organic 
liquids used for the experiment. 
Our current R$\&$D is focused on a new liquid scintillator using LAB.
Comparison of some chemical properties between LAB and PC is summarized 
in Table~\ref{PC-vs-LAB}.  
LAB can be obtained domestically in Korea from ISu Chemical Ltd~\cite{LS5}. 

\begin{table}
\begin{center}
\begin{tabular}{ccc}\hline
        &PC &LAB
\\\hline
Molecular formula         &C$_{9}$H$_{12}$ 
                          &C$_n$H$_{2n+1}$-C$_{6}$H$_{5}$, n = 10 $\sim$ 13 \\  
Molecular weight (g/mol)  &120.19            &233$\sim$237   \\
Flash point ($^o$C)       &48                &130  \\
Density (g/ml)            &0.89              &0.86 \\
Compatibility(acrylic)    &Bad, need diluent &Good \\
Cost                      &Moderate          &Low \\
Fluor dissolution         &Very good         &Moderate \\
Domestic availability     &No		     &Yes \\
Toxicity                  &Toxic fume        &Non toxic \\\hline
\end{tabular}
\end{center}
\caption{Comparison between PC and LAB.}
\label{PC-vs-LAB}
\end{table}

\subsection{Purification of LAB}\label{Purification of LAB}
Purification of liquid scintillator is performed to remove chemical 
impurities as well as particulates containing radioactive isotopes. 
It enhances light transmittance and long term stability of liquid 
scintillator. 
Various purification methods were investigated by neutrino experiments;
adsorption by activated Al$_2$O$_3$ or silica gel, water extraction,
vacuum distillation, and filtration by a particulate filter.  
Adsorption method has many advantages. Specific separation can be done 
based on high selectivity of adsorbents. A large amount of liquid can 
be continuously purified. Drawbacks include need for periodic 
replacement of adsorbent and slow purification speed for high viscosity
liquids. 

LAB was purified by adsorption by passing LAB through a column of 
activated Al$_2$O$_3$.
Unpurified LAB samples obtained from ISu Chemical Ltd have an attenuation 
length $7\sim10$~m at 420 nm, depending on production batches. These samples 
were delivered in 200~$l$ steel drums. 
The absorbance of purified LAB started to decrease after passing LAB
five times the volume of Al$_2$O$_3$ used through the column.
Therefore, we replaced Al$_2$O$_3$ in the column with fresh one 
for each volume of LAB five times that of Al$_2$O$_3$ used.

\begin{figure}
\begin{center}
\leavevmode
\includegraphics[width=3.5in]{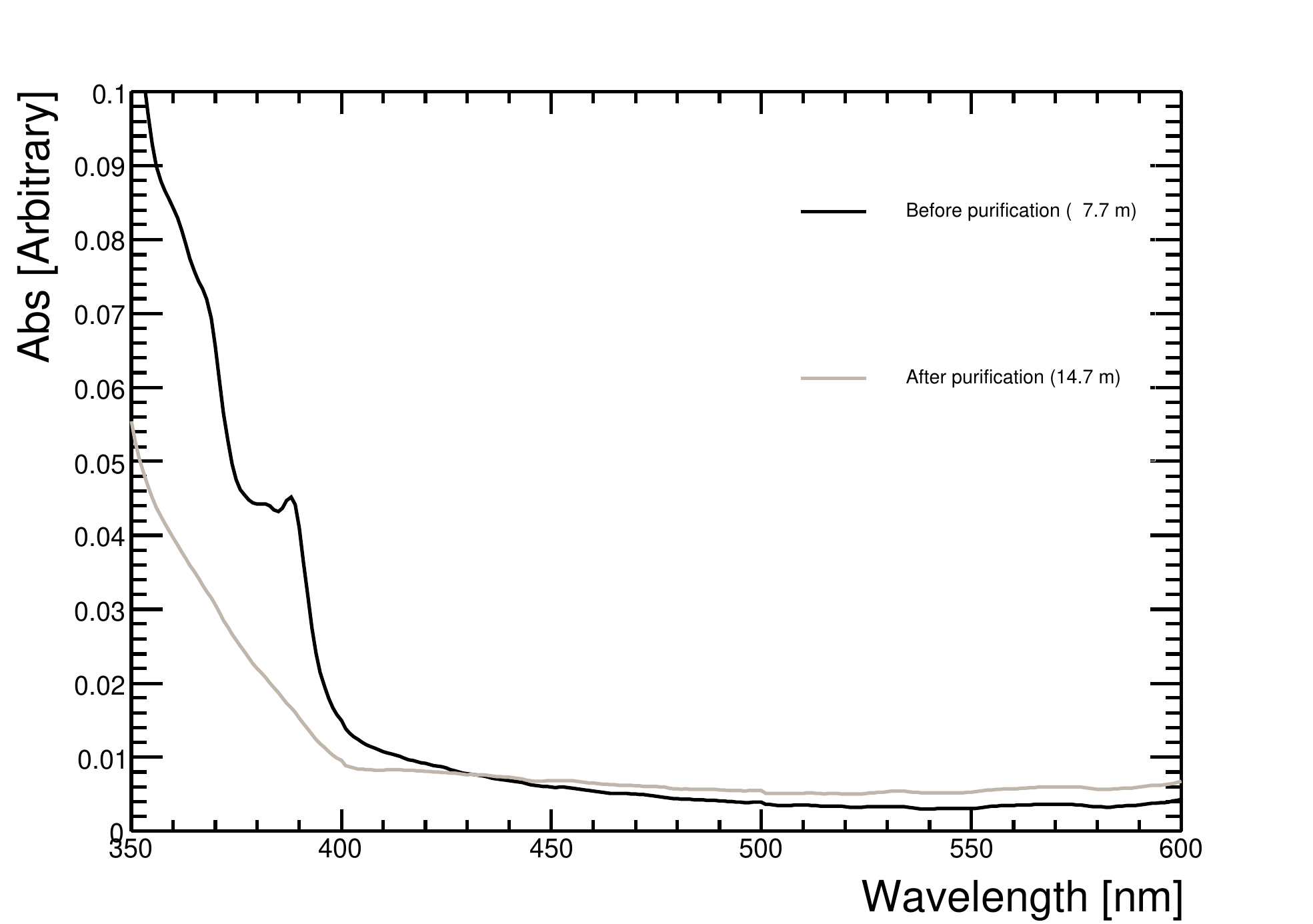}
\end{center}
\caption{Absorption spectra of a LAB sample before and after purification
by Al$_2$O$_3$ column separation. Absorbance is 
measured using Jasco UV/VIS 530 spectrometer. Two curves have different
baseline values.} 
\label{LAB-B-A}
\end{figure}

Figure~\ref{LAB-B-A} shows UV-visible spectra of LAB before and after 
Al$_2$O$_3$ purification. 
The attenuation
lengths at 420~nm calculated from absorption data before and after purification 
are $7.7\pm0.3$~m and $14.7\pm 0.3$~m, respectively.
This shows that Al$_2$O$_3$ purification method could be used to improve 
optical properties of LAB.

ISu Chemical Ltd. kindly provided a high quality LAB sample obtained from 
the upstream in their production line specially for this test.
This sample was handled with care and delivered
in very clean plastic containers. First, we measured its attenuation 
length without any purification. The result showed that the attenuation length 
is greater than 20~m. Based on this, we decided that purification with 
Al$_2$O$_3$ is not necessary to improve the attenuation length. 
Then we tried filtration on the sample with a Teflon membrane filter 
with 0.2~$\mu$m pore size, but it also did not improve the attenuation 
length. From these results, it is evident that 
a high quality LAB does not need to go through purification processes
to improve the attenuation length.
Table~\ref{lambda_LAB} lists the attenuation lengths for several LAB 
samples. 

\begin{table}
\begin{center}
\begin{tabular}{cc}\hline
Conditions  & Attenuation Length (m) at 420~nm
\\\hline
Steel drum, unpurified                 & 7 $\sim$ 10 \\
Steel drum, purified by  Al$_2$O$_3$   & 15 \\
Best quality, unpurified               & $>$ 20 \\
Best quality, filtered                 & $>$ 20 \\\hline
\end{tabular}
\end{center}
\caption{The attenuation lengths of four LAB samples 
provided by ISu Chemical Ltd. in different conditions.}
\label{lambda_LAB}
\end{table}

\subsection{LAB Composition Measurement with GC-MS}
\label{LAB composition}
\begin{figure}
\begin{center}
\leavevmode
\includegraphics[width=5in,clip=true,trim=0mm 0mm 0mm 50mm]{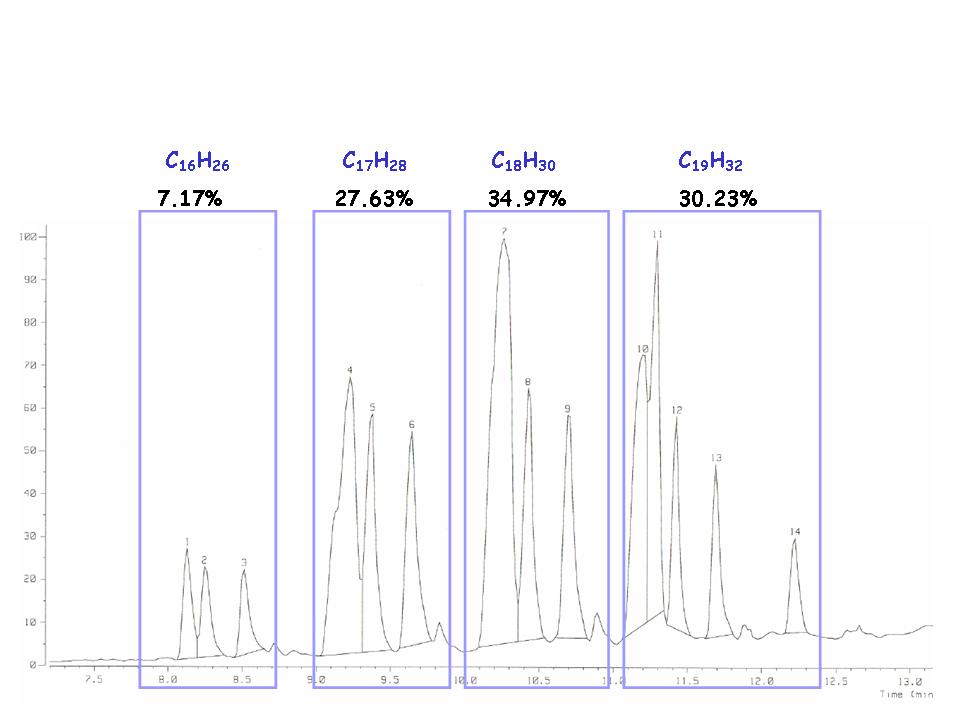}
\end{center}
\caption{LAB components identified with GC-MS. This LAB sample consists of 
four main components; C$_{16}$H$_{26}$, C$_{17}$H$_{28}$, C$_{18}$H$_{30}$, 
and C$_{19}$H$_{32}$. 
}
\label{GC-MS}
\end{figure}

To reduce the systematic uncertainty between near detector and far 
detectors at RENO experiment, it is important to know the exact 
composition of LAB. The composition of a LAB sample was measured by Gas 
Chromatography with Mass Spectrometry (GC-MS) at Korea Basic Science 
Institute~\cite{LS6}.
The results are shown in Table~\ref{comp} and Fig.~\ref{GC-MS}. Based on this 
measurement, we can calculate the number of protons and H/C ratio of the 
sample. The sample has a H/C ratio of 1.66. Table~\ref{table_proton} shows 
the proton densities of PC, PXE, dodecane, and LAB.  

\begin{table}
\begin{center}
\begin{tabular}{ccc}\hline
   n &Composition ($\%$)
\\\hline
10     &7.17   \\
11     &27.63  \\
12     &34.97  \\
13     &30.23 
\\\hline
\end{tabular}
\end{center}
\caption{LAB composition measured with GC-MS method. Molecular formula of 
LAB is C$_n$H$_{2n+1}$-C$_{6}$H$_{5}$, $n = 10 \sim 13$.}
\label{comp}
\end{table}
\begin{table}
\begin{center}
\begin{tabular}{cc}\hline
Solvent  & Proton Density (/m$^3$)       \\ \hline
LAB(C$_n$H$_{2n+1}$-C$_{6}$H$_{5}$, $n = 10\sim 13$)  &$0.631\times 10^{29}$   \\
PC(C$_{9}$H$_{12}$)                          &$0.530\times 10^{29}$   \\
PXE(C$_{16}$H$_{18}$)                        &$0.512\times 10^{29}$   \\
Dodecane(C$_{12}$H$_{26}$)                   &$0.694\times 10^{29}$   \\\hline
\end{tabular}
\end{center}
\caption{Numbers of hydrogen atoms per m$^3$ of various solvents 
for liquid scintillators. 
The composition of LAB is given in Table~\protect\ref{comp}.
}
\label{table_proton}
\end{table}
\subsection{Optical Properties}
The acrylic that target and $\gamma$-catcher vessels are made of
becomes opaque rapidly for light with wavelength below 400~nm. 
And the PMTs being used (R7081, Hamamatsu) is most sensitive to
light at $\sim 390$~nm but retains good sensitivity at $400\sim430$~nm. 
Therefore, it is imperative for the liquid scintillator emitting 
light at $\sim 420$~nm.

\begin{figure}[htbp]
\begin{center}
\includegraphics[width=11.8cm,clip=]{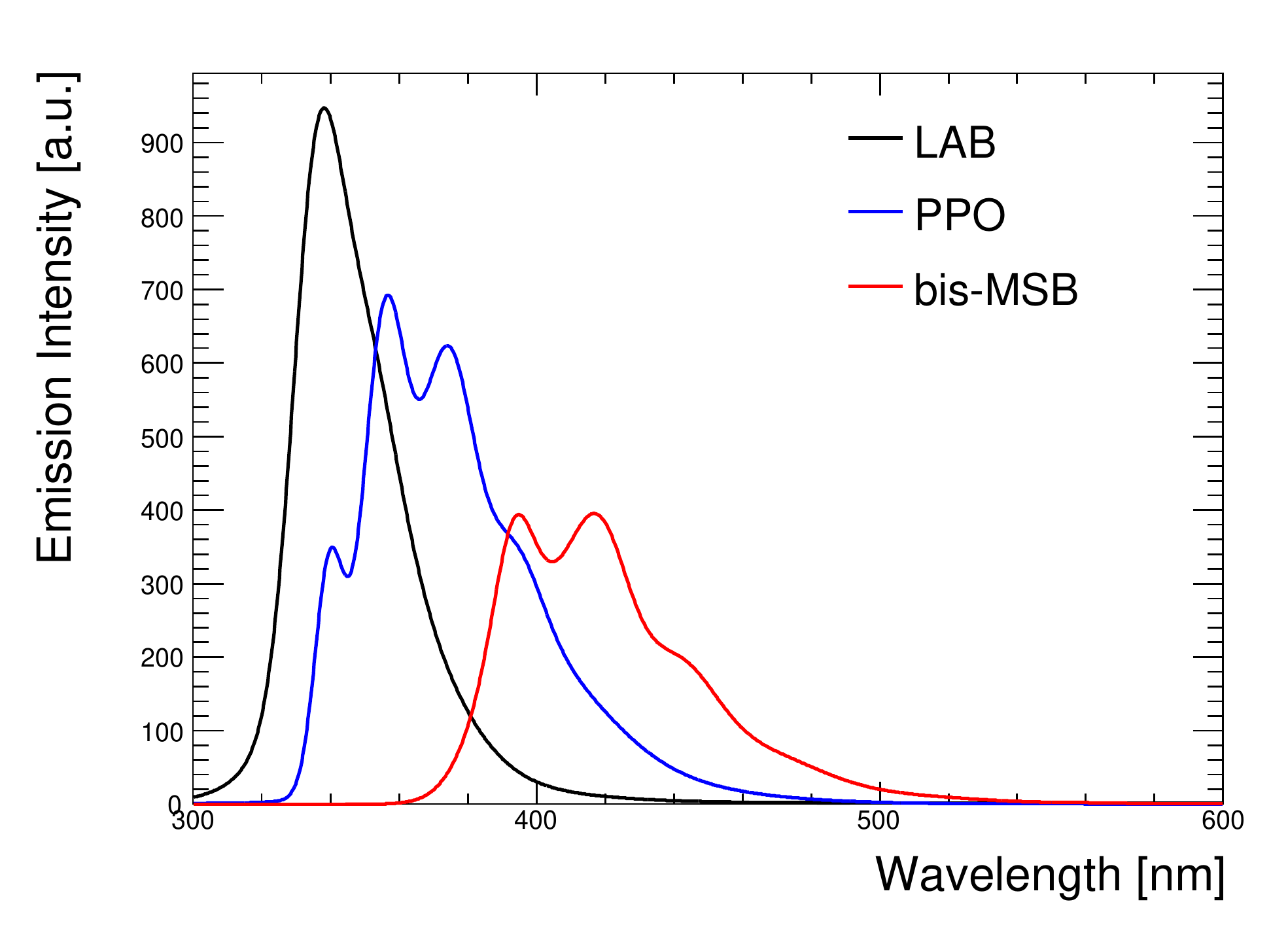}
\caption{Emission spectra of the solvent LAB (black), the primary 
fluor PPO (blue), and 
the wavelength shifter bis-MSB (red).\label{emissionAll}}
\end{center}
\end{figure}

The optical and scintillation properties of the pure LAB solvent 
and of LAB-fluor mixture have been investigated by UV/VIS spectrometry 
The emssion spectra of LAB, PPO, and bis-MSB are shown in 
Fig.~\ref{emissionAll}.
The pure LAB solvent shows an emission maximum at 340~nm.
Therefore, the wavelength of the scintillation light from LAB needs 
to be shifted above 400~nm. This is achieved by using PPO as a 
primary solute and bis-MSB as a secondary wavelength shifter.
As shown in Fig.~\ref{emissionAll}, PPO and bis-MSB emit photons   
at $340\sim 440$~nm and $380\sim460$~nm, respectively. 

\begin{figure}[htbp]
  \begin{center}
    \includegraphics[width=11.8cm,clip=]{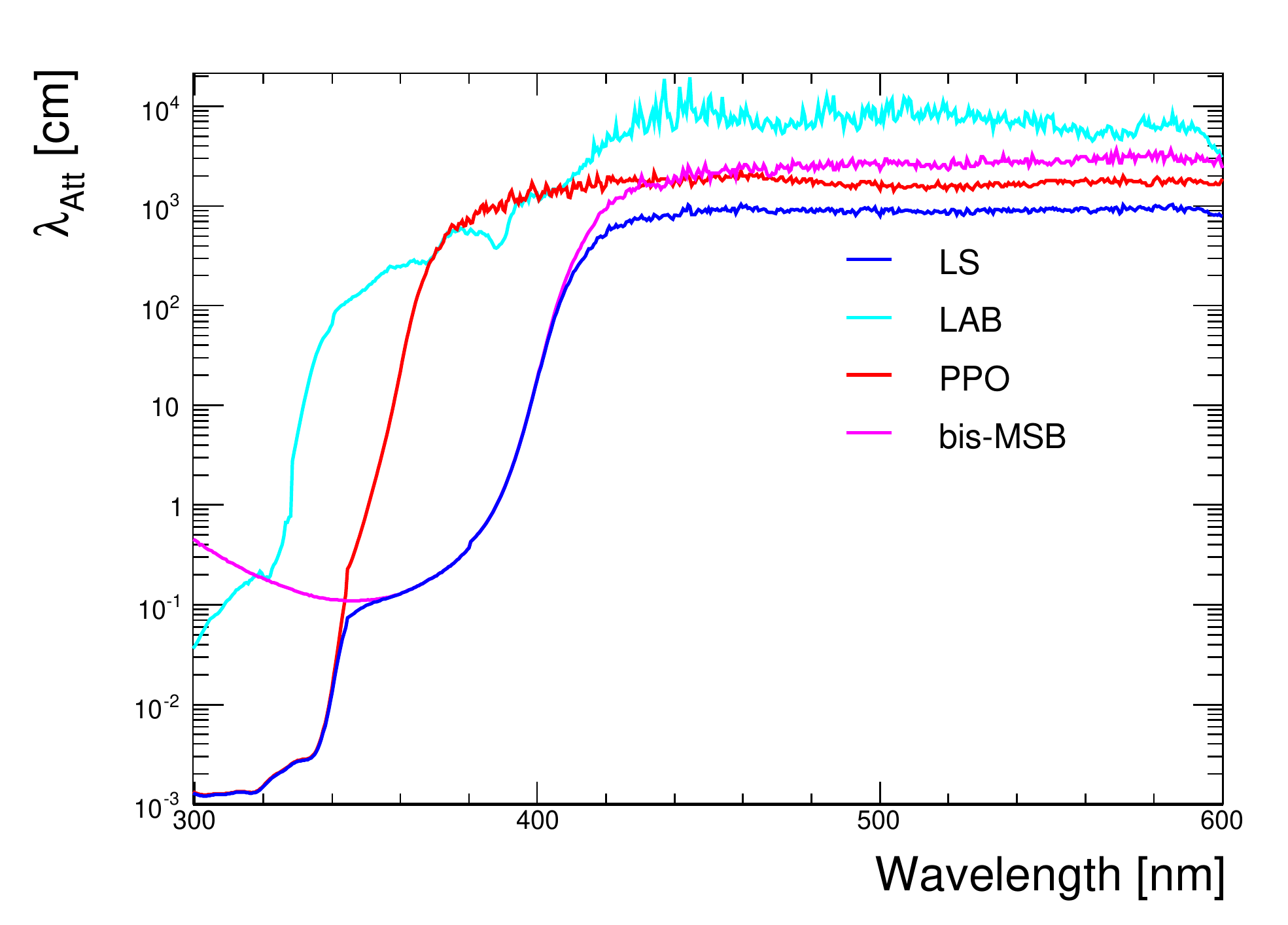}
    \caption{Attenuation lengths of the LAB based LS with 3~g/l PPO
             and 30~mg/l of bis-MSB (black), LAB (light blue), 3~g/l PPO
             (red), and 30~mg/l bis-MSB (light purple).
             Cyclohexane was used as the solvents when measuring PPO and 
             bis-MSB attenuation lengths.}
    \label{attlength}
  \end{center}
\end{figure}

The LAB based liquid scintillator has 3~g/l PPO and 30~mg/l bis-MSB
dissolved in LAB. 
The attenuation lengths of the LAB based scintillator and its individual 
components
are shown in Fig.~\ref{attlength}. 
The sample's attenuation lengths were measured
using a JASCO V530 UV/VIS spectrophotometer
with a sample in 10-cm quartz cuvettes.
The attenuation length of LAB was measured to be 25.2~m
at 420~nm.
The attenuation length of the liquid scintillator has been measured to 
be 7.1~m at 420~nm.
Based on the comparison of the attenuation lengths of LAB and LAB based
liquid scintillator, it is evident that the absorption of photons with 
wavelength shorter than 420~nm is mostly by fluor. 

\begin{figure}[htbp]
\begin{center}
\includegraphics[width=11.8cm,clip=]{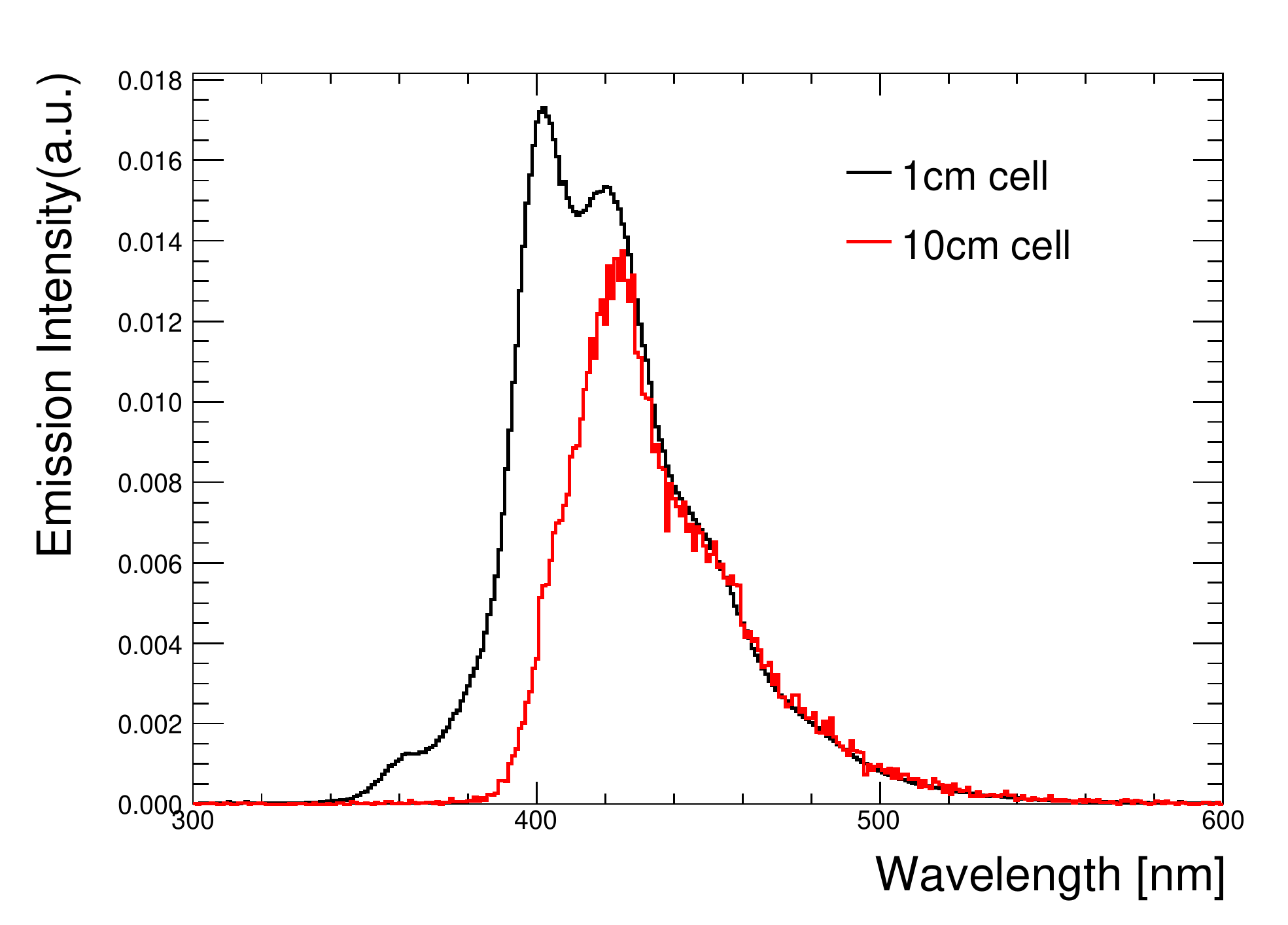}
\caption{Fluorescent spectra of LS measured using 1~cm (black) and 10~cm 
(red) sample cells.
The spectrum shapes are alike above $\sim 430$~nm but quite different
below, where absorption by bis-MSB is dominant. This indicates the 
short wavelength photon escaping scintillation volume before 
fully red-shifted through absorption-reemission process.}
\label{comp101}
\end{center}
\end{figure}

Below the wavelength $\sim 340$~nm, photons are absorbed mostly by 
PPO and $340\sim 400$~nm by bis-MSB.
PPO and bis-MSB molecules absorb a photon emitted by LAB and 
themselves and emit longer wavelength photons. 
A reemitted photon has
random direction with respect to the absorbed photon.
This absorption-reemission process could occur multiple times until 
either the photon escapes the scintillator volume or its wavelength 
falls in a region where PPO and bis-MSB absorption probability 
is negligible. 
Since the absorption probability of PPO and bis-MSB is much higher for 
the shorter wavelength photons than for the longer ones, successive 
absorption and reemission result in progressive red-shift of the 
spectrum (Fig.~\ref{comp101}). 
The small attenuation length of liquid scintillator below 400~nm 
indicates that most of the absorption and reemission processes
happen close to the location of the initial scintillation process.

\begin{figure}[htbp]
\begin{center}
\includegraphics[width=11.8cm,clip=]{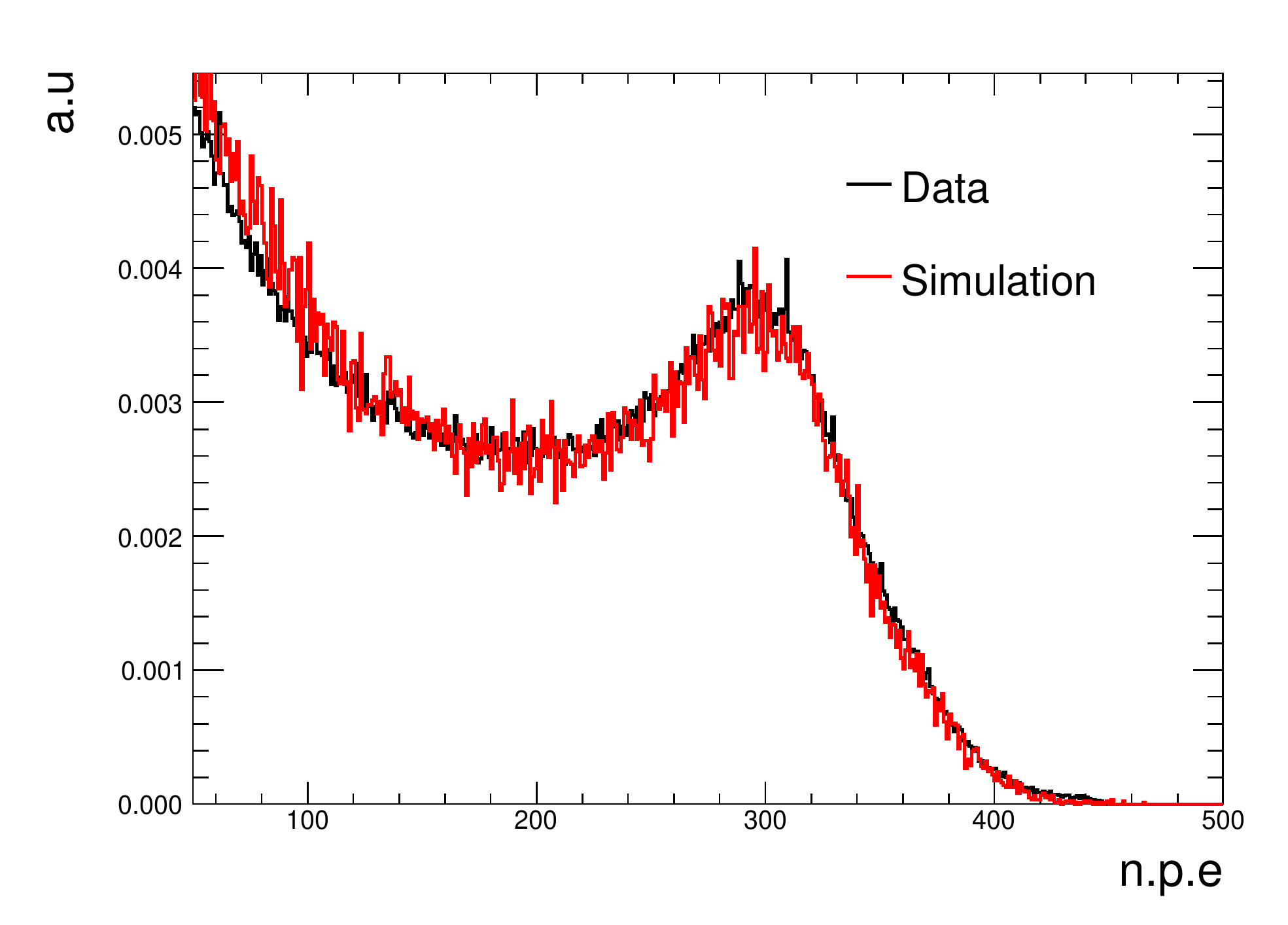}
\caption{The number of photoelectrons distribution of $^{137}$Cs measured with 
the light yield measurement setup. Data is shown in black and simulation 
in red. The light yield value used for the simulation is 10\,000/MeV.
\label{LY6300_vs_MC}}
\end{center}
\end{figure}

The light yield is given as the number of optical photons emitted 
per one MeV deposited in the scintillator.
A gamma ray source of $^{137}\mbox{Cs}$ is used to measure the light 
yield of a scintillator in a dark box with 5-inch PMT (H6527, Hamamatsu). 
Liquid scintillator is filled in an 1-liter Teflon bottle. A GEANT based 
simulation shows the 1-liter Teflon bottle is too small for a gamma to 
deposit all of its energy.
Thus Compton edge at 0.477~MeV from scattering of photons from 
$^{137}$Cs is used as a reference in the light yield measurement instead. 
After considering the PMT quantum efficiency and coverage of the PMT cathode, 
light yield of the our sample scintillator was measured to be 10\,000 optical 
photons per MeV. 
Figure~\ref{LY6300_vs_MC} shows comparison of data with the simulation 
result in which 10\,000 optical photons per MeV was used as the input to 
the simulation.

\subsection{Titration of Gd Concentration}
\begin{figure}
\begin{center}
\leavevmode
\includegraphics[width=3.5in]{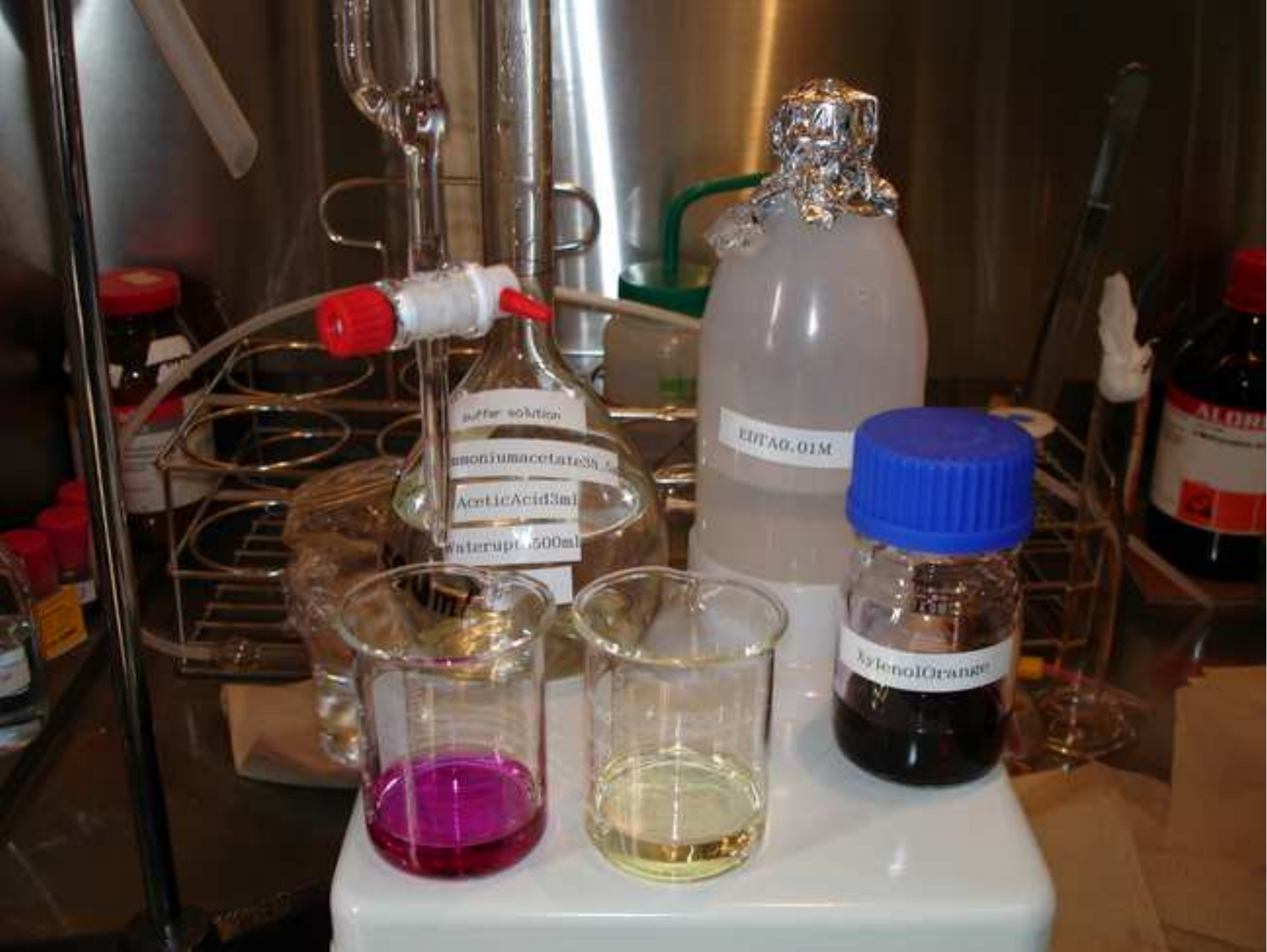}
\end{center}
\caption{EDTA, Xylenol orange indicator, and buffer solution. 
Microburette with bang pattern and lateral stopcock are used 
for titration.}
\label{violet}
\end{figure}

To measure the Gd concentration, a titration method was tested.
We prepared Xylenol orange indicator, buffer solution, 
and EDTA (Ethylene Diamine Tetra Acetic acid). 
At pH$>$5, EDTA reacts with Gd until Gd is depleted in the Gd loaded
solvent. 
But when more EDTA is added into solution, EDTA then reacts with the 
indicator and color of the solvent changes from violet to yellow as shown 
in Fig.~\ref{violet}.  We can calculate the concentration of Gd from the 
following equilibrium equation,
\begin{equation}
V_{EDTA} \times C_{EDTA} = V_{sample} \times C_{sample},  
\end{equation}
where $V_{EDTA}$ is the volume of EDTA, $C_{EDTA}$ is the concentration of 
EDTA (mol), $V_{sample}$ is the volume of the sample, and $C_{sample}$ is 
the concentration of Gd in the sample to be measured. 

\section{Radiopurity}
The main radioactive sources of low energy gamma rays 
are $^{238}$U, $^{232}$Th, and $^{40}$K. Double Chooz and Daya Bay experiments 
require the concentrations of each of these radioactive impurities in liquid 
scintillator to be less than $10^{-13}$~g/g. 
An acceptable single event rate due to such radioactivity in the detector is 
estimated by Monte Carlo simulations. This is described in 
Sect.~\ref{sect: radioactive backgrounds}. 

The radioactivities were measured by Inductively Coupled Plasma Mass 
Spectrometry (ICP-MS). ICP-MS is the simplest and fastest method but 
it cannot differentiate $^{40}$K from $^{40}$Ar because they have 
similar masses. In addition to ICP-MS, independent measurements of 
radioactivity are also available; High Resolution ICP-MS, Neutron 
Activation Analysis (NAA), and High Purity Germanium detector (HPGe).

Gammas and electrons from radioactive isotopes in the detector materials 
together  
with a neutron-like signal within a given time window can make a uncorrelated 
background. Assuming $10^{-13}$~g/g of $^{238}$U, $^{232}$Th, and 
$^{40}$K in the $\gamma$-catcher and target, we obtain 0.7~Hz of 
the single event rate using {\sc geant4} based simulation. This rate is 
quite smaller than the single event rate 
due to radioactivity of the PMT glass. 

\section{Liquid Handling and Purification System}
\begin{figure}[htp]
  \centering
  \includegraphics[width=14cm]{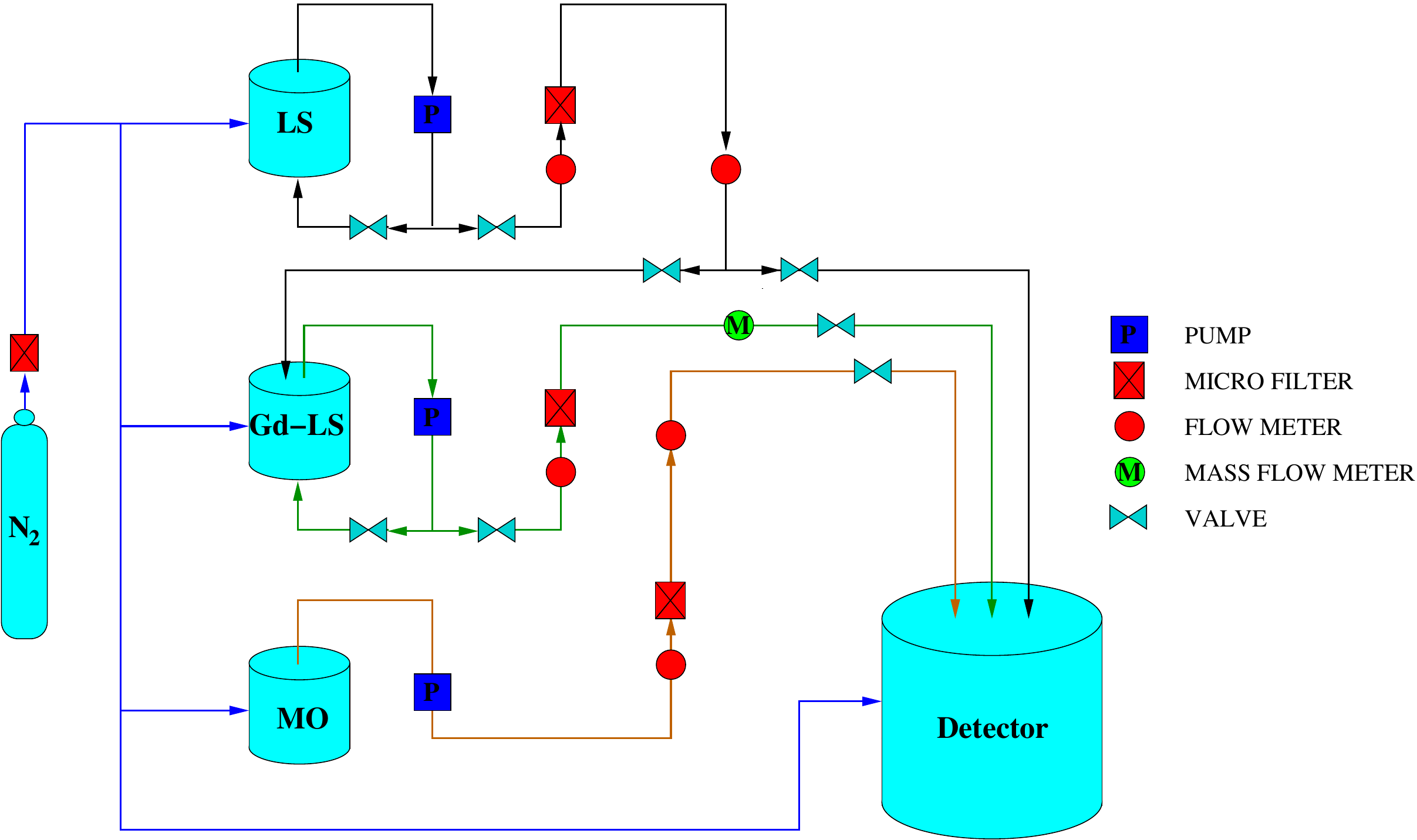}
  \caption{Schematic view of liquid handling and purification system.}
\label{handling}
\end{figure}

The liquid system consists of three sets of liquid storage tanks, pumps, 
and 0.05~$\mu$m filters, each for target, $\gamma$-catcher, and buffer, 
as shown in Fig.~\ref{handling}.
The speed of filling the detector with liquids should be carefully controlled 
to keep the equal liquid levels in target, $\gamma$-catcher, and buffer so as 
to avoid any stresses on acrylic vessels. The amount of liquid is measured by 
mass flow meters. Nitrogen purging is done during the liquid filling.

\chapter{Front-End Electronics, DAQ, Trigger, and Slow Control}
%

\section{Front-End Electronics}
The antineutrino interaction in the RENO detector produces scintillation
lights, and a part of them are converted  into photoelectrons at the PMT.
To detect the antineutrino event, the RENO detector has 354 inner PMTs 
and 67 outer PMTs.
A RENO readout system is designed to record the charge and 
arrival time of PMT hits.
Based on the energy
and timing information we can select the neutrino events,
reject background events, and reconstruct the vertex of
antineutrino interaction.  The near and far detectors are designed to 
have the same PMT configuration and readout system.

The RENO DAQ employs a new electronics developed for Super-Kamiokande 
experiment which uses charge-to-time conversion chips
to record hits at 60~kHz with no no dead time.  The following 
section describes the descriptions of the RENO DAQ electronics.

\subsection{Specification of RENO Electronics}
The readout electronics system records the charge and arrival time of
PMT hits to measure the energy and reconstruct the neutrino interaction.
Characteristics of RENO electronics are summarized below.
\begin{itemize}
\item PMT gain: $\sim 10^7$
\item Time window: $\sim 300$~ns
\item Dynamic range of PMT signals: 1 $\sim$ 1000 photoelectrons
\item Time resolution of each PMT signals: $1\sim 1.5$~ns
\item Data size: $\leq 200$~kbyte/s for each detector
\item No electronic deadtime
\item Time resolution between $e^+$ signal and neutron-like signal : $\sim 10$~ns
\end{itemize}

%
%
\subsection{QBEE Board}
The QBEE board is a charge-to-time conversion chip (QTC) based electronics 
with an onboard ethernet card, developed for Super-Kamiokande experiment
and used since Sept. 2008. 
The new electronics system are fast enough to record every PMT hits 
and are guaranteed for stable data acquisition over ten years. 
Each QBEE board has an 100~Mbps ethernet card,
which is fast enough to transfer every hit information to online computer
without any loss.
The hit information is stored in the online storage and then software 
triggers are applied. 

The PMT pulse generated by a photon hit is fed to a QTC chip. The QTC chip 
measures the hit time and charge of the PMT pulse and convert 
them into a form that can be easily read and stored by the TDCs. 
The output of the QTC chip is a logic pulse with its leading edge marking the 
hit arrival time and width representing the integrated charge of the PMT pulse.
The characteristics of the QTC chip are summarized in Table~\ref{fe-parameters}.

\begin{figure}
\begin{center}
\includegraphics[width=13cm]{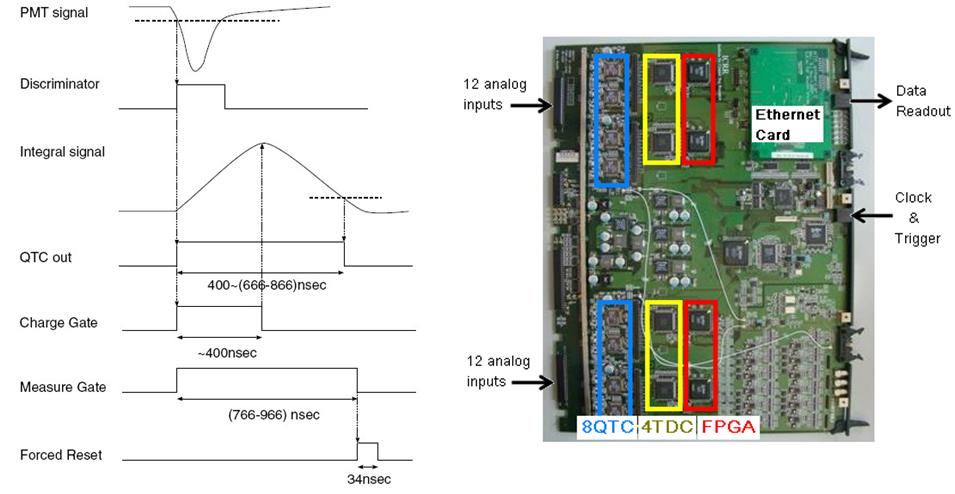}
\end{center}
\caption{Operation logic diagram of QTC chip and QBEE board}
\label{qbee}
\end{figure}

The operation logic diagram of QTC chip is shown in Fig.~\ref{qbee}. 
The QTC chip integrates charge of a PMT pulse fed to the chip and
outputs a pulse with a width proportional to the integrated charge.
The QTC chip produces two gates for its charge integration operation,
one for charging the capacitor (charge gate) in the QTC chip and the 
other for discharging the capacitor for measuring the charge in the 
capacitor (measure gate). If an incoming PMT pulse exceeds a 
current threshold, the 400~ns wide charge gate and 966~ns wide measure gate
are generated. 
Therefore, the width of output pulse from a QTC chip is between 400 and 966~ns 
which is proportional to the size of the integrated charge.
A reset signal of 34~ns is generated after the measure gate. So the processing 
time of a QTC chip is ~1~$\mu$s per cycle. The output pulse from the QTC is fed 
into a multi-hit TDC where the timing information of all leading and trailing 
edges are recorded.

A QTC chip receives three analog inputs and processes each input with one of
three gains of 1, 7, and 49. The charge resolution is about 0.1~pC and the
dynamic range is 0.2 to 2\,500~pC. The timing resolution is 0.3~ns for one 
photoelectron and 0.2~ns for more than five photoelectrons.

A QBEE board accommodates eight QTC and four TDC chips to process 24 analog 
inputs.
The QBEE board receives an external clock of 60~MHz and a periodical trigger
signal of 60~kHz from a master clock. The 60~kHz periodical trigger signal 
initializes TDC and 
comes with a timing tag and an event number, which are used
to identify the PMT hits in the same trigger. After collecting all the hits, 
an event is built and selected by software triggers. The adjustable
QTC parameters for RENO are 1) the threshold level for a single photoelectron
signal, 2) the length of charge gate and measure gate.

\begin{table}
\begin{center}
\begin{tabular}{cc}      \hline\hline
  Dynamic range            & $0  \sim 2500$~pC      \\
  Self trigger             & Built-in discriminator \\
  Number of input channels & 3                      \\
  Processing speed         &  $\sim 500$~ns/cycle  \\
  Gain                     & 1/7/49 (3 settings) \\
  Charge resolution        & 0.05~p.e. ($<25$~p.e.)    \\
  (Non-) Linearity (Q)     &   $< 1$\%        \\
  Timing resolution        &  0.3~ns (1~p.e.$=-3$~mV), 0.2~ns($>5$~p.e.)  \\
  Power dissipation        & $< 200$~mW/channel           \\\hline\hline
\end{tabular}
\end{center}
\caption{Characteristics of QTC chips. p.e. is photoelectron.} 
\label{fe-parameters}
\end{table}

%
%
\section{DAQ}
\begin{figure}
\begin{center}
\includegraphics[width=13cm]{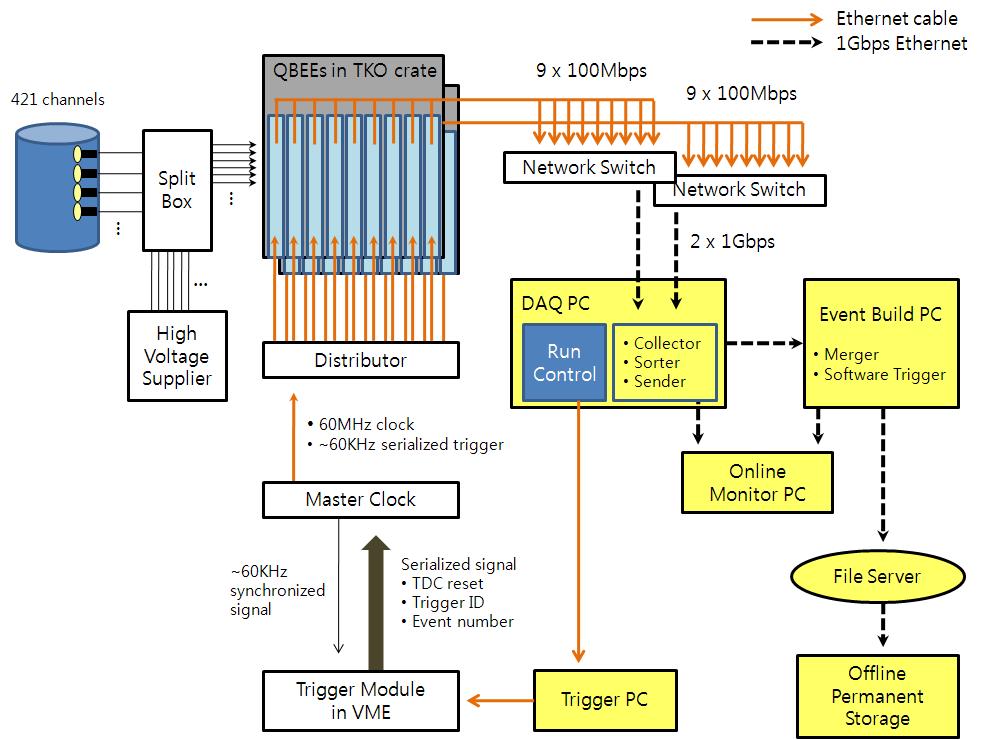}
\end{center}
\caption{Diagram of DAQ system for RENO. 
         There are 18 QBEE boards in two TKO crates collecting
         the hit signals from 421 PMTs (354 PMTs in inner
         detector and 67 PMTs in veto). The near and far
         detectors have the same DAQ architecture. 
        }
\label{daq-diagram}
\end{figure}

\begin{figure}
\begin{center}
\includegraphics[width=11cm]{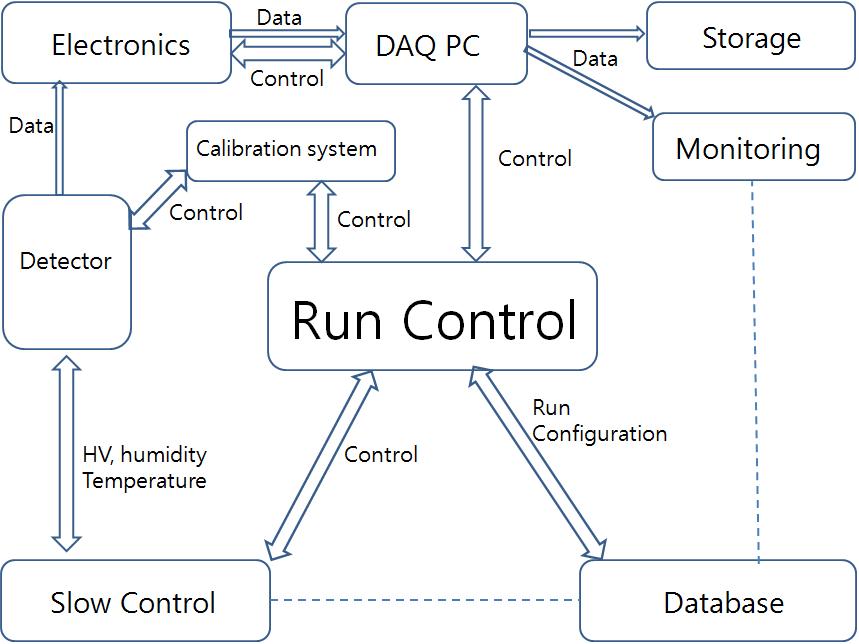}
\end{center}
\caption{Flow diagram of run control for RENO.
         The run control sends commands to DAQ component and makes
         run conditions. Shift crew will use integrated GUI.
        }
\label{run-control}
\end{figure}

The RENO data acquisition consists of data readout using front-end electronics,
event builder, software triggers, data logger and run control. A schematic
diagram of the RENO DAQ system is shown in Fig.~\ref{daq-diagram} and 
Fig.~\ref{run-control}.

\subsection{Data Readout and Run Control}
The front-end electronics for data readout are based on QBEE boards in the TKO
crate and ethernet cards on QBEE. A QBEE board receives 24 analog PMT inputs,
digitizes them, and sends the signal outputs to the online computer through an
100~Mbps ethernet card. RENO uses 18 QBEE boards for 421 channels per detector,
and data throughput rate is about 1.8~Gbps per detector. The near and far 
detectors have the same DAQ architecture. 

The run control sends command to DAQ components and makes run conditions. 
Shift crew will use an integrated GUI, which can be used to select run mode, 
trigger type, and detector parameter. The possible run modes are data taking 
and calibration. The trigger type can be one of predefined trigger sets. 
The detector parameters are high voltage setting for PMTs. 

\subsection{Event Builder}
All the QBEE boards are driven by a common 60~MHz master clock (MCLK). A 60~kHz
periodical trigger and a serialized 32-bit event number are generated by a
trigger module, and fanned out via a distributor to all the QBEE boards through
network cables. All the hit data are sorted and merged according to the trigger
event number and the timing information.

A periodic trigger of 60~kHz makes a data block of hits. The order of the data 
blocks is made according to the event number. The hits in a data block are sorted 
by their hit time and merged. The hit data in the same block are merged, sorted 
by hit time, and stored with an event number.
An event builder constructs events by applying software trigger to the merged 
hit data. The  merged data before the software trigger are stored 2 to 3 days 
and will be used for monitoring purposes. 

\subsection{Software Trigger}
The software triggers are applied to the events constructed by the merger 
for identifying neutrino candidate events, cosmic muon events, or calibration
events. 
The software trigger calculates the total number of hits (multiplicity) 
within a certain time window and constructs event if the hit sum exceeds 
a certain threshold number. The first hit time in an event is set
to $T_0$, and 
time windows before and after $T_0$ determine an event gate
by software triggers. 
All of PMT hits within this event gate makes an event with ``in gate event'' 
flag and the hits are set with ``in gate hit'' flag.
In analyzing data hits with ``in gate hit'' flag will be used.
Some of QBEE bits are assigned for calibration trigger ID. If there is
even a hit with those bits, a calibration trigger is generated. The 
software trigger rates are monitored in the online, and those events with ``in gate event'' 
flag are stored in the long-term storage.

\section{Trigger}
The triggering strategy at RENO is to record all the hits having signal 
over a given threshold, and then to select events by software triggers. 
The software triggers make decisions for neutrino-like events, cosmic 
ray muon events, background events, and calibration events.

\subsection{Energy Threshold}
The signature of a neutrino interaction in the RENO detector is
a prompt signal from a positron with a minimum energy of 1.022~MeV followed by a 
delayed signal from a neutron.
About 90\% of the neutrons are captured on Gadolinium at the target of the RENO 
detector, yielding an $\sim 8$~MeV gamma cascade with a capture time of $\sim 30~\mu$s. 
The main backgrounds to the signal
in the antineutrino detectors are fast neutrons produced by cosmic muon
interactions in the rock, $^8$He/$^9$Li, which are also produced by cosmic muons and
accidental coincidences between natural radioactivity and neutrons produced by
cosmic muons. Dominant backgrounds to the delayed event are related to cosmic muons.

The energy threshold should be set for the software trigger to accept both 
the prompt positron signal above 1.022~MeV and the delayed neutron capture 
event with a photon cascade of 8~MeV with a very high efficiency.
This requirement for energy threshold expects the DAQ system 
to record all prompt positron signals produced from the antineutrino interactions 
and to make a complete energy spectrum analysis to be possible for increasing the
sensitivity of $\sin^2(2\theta_{13})$.  
Furthermore, it requires the DAQ to register uncorrelated background
events due to either PMT dark noise or low energy radioactivity for
a detailed background analysis in the offline.
 
\subsection{Event Building}
After calculating the global time of hits,
all the recorded hits are sorted and merged according to the timing information,
and then software triggers are applied to select an event.

\subsection{Online Software Trigger}
There are two groups of triggers, one is for selecting neutrino 
events and the other is for calibration and monitoring backgrounds.
\begin{itemize}
  \item neutrino triggers: energy sum trigger and multiplicity trigger.
  \item calibration triggers: LED, laser, and radioactive sources.
  \item random triggers: pedestal run and random background.
  \item cosmic trigger: cosmic ray muons.
\end{itemize}
%
%
\section{Slow Control and Monitoring}
\begin{figure}
\begin{center}
\includegraphics[width=11cm]{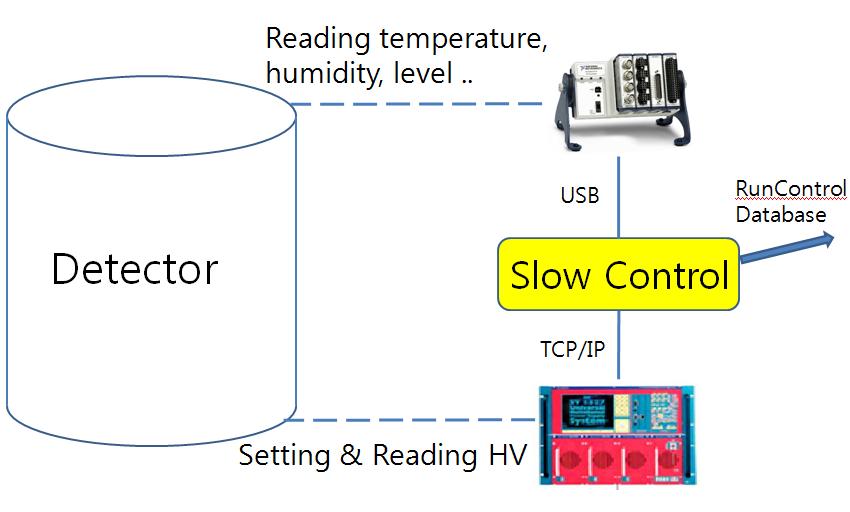}
\end{center}
\caption{Diagram of slow control.
         The slow control system monitors detector conditions and 
         controls PMT HV power supplies.
        }
\label{slow-control}
\end{figure}

An online monitoring computer located in the control room reads data from
the DAQ host computer via network. It provides event display, online history 
histograms to monitor detector performance, and variety of additional tasks 
needed to efficiently monitor detector performance parameters and diagnose 
troubles of detector and DAQ system. 
The slow control monitors 
the status of the HV systems, the temperatures of the electronics crates and
detectors, the fluids levels, and humidity. And the slow control is able to 
set up high voltage for each channel and turn on and off HV remotely.
The slow control scheme is shown in Fig.~\ref{slow-control}.
The data collected by the slow control system are sent to online monitoring 
system and database.

\newpage

\bibliographystyle{plain}
\chapter{Monte Carlo Simulation}\label{Simulation Chapter}
\section{Overview}
As with other particle experiments, extensive studies using Monte Carlo
simulation have been performed for RENO experiment. The Monte Carlo (MC)
studies provide valuable guidance to optimize and determine various 
design parameters of the detector. The most cost effective design without 
much compromising the sensitivity of the experiment is attempted based 
on the studies using MC simulation.
The MC simulation also helps the development of analysis 
tools to be used for the actual data from the experiment. In 
addition, some of systematic uncertainties can be estimated from 
simulation studies as well.

The RENO detector simulation is modified from {\sc glg4sim}, a 
{\sc geant4} based program for liquid scintillator neutrino detectors. 
The ``generic'' program has been customized for the RENO detector 
with a new event generator which provides better physics models. 
More details on the simulation can be found later in this chapter.

Studies on backgrounds is also performed using MC simulations. 
Major background sources are cosmic muon induced events and radioactivities 
from the surrounding rocks and detector itself. The rates of these 
background events have been independently estimated and appropriate 
event generators have been added to the main simulation program so 
that we can obtain realistic results from the MC simulations.

This chapter describes the generation of inverse beta decay (IBD) 
events, calculations of the detector performance, and the 
background simulations. It also includes the description of 
the systematic uncertainties estimated from MC studies.

\section{Event Generation}\label{Event Generation}
This section describes the event generator of inverse beta
decay for reactor neutrinos.

\subsection{Inverse Beta Decay Events}
\begin{figure}
\begin{center}
\includegraphics[width=3in]{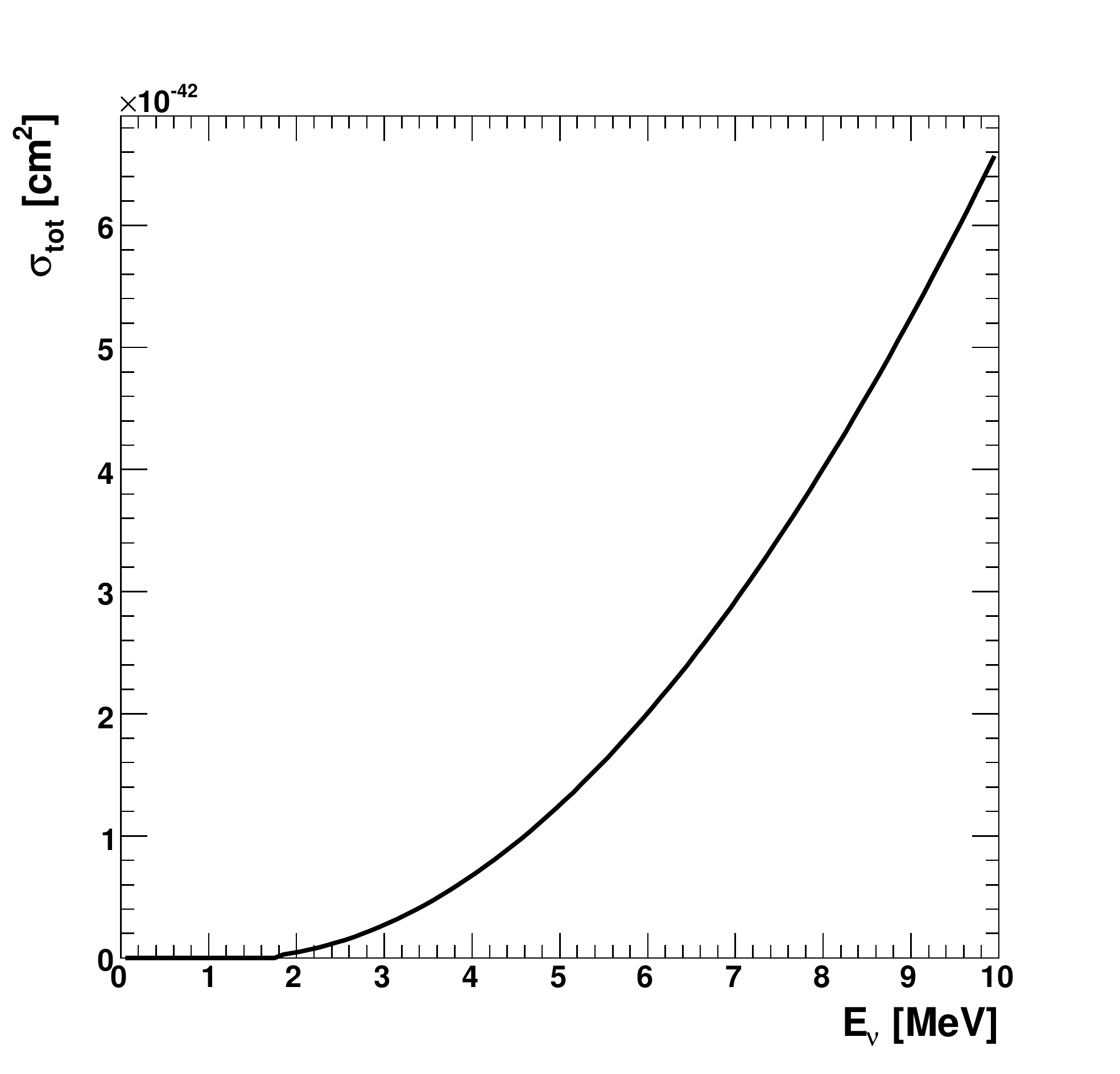}
\includegraphics[width=3in]{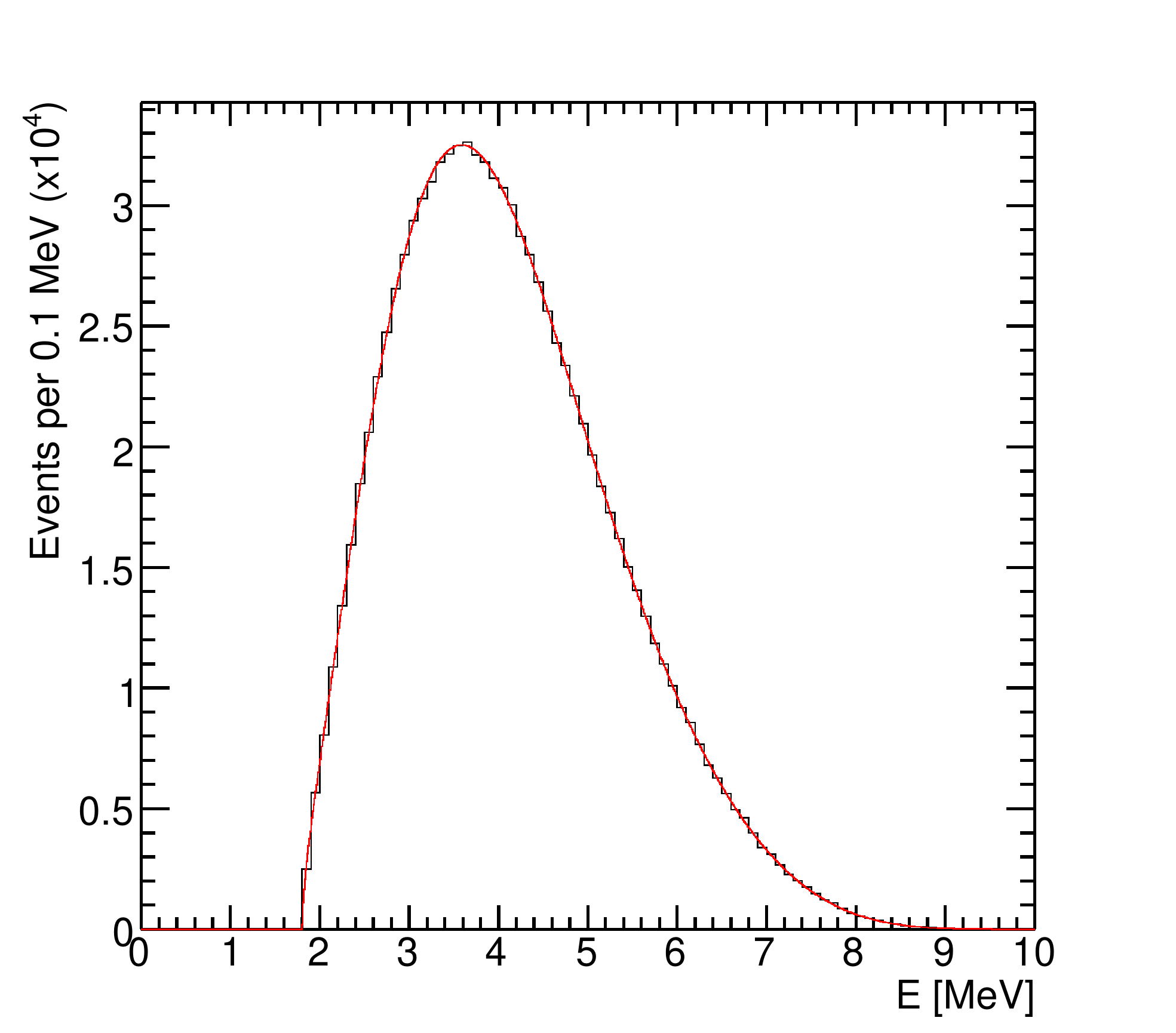}
\end{center}
\caption{
Energy dependence of anti-neutrino cross section (left) and observed
spectrum of reactor neutrinos (right). In the observed spectrum,
a theoretically calculated spectrum is shown as a red curve compared
with that of the MC simulation shown as a black histogram.
}
\label{IBDevents}
\end{figure}

In reactor based neutrino experiments, the neutrino detection is
made via the inverse beta decay (IBD) process that an incoming neutrino
interacts with a free proton in the detector material, that is,
\begin{equation}
\bar{\nu}_e + p \to n+e^+.
\end{equation}
In the event generator, the four-momenta are specified for all four
particles involved in the IBD process. The energy dependence of 
anti-neutrino's cross section and the observed spectra of reactor
neutrinos are shown in Fig.~\ref{IBDevents}.

The cross sections for quasi-elastic neutrino scattering on proton at
energies greater than 1~GeV have been calculated by 
Llewellyn-Smith~\cite{Smith}.
Vogel and Beacom~\cite{Vogel99} presented a simple and precise description 
of neutrino--nucleon scattering at a low energy region of a few MeV,
relevant to reactor, solar, and supernova neutrinos.

The IBD event generator includes Vogel and Beacom cross 
section~\cite{Vogel99} and the total cross section at the first 
order in $1/M$ is defined by  
\begin{eqnarray}
\label{xc eqn4}
\sigma^{(1)}_{tot} = \sigma_{0}\left(\alpha_{1}+\beta_{1}{\Delta\over M}+\gamma_{1}{E^{(0)}_{e}\over M}\right)E^{(0)}_{e}E^{(0)}_{e},
\label{Simulation:eq2}
\end{eqnarray}
where $E_e^{(0)}=E_{\nu}-\Delta$, $\Delta=(M_{n}-M_{p})$, and $M$ is the 
nucleon mass.
The parameters in Eq.~\ref{Simulation:eq2} are 
$\alpha_{1}=f^{2}+3g^{2}$, $\beta_{1}=-2(f+f_{2})g-2f^2-8g^2$, and 
$\gamma_{1}=-4(f+f_{2})g-2f^{2}-10g^{2}$, where $f = 1$, $f_{2}=3.706$, 
and $g=1.26$. 
The normalization constant $\sigma_{0}$ is given as
\begin{eqnarray}
\sigma_{0} = {G^{2}_{F}\cos^{2}\Theta_{c}\over \pi}
\left(1+\Delta^{R}_{inner}\right),
\end{eqnarray}
where $G_{F}$ is the Fermi coupling constant, $\Theta_{c}$ is the 
Cabibbo angle, and $\Delta^{R}_{inner}\simeq0.024$.
In the laboratory frame the threshold energy is 
$E^{thr}_{\nu}=((M_{n}-M_{e})^{2}-M^{2}_{p})/2M_{p}=1.806$~MeV
as shown in Fig.~\ref{IBDevents}. 
We calculated the neutrino spectrum at the detector using the 
total cross section weighted 
by the neutrino flux from the Yonggwang nuclear power plant, Korea.

\subsection{Kinematics for Positron and Neutron}
\begin{figure}
\begin{center}
\includegraphics[width=3in]{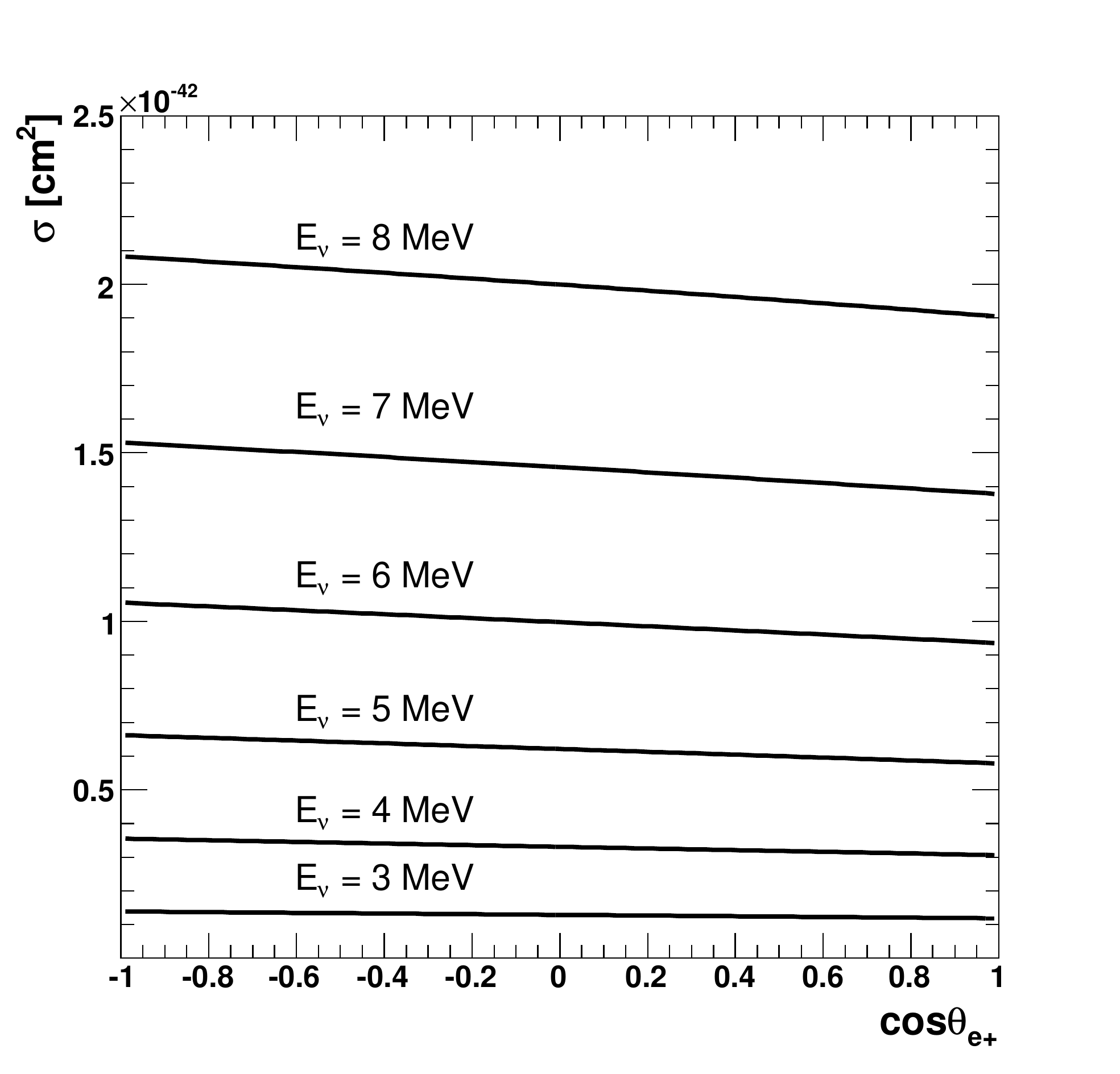}
\includegraphics[width=3in]{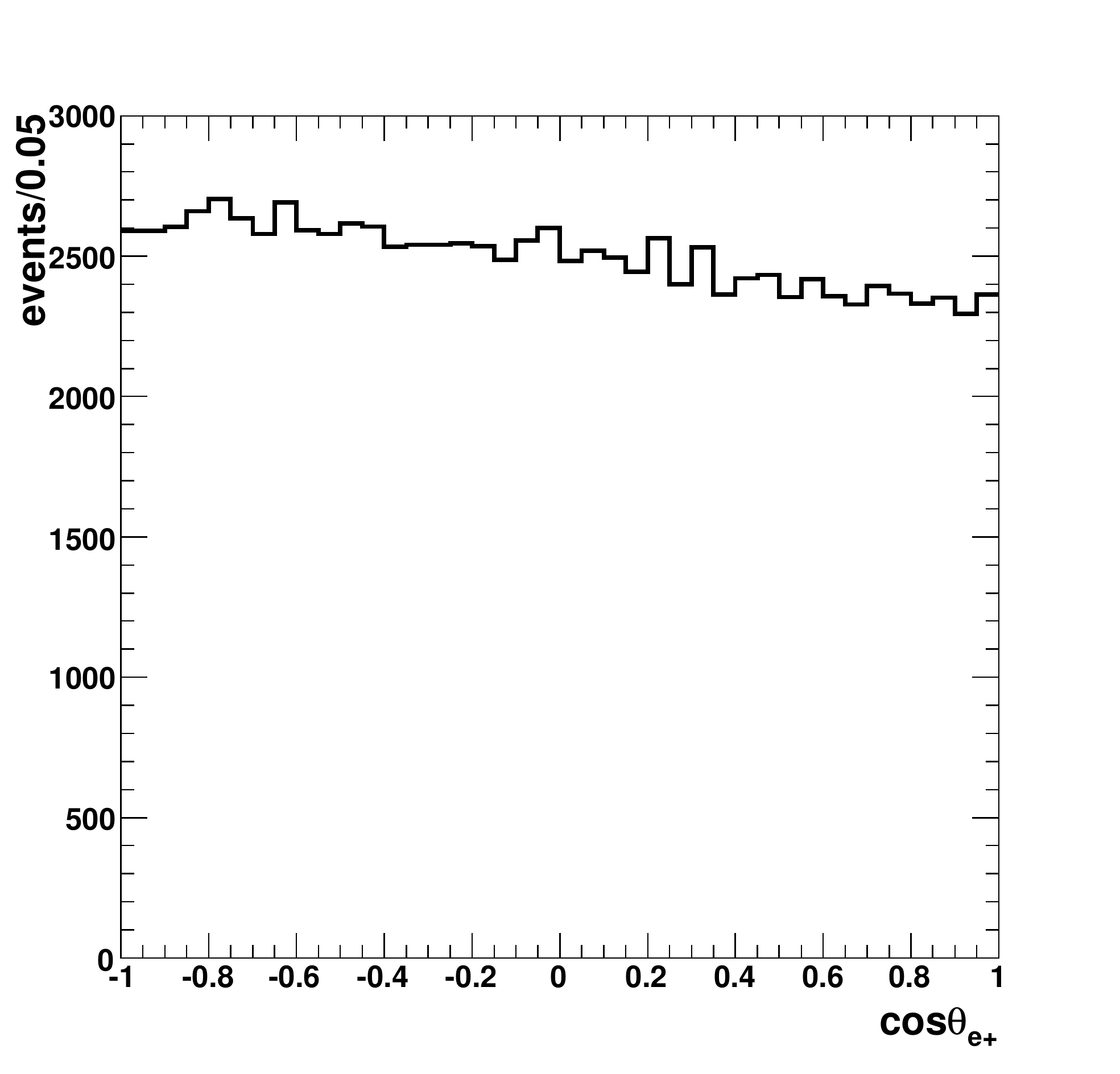}
\end{center}
\caption{
Differential cross section as a function of the angle between the positron
direction and the incident neutrino direction for several neutrino energy
values (left). Angular distribution of the positron from the IBD 
event generator (right).
}
\label{IBDevents-properties1}
\end{figure}

\begin{figure}
\begin{center}
\includegraphics[height=2.8in]{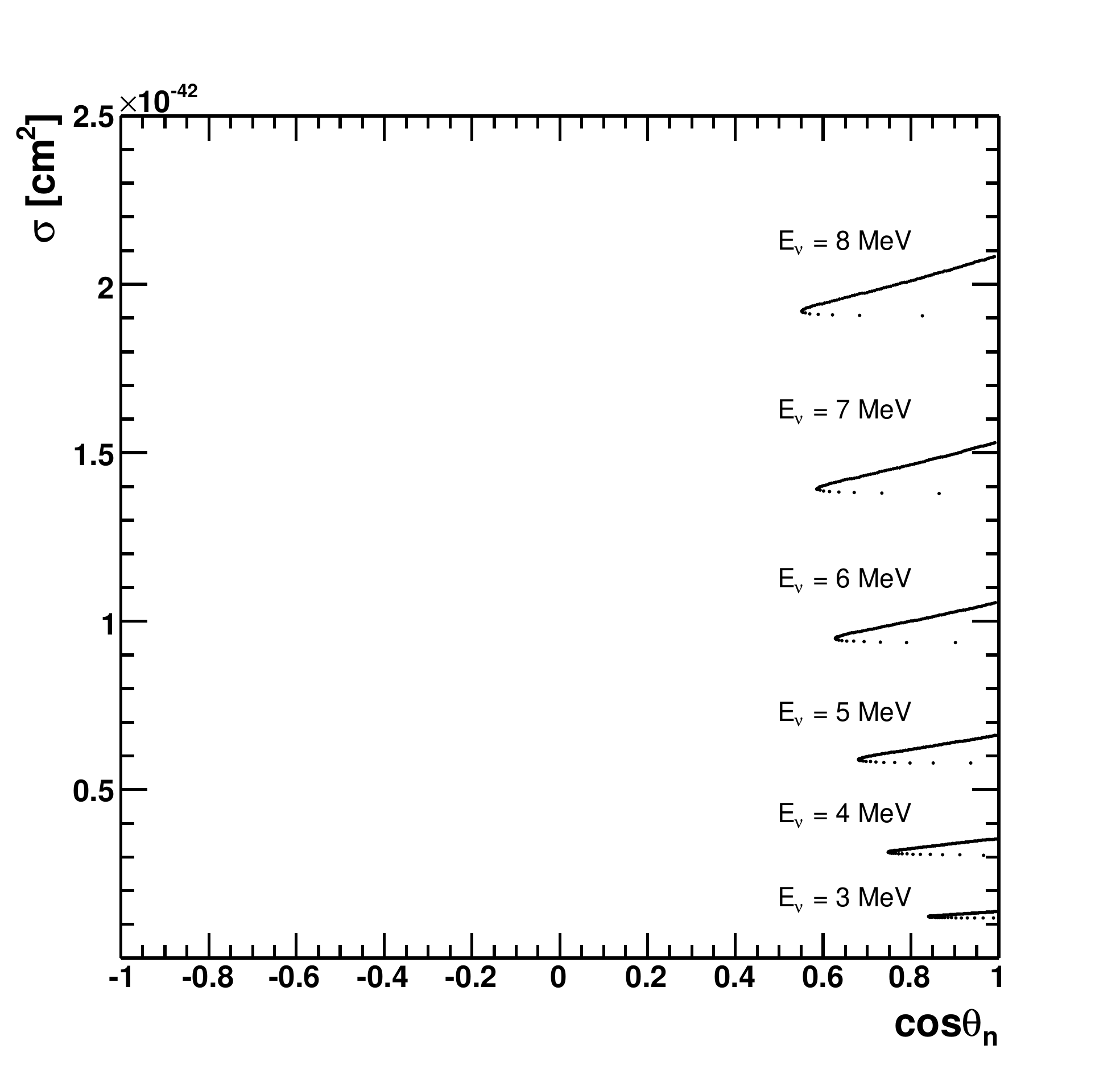}
\includegraphics[height=2.75in]{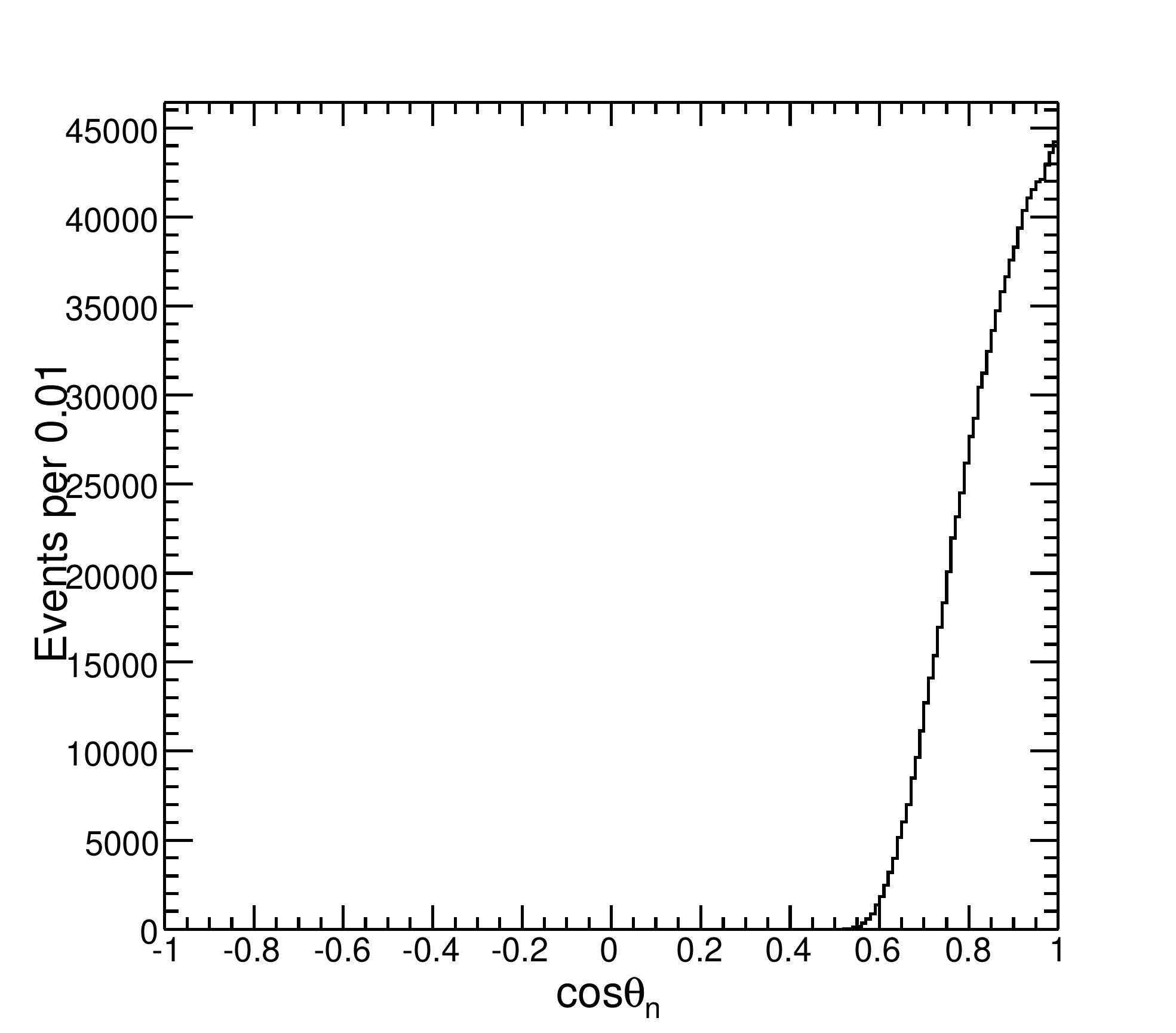}
\end{center}
\caption{Cross section versus the cosine of neutron angles for
several neutrino energies (left) and neutron angular distribution
(right) from the IBD process.}
\label{IBDevents-properties2}
\end{figure}

The event generator gives the energy and scattering angle 
distributions of a positron and a neutron coming from an IBD 
event. 
The generator is based on Vogel and Beacom's 
IBD differential cross section calculations~\cite{Vogel99}.
The angular distribution of the positron is almost uniform 
as shown in Fig.~\ref{IBDevents-properties1}. 
Unlike the positron from the IBD process, the neutron recoiling from
the IBD process has a strong angular correlation with respect to
the incoming neutrino direction.  
Since the proton is at rest in the laboratory 
frame, the neutron is
essentially emitted in the forward directions with the maximum angle 
$\theta^{\rm{max}}_n$:
\begin{eqnarray}
\label{nCosThetaMax eqn}
\cos(\theta_{n}^{\rm{max}}) 
= {\sqrt{(2E_{\nu}\Delta-(\Delta^{2}-m^{2}_{e}))}
\over E_{\nu}}.
\end{eqnarray}
Figure~\ref{IBDevents-properties2} shows the differential cross section 
of IBD process as a function of the angle between the recoiling neutron 
direction and the incident neutrino direction for various incident neutrino
energies. Also the angular distribution 
of the neutron with respect to the incident neutrino direction
is shown.
The kinetic energy of the neutron is given by
\begin{equation}
T_{n} = {E_{\nu}E_{e}^{(0)} \over M}(1-v_{e}^{(0)}\cos\theta_{e+})+{y^{2} 
\over M},
\end{equation}
where $y^{2}=(\Delta^{2}-m_{e}^{2})/2$ and $v_{e}=p_{e}/E_{e}$.


\begin{figure}
\begin{center}
\includegraphics[width=3in]{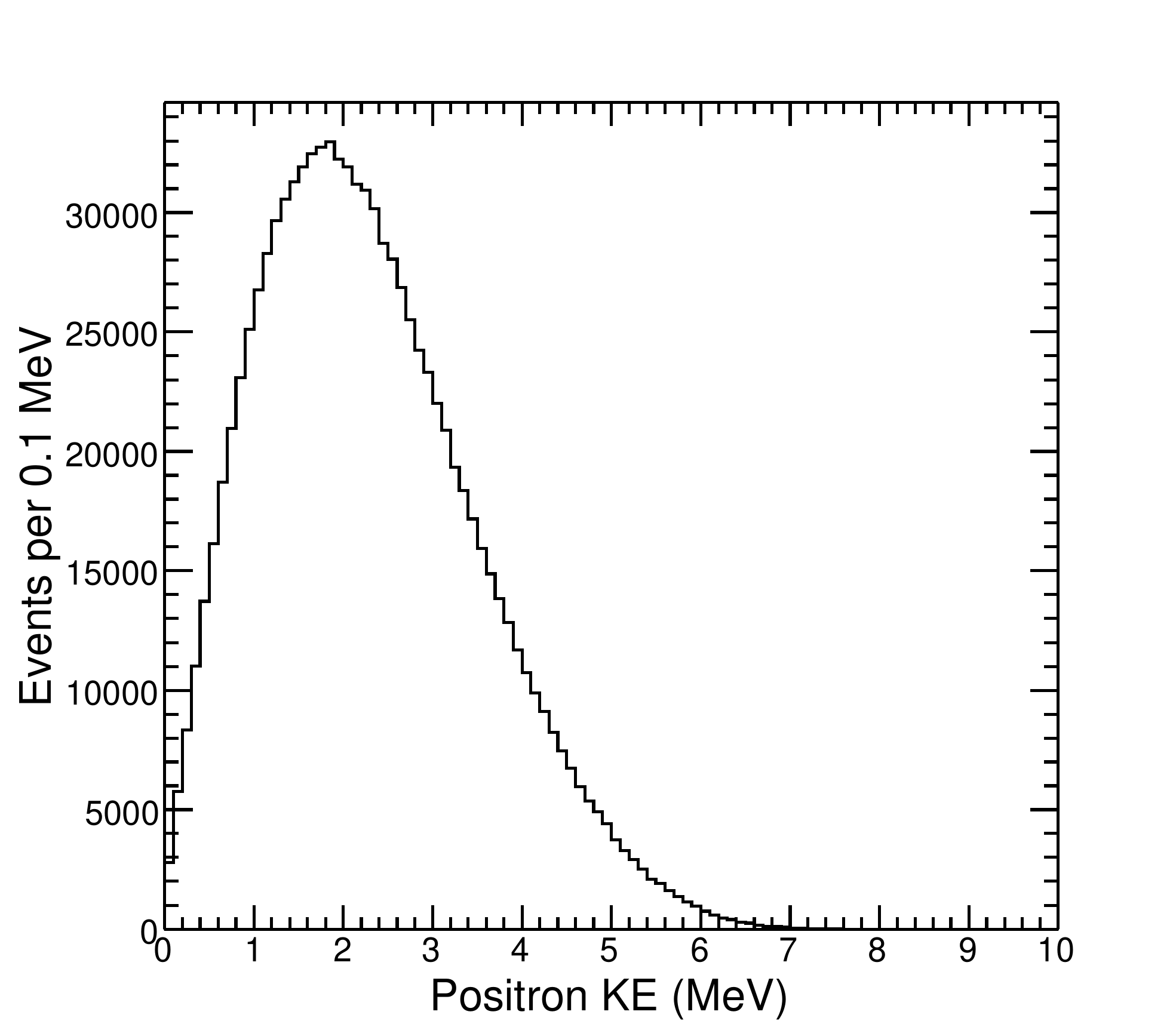}
\includegraphics[width=3in]{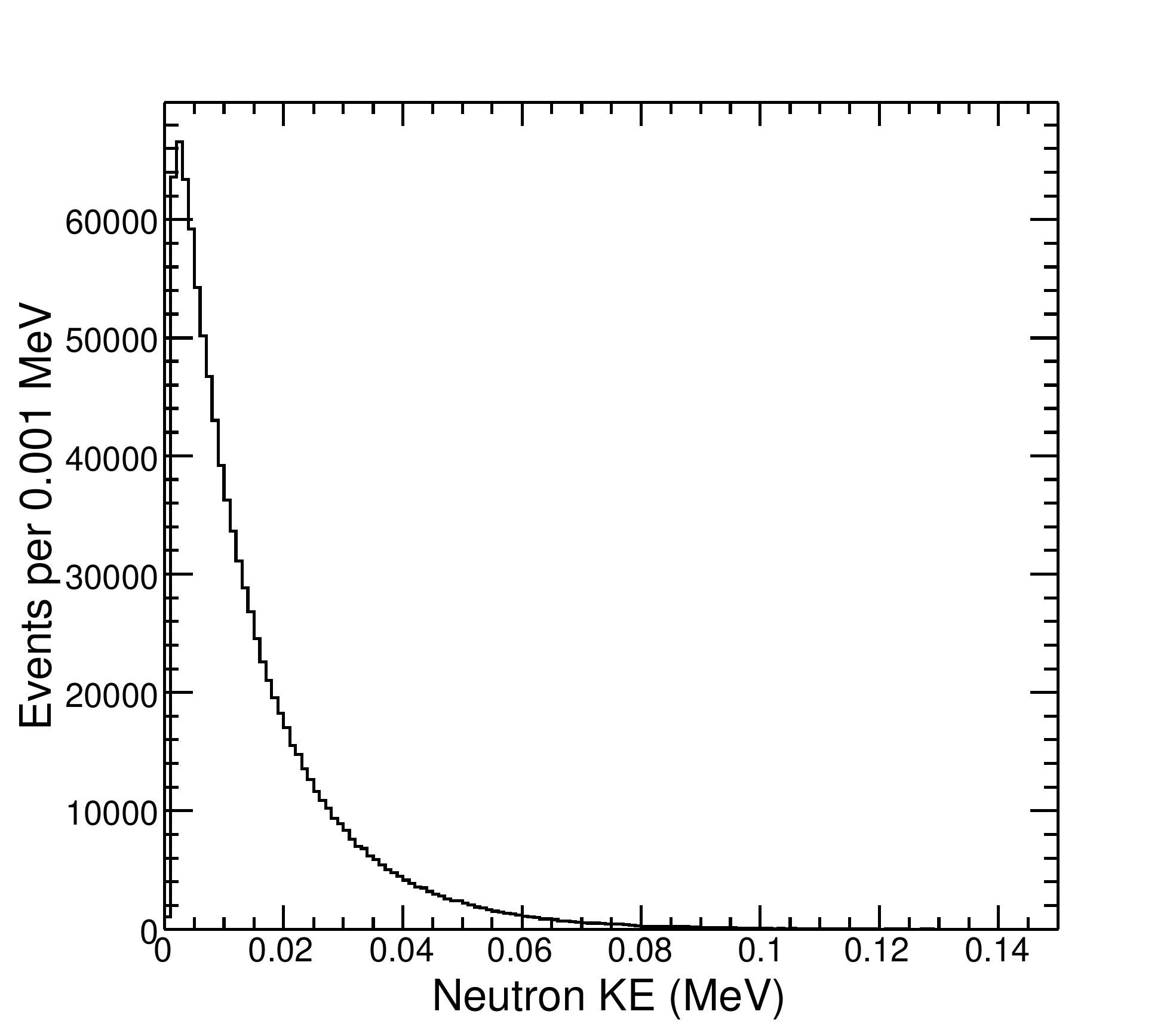}
\end{center}
\caption{Positron kinetic energy spectrum (left) and neutron 
kinetic energy spectrum (right) from the IBD event generator.
}
\label{IBDevents-properties3}
\end{figure}

The kinetic energy spectrum of the positron from IBD process is given by 
\begin{eqnarray}
\label{positron eqn}
{d\sigma\over dT_{e+}} = {K\over ft}\left({T_{e}\over m}+1\right)
\left[\left({T_{e}\over m}+1\right)^{2}-1\right]^{1/2}\rho(T_{e}-\Delta),
\end{eqnarray}
where $\Delta=M_{n}-M_{p}+m=1.804$~MeV, $ft=1087$~s,
$T_{e}$ is the kinetic energy of the positron,
$K=2.63\times 10^{-41}$~cm$^{2}\cdot$s, and
$\rho$ is the energy spectrum of the incident $\bar\nu_{e}$~\cite{fayans79}.
The kinetic energy spectra of the positron and 
the neutron are shown in Fig.~\ref{IBDevents-properties3}.

\subsection{Neutrino Flux}
When fissile isotopes undergo fission, neutrinos are emitted 
isotropically. There are four dominant fissile isotopes in 
the nuclear fuel cycle; $^{235}$U, $^{238}$U, $^{239}$Pu, 
and $^{241}$Pu. 
On the average six neutrinos are emitted per fission with their 
energy peaked at around 1~MeV~\cite{Bempo02}. But only neutrinos 
with energies above the threshold of 1.806~MeV contribute to the 
inverse beta decay.  
The mean energy and the number of neutrinos above $E_{\nu}=1.8$~MeV
released from the fission of these isotopes are shown in 
Table~\ref{fission summary}.

\begin{table}
\begin{center}
\begin{tabular}{ccc}\hline
Isotope &$N_{\nu}$~($>1.8$~MeV) &$E_{f}$~(MeV) \\\hline
$^{235}$U    &$1.92(1\pm 0.019)$     &$201.7\pm0.6$ \\
$^{238}$U    &$2.38(1\pm 0.020)$     &$205.0\pm0.9$ \\
$^{239}$Pu   &$1.45(1\pm 0.021)$     &$210.0\pm0.9$ \\
$^{241}$Pu   &$1.83(1\pm 0.019)$     &$212.4\pm1.0$ \\
\hline
\end{tabular}
\caption{The total number of ${\bar\nu}_e$ per fission above 1.8~MeV
and energy released per fission from Ref.~\protect\cite{Apollonio03}.}
\label{fission summary}
\end{center}
\end{table}

\begin{figure}
\begin{center}
\leavevmode
\includegraphics[width=4.5in]{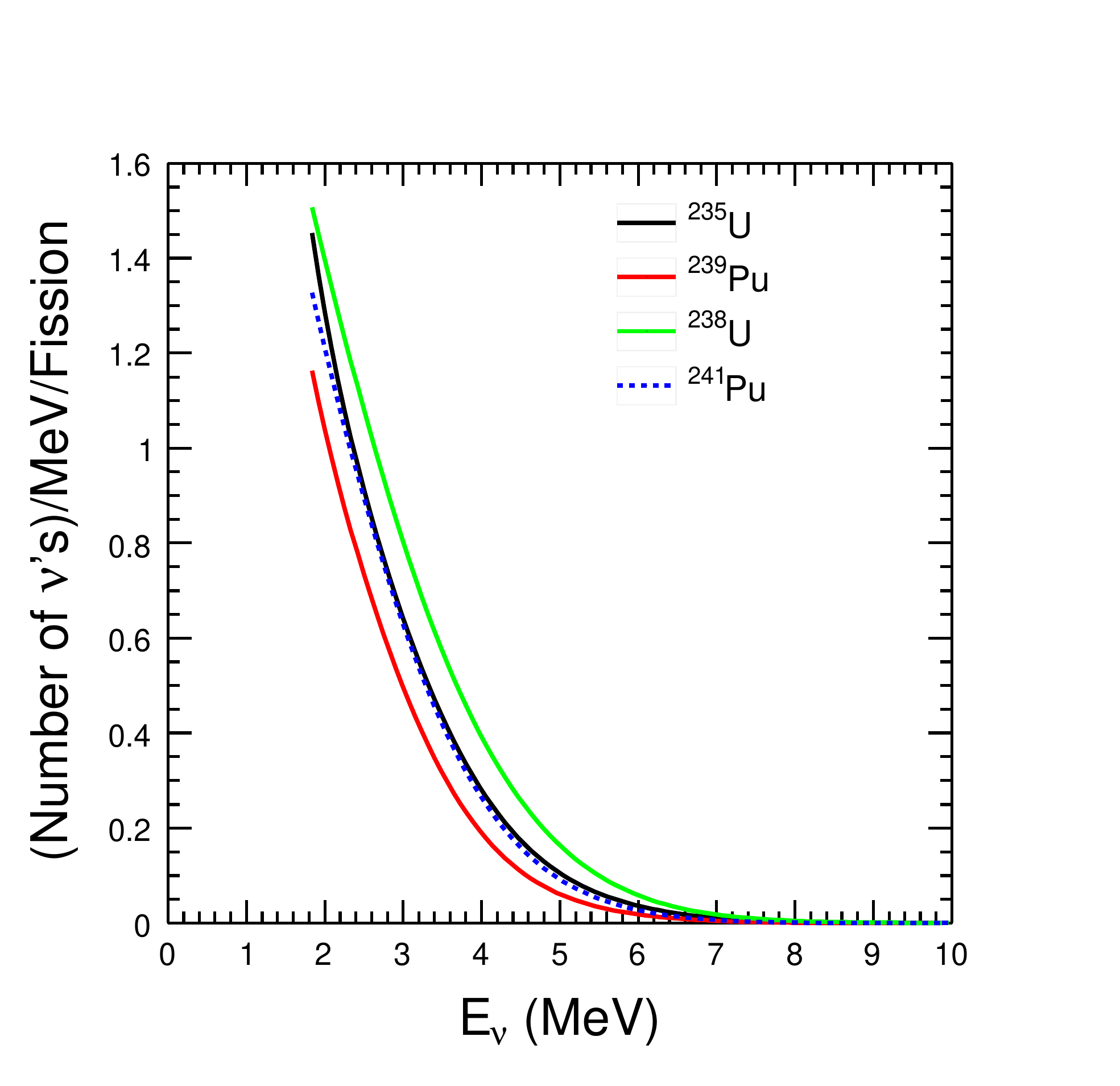}
\end{center}
\caption{Neutrino flux of four main isotopes in the nuclear 
fuel using parametrization in Table~\protect\ref{oink} given
in Refs.~\protect\cite{Huber04,Vogel89}.}
\label{neutrino flux}
\end{figure}

The neutrino energy spectra from fission processes are 
parametrized in Refs.~\cite{Huber04,Vogel89} using 
\begin{equation}
{dN_{\nu}^{(j)}\over dE_{\nu}} = 
\exp\left(\sum_{i=0}^{5} a_i^{(j)} E_{\nu}^i\right),
\end{equation}
where $a_i^{(j)}$ are the fit parameters for the $j$th 
isotope and $E_{\nu}$ is a neutrino energy in MeV.
The results are shown in Table~\ref{oink} and 
Fig.~\ref{neutrino flux}.
\begin{table}
\begin{center}
\begin{tabular}{cccccc}\hline
Parameter       &$^{235}$U   &$^{238}$U$^{(*)}$ 
                &$^{239}$Pu  &$^{241}$Pu  \\\hline
$a_0$           &$3.519$        &$0.976$        &$2.560$        &$1.487$ \\
$a_1$           &$-3.517$       &$-0.162$       &$-2.654$       &$-1.038$ \\
$a_2$           &$1.595$        &$-0.790\times 10^{-1}$
                &$1.256$        &$4.130\times 10^{-1}$ \\
$a_3$           &$-4.171\times 10^{-1}$ &$-$ &$-3.617\times 10^{-1}$ 
                &$-1.423\times 10^{-1}$ \\
$a_4$           &$5.004\times 10^{-2}$ &$-$ &$4.547\times 10^{-2}$ 
                &$1.866\times 10^{-2}$ \\
$a_5$           &$-2.303\times 10^{-3}$ &$-$ &$-2.143\times 10^{-3}$ 
                &$-9.229\times 10^{-4}$ \\
\hline
\end{tabular}
\end{center}
\caption{Parameters of the polynomial of order of 5 for the neutrino flux
from dominant isotopes in nuclear fuel. Parameters for isotopes $^{235}$U, 
$^{239}$Pu, and $^{241}$Pu are from Ref.~\protect\cite{Huber04} and
$^{238}$U from Ref.~\protect\cite{Vogel89}. The resulting distributions
are shown in Fig.~\protect\ref{neutrino flux}.}
\label{oink}
\end{table}

The fission rate in a reactor with a power $P_{th}$ is   
\begin{equation}
n_{fis} = {P_{th}\over \sum_i f_i \bar E_{fi}},
\label{number of fissions}
\end{equation}
where $f_i$ and $\bar E_{fi}$ are the fission fraction in the nuclear 
fuel and the mean energy released per fission of isotope $i$ given
in Table~\ref{fission summary}, respectively, and
$P_{th}$ is the reactor power.
Then the number of fissions per second is related to the reactor power 
by $(6.24\times 10^{18})\cdot n_{fis}$, 
where $P_{th}$ is given in Watts and $E_{fi}$ in eV
in calculating $n_{fis}$ in Eq.~\ref{number of fissions}.

The number of neutrinos with energy between $E_{\rm min}$ and 
$E_{\rm max}$ from 
the fission process of the $i$th isotope is 
\begin{equation}
N_{\nu} = n_{fis} \cdot \sum_i f_i 
\int_{E_{\rm min}}^{E_{\rm max}}{dN_{\nu}^{(i)}
\over dE_{\nu}} dE_{\nu}
\end{equation}
The neutrino flux is isotropic about the source and the neutrino flux
at distance $r$ is
\begin{equation}
n_\nu(r) = {1\over 4\pi r^2} N_{\nu}. 
\end{equation}

\section{Detector Simulation}
A major goal
of the detector simulation, besides 
being used as a data analysis tool, is to optimize 
the detector design in the early stage of the experiment.
To achieve the sensitivity goal of $\sin^2(2\theta_{13})>0.02$ 
within the budget and time constraints, the detector design 
needs to be carefully studied and optimized.

\begin{figure}
\begin{center}
\includegraphics[width=3.0in]{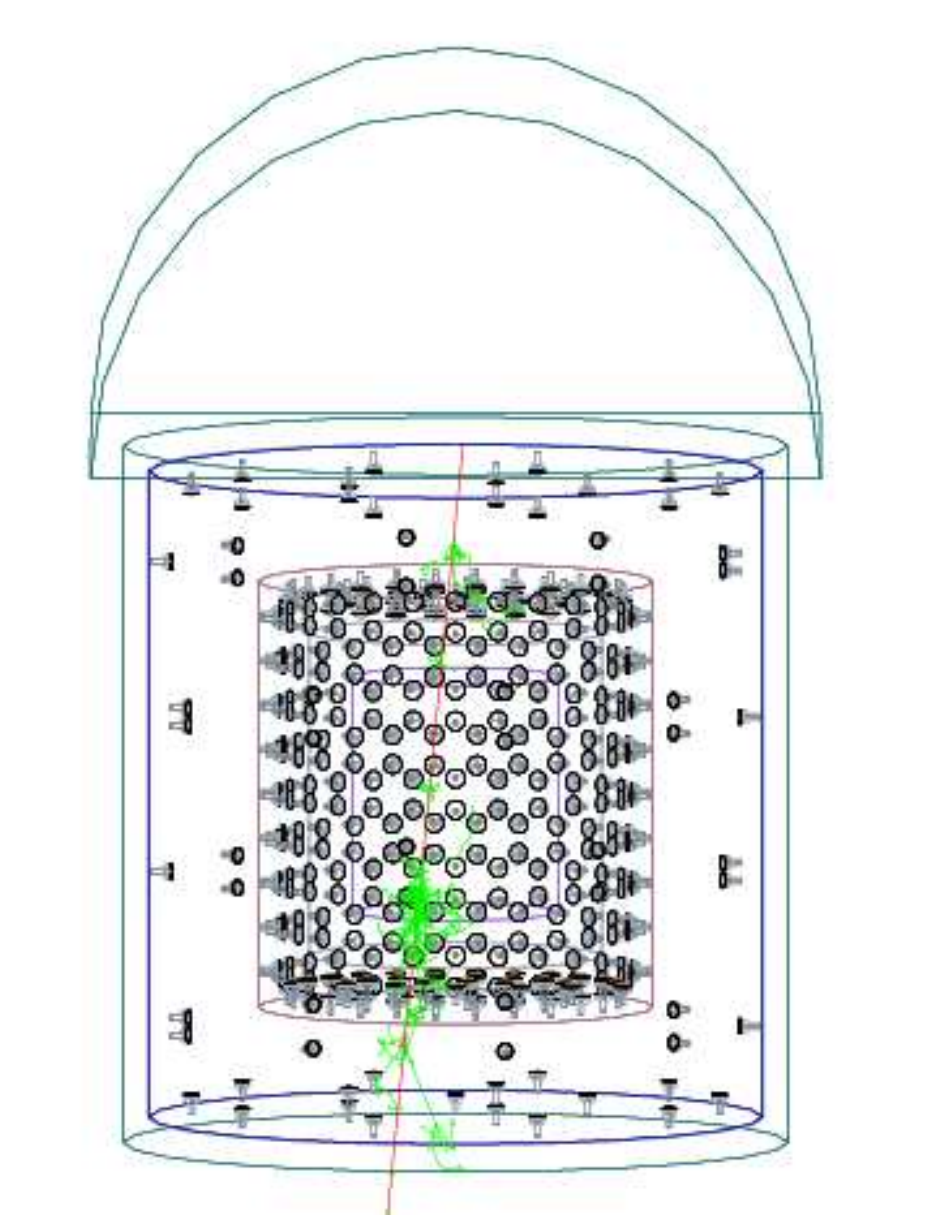}
\includegraphics[width=2.5in]{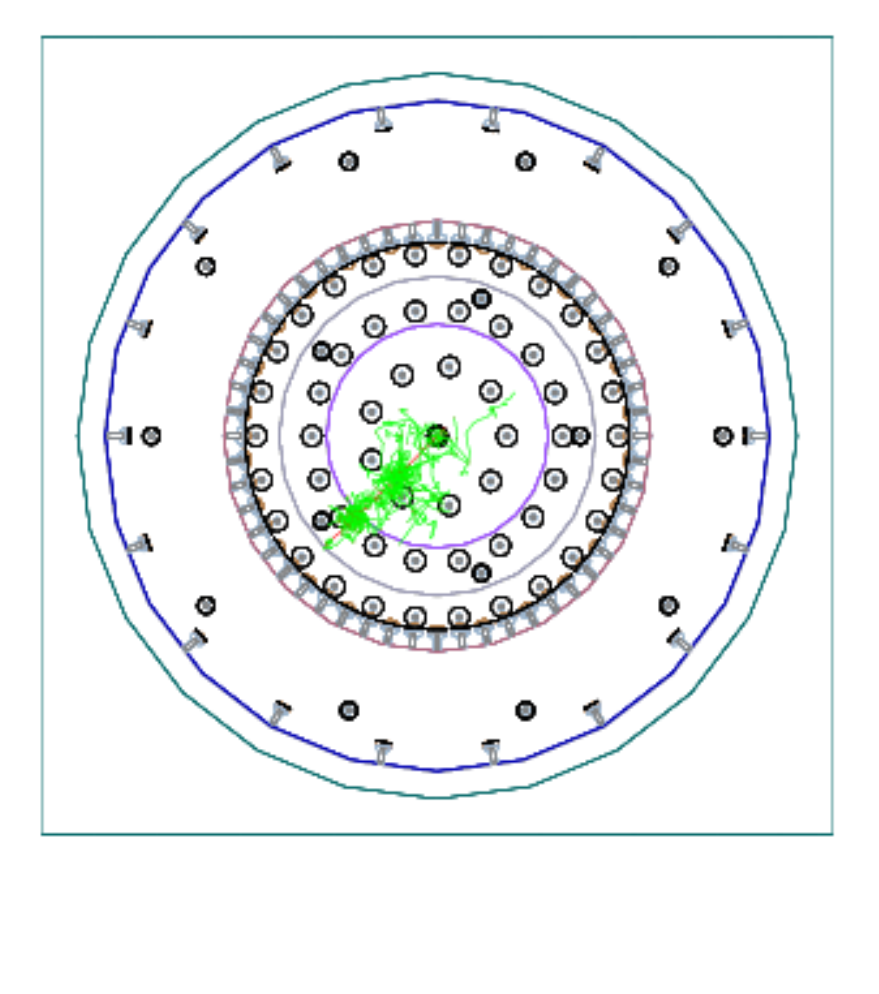}
\end{center}
\caption{Side and top view of the RENO detector simulation with a
muon (red line) passing through the target and leaving showers (green lines). 
}
\label{detectorSim-layout}
\end{figure}

The RENO detector is designed to have four concentric cylindrical 
modules, two active inner modules called target and $\gamma$-catcher 
and two inert outer modules called buffer and veto as shown in
Fig.~\ref{detectorSim-layout}.
Compared to the past reactor neutrino experiments, an additional
active layer, $\gamma$-catcher, is added to the detector design,
surrounding the target, to contain gamma rays escaping from target.
There are 354 and 67 10-inch PMTs mounted on the buffer vessel wall 
and veto wall, respectively, pointing inward normal to the wall surfaces. 

The geometrical parameters of the detector modules and the number of 
PMTs are determined based on the MC simulation studies as shown 
later in this chapter.

\subsection{Software Tools}
The primary software tool for modelling the RENO detector response 
is {\sc glg4sim}, a {\sc geant4}-based simulation package for liquid
scintillator detectors derived from {\sc klg4sim} of KamLAND collaboration.
This software is designed for simulation of the detailed detector 
response to particles moving through and interacting with 
a large volume of liquid scintillator detector.

\subsubsection{GEANT4 Simulation}
The RENO detector has four concentric cylindrical sub-detectors 
each filled with Gd-loaded liquid scintillator, liquid scintillator 
without Gd, mineral oil, and water, respectively.   
The \geant4 toolkits are used for simulating the physics processes
involving particles with energies above a few keV
propagating through the materials in the sub-detectors. 
However, the optical photon production 
and propagation through liquid scintillator, including 
processes like absorption, re-emission, and elastic collisions, are handled 
by specifically written codes in \glg4sim. 

\begin{figure}
\begin{center}
\includegraphics[width=3in]{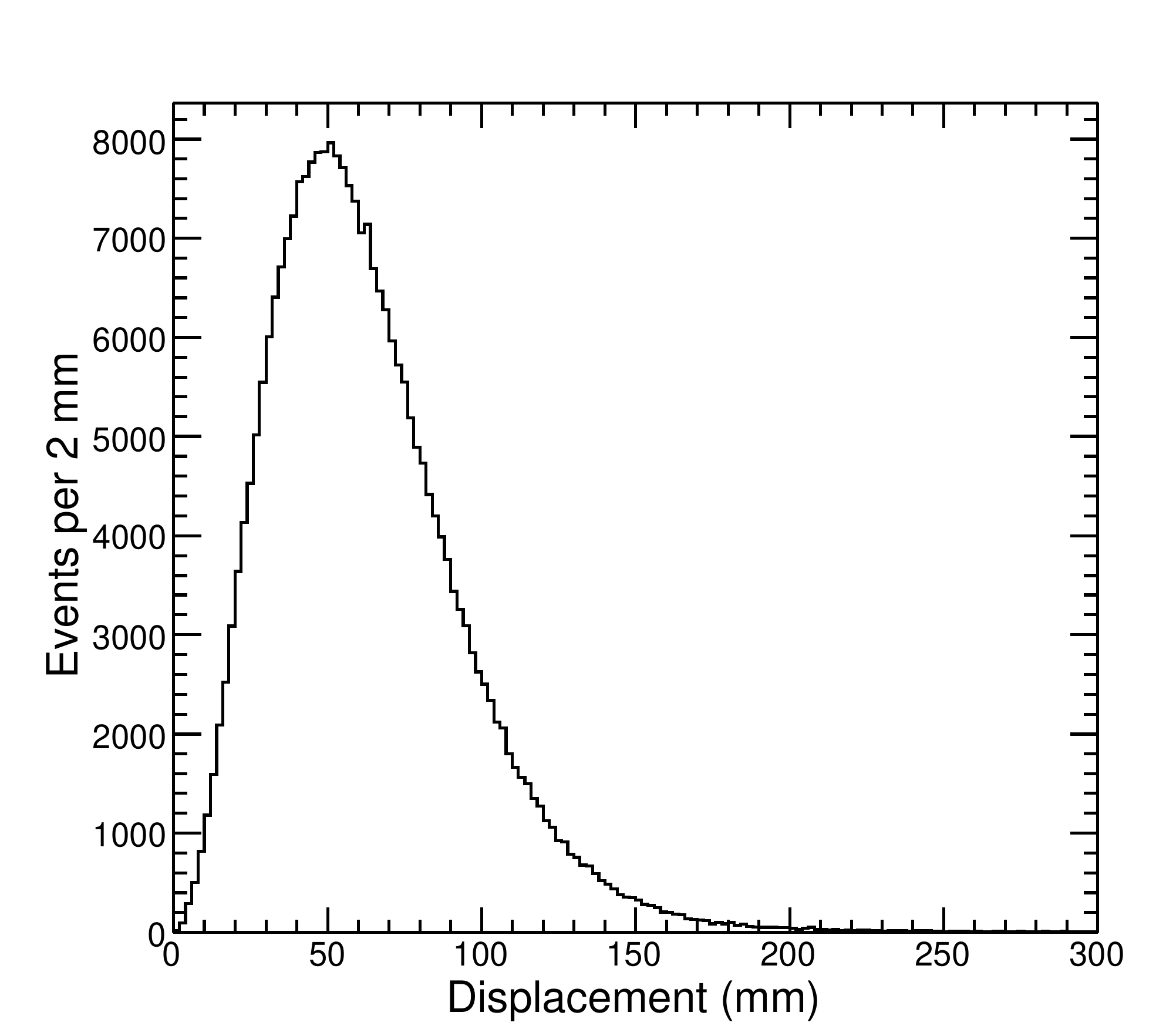}
\includegraphics[width=3in]{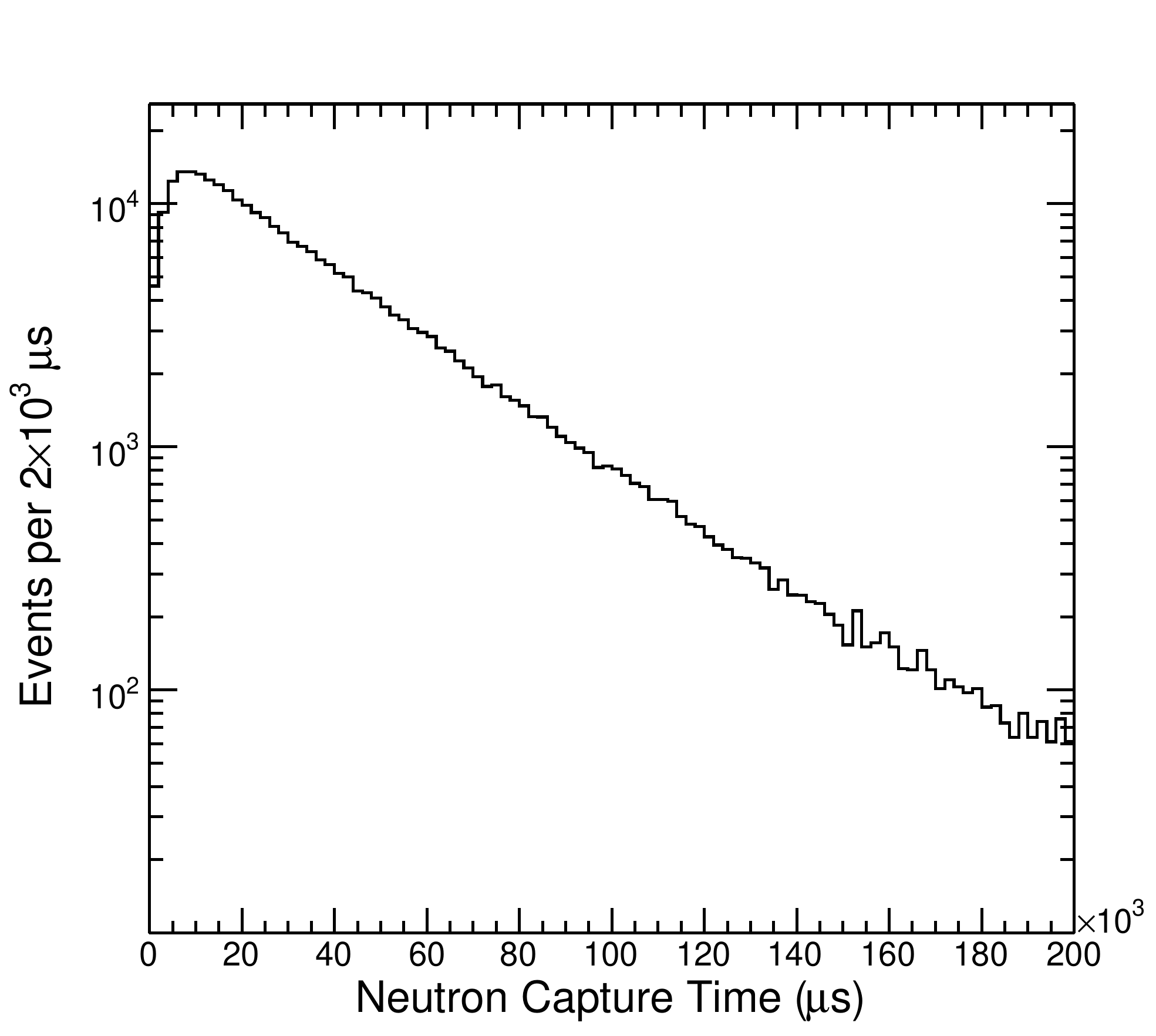}
\end{center}
\caption{Neutron capture distance from inverse beta decay events (left) and
neutron capture time (right).}
\label{detectorSim-captureinfo}
\end{figure}

In the detector simulation, 
the liquid scintillator consists of LAB
for the organic solvent, 1.5~g/l of PPO as a fluor, and 0.3~mg/l of 
Bis-MSB as a secondary wavelength shifter. 
In the target region, 0.1\% 
Gadolinium (Gd) is loaded. 
\geant4 Neutron Data Library (NDL) version 3.8 gives a reasonable 
approximation for the continuum gamma spectrum 
after neutron capture on Gd.
However, the discrete lines of high-energy gammas are not included in the 
NDL version 3.8. 
Fortunately, an update is available for {\sc glg4sim} for an additional 
Gd support for a proper modelling of discrete lines of high energy gamma.
The resulting distributions of the neutron capture distance and capture 
time are shown in Fig.~\ref{detectorSim-captureinfo}.

\glg4sim uses a custom written simulation code for PMT with 
detailed PMT geometries. This PMT simulation handles 
transmission, absorption, and reflection of optical photons at the 
photocathode. The PMT modelling includes a finite photocathode 
thickness and a wavelength dependent photocathode efficiency 
supplied by the PMT manufacturer. 

\subsection{Optical Photon Processes}
Each photon generated in the simulation is tracked in the 
detector until it either reaches a PMT or is lost. The simulation 
accounts for several light propagation phenomena while tracking 
the photons. In the scintillator, photons can undergo absorption 
or elastic scattering (Rayleigh scattering) by solvent and fluor molecules. 

Attenuation length, $\lambda_{att}$, of the liquid scintillator 
is defined as 
\begin{equation}
\frac{1}{\lambda_{att}}=\frac{1}{\lambda_{scat}}+\frac{1}{\lambda_{abs}},
\end{equation}
where $\lambda_{scatt}$ and $\lambda_{abs}$ are the scattering length 
and the absorption length, respectively. The reciprocal value of the 
liquid scintillator attenuation length, $1/\lambda_{att}^{LS}$, is equal 
to the sum of those of scattering lengths and absorption lengths, 
\begin{equation}
\frac{1}{\lambda_{att}^{LS}}=\frac{1}{\lambda_{scat}^{LS}}+\frac{1}{\lambda_{abs}^{LS}} 
= \frac{1}{\lambda_{scat}^{LS}}+\frac{1}{\lambda_{abs}^{solvent}}+\frac{1}{\lambda_{abs}^{fluors}}.
\end{equation}

\begin{figure}[htbp]
  \begin{center}
    \includegraphics[width=10.0cm,clip=]{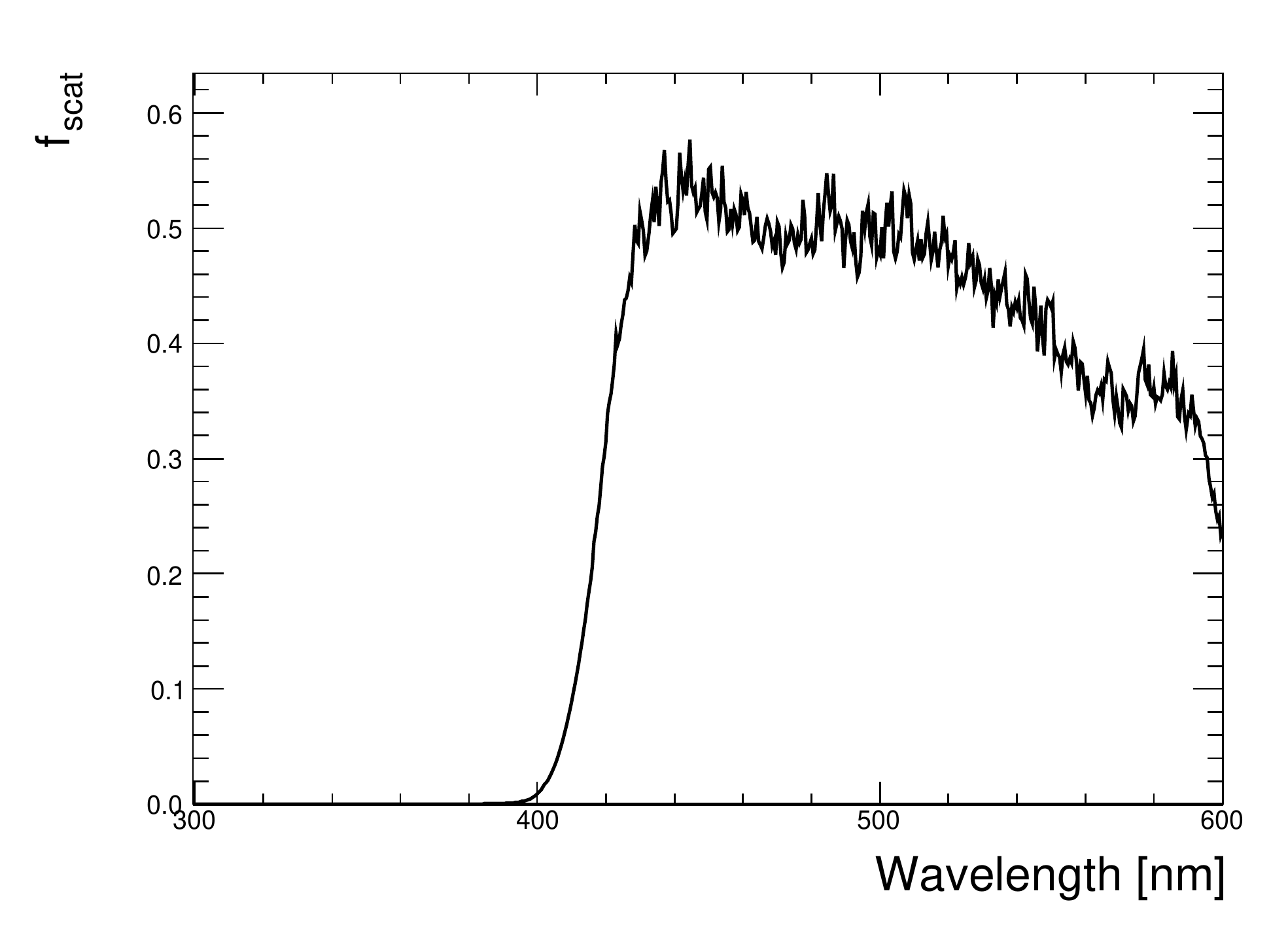}
    \caption{Measured scattering fraction of the LAB based liquid scintillator. 
\label{scatfrac}}
  \end{center}
\end{figure}

In the simulation photons can be either scattered or absorbed by the solvent
and fluors according to appropriate fractions. Because a large fraction of
liquid scintillator is the solvent, photons are scattered mostly by LAB.
It has to be noted that the band gap for the lowest-energy electronic 
transitions in the LAB molecules is at 320~nm, and thus the absorption 
by LAB below 320 nm is strong. At wavelengths longer than 320~nm the 
absorbance by LAB drops rapidly and the measured extinction coefficient 
roughly obeys a $\lambda^{-4}$ dependence, as expected in Rayleigh 
scattering. The scattering fraction, $f_{scatt}$, can be obtained from 
\begin{equation}
f_{scatt} = {\frac{\lambda_{att}^{LS}}{\lambda_{scatt}^{LAB}}}.
\end{equation}

\begin{figure}[htbp]
  \begin{center}
    \includegraphics[width=10.0cm,clip=]{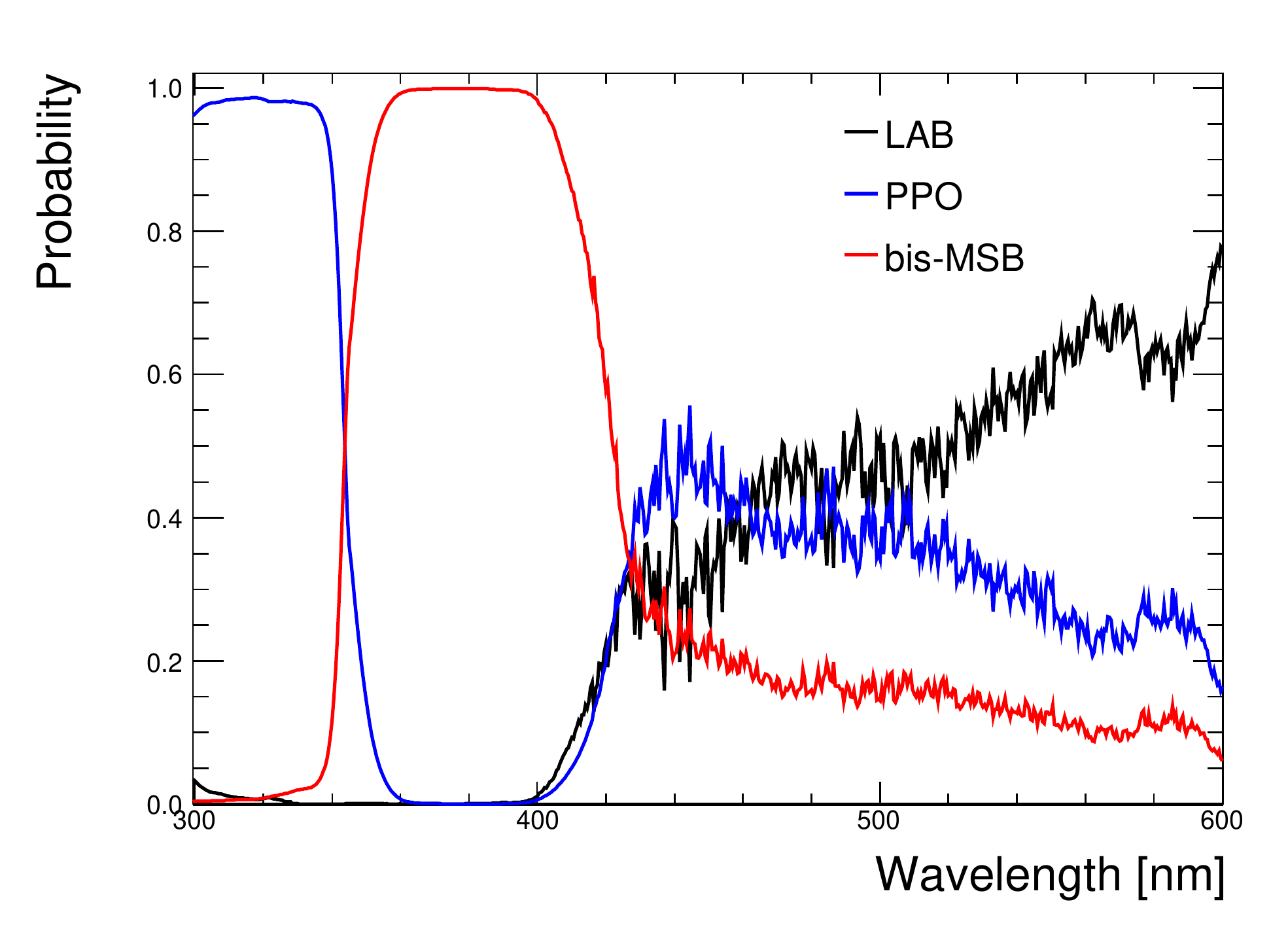}
    \caption{Measured absorption probabilities of LAB, PPO, and bis-MSB. These
             are used in the detector simulation.
             \label{absorptionpro}}
  \end{center}
\end{figure}

Figure~\ref{scatfrac} shows the measured scattering fraction of an optical 
photon in the liquid scintillator.
If a photon undergoes elastic scattering, 
its wavelength remains unchanged but its direction is altered. The direction
of a photon after elastic scattering has an $(1+\cos^2\theta)$ dependence,
where $\theta$ 
is the photon scattering angle.
Absorption of a photon by fluors can be followed by their re-emission,
but there is a chance for an absorbing molecule undergoing non-radiative 
relaxation process depending on its quantum yield efficiency.
The non-radiative relaxation results in the loss of the photon, and 
tracking in the simulation is terminated in that case. The absorption 
probability of LAB, PPO, and bis-MSB can be calculated by 
\begin{equation}
P_{abs}^i = {\frac{\lambda_{abs}^{LS}}{\lambda_{abs}^{i}}},
\end{equation}
where $i$ represents LAB, PPO, or bis-MSB.
Figure~\ref{absorptionpro} shows the measured absorption probability 
for each component in the liquid scintillator. 
Re-emission occurs isotropically and a longer wavelength than 
that of the absorbed photon, based on the emission spectrum, is 
assigned to the re-emitted photon. 

\begin{figure}[htbp]
  \begin{center}
    \includegraphics[width=10.0cm,clip=]{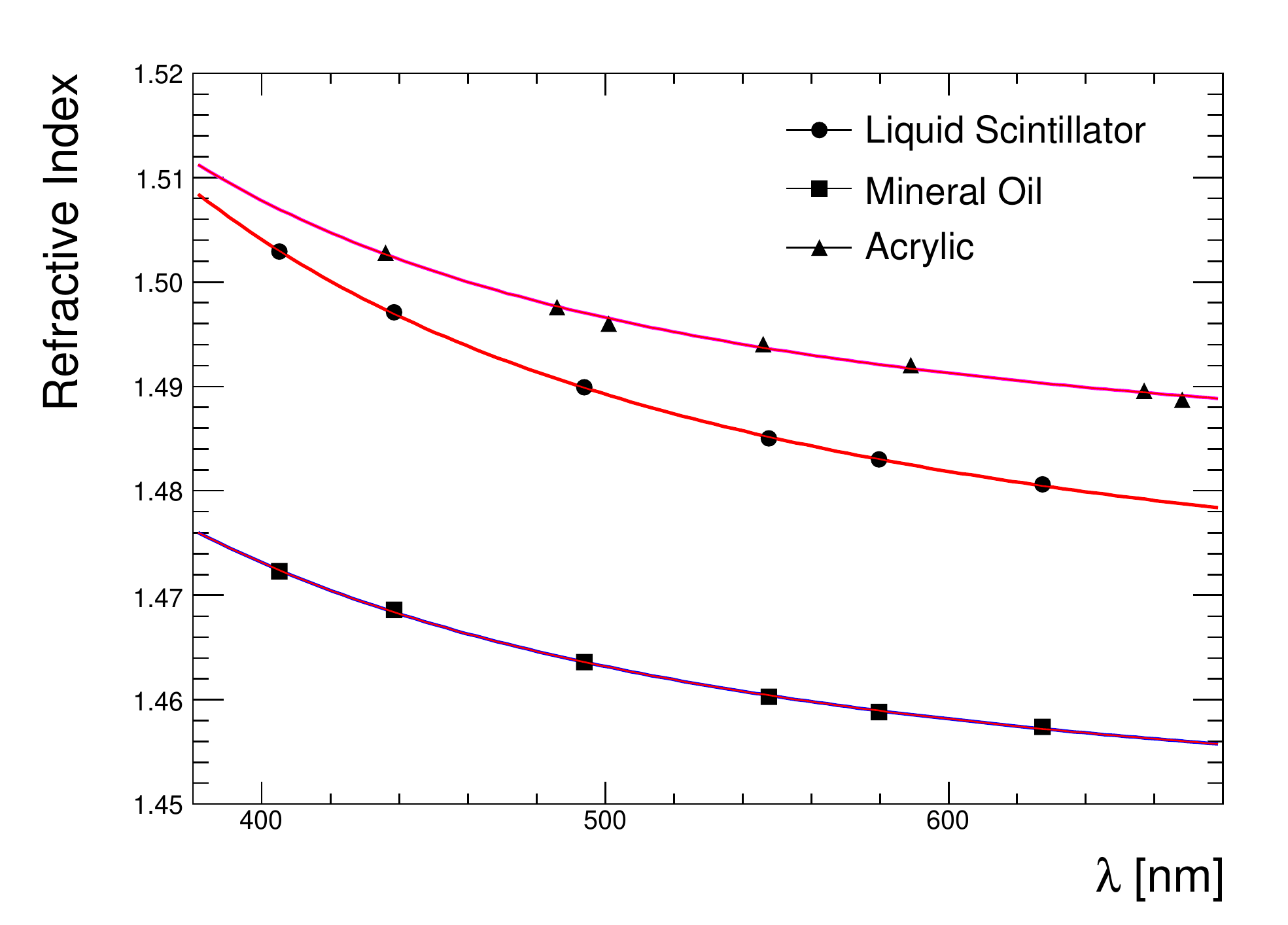}
    \caption{Measured refractive indices of liquid scintillator, mineral oil, 
             and acrylic. \label{refindex}}
  \end{center}
\end{figure}

The absorption of photons within the acrylic medium (vessel walls) 
is simulated according to the absorption probability calculated with
medium's attenuation length.
Also, the reflection and refraction of photons at the surface of the 
acrylic vessel are simulated using the Fresnel's law.
The refractive indices of all dielectric materials in the detector  
are measured at different wavelengths and implemented in the simulation. 
Figure~\ref{refindex} shows the measured refractive 
indices of some of detector materials. 
After a photon enters a PMT and is absorbed by the photocathode, 
tracking is terminated, and a hit is made depending on the quantum 
efficiency of the photocathode.


\subsection{Event Reconstruction}
\subsubsection{Vertex Reconstruction}
For vertex reconstruction, two independent algorithms,
``charge weighting method'' and ``likelihood method,'' have been 
used.
The charge weighting method is simple and fast, and is suitable for 
the online event display or as a filter to extract interesting 
events to apply more sophisticated event selection criteria. 
The likelihood method has a better vertex position
resolution than the charge weighting method but requires more CPU
time and therefore to be used as an offline reconstruction method.

\begin{figure}
\begin{center}
\includegraphics[width=10cm]{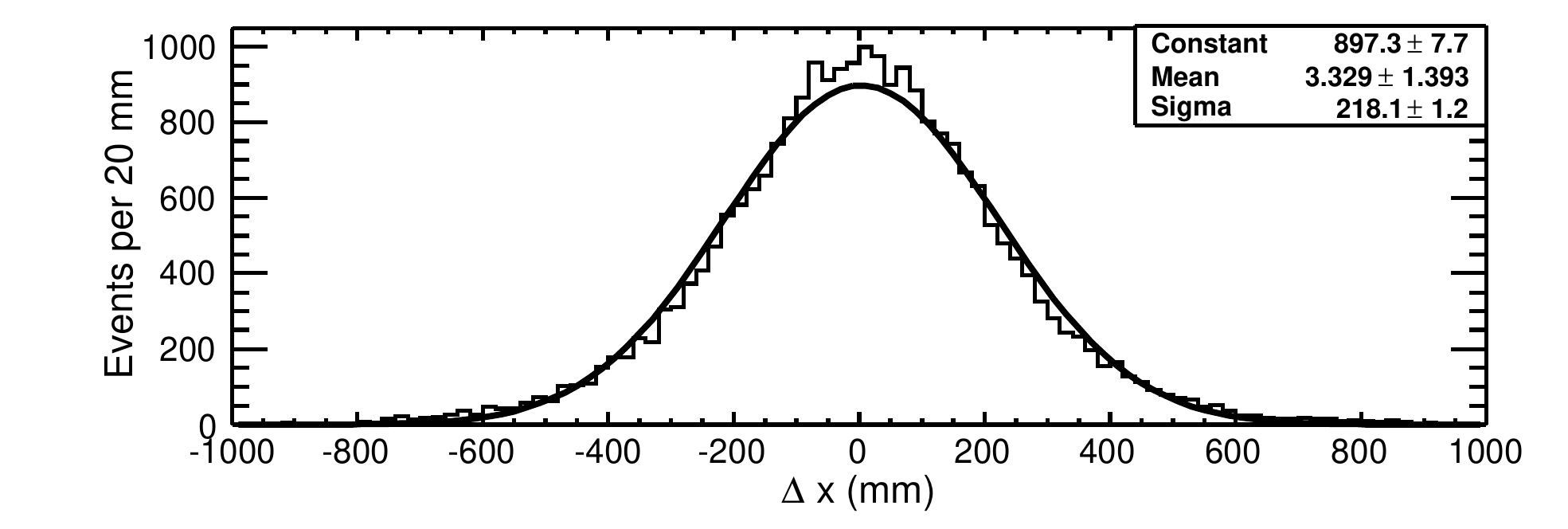}
\includegraphics[width=10cm]{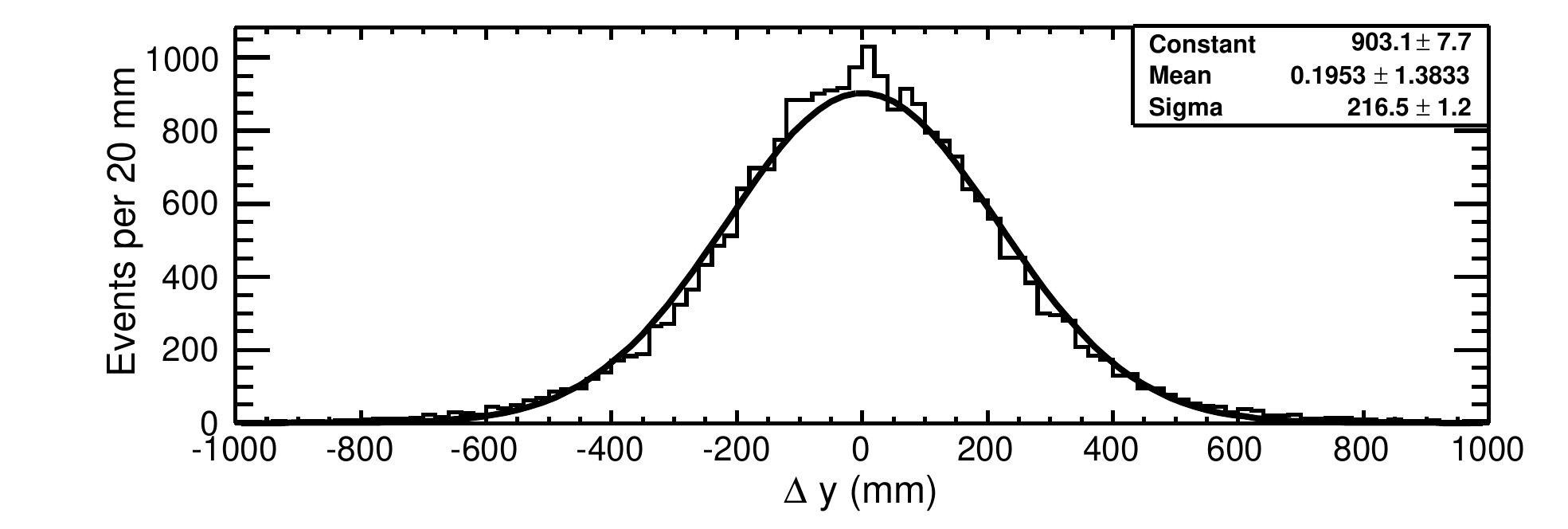}
\includegraphics[width=10cm]{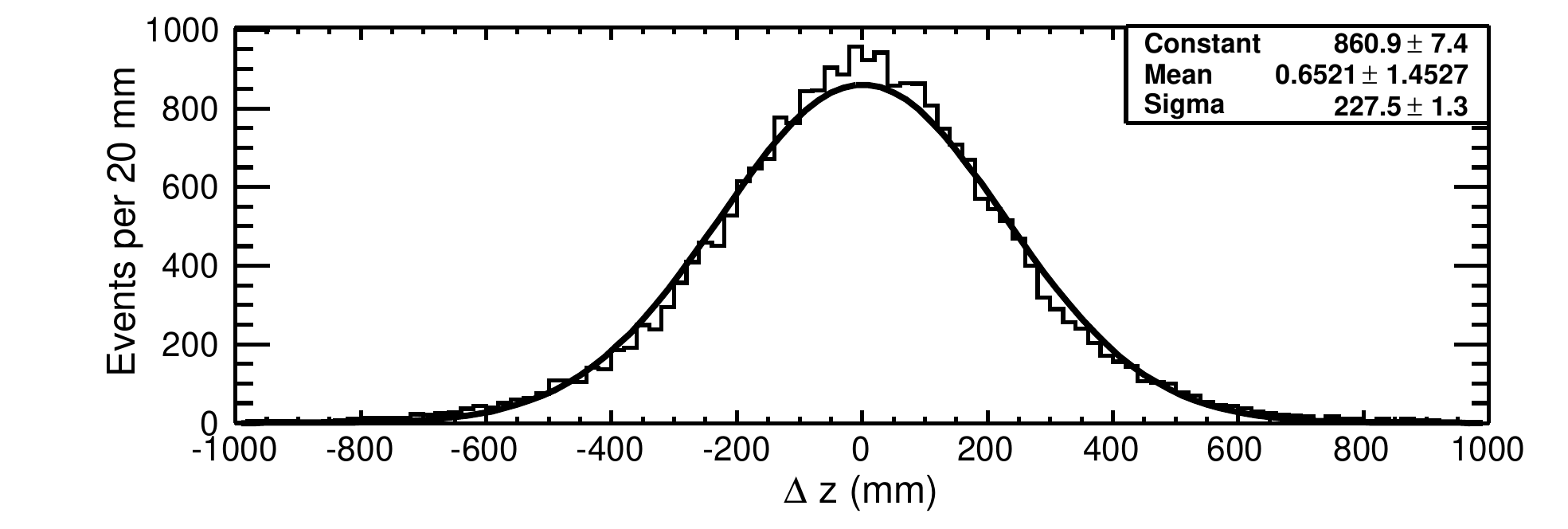}
\caption{Difference between reconstructed and the generated vertex positions
for an 1~MeV $\gamma$ rays in random direction in the target using a 
simple weighting method. 
}
\label{vertex reconstruction}
\end{center}
\end{figure}

The event vertex in the charge weighting method is calculated as  
\begin{equation}
\vec{r}_{vtx}
=\frac{\displaystyle \sum_{i=PMT}n_i \vec{r}_i}
      {\displaystyle \sum_{i=PMT}n_i },
\end{equation}
where $n_i$ is the number of photoelectrons on the $i$th 
PMT and $\vec{r}_i$ is the vector pointing from the center of the
detector to the $i$th PMT. 
The number of photoelectrons is calculated by $n_i = c_i q_i$, where
$q_i$ and $c_i$ are the amount of charge measured on the $i$th
PMT and the charge-to-number of photoelectron conversion factor
on that PMT.
Because the reconstructed vertex 
position calculated with the charge weighting method is inherently closer 
to the center of the detector than the actual vertex position, linear 
corrections are applied based on the detector simulation. The 
position resolution is found to be $\sim 38$~cm for a 1~MeV gamma ray
as shown in Fig.~\ref{vertex reconstruction} and improves 
for a higher energy gamma.

The likelihood method uses not only the number of scintillation
photons detected by PMTs but also the arrival time of those photons. 
The expected number of photoelectrons on the $i$th PMT can be written as
\begin{equation}
\nu_i = N_{tot}{A_i\cdot f(\cos\theta_i)\over 4\pi R_i^2}
\epsilon_i\cdot\prod_{j} e^{-R_{ij}/\lambda_j},
\label{expected number of photoelectrons}
\end{equation}
where $N_{tot}$ is the total number of optical photons generated, $A_i$ and 
$\epsilon_i$ are the frontal area of the cathode and quantum efficiency 
of the PMT, respectively, $R_{ij}$ is the distance from the
vertex to the PMT in medium $j$, and $\lambda_j$ is the 
attenuation length of the $j^{\rm th}$ medium in between the vertex 
and the PMT.  
The effective area of photocathode of the PMT seen from 
the incident angle $\theta_i$ is accounted for in function 
$f(\cos\theta_i)$.

The likelihood is then written as
\begin{equation}
{\cal L} = \prod_{i=PMT} {\cal G}(n_i, \vec{r}; \nu_i, \sigma_i)
\cdot {\cal T}(t_i;n_i, R_i),
\label{likelihood}
\end{equation} 
where ${\cal G}(n_i;\nu_i,\sigma_i)$ is the Gaussian probability
with its mean of $\nu_i$ and width of $\sigma_i$, and
${\cal T}(t_i;n_i,R_i)$ is the probability of having the first
hit of $n_i$ hitting the $i$th PMT to have the hit time of 
$t_i$. 
The number of observed photoelectrons, $n_i$, is calculated from
the charge output from PMT using the charge--to--photoelectron
conversion factor from calibrations.
The negative log likelihood is then minimized using {\sc minuit} to
find the vertex position and the total number of optical photons 
created.

\subsubsection{Energy Reconstruction}
Energy is calculated from the number of reconstructed optical photons.
Since the PMT coverage depends on the position of event vertex, 
the reconstructed energy is dependent on 
the vertex position. This is naturally incorporated in the
expected number of photoelectrons in 
Eq.~\ref{expected number of photoelectrons}.
Energy can be written as $E= \xi N_{tot}$
where $\xi$ is a constant determined from simulation and source 
calibration. 

\begin{figure}
\begin{center}
\includegraphics[width=10cm]{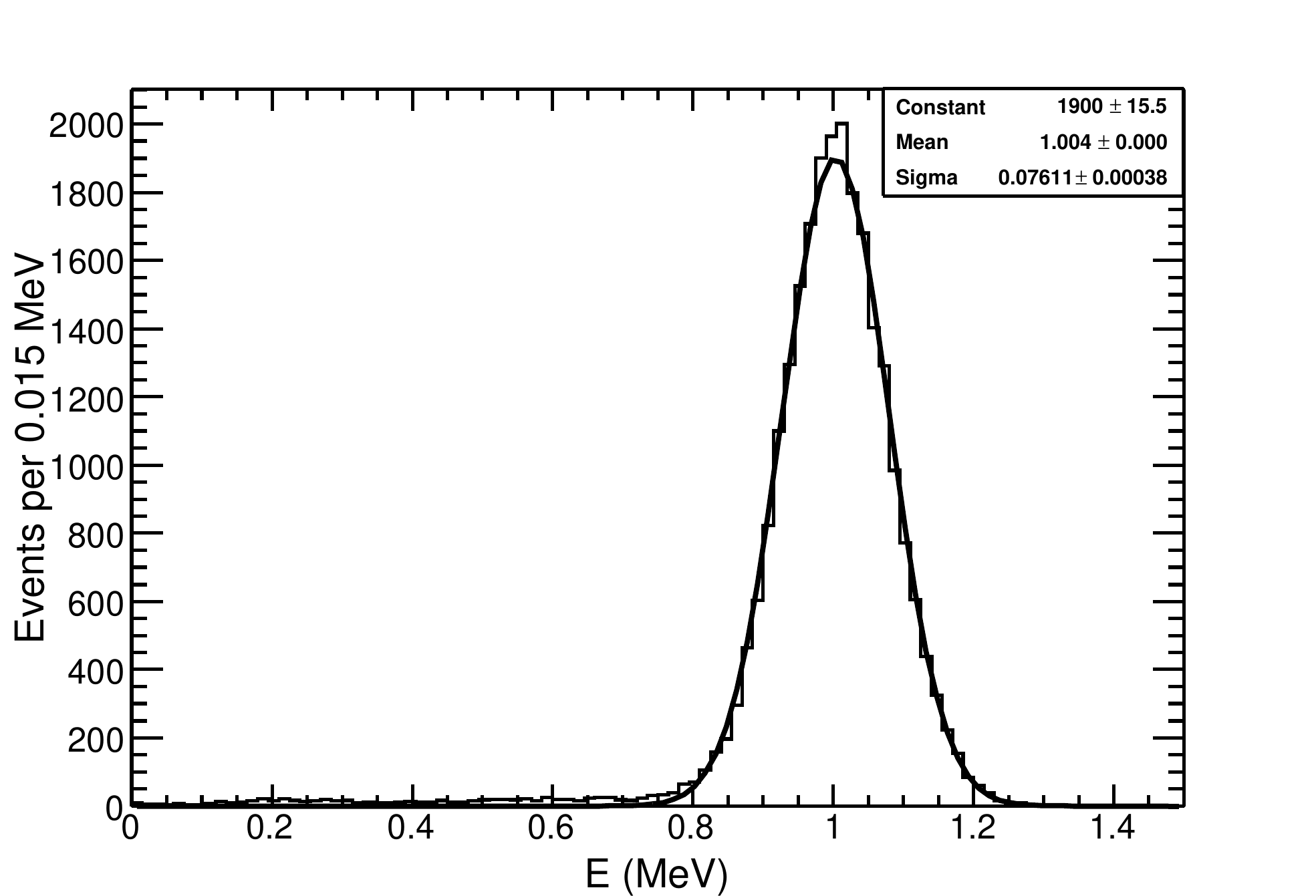}
\caption{Reconstructed energy distribution of an 1~MeV $\gamma$ in the 
target.}
\label{energy reconstruction}
\end{center}
\end{figure}

\begin{figure}
\begin{center}
\includegraphics[width=3.0in]{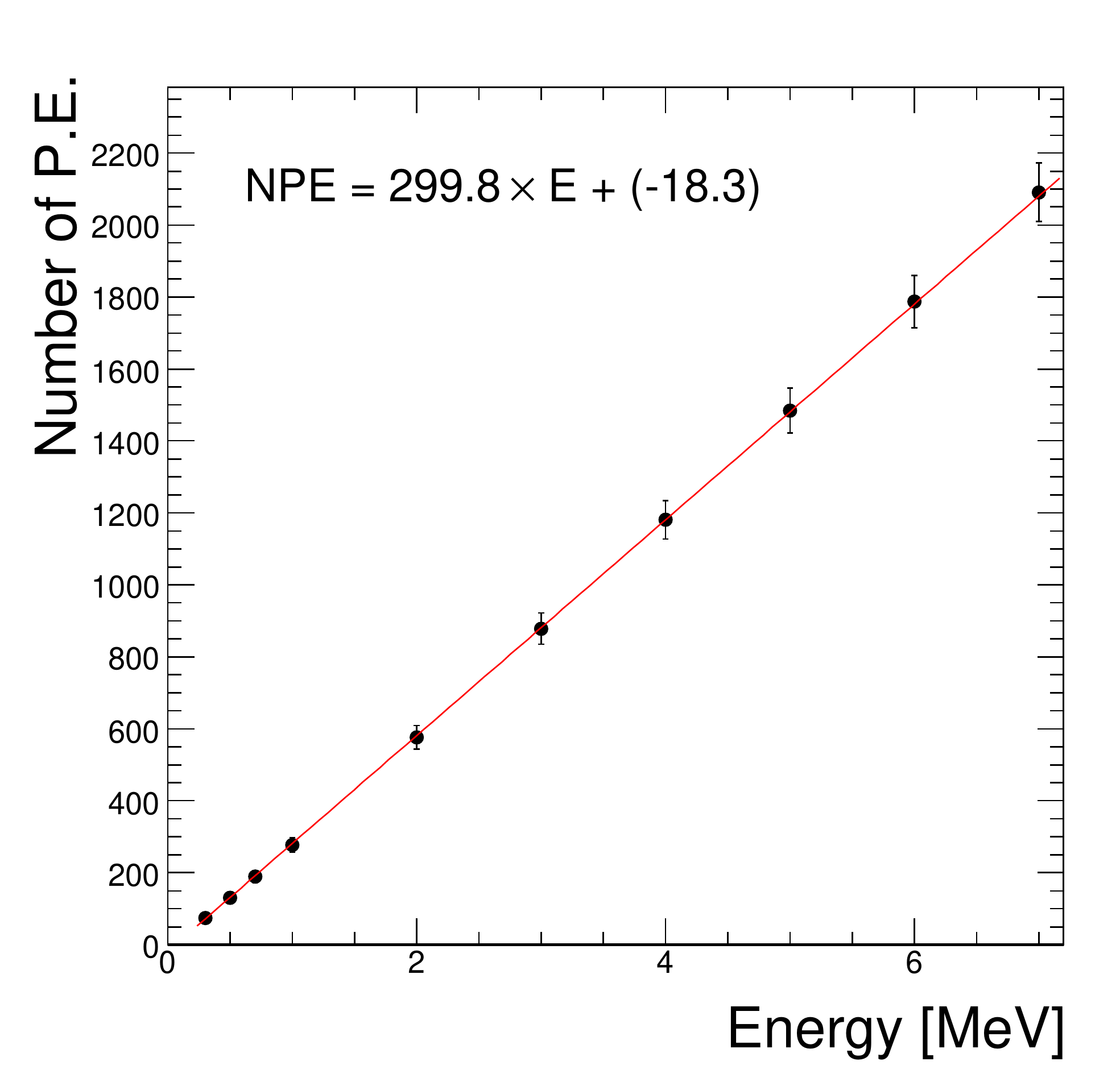}
\includegraphics[width=3.0in]{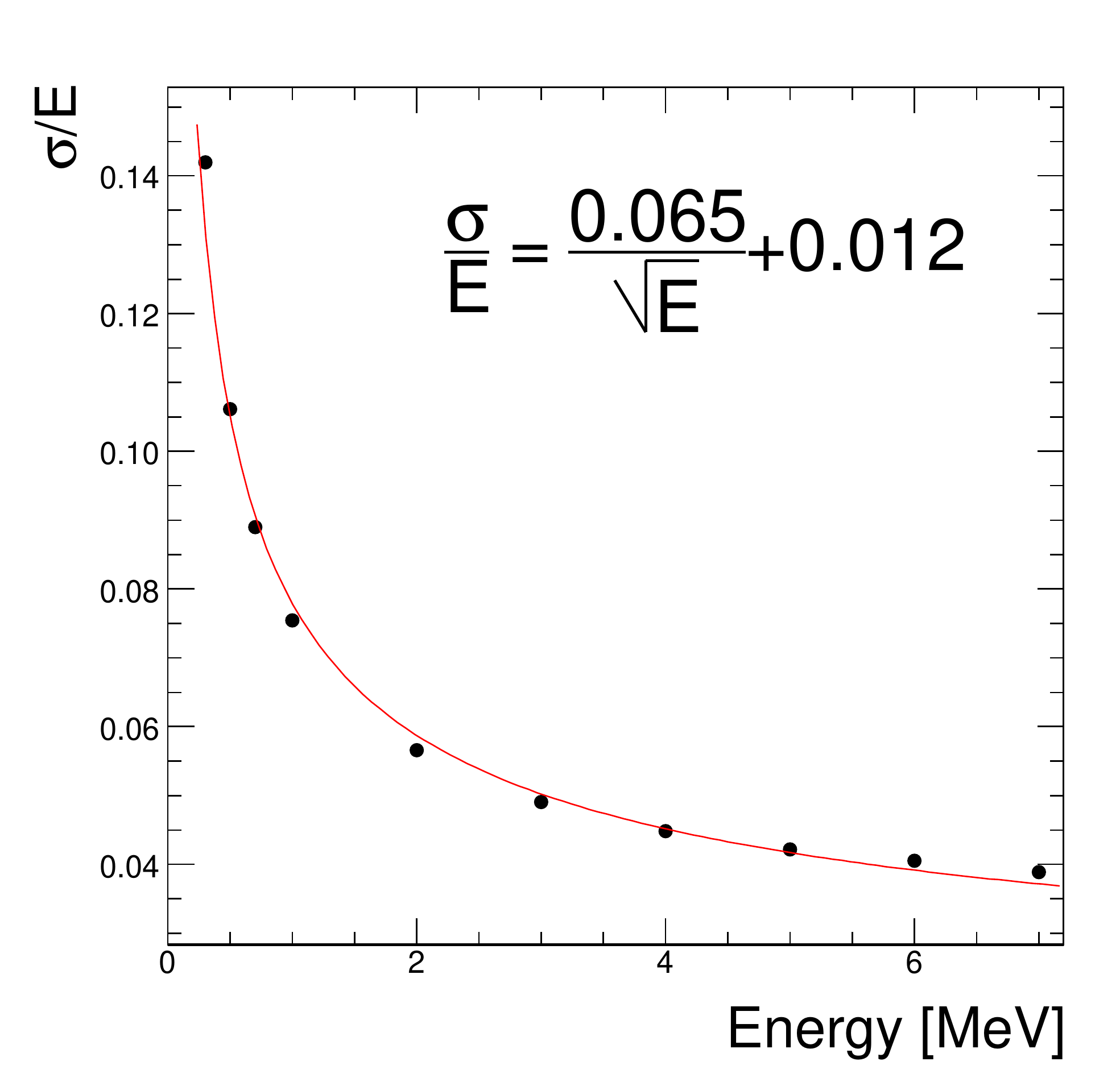}
\end{center}
\caption{Energy linearity (left) and energy resolution (right). 
}
\label{escale}
\end{figure}

The energy reconstruction has a good linearity and has the
resolution of ${\delta E\over E}={0.065\over\sqrt{E}}\oplus 0.012$, 
where $E$ is given in MeV. 
Figure~\ref{energy reconstruction} shows the reconstructed energy
distribution for an 1~MeV gamma ray and Fig.~\ref{escale} shows energy
linearity and resolution.

\subsubsection{Muon Tracking}
\begin{figure}
\begin{center}
\includegraphics[width=4in]{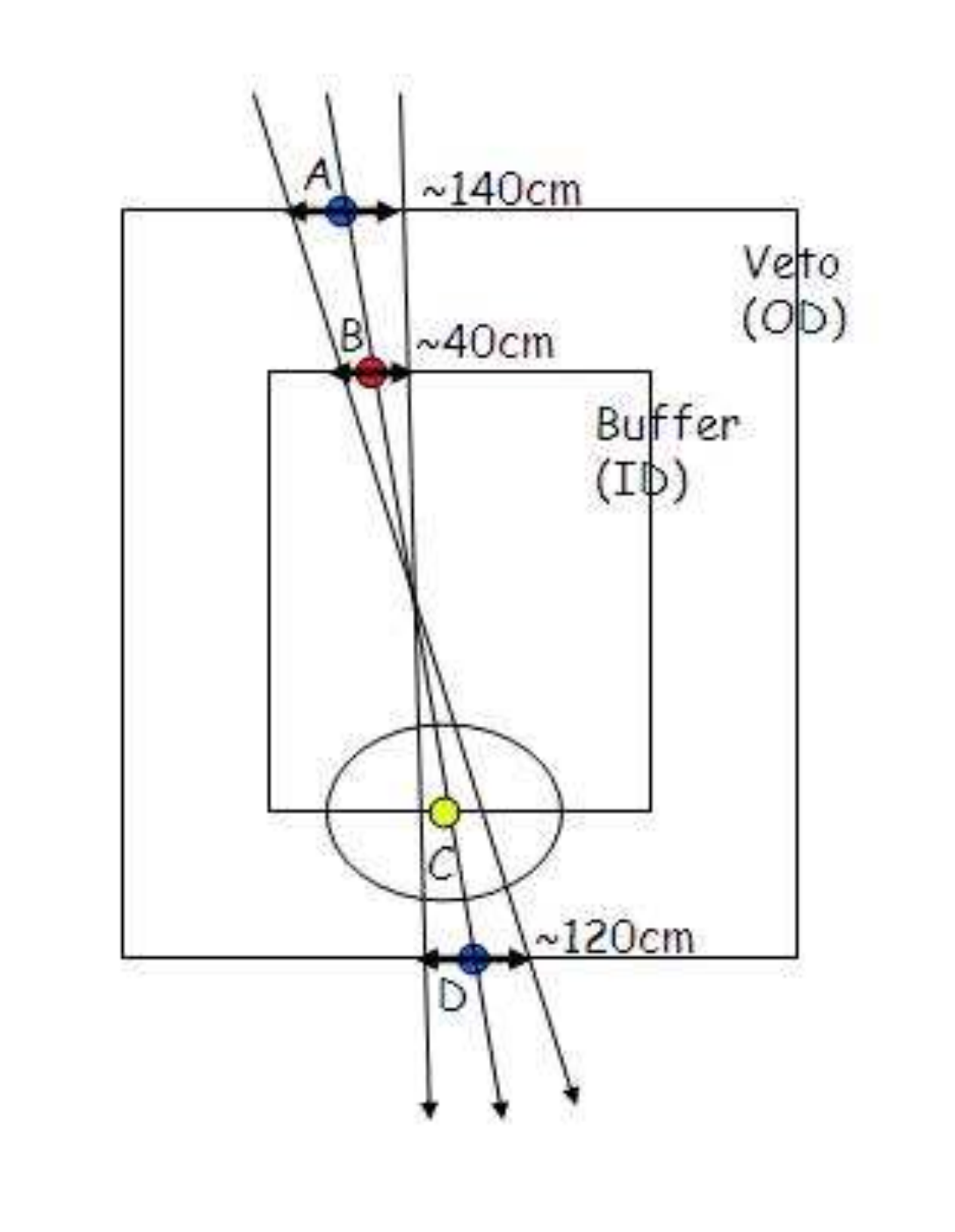}
\caption{Schematic of the reconstructed through-going muon track. 
The best fit muon track is constructed with hit information from
the inner detector PMTs for the entrance point and the outer 
detector PMTs for the exit point ($\bar{BD}$).}
\label{muonTrack}
\end{center}
\end{figure}

\begin{figure}
\begin{center}
\includegraphics[width=4in]{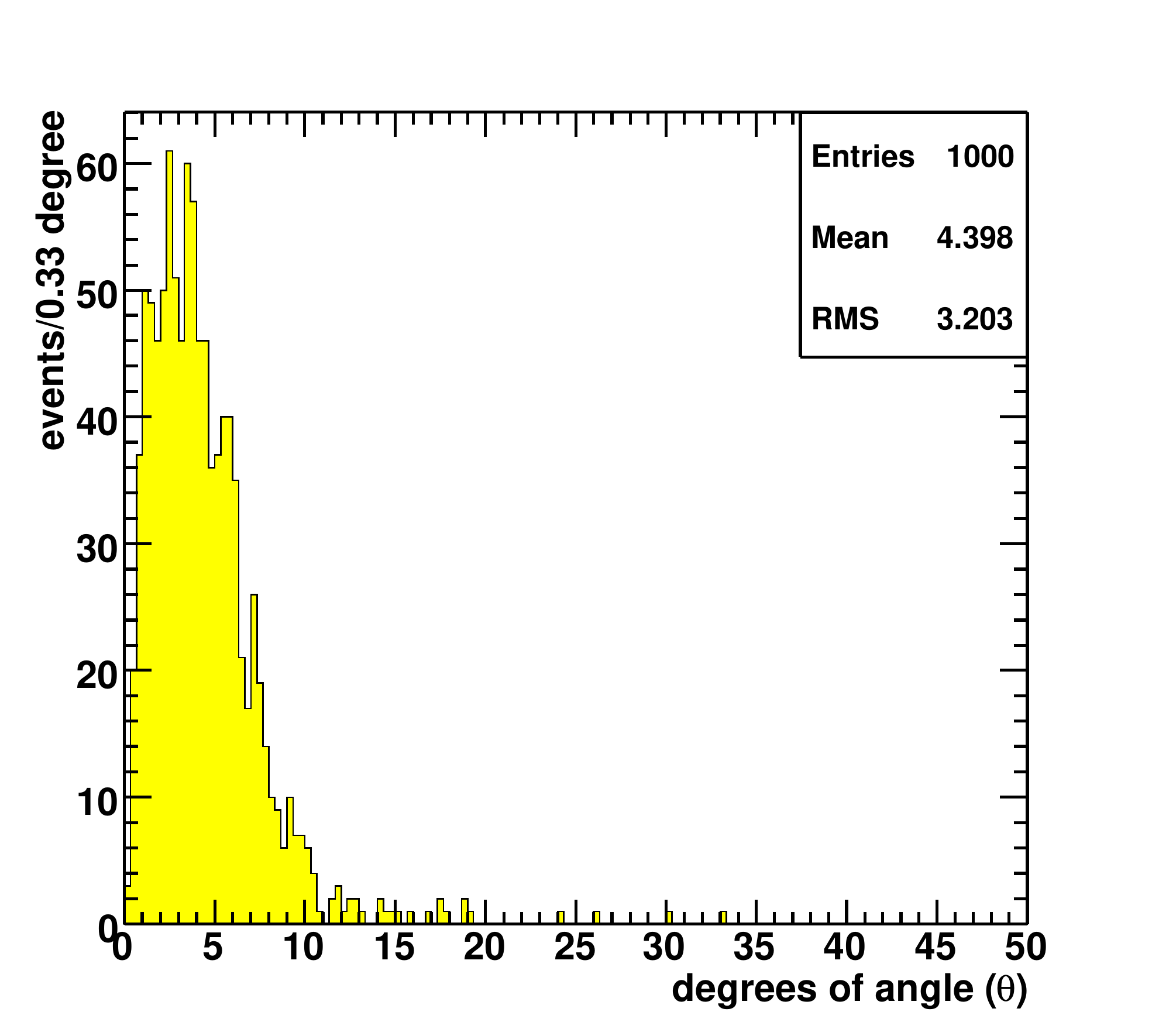}
\caption{Angular difference between the reconstructed muon 
track and the simulated muon track.}
\label{angleDiff}
\end{center}
\end{figure}

Even though the detectors are installed underground, the cosmic muons
can still penetrate into the surrounding rocks and generate fake signals
similar to the inverse beta decay from the anti-neutrinos. To reduce
the background events from these cosmic muons, we have developed muon
tracking algorithm.
\begin{itemize}
\item{Muons passing through the detector} \\
The RENO detector consists of four concentric cylindrical modules.
Those modules are filled with different liquids; Gd loaded liquid 
scintillator in target, liquid scintillator in $\gamma$-catcher,
mineral oil in buffer, and water in veto.
Therefore, the muons can generate photons in two ways through the 
detector; Cerenkov radiation and scintillation.
\item{Test of the earliest photon hit time} \\
To study the arrival time of photons from incident muons, we have 
generated muons events with $20~\GeVc$, going down vertically through 
the center of the detector. The relationship between the photon hit 
time and the number of photoelectrons was investigated for each PMT.
It was found that the earliest photon hit was always registered 
in a PMT within a group of neighboring PMTs with photon hits (PMT cluster),
which is true regardless of photon generation mechanism of a muon event. 
We also investigated the relationship between the distance of closest 
approach to a PMT and the amount of light it sees relative to the 
neighboring PMTs for the exit point of muon track.
\item{Muon tracking algorithm} \\
The muon tracking algorithm finds the incident muon's entrance point on 
the detector by taking the PMT with the earliest time that has at least 
two nearest neighboring PMTs in a cluster. 
And the exit point of the muon track is 
determined using the PMT with the highest
number of photoelectrons that also has at least two nearest neighboring 
PMTs with hits in a cluster.
The muon track is obtained by tracing the straight path from the entrance and exit positions.
\end{itemize}
From the simulation, when muons pass through the active layers (target
and $\gamma$ catcher), muons are tagged with an 100\% efficiency.
To reconstruct the trajectory of each through-going muon, we combine 
information obtained from inner detector PMTs, which are mounted on the buffer 
vessel, as well as outer detector PMTs, which are mounted on veto wall. 
The best fit muon track is reconstructed by tracking the path from 
entrance point from inner detector PMTs and the exit point from outer 
detector PMTs (Fig.~\ref{muonTrack}).
The angular difference between the reconstructed muon track and the 
simulated muon track is shown in Fig.~\ref{angleDiff}.


\subsubsection{Event Display}
\begin{figure}[htbp]
\begin{center}
\includegraphics[width=15cm]{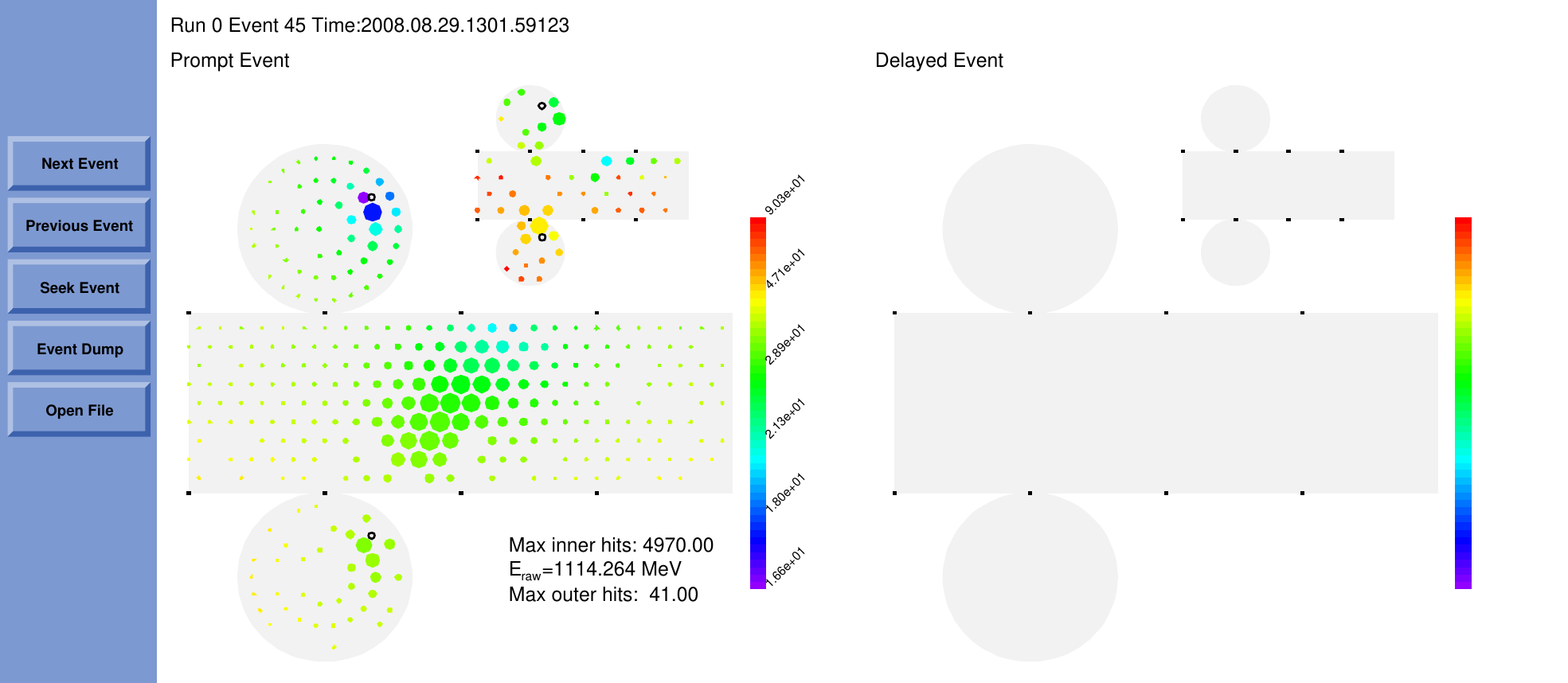}
\includegraphics[width=15cm]{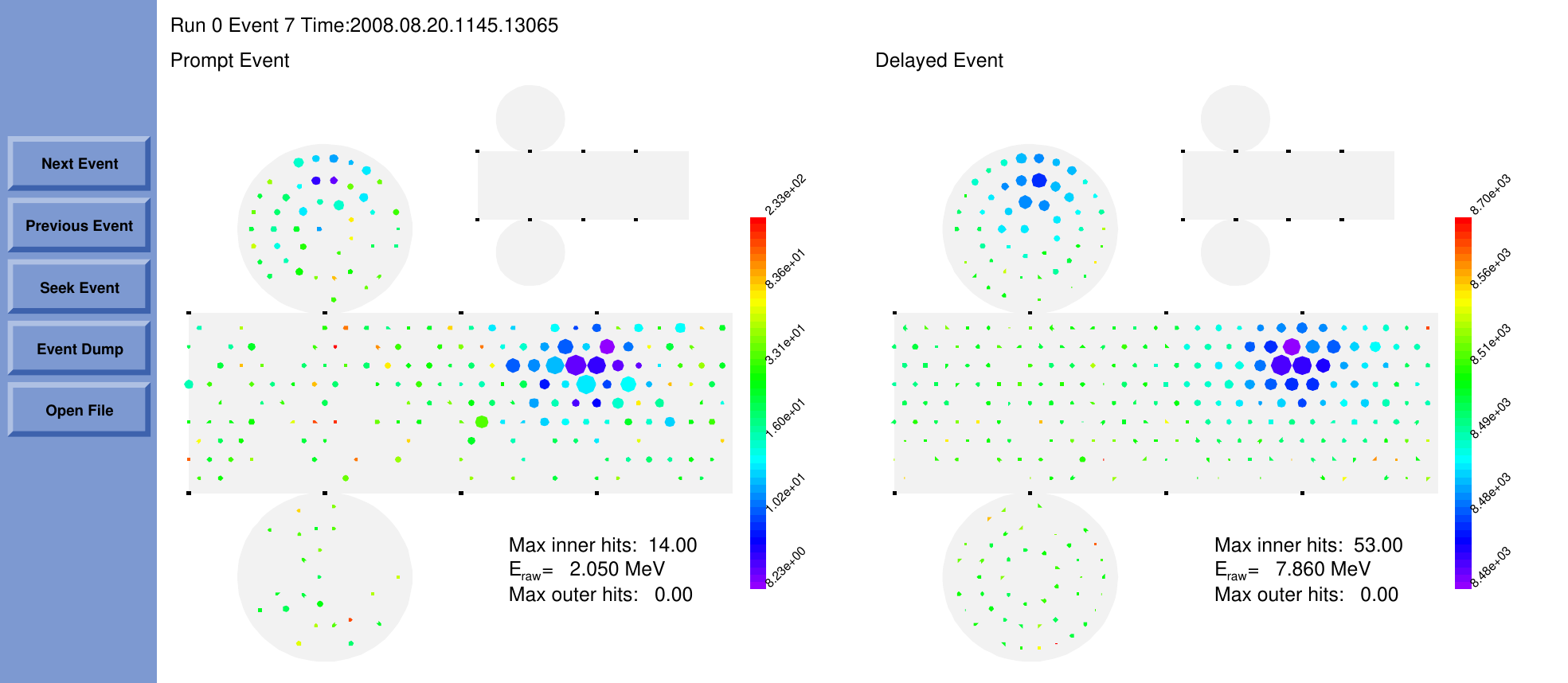}
\caption{Examples of 2D--event display of a simulated cosmic ray event (top) 
and of a simulated inverse beta decay event (bottom). The large (small) 
cylinder development represents the buffer (veto) vessel surface where 
The colored circles shows the hits on PMTs with their radius 
proportional to the number of hits. The first hit time is 
color coded. 
Note that the PMT locations are not for the latest design.
}
\label{event display}
\end{center}
\end{figure}

An event display tool based on {\sc root} is under development. 
Figure~\ref{event display} shows a rudimentary event display of 
sample events.

\section{Optimization of the Detector Design}
One of the important goals of the Monte Carlo simulation studies 
is to design a cost effective detector with a good detection efficiency
and performance.
%
%
This section describes how the optimization of the RENO detector
was done.

\subsection{Target Mass}
\begin{figure}
\begin{center}
\includegraphics[width=4.0in]{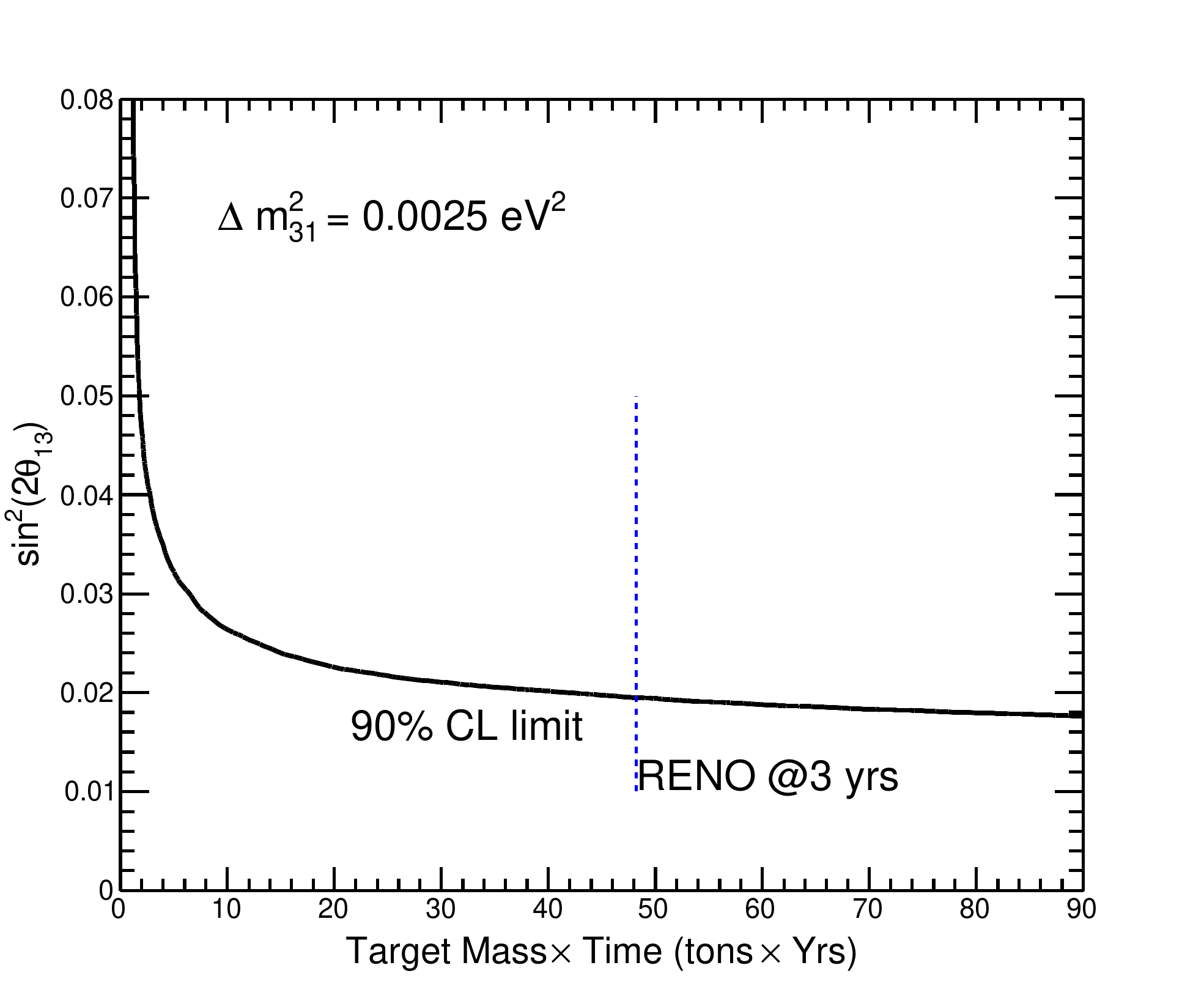}
\end{center}
\caption{%
The 90\% CL limit of $\sin^2(2\theta_{13})$ as a function of number of years 
of data taking times target mass at $\Delta m^2_{31}=0.0025~\hbox{eV}^2$. The 
intercept in the limit by the dashed line shows the expected limit by
RENO experiment with target mass of 16 tons in three years. 
}
\label{detectorPerformance-targetMass}
\end{figure}

With the goal of having systematic uncertainty less than 1\%,
we estimated the 90\% confidence level (CL) limit on $\sin^2(2\theta_{13})$ as a function 
of target mass times number of years of data taking as shown in 
Fig.~\ref{detectorPerformance-targetMass} (see Sect.~\ref{section: sensitivity}). 
The target mass is determined to be 16 tons as a result of compromise 
between construction cost and measurement sensitivity. With three years 
of data taking and the target mass of 16 tons we expect to set a 
limit on the neutrino mixing angle down to $\sin^2(2\theta_{13})=0.02$. 


\subsection{$\gamma$-Catcher Thickness}
\begin{figure}
\begin{center}
\includegraphics[width=4in]{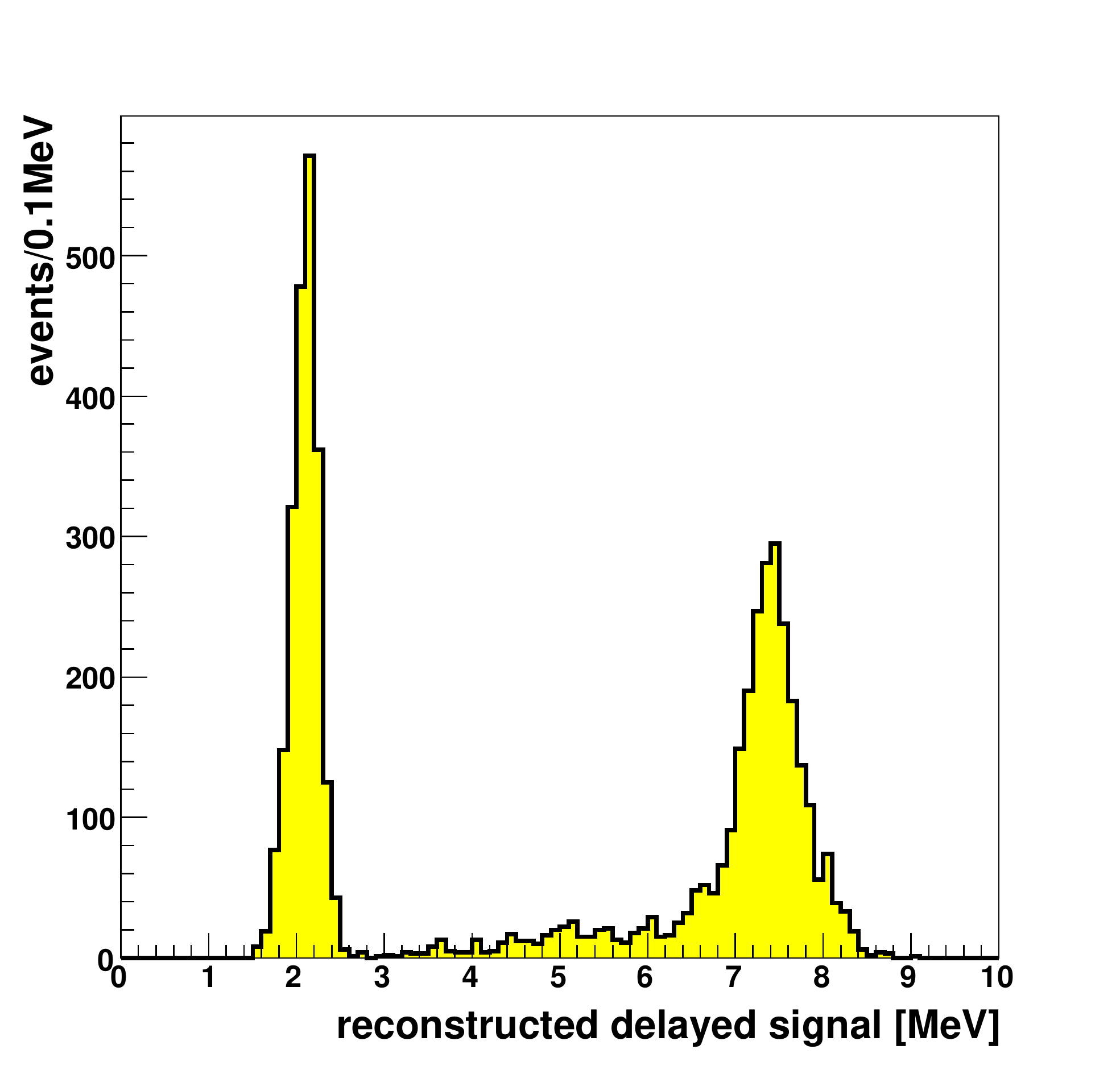}
\end{center}
\caption{The reconstructed energy spectrum of neutrons captured 
in the target volume. The peaks at $\sim 2.0$~MeV and $\sim 7.5$~MeV
are from neutrons captured by hydrogen and Gd, respectively.
}
\label{detectorPerformance-catcher-recE}
\end{figure}

The purpose of $\gamma$-catcher is to contain energy of gamma rays 
escaping from the target vessel.
In target, $\gamma$ rays are produced by the positron-electron 
annihilation and by the neutron capture. 
It is found that some of gamma rays from the neutron capture can escape
even from $\gamma$-catcher, and cannot be fully contained if
$\gamma$-catcher is not thick enough. 
Using the simulation, neutron detection efficiency
has been obtained for several different thicknesses of $\gamma$-catcher
as shown in Fig.~\ref{detectorPerformance-catcher-recE}.
Here, again a compromise has been made between
detection efficiency and the construction cost. 
The thickness of $\gamma$-catcher is chosen to be 60~cm where 93\% of
Gd captured neutron events passing $E>6$~MeV requirement.

\begin{figure}
\begin{center}
\includegraphics[width=4in]{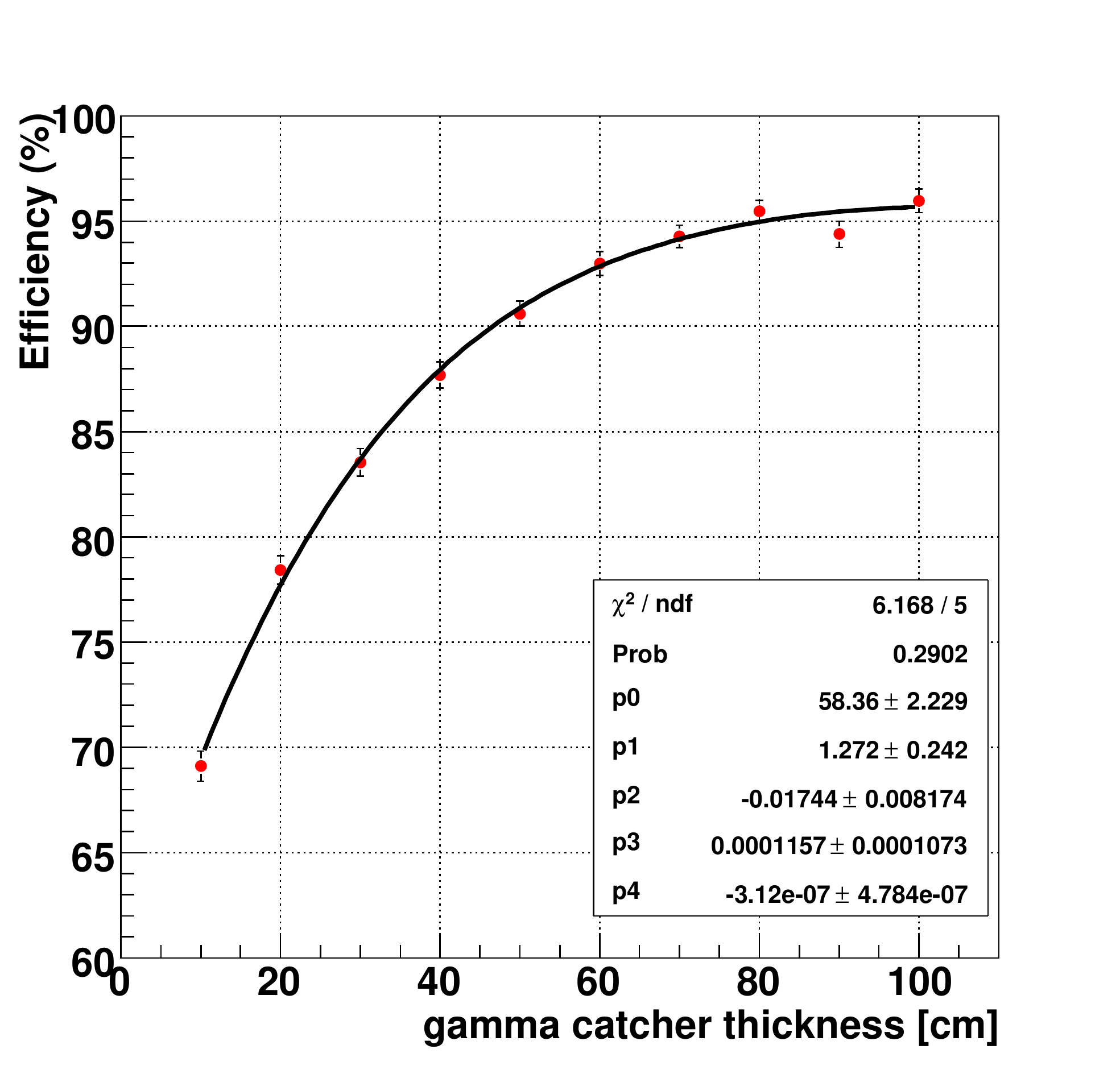}
\end{center}
\caption{The Gd captured neutrino detection efficiency as a 
function of the $\gamma$-catcher thickness. The cut used is
$E>6$~MeV.}
\label{detectorPerformance-catcher-eff}
\end{figure}

\subsection{Buffer Thickness}
The buffer layer is used to shield the active area of target and
$\gamma$-catcher against photons from radioactivities in the PMTs 
and the surrounding rocks. 
Increasing the thickness of the buffer layer reduces these photon
backgrounds while increasing the construction cost as well. Therefore, 
the thickness of the buffer layer is determined to be large enough 
to reduce the photon backgrounds to an acceptable level from
Monte Carlo studies.

The gammas from the radioactive decays of $^{40}$K, $^{232}$Th and
$^{238}$U in PMTs are simulated and traced through the whole detector
media. The optical photons generated in the liquid scintillator due to
the gamma rays are traced in the detector until they reach the
PMTs. The simulation was repeated for six buffer
thicknesses from 50~cm to 100~cm in steps of 10~cm and the results are
shown in Table~\ref{buffer thickness}.
We chose the buffer thickness of 70~cm to limit the rate of events above
1~MeV coming from the radioactivities.



\begin{table}
\begin{center}
\begin{tabular}{ccccc} \hline
Buffer Thickness (cm) & $^{40}$K & $^{232}$Th & $^{238}$U & Total \\ \hline
50 & 3.4 & 8.0 & 11.2 & 22.6 \\ 
60 & 2.0 & 5.3 & 7.1 & 14.4 \\ 
70 & 1.4 & 3.9 & 5.1 & 10.4 \\ 
80 & 0.8 & 2.7 & 3.6 & 7.1 \\ 
90 & 0.5 & 1.6 & 1.9 & 4.0 \\ 
100 & 0.3 & 1.4 & 1.1 & 2.8 \\ \hline
\end{tabular}
\end{center}
\caption{Event rates from radioactivities in the PMTs with energy above 
1~MeV. Rates are in Hz.}
\label{buffer thickness}
\end{table}

\section{Detection Efficiencies and Uncertainties}
\subsection{Introduction}
Based on the various studies we have attempted to design the RENO
detector with a good efficiency and small systematic uncertainties.
Due to addition of $\gamma$-catcher and buffer modules, several geometrical
requirements used by CHOOZ experiment are not needed at RENO experiment;
fiducial cuts on positron and neutron vertices, and the cut on the
positron-neutron vertex distance.   

\subsection{Positron Detection Efficiency}\label{positron efficiency}
\begin{figure}[htbp]
  \begin{center}
    \includegraphics[width=4in]{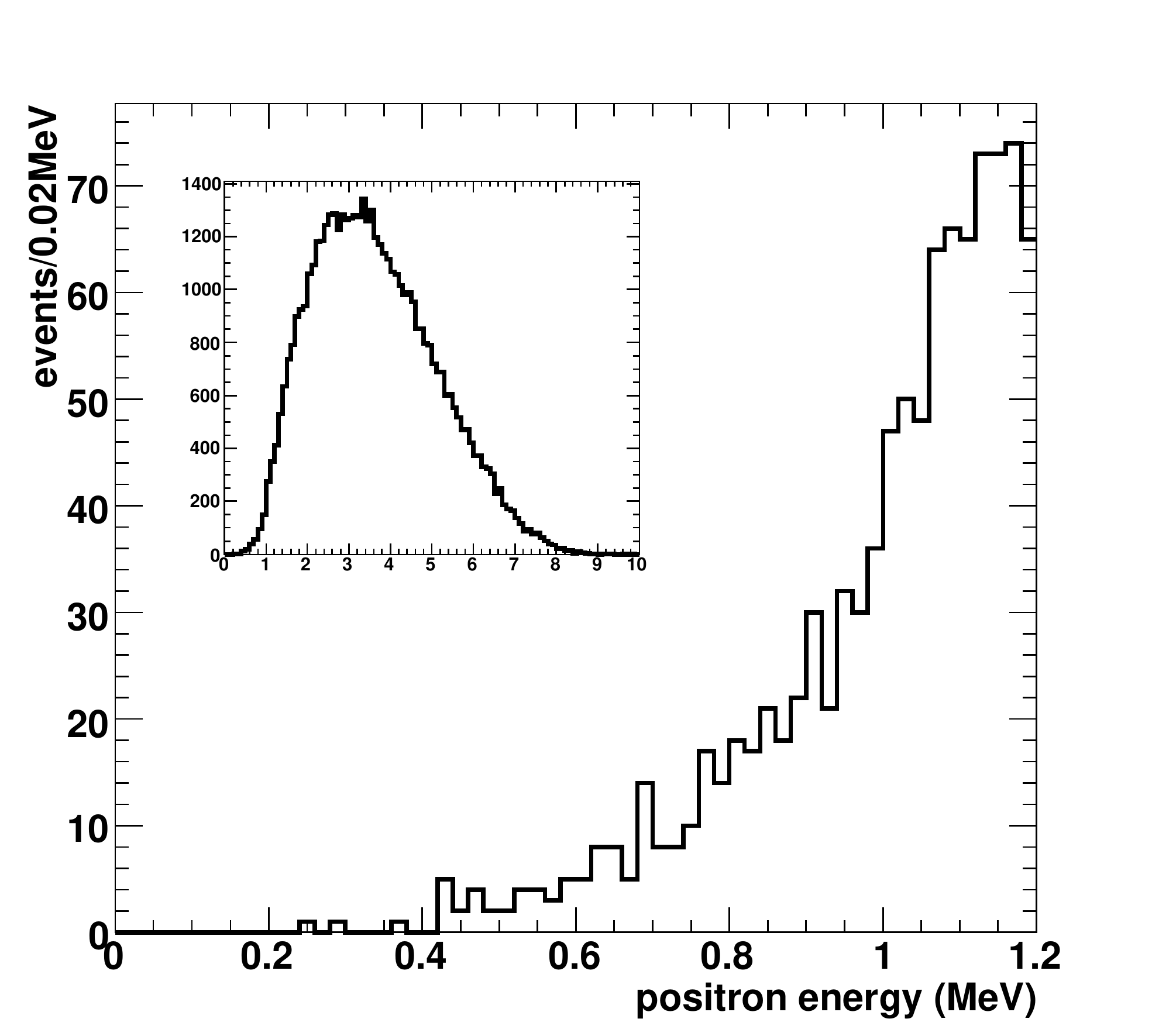}
     \caption{Simulated measured positron energy spectrum near the threshold from reactor neutrinos. 
      The full range of the spectrum is shown in the inset.}
\label{prompt}
  \end{center}
\end{figure}

An IBD event is identified by a prompt signal from the positron
followed by the neutron capture on Gd producing gamma rays with
total energy of about 8~MeV.
The positron deposits its kinetic energy through scintillation and 
then annihilates to yield two gammas with 0.51~MeV each. Thus
the minimum energy from the prompt signal is 1.02~MeV. 
Therefore, the energy requirement of $E>1$~MeV for the prompt 
signal should be fully efficient.
However, the
reconstructed energy could be less than 1~MeV as shown in 
Fig.~\ref{prompt} because of a finite energy resolution of the detector,
which is 6.5\% at 1~MeV.
We studied the change of the reactor neutrino event rate depending on
the positron energy threshold assuming the energy scale uncertainty
of 2\% using Monte Carlo calculations.
The detection efficiencies are found to be 99.23, 99.14, and 99.05\%
at the positron energy thresholds of 0.94, 1.00, and 1.02~MeV,
respectively.
Therefore, we estimate the positron detection efficiency with the
1~MeV threshold requirement as $99.1\pm 0.1$\%.

\subsection{Neutron Detection Efficiency}
The efficiency for detecting a neutron is given by
\begin{equation}
{\varepsilon}=P_{Gd}{\varepsilon_{E}}{\varepsilon_{T}},
\end{equation}
where $P_{Gd}$ is the probability of a neutron being captured by Gd,
$\varepsilon_{E}$ is the efficiency of the $E>6$~MeV cut for gamma
rays from the neutron Gd capture, and $\varepsilon_{T}$ is the efficiency 
of the delayed time cut, $0.3 < T < 200~\mu$s.
The combined neutron detection efficiency is estimated to be $79.3\pm0.5$\%.

\subsubsection{Neutron Capture Fraction on Gd}
\label{Neutron Capture Fraction on Gd}
\begin{figure}[htbp]
  \begin{center}
    \includegraphics[width=4in]{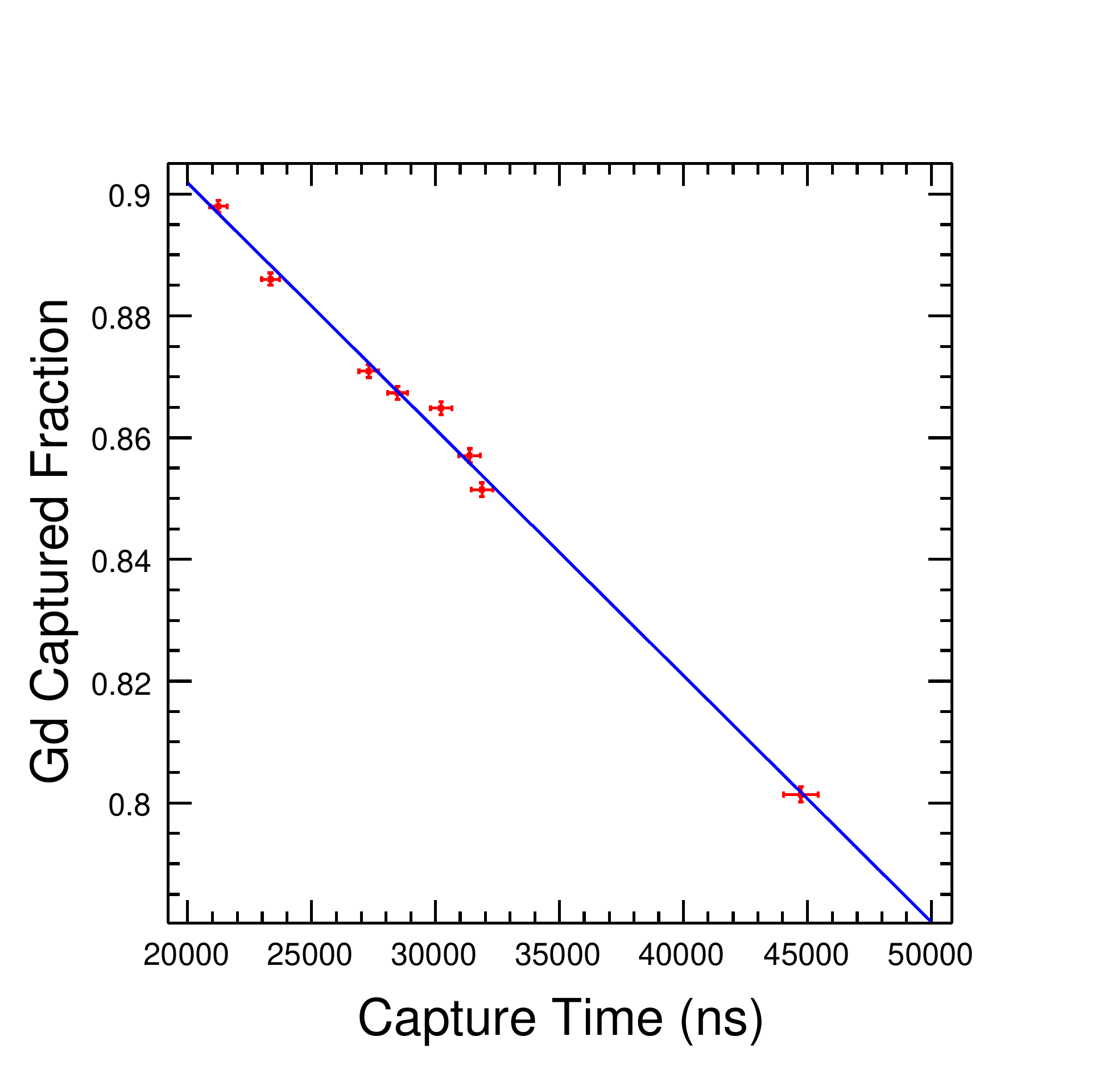}
    \caption{Fraction of neutrons captured by Gd vs. capture time for various
    Gd concentrations in the target liquid scintillator. The Gd concentration 
    was varied from 0.06 to 0.15\%.\label{capture}}
  \end{center}
\end{figure}

The neutron capture fraction on Gd is expected to be 85.3\% in 
a liquid scintillator with a 0.1\% Gd concentration.
We studied the relation between the neutron capture fraction on Gd and 
the capture time with the various concentration of Gd using Monte Carlo 
calculations. In Fig.~\ref{capture} the result shows that the capture 
time difference of 1~$\mu$s gives about 0.4\% change in the Gd capture 
fraction.
Our goal is to measure the relative Gd captured neutron fraction to 0.4\%
precision and ultimately to 0.1\%. Since we expect to measure the
capture time to better than $1~\mu$s uncertainty with less than $10^4$ 
neutron capture events, which can be obtained in a few minutes of 
data taking with a neutron source,
we take 0.4\% to be the neutron capture fraction uncertainty.

\subsubsection{Efficiency of Energy Cut}
The gammas Gd captured neutron event have a total energy of $\sim 8$~MeV,
whereas low energy backgrounds mainly coming from radioactive sources mostly
have energy below 1~MeV. To discriminate low energy backgrounds as well
as prompt signals and hydrogen captured neutron signals, $E>6$~MeV
requirement is imposed on the delayed signal. 
This cut was studied using Monte Carlo simulation (see 
Fig.~\ref{detectorPerformance-catcher-eff}).
The efficiency of the energy cut is found to be $93.0\pm 0.4\%$,
assuming the energy scale uncertainty of 2\%.
\subsubsection{Efficiency of Delayed Time Cut}
The delayed time cut of $1.0<T<200~\mu$s was studied using an MC simulation
with a 0.1\% Gd concentration in the liquid scintillator.
The efficiency of the delayed time cut is found to be 95\%. 
Assuming an 1~ns precision of the electronics, the systematic 
uncertainty of the efficiency of this cut is negligible. 

\subsubsection{Neutron Multiplicity}
In CHOOZ experiment the muon-induced spallation neutron events were 
rejected with an efficiency of $97.4\pm 0.5\%$~\cite{Apollonio03}.
In RENO these events will be identified with a very high efficiency
due to having the veto layer unlike CHOOZ.
Therefore, we consider the cut efficiency on the neutron multiplicity
fully efficient and its systematic uncertainty to be negligible.

\subsection{Dead Time}
In RENO experiment the electronics does not introduce dead time. 
However, vetoing cosmic muon events can introduce dead time.
We plan to use a 0.5~ms veto duration after each muon event.
In RENO experiment the muon fluxes are estimated to be 5.5~Hz/m$^2$
and 0.85~Hz/m$^2$ for near and far detectors, respectively.
Therefore, we estimate the veto efficiencies associated
with dead time to be 75.4\% and 95.8\% for near and far detectors,
respectively.

\subsection{Summary}
The detector related efficiencies with respect to inverse beta decay events 
are summarized in Table~\ref{detectionEffsummary}.
The total detection efficiencies are expected to be $56.3\pm 0.5$\% and 
$71.6\pm 0.5$\% for near and far detector, respectively.

\begin{table}[htbp]
\begin{center}
\begin{tabular}{lcc}\hline
 Cut	&\multicolumn{2}{c}{Efficiency (\%)}\\
 &CHOOZ &RENO  \\\hline
 Positron energy 	&$97.8\pm0.8$ 	&$99.1\pm 0.1$ \\
 Positron-geode distance 	&$99.9\pm0.1$ 		&--  \\
 Neutron capture 	&$84.6\pm1.0$ 	&$85.3\pm 0.4$  \\
 Capture energy containment &$94.6\pm0.4$ 	&$93.0\pm 0.4$  \\
 Neutron-geode distance	&$99.5\pm0.1$ 	&--  \\
 Neutron delay 	&$93.7\pm0.4$ 	&$95.0\pm 0.0$  \\
 Positron-neutron distance 	&$98.4\pm0.3$ 	&--  \\
 Neutron multiplicity 	&$97.4\pm0.5$ 	&--  \\
 Dead time &-- &$75.4$(near) \\
 & &$95.8$(far) \\\hline
Total &$69.8\pm1.5$ &$56.3\pm 0.5$(near) \\
 & &$71.6\pm 0.5$(far) \\\hline
\end{tabular}
\end{center}
\caption{Event selection efficiencies.}
\label{detectionEffsummary}
\end{table}

\section{Background Simulation}
\label{back sim}
\subsection{Simulation of Muon Background}
An accurate and efficient method of calculating the muon intensity 
and energy at the underground site requires the detailed profile of 
surrounding terrain and the parameterization of muon intensity at 
its surface. 
A standard atmospheric muon parameterization given by 
Gaisser~\cite{gaisser} is well known for  describing the muon intensity 
at the surface in the high energy region but poor in the low energy 
region. 
Since detector sites are at relatively shallow depth ($\sim$200 m), 
low energy muons can survive at the depth. 
To consider the effect, we use the modified Gaisser 
parameterization~\cite{mgaisser} in the low energy region. 
The Gaisser parameterization is written as 
\begin{equation}
\frac{dN_{\mu0}}{dE_{\mu0}d\Omega} 
\simeq A\frac{0.14 E^{-\gamma}_{\mu0}}{\mbox{cm}^{2}\mbox{ sr \textit{s} GeV}}
\left\{\frac{1}{\displaystyle 1+\frac{1.1 \tilde{E}_{\mu0}\cos\theta^{*}}{115}}
+
\frac{0.054}{\displaystyle 1+\frac{1.1 \tilde{E}_{\mu0}\cos\theta^{*}}{850}}
+r_{c} \right\},
\end{equation}
where muon energy $E_{\mu0}$ at the surface is measured in GeV. 
The standard Gaisser parameterization has $A=1$, $\gamma=2.7$, 
$\tilde{E}_{\mu0}=E_{\mu0}$, and $r_{c}=0$. 
The detailed parameters for the modified Gaisser parameterization are 
\begin{eqnarray}
\gamma&=&2.70,\nonumber
\\
r_{c}&=&10^{-4},\nonumber
\\
\Delta&=&2.06\times10^{-3}\left(\frac{950}{\cos\theta^{*}}-90\right),
\nonumber
\\
\tilde{E}_{\mu0}&=&E_{\mu0}+\Delta, 
\nonumber
\\
\hbox{and} \nonumber
\\
A&=&1.1\left(\frac{90\sqrt{\cos\theta+0.001}}
{1030}\right)^{\frac{4.5}{\tilde{E}_{\mu0}\cos\theta^{*}}},\nonumber
\end{eqnarray}
where $\cos\theta^{*}$ is calculated using a simple geometrical 
extrapolation as 
\begin{equation}
\cos\theta^{*}=
\sqrt{\frac{x^{2}+p_{1}^{2}+p_{2}x^{p_{3}}+p_{4}x^{p_{5}}}
           {1+p_{1}^{2}+p_{2}+p_{4}}}.
\end{equation}
Here $x$ denotes $\cos\theta$ and $\theta$ is the angle subtended 
between the incoming cosmic ray particle 
and the normal to the upper atmospheric layer, 
and $p_{1}=0.102573$ ,$p_{2}=-0.068287$, $p_{3}=0.958633$, 
$p_{4}=0.0407253$, and $p_{5}=0.817285$. 

\begin{figure}[htbp]
\begin{center}
\includegraphics[width=12cm]{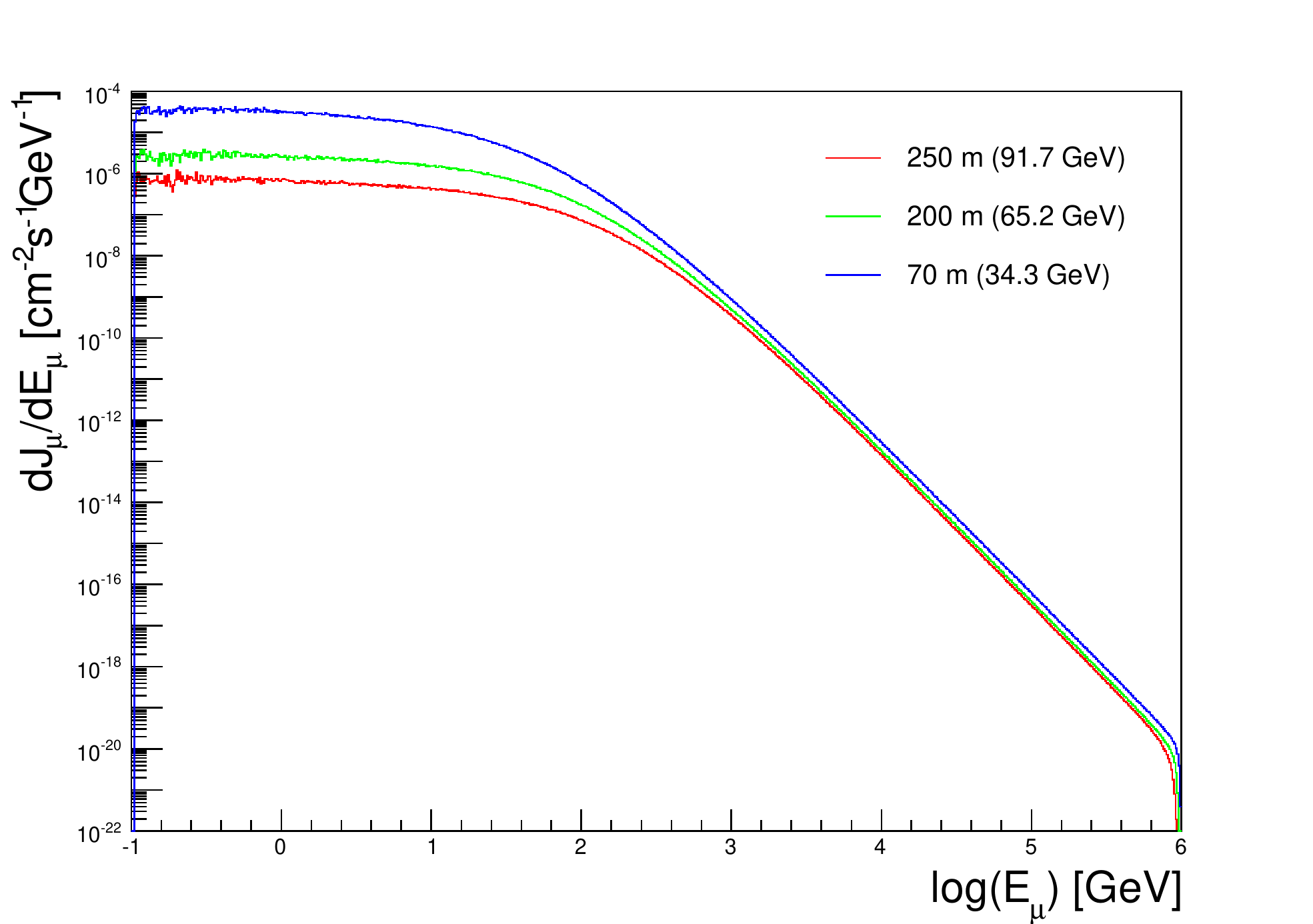}
\caption{\label{muonspectrum} 
The differential muon intensity at 
depths of 70, 200, and 250~m underground.
far detector candidates and near detector candidate.
The numbers in the parenthesis are average muon energy.} 
\end{center}
\end{figure}

Muon generated uniformly in the energy range, $\theta$, and azimuthal 
angle are propagated through the rocks using \textsc{music}. 
\textsc{music} is a code that simulates the 3-dimensional 
transportation of muons through a slant depth of a material,
taking into account the energy loss due to the ionization, 
pair production, Bremsstrahlung, and inelastic 
scattering~\cite{music1,music2}. 
We calculated the integrated muon intensity and average energy 
at the near and far detector sites, 70~m and 200~m in depth, respectively,
as given in Table~\ref{resultmuonsim}.
The muon energy spectra of the detector sites are shown in 
Fig.~\ref{muonspectrum}. 
\begin{table}
\begin{center}
\begin{tabular}{ccc}\hline
Depth	& Integrated intensity ($\mbox{cm}^{-2}s^{-1}$) & Average energy (GeV) \\\hline
70 m	& $5.5\times 10^{-4}$& 34.3\\
200 m	& $8.5\times 10^{-5}$& 65.2\\
\hline
\end{tabular}
\caption{Result of muon transport simulation for the detector sites.
}
\label{resultmuonsim}
\end{center}
\end{table}

%
%
\subsection{External Neutron Background}
The fast neutron produced by a cosmic muon in surrounding rock or 
in the detector can mimic an inverse beta decay signal in the detector
in two ways.
First, the fast neutrons with energy over 10~MeV can be scattered 
elastically from protons in the target or $\gamma$-catcher many 
times and be captured in the target volume. The quenched proton 
recoil scintillation signals can be the prompt signal with an energy 
between 1 and 10~MeV and the moderated neutrons can be captured with 
the same time distribution as the neutrino signal.
Second, 
some neutrons are produced without recoil protons by cosmic muons and
are captured in the detector.
A single neutron capture signal has some
probability to fall accidentally within the time window of a preceding 
signal 
caused by
natural radioactivity in the detector, producing an accidental 
background. In this case, the prompt and delayed signals are 
from different sources, forming an uncorrelated background. 
A detailed neutron 
production simulation will give the estimation of the neutron background 
originating from cosmic muons.

A detailed simulation of the cosmogenic background requires accurate
information of the overburden profile and rock composition.  We assume an 
uniform rock density of 2.70~g/cm$^3$ in the present background simulation. 
The modified Gaisser formula with corrected polar angle is used to describe 
muon flux. 

At muon energies of several tens of GeV, the standard Gaisser formula 
has large discrepancies with data while the modified formula agrees 
with data in the whole energy range.
Using the mountain profile data, the cosmic muons are transported from 
the atmosphere to the underground detector sites using the {\sc music}
package. The simulation results are shown in Table~\ref{resultmuonsim}
for the detector sites.

\begin{figure}
\begin{center}
\includegraphics[width=13cm]{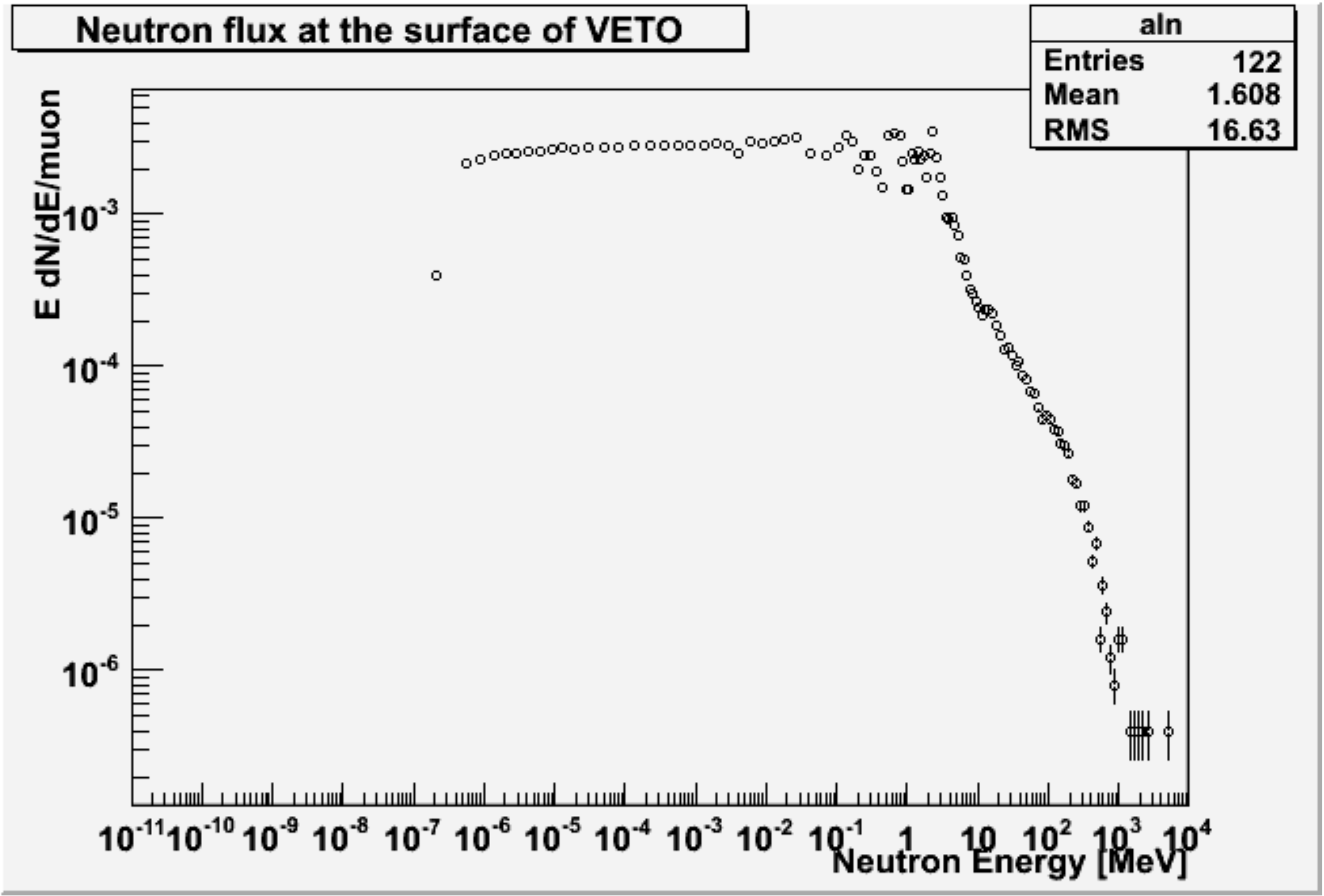}
\end{center}
\caption{Neutron energy distribution coming from rock by cosmic muon}
\label{neutenergy}
\end{figure}

The production rate of the cosmogenic neutrons originating from 
surrounding rocks depends not only on the cosmic muon flux and 
its average energy at the detector but also on the detector 
shielding. The external neutron background rate is calculated 
using {\sc fluka} package, in which the detailed detector site
geometries are accounted for.
The resulting neutron energy distribution coming from the 
surrounding rocks is shown in Fig.~\ref{neutenergy}.

Mei and Hime parametrized the total neutron flux as a function of mean 
muon energy at various depths at underground~\cite{mei06}.
The neutron rate at far detector site is estimated using Mei's 
parameterization to be about $10^{-2}$ neutrons/m$^{2}/$s. 
In addition, the energy spectra of muon induced neutrons at rock is 
simulated again according to Mei's parameterization of the neutron 
energy spectra as a function of muon energy. We adopted the mean muon 
energy at far site for the energy spectra. Then {\sc geant4} simulation 
was done with $5\times 10^6$ neutrons produced at rock surrounding the 
far detector, and selection cuts for the inverse beta neutrino event 
were applied; 1--10~MeV for the prompt signal and 6--10~MeV for the 
delayed signal.
The rate of coincident background events due to fast neutrons passing 
the selection cuts is 2.5 events per day at the far detector. 
A further reduction larger than a factor of two on the background rate is
expected by rejecting the events with neutron induced signals at 
the veto detector.
Therefore, the expected fast neutron background at the far detector
is estimated to be about 1 event per day.
We plan to understand the background rate with an uncertainty better
than 50\%.

%
%
\subsection{Radioactive Isotopes Induced by Cosmic Ray Muons}
From the experiences of previous reactor neutrino experiments such 
as KamLAND and CHOOZ, we know there are irreducible background events 
from the decay of cosmogenic isotopes. Most prominent radioactive isotopes 
are $^{8}$He and $^{9}$Li. $^{8}$He decays by 
$\beta^{-}+n$ (16\%, $Q_{\beta^{-}}=10.653$~MeV) 
with a half-life time of 119~ms. 
$^{9}$Li decays by 
$\beta^{-}+n$ (49.5\%, $Q_{\beta^{-}}=13.606$~MeV) 
with a half-life time of 178.3~ms. 
The $\beta^{-}+n$ decay gives a prompt and delayed signal similar to 
an IBD events.

The production rates of these long-lived cosmogenic isotopes are 
studied initially by rock dating group. 
There are a lot of data available on the production rates of $^{10}$Be 
and $^{26}$Al from silicon and oxygen by cosmic muons.  
The production 
cross section of $^{8}$He and $^{9}$Li in the carbon has been measured 
with accelerator muon beams
at an energy of 190~GeV at CERN\cite{hagner00}. 
Their combined cross section is
$\sigma (^9\hbox{Li} +^8\hbox{He}) = (2.12 \pm 0.35)~\mu$b.
The energy dependent production cross section is estimated as  
$\sigma_{tot} = E_\mu^{0.73}$, where $E_\mu$ is the muon energy.
The $^8$He+$^9$Li background cross section and rate are 
estimated by using the average muon energies at near and
far detector sites and the results are  
listed in Table~\ref{isotoperate_1}.

\begin{table}
\begin{center}
\begin{tabular}{ccc} \hline
                        & near & far  \\\hline
cross section ($\mu$b)  & 0.61 & 0.97 \\
($^9$Li + $^8$He)/day   & 18.6 & 4.6  \\\hline
\end{tabular}
\end{center}
\caption{Cosmogenic isotope production rates in RENO experiment 
estimated from the CERN cross section.}
\label{isotoperate_1}
\end{table}

In addition, we can use the measured data in the previous reactor 
neutrino experiments to estimate the cosmogenic isotope production
rates. The CHOOZ collaboration presented the $^{8}$He and $^{9}$Li 
background rate as $0.7\pm0.2$ events per day for their 5 ton target 
mass~\cite{Apollonio03}. If we scale this rate to RENO detector 
mass and use the dependence of production rate on the muon energy, then
we have 9.3 and 2.3 events per day for near and far detectors,
respectively, which are half of the event rates estimated with the
CERN cross section data.

KamLAND experiment reported that their $^{8}$He and $^{9}$Li backgrounds 
are correlated with showering muons, which has more than $10^{6}$ 
photoelectrons, and they applied two second veto for the muons. 
For RENO experiment, we plan to apply 1~ms veto for 
non-showering muons and 5~ms for showering muons. This veto on the 
showering muons will further reduce the $^{8}$He and $^{9}$Li background 
rate by about 70\%. The final expected $^{8}$He and $^{9}$Li background 
rates will be 2.8 and 0.7 events per day for near and far detectors,
respectively. 

\subsection{Radioactive Backgrounds}
\label{sect: radioactive backgrounds}
Radioactive isotope (\textit{i.e.} radionuclide) is an atom with an unstable nucleus. 
They undergo radioactive decay, and emit a gamma(s) and/or subatomic particles 
like alpha, electron and positron, \textit{etc}. Some radioactive isotopes, 
like $^{40}$K, $^{60}$Co, $^{232}$Th, and $^{238}$U, are naturally abundant in detector 
materials and rocks around the detector hall. Energetic gammas and particles coming 
from radioactive decay of isotopes in the inner detector can create accidental and
correlated backgrounds. Especially, the signal made by radioactive gamma and electron can 
mimic the prompt signal of the inverse beta decay. This background can form an 
accidental (uncorrelated) background with the single neutron events induced by cosmic 
muons.

Radioactive background can come from a variety of sources, mainly inner detector 
materials and the rock. In this study, we consider K, Th, and U isotopes as the radioactive 
sources existing in the following materials.

\begin{itemize}
\item rocks surrounding the detector hall
\item liquid scintillators
\item acrylic vessels
\item buffer oil
\item stainless steel vessel
\item PMT glasses
\end{itemize}

\begin{figure}[hbtp]
\begin{center}
\includegraphics[width=4.5in]{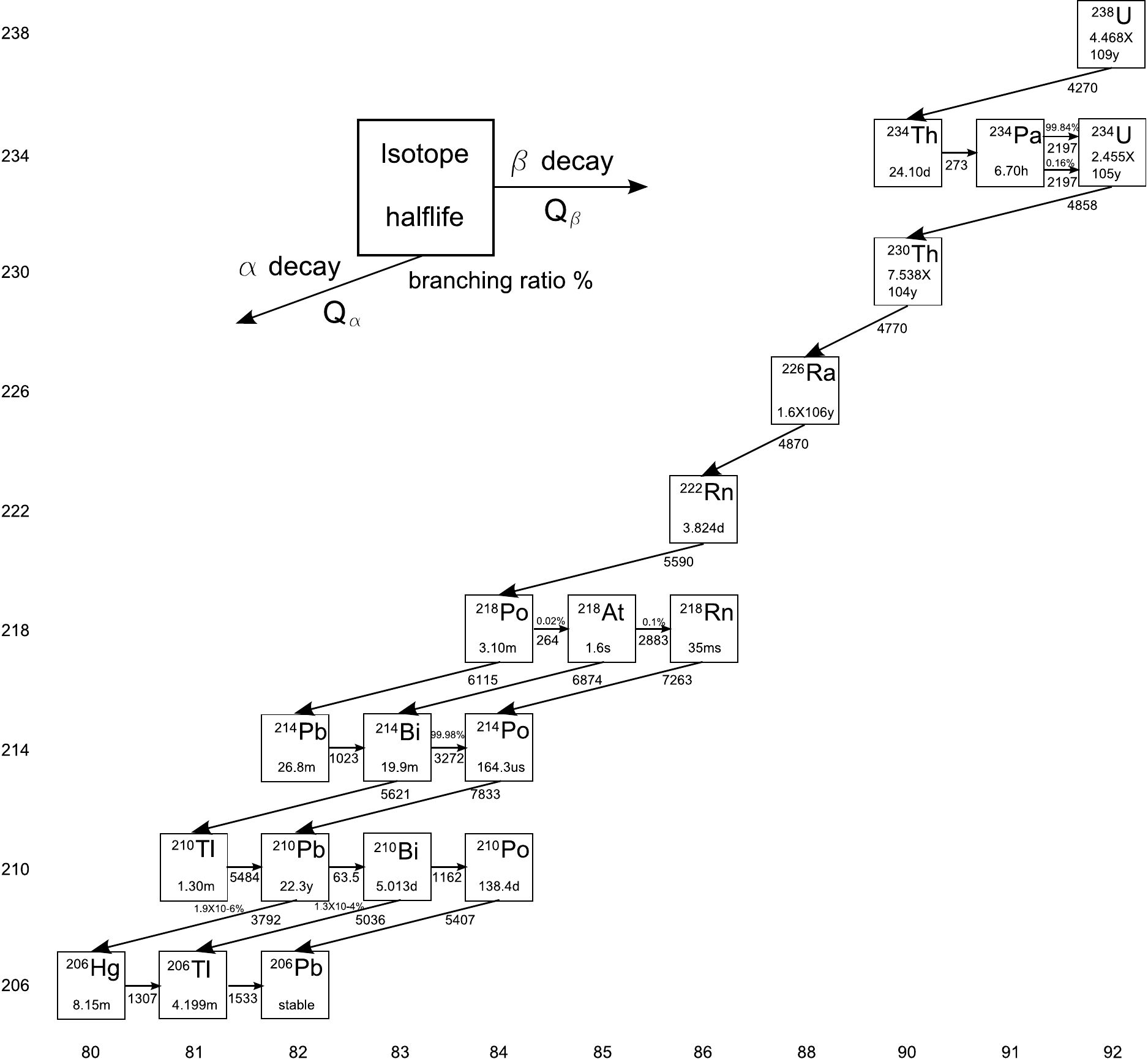}
\includegraphics[width=3.2in]{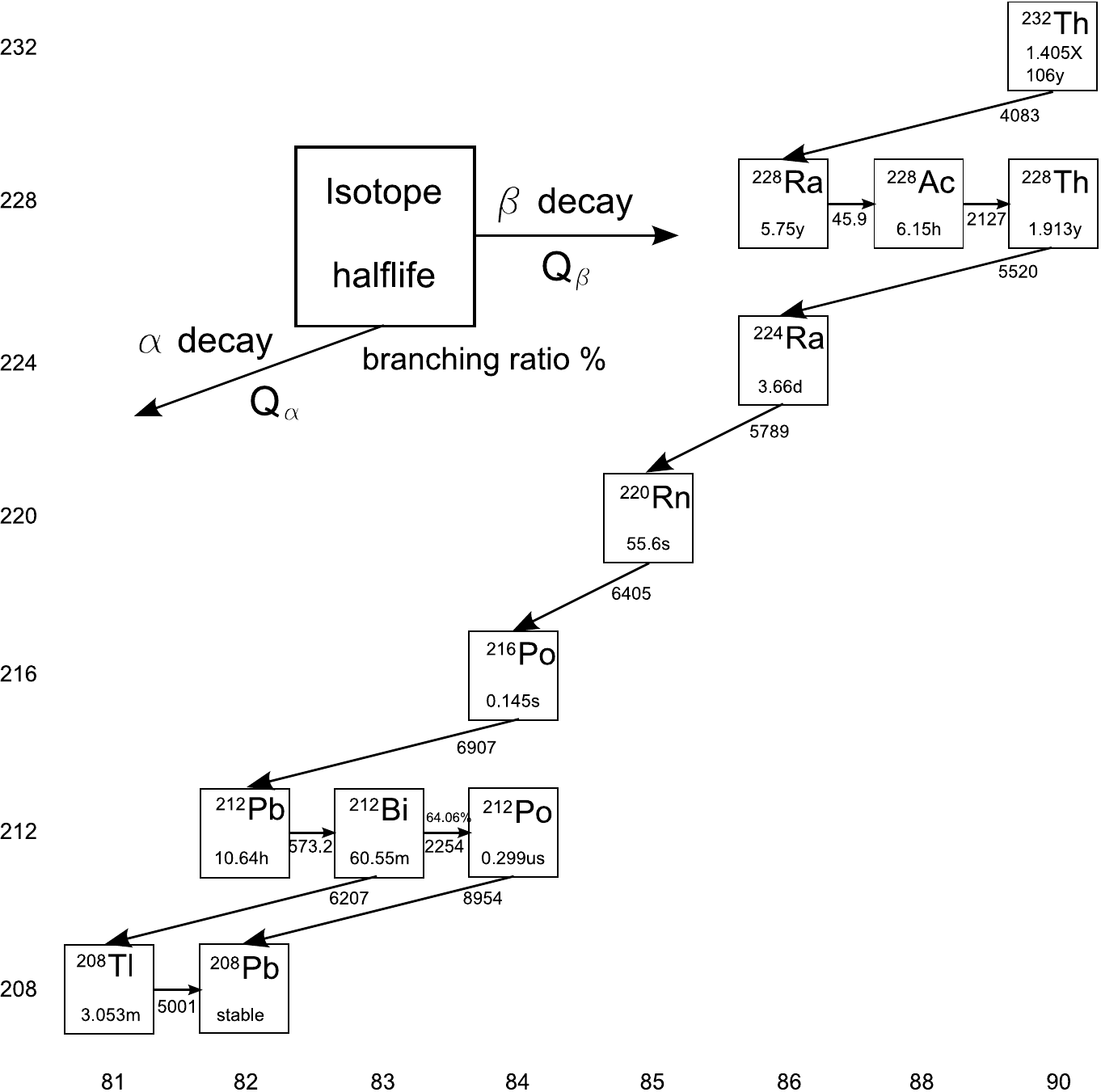}
\end{center}
\caption{Decay chains of $^{238}$U and $^{232}$Th series}
\label{UThchain}
\end{figure}

\begin{figure}[hbtp]
\begin{center}
\includegraphics[width=4in]{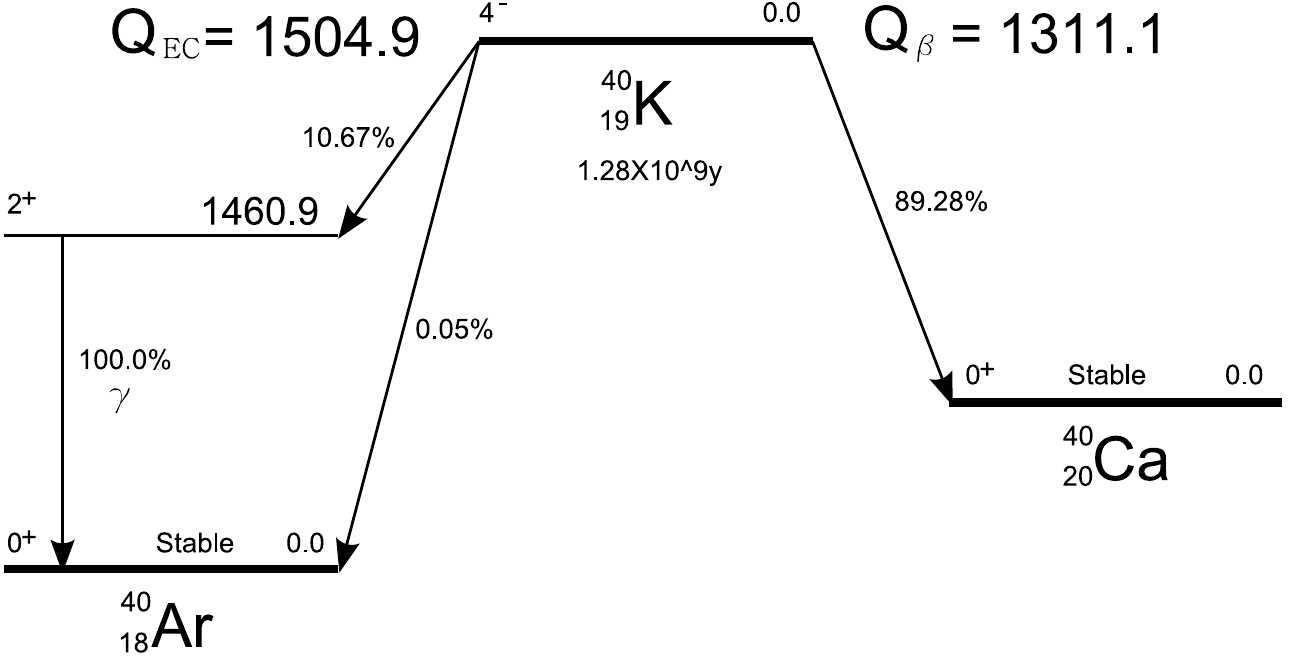}
\end{center}
\caption{Level diagram for $^{40}$K}
\label{Klevel}
\end{figure}

Most radioactive isotopes do not decay directly to a stable state, but rather undergo 
a series of decays until a stable isotope is reached eventually. Figures~\ref{UThchain} 
shows the decay chains of $^{238}$U and $^{232}$Th. 
Figure~\ref{Klevel} shows the energy level diagram for $^{40}$K decay.
Gamma emission occurs not only when a radioactive isotope undergoes gamma 
decay, but also when a beta particle from an isotope decay emits gammas. 
To take account of probability of decay and coincidence of generated particle, a 
decay chain generator is developed.

Using the radioactive decay chain generator, radiations from three 
dominant decay chains are generated at the detector subsystem. We 
apply 1 MeV threshold to the reconstructed energy. Table~\ref{daccept} 
shows the detector acceptance of the radiation from each radioactive
isotope decay chain.

\begin{table}
\begin{center}
\begin{tabular}{cccc}\hline\hline
		& $^{40}$K	& $^{232}$Th	& $^{238}$U	 \\ \hline
Target LS	& 0.20		& 1.34		& 1.25		\\
Target Acrylic	& 0.09		& 0.67		& 0.79		\\
GC LS		& 0.23		& 1.29		& 1.25		\\
GC Acrylic	& 0.04		& 0.28		& 0.35		\\
Buffer Oil	& 0.007		& 0.05		& 0.05		\\
Buffer Vessel	& 0.0005	& 0.007		& 0.004		\\
PMT		& 0.01		& 0.04		& 0.04 		\\ \hline\hline
\end{tabular}
\end{center}
\caption{Detector acceptance of the radioactive radiations from each 
detector subsystem with 1 MeV threshold.}
\label{daccept}
\end{table}

\begin{figure}[t]
\begin{center}
\includegraphics[width=6.0in,clip=true,trim=0mm 0mm 0mm 145mm]{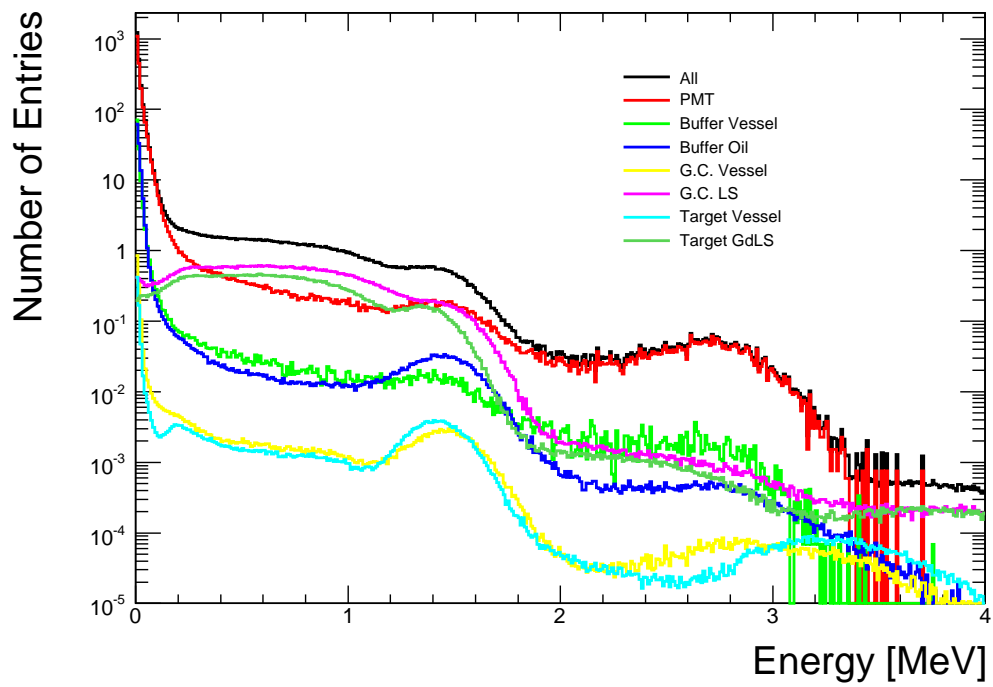}
\end{center}
\caption{The energy distributions of backgrounds due to radioactive material 
in each subsystem.} 
\label{we}
\end{figure}

The concentration of isotopes in the inner detector material is measured 
using ICP-MS and HPGe detector. With the concentration and the detector 
acceptance from full simulation, the single event rates caused by each 
detector subsystem are calculated. Table~\ref{SER} shows the results. 
The most dominant contribution comes from the liquid scintillator in target 
and gamma catcher. But if we control the concentration of the isotopes 
under $10^{-12}$g/g, these rates are negligible compared to that from PMT 
glass. Figure~\ref{we} shows the weighted energy spectrum from the 
radioactive isotopes in each detector subsystem. 

\begin{table}
\begin{center}
\begin{tabular}{ccccc}\hline\hline
		& $^{40}$K     & $^{232}$Th   & $^{238}$U    & Single Event Rate \\
		& ppt	       & ppt	      & ppt	     & Hz    \\ \hline
Rock            & 4.33(ppm)    & 7.58(ppm)    & 2.32(ppm)    & 9.2               \\		        
Target LS	& $< 0.32$     & 17.7	      & 13.9	     & $< 5.6$           \\
Target Acrylic	& 8 	       & 206.8	      & 167.5	     & 0.95              \\
GC LS		& $< 0.32$     & 17.7	      & 13.9	     & $< 8.4$           \\
GC Acrylic	& 8 	       & 206.8	      & 167.5	     & 0.87              \\
Buffer Oil	& 10 	       & 19.7	      & 5.0	     & 1.07              \\
Buffer Vessel	& 60 	       & 900	      & 900	     & 0.33              \\
PMT		& 10.8	       & 125.9	      & 50.3	     & 8.19              \\ \hline
Total		&	       &	      &		     & $<34.6$           \\ \hline\hline
\end{tabular}
\end{center}
\caption{Concentrations of $^{40}$K, $^{232}$Th, and $^{238}$U in surrounding rock 
and the main components of RENO detector, and their event rates.}
\label{SER}
\end{table}

\section{Summary of Backgrounds}
The background event rates are summarized in the Table~\ref{background summary}.
The single background rate over 1 MeV energy is about 45~Hz for both 
near and far detectors. The correlated background rates will be
1.7 and 5.8 events per day at far and near detector, respectively.

\begin{table}
\begin{center}
\begin{tabular}{ccc}\\\hline
	&Near	&Far \\\hline
Gamma Single Rates(Hz)	&$\sim 30$	&$\sim 30$	\\
$^{8}$He + $^{9}$Li (/day)     &  2.8 & 0.7    \\ 
Correlated Neutron Backgrounds (/day)	& 3.0  & 1.0	\\\hline
\end{tabular}
\end{center}
\caption{Summary of backgrounds for single hit events over 1~MeV and 
correlated neutrino-like events at both near and far detector.}
\label{background summary}
\end{table}

\newpage

\chapter{Calibration}
\section{Overview}
Since the $\theta_{13}$ measurement depends on the systematic uncertainties 
in the relative parameters between near and far detectors, it is important 
to understand detector performance in great detail. 
There are three major motivations for having the calibration system.
First, the characteristics of the events in the energy range of 1--10~MeV 
depend on the positions of the event vertex since the scintillation lights 
propagate through the liquid scintillator, acrylic vessel, and buffer oil. 
The vertex position dependence of energy measurement can be understood by 
placing a radioactive source at various locations inside the liquid 
scintillator and measuring the energy deposit.
The detail optical parameters of the liquid scintillator and acrylic vessel, 
and stainless steel tank can be obtained and compared between two detectors. 
Second, liquid scintillators may change its scintillation and optical properties 
during the long-term data taking period. Therefore, it is crucial to monitor 
the detector response throughout the duration of the experiment. Also, the 
day and night oscillation in the energy measurement due to temperature and 
other environmental factors inside and outside of the detector could occur, 
so the monitoring should be done at all times.
Finally, the calibration system could be used to calculate the dead time for
inverse beta decay events. Any difference of the dead time between near and
far detector should be understood.

\section{Radioactive Sources}
The purpose of using radioactive sources is to calibrate the detector 
response for the inverse beta decay of reactor antineutrinos. 
A neutrino event generates two signals 
separated by several tens of $\mu$s. The first signal is from $e^{+}$ 
annihilation with $e^{+}$ kinetic energy up several MeV, and
the second signal from a neutron with kinetic energy of tens of keV.
The neutron capture by Gadolinium isotopes produces several gammas of 
1--2~MeV with their total energy of about 8 MeV. Table~\ref{sources} 
shows the characteristics of radioactive sources to be used for calibrating 
the detector for both the $e^{+}$ and neutron signals.

\begin{table}
\begin{center}
\begin{tabular}{ccccc}\hline
type & sources	& energy (keV) & calibration \\ \hline
$e^{+}$  & $^{22}$Na  & 511(2)+1274.5 & positron\\
         & $^{68}$Ge  & 511(2) & position E threshold \\ \hline
$\gamma$ & $^{137}$Cs & 662 & gamma \\
         & $^{54}$Mn  & 835       & \\ 
         & $^{60}$Co  & 1173+1333 & multiple gamma \\ \hline
neutron  & $^{252}$Cf & neutron + $\sim$10~MeV & neutron efficiency\\
         & $^{241}$Am+Be & neutron + 4.4~MeV & \\ \hline
\end{tabular}
\caption{A list of radioactive sources for RENO detector calibration.}
\label{sources}
\end{center}
\end{table}

The size of the radioactive source is limited by the attenuation length of 0.511~MeV gamma ray
in case of positron source. The attenuation length of 0.511~MeV gammas in the liquid scintillator
is about 10~cm. Therefore, the overall size of the source should be smaller than the attenuation
length by several times to minimize the amount of the scintillation light scattering by the source
itself. The overall size of the source will be $2~{\rm cm}\times 3~{\rm cm}$. 
The material encapsulating the source should be compatible with the scintillator materials, and
PTFE could be one of the best candidate materials.

\section{Light Sources}
\begin{figure}
\begin{center}
\includegraphics[width=5.3in]{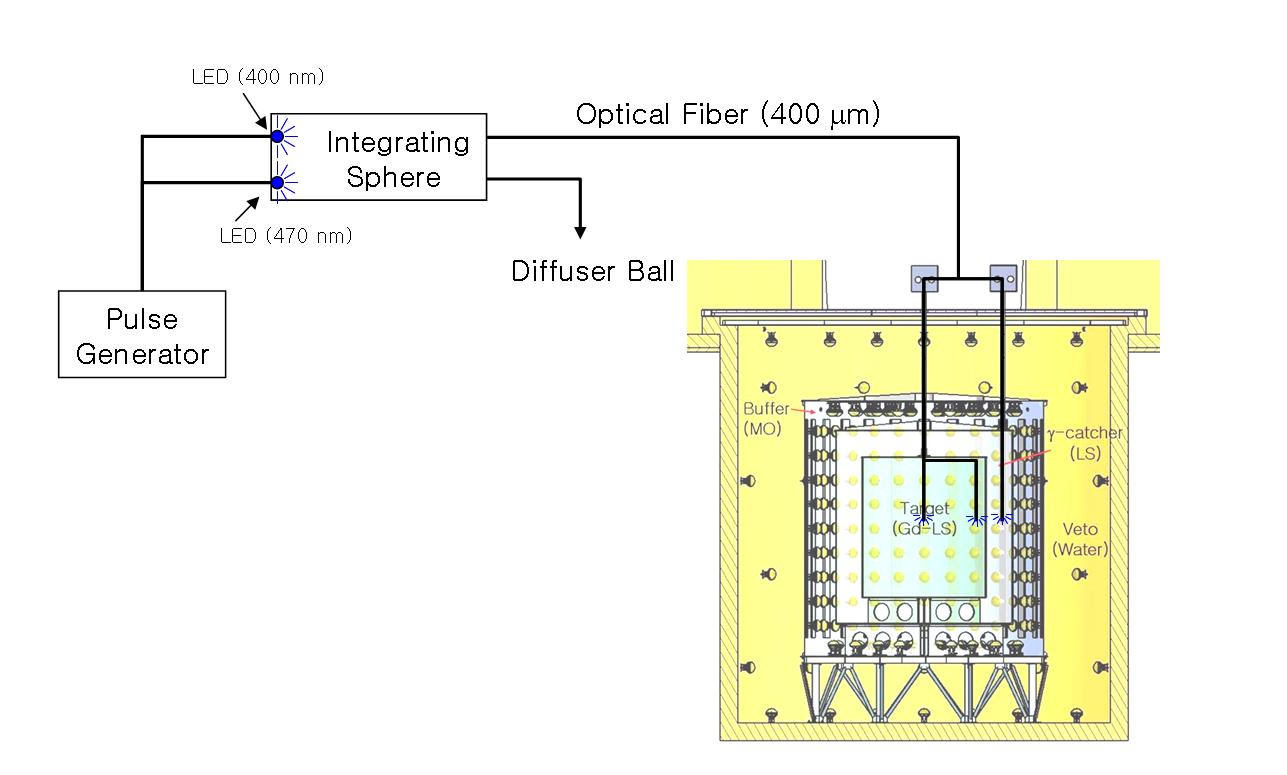}
\end{center}
\caption{A conceptual design of light source system.}
\label{lightsourcesystem}
\end{figure}

Low energy background events may be analyzed to monitor the PMT gain and 
single photoelectron energy resolution. Additional light sources can be
used with a definite external trigger for the same purpose. These two 
continuous monitoring methods are complementary. 
The light source calibration system can be also used to check the time and position
resolutions at various energy ranges.
Figure~\ref{lightsourcesystem} shows a conceptual design of the light source and
fiber optic system. Ultraviolet (UV) and blue LEDs are used as light sources. 
A pulse generator selectively fires one of the LEDs with a capability of 
generating double pulse with variable amplitudes. The pulse width is a few ns.
The LED lights are fed into an integrating sphere to make a stable light source
and transmitted through the optical fiber. Several different LED can 
be mounted on the integrating sphere. 

In case of UV light the output fiber from the integrating sphere is fanned out 
into three identical optical fibers. There is an optical shutter for each fiber 
to be activate only one fiber at one time. The three optical fibers are fixed 
at the center and edge of target vessel and at the center of $\gamma$-catcher 
vessel. 
The UV light, when radiated into the target liquid scintillator, is absorbed 
within a few centimeters from the end of the optical fiber and generates 
isotropic blue light. 
The attenuation length is a few centimeters for the light with 400~nm wavelength. 
This system has advantages of a diffuser ball not being necessary and generating
isotropic emission of similar wavelengths to real events. A 400~$\mu$m diameter bare 
optical fiber is used for light transmission. If a blue LED (470~$\mu$m) is used,
then the attenuation length of the liquid scintillator can be measured directly, 
since the liquid scintillator does not undergo scintillation processes and, 
therefore, is transparent to the emission light. 

Alternatively, a diffuser ball with a blue LED can be used. A diffuser ball can 
be made of a solid Teflon ball or a spherical acrylic shell filled with diffuser 
material such as Ludox.

The light source calibration system can be used to measure dead time for both near 
and far detectors by generating double pulses in light intensity corresponding to 
various neutrino energies. For this purpose, at least a set of fixed light source 
system will be set up for calibrating each detector for whole data taking period.

\section{Source Driving System}\label{Source Driving System}
It is a challenging work to load the various sources at desired locations 
within the liquid scintillator regions of the detector.
Signals from a radioactive source should be similar to that of the 
inverse beta decay events and background events. Therefore, the source 
driving system itself should affect minimally the propagation of the 
scintillation lights throughout the detector. 
The main goals of the calibration using radioactive sources are
to measure energy scale and resolution, and the vertex position dependence 
of energy measurement.
 
The source driving system consists of a stepping motor driven pulley with
a polyethylene wire attached. The radioactive source container made of Teflon-PFA 
is connected at the end of the wire with a weight countering buoyant force when 
submerged in liquids. The pulley has a spiral groove to avoid entanglement of wire.
The system is expected to have the z-position accuracy of an order of a few mm, 
which is much better than the the intrinsic vertex position resolution of the 
detector.

\begin{figure}[tbh]
\begin{center}
\includegraphics[width=10cm]{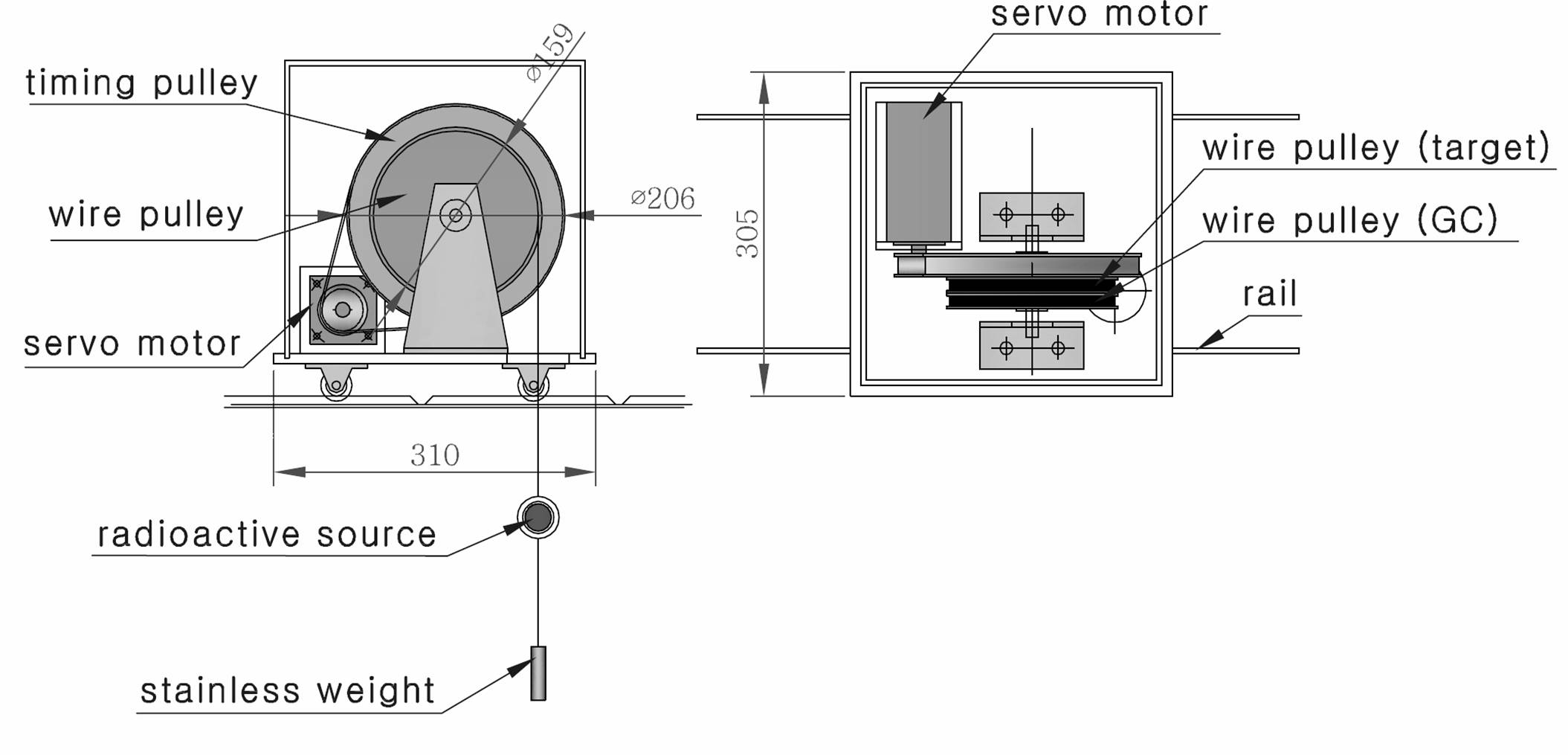}
\end{center}
\caption{Design of RENO source driving system. The wire pulley is driven by 
a stepping motor via a belt. The radioactive source is encapsulated in a 
Teflon-PFA capsule attached to polyethylene wire.}
\label{wire_system_1}
\end{figure}

\begin{figure}
\begin{center}
\includegraphics[width=8cm]{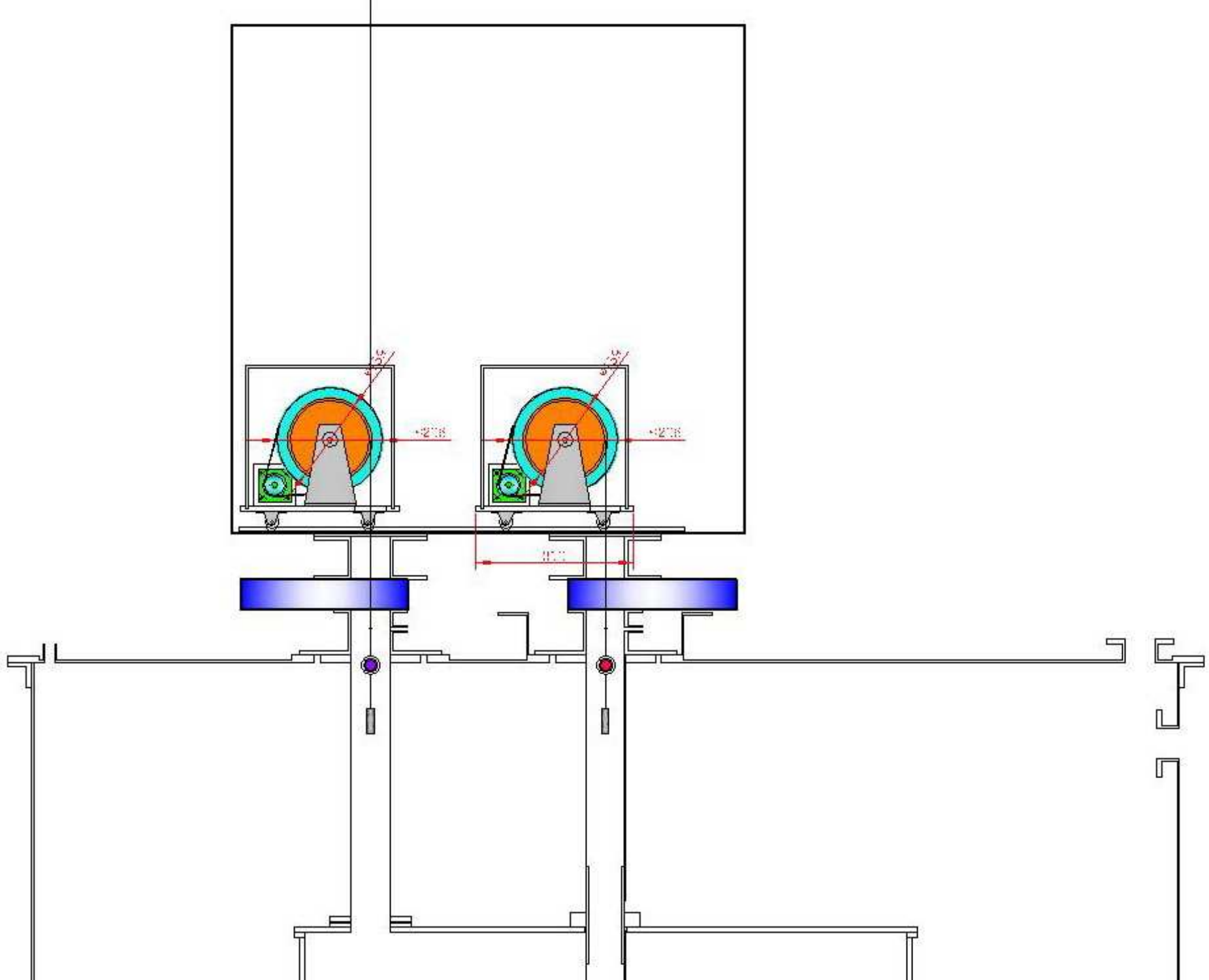}
\end{center}
\caption{The source driving system for the mockup detector. A similar 
system will be used for the RENO detector.}
\label{wire_system_2}
\end{figure}

Figure~\ref{wire_system_1} shows the design of the source driving system.
The system is enclosed in an air-tight stainless steel glove box and is located
on top of the detector as shown in Fig.~\ref{wire_system_2}.
The source driving system is remotely controlled by a computer located
in a main control room. The current system can move the radioactive source
in z-direction only and a 3-dimensionally deployable system is under development.

\newpage

\bibliographystyle{plain}

\begin{thebibliography}{9}
\bibitem{Systematics:powerUncertainty} S.~K.~Ji (Korea Power Engineering Co.), private communications. 
The power output 
under a normal reactor operation at Yonggwang is well over 90\% 
of the rated power.
\end{thebibliography}
\chapter{Systematic Uncertainties and Sensitivity}
\section{Systematic Uncertainties}\label{systematic uncertainties}
The RENO experiment is expected to record $\sim 5\times 10^4$ 
inverse beta decay events at the far detector 
over the course of three years of data-taking, yielding 
a statistical uncertainty of $\sim 0.4$\%.
Therefore, our goal for the systematic uncertainty is to keep the 
uncertainty comparable to 0.4\% or less to achieve 
our target sensitivity of $\sin^2(2\theta_{13}) \lesim 0.02$
at 90\% confidence level (CL).

As a comparison, the CHOOZ experiment, a previous reactor neutrino 
experiment has set a limit on $\sin^2(2\theta_{13})$ at 
$\lesim 0.2$ at 90\% CL, with 2.8\% and 2.7\% statistical and 
systematic uncertainties, respectively~\cite{Systematics:Apollonio03}. 
The uncertainties in neutrino flux was dominant source of systematic
uncertainty accounting 1.9\%.
The systematic uncertainty goals for RENO are shown in 
Tables~\ref{reactor systematic uncertainty summary} 
and \ref{detector systematic uncertainty summary}.

The sources of systematic uncertainties are reactor related, 
detector related, and background related.
By using two identical detectors, the detector related systematic
uncertainties could be mostly cancelled out and, in addition, the 
effects of reactor related uncertainties are greatly reduced.  

\begin{table}
\begin{center}
\begin{tabular}{lc}\hline
\hline
Uncertainty Source &RENO \\\hline
\hline
Reactor Power			&0.4	\\
Energy Released per Fission	&$<0.1$ \\
Reactor/Detector Distances	&0.06 \\\hline
Combined	 		&$<$0.5 \\\hline\hline
\end{tabular}
\end{center}
\caption{Reactor related systematic uncertainties for RENO.}
\label{reactor systematic uncertainty summary}
\end{table}

\begin{table}
\begin{center}
\begin{tabular}{lcc}\hline
\hline
Uncertainty Source &CHOOZ (\%) &RENO (Goal \%) \\\hline
\hline
        H/C Ratio                      &0.8    &0.1 \\
        Target Mass                    &0.3    &0.1 \\
        Gd Capture Fraction            &1.0    &0.4 \\
        Positron Energy                &0.8    &0.1 \\
        Positron Geode Distance        &0.1    &-- \\
        Neutron Energy                 &0.4    &0.2 \\
        Neutron Geode Distance         &0.1    &-- \\
        Neutron Capture Time           &0.4    &$<0.1$\\
        Positron--Neutron Distance     &0.3    &-- \\
        Dead Time                      &0      &$<0.1$ \\
        Neutron Multiplicity           &0.5    &$<0.1$ \\\hline\hline
	Combined        	       &1.7    &$<0.5$\\\hline\hline
\end{tabular}
\end{center}
\caption{Detector related absolute systematic uncertainty goals for 
RENO compared with CHOOZ experiment.}
\label{detector systematic uncertainty summary}
\end{table}

\subsection{Reactor Related Uncertainties}\label{section:reactor related uncertainties}
At CHOOZ experiment, the dominant source of systematic uncertainty 
was the neutrino flux, which amounts to 1.9\%~\cite{Systematics:Apollonio03}.
The expected neutrino fluxes at the near and far detectors depend 
on various factors; the fission rate, the number of neutrinos per 
fission, the composition of fissile materials in the fuel, and the 
distance between the reactor and detector and so on.
For an experiment with a single reactor, uncertainties from the 
neutrino flux can be completely removed by using two identical 
detectors through normalization of the neutrino flux, if the 
distances between the reactor and detectors are precisely known. 
However, the Yonggwang power plant has six roughly equally spaced 
reactors with its span comparable to the distance between the reactor
baseline and the far detector. The neutrino flux from each reactor 
is comparable at the far detector but quite different at the 
near detector, where the nearest two reactors contributing $\sim 60$\% 
of the total neutrino flux. Therefore, uncorrelated uncertainties 
among the reactors can be reduced but not as much as 
the case for having one or two reactors.

\subsubsection{Reactor Power Uncertainties}
At Yonggwang power plant, the thermal power of each reactor can be
measured better than 1.6\% when the power output is above 90\% of
the rated power output of the reactor~\cite{Systematics:powerUncertainty}.
The correlations in the power output measurements among reactors are
not yet explicitly investigated. 

To be conservative, we assumed no correlations in the reactor power
output uncertainty among the reactors and estimated the effects of 
the uncertainty.
Each reactor was 
assumed to have the same power output as well as the same amount of 
uncertainty on the power output. The change in the the ratio of the number of 
events detected at the near detector to that of the far detector was 
numerically calculated when the power output is fluctuated. 

\begin{figure}
\begin{center}
\includegraphics[width=4in]{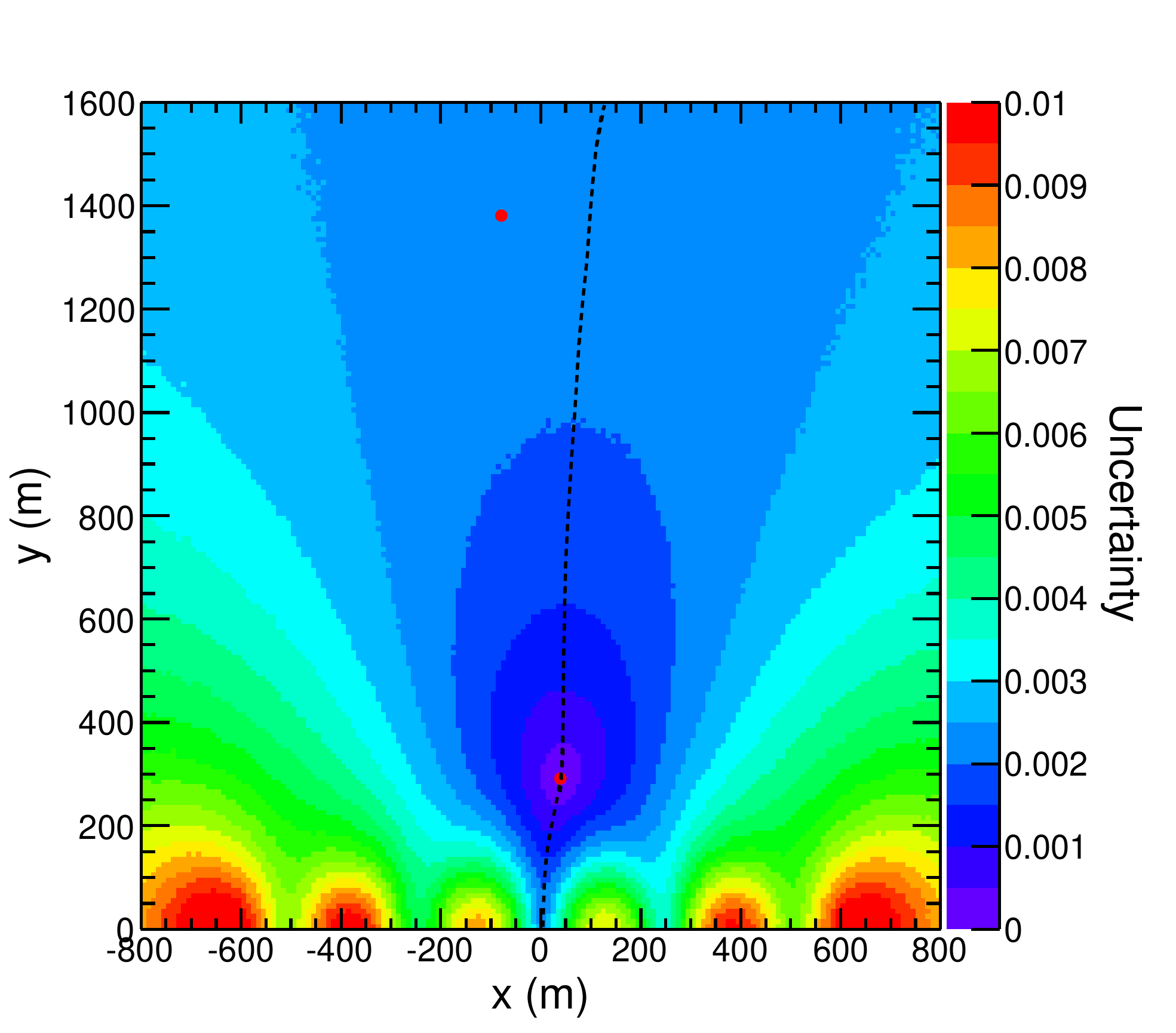}
\end{center}
\caption{The uncertainty of ratio of numbers of events in the near detector to
far detector for uncorrelated power uncertainty. The reactor power is assumed
to be the same and has an 1\% uncertainty as a reference.
The red dots represent the locations of the near and far detectors. The 
dotted line represents the minimum uncertainty at a given distance from the 
reactor baseline (y-axis).}
\label{power uncertainty ratio}
\end{figure}

Figure~\ref{power uncertainty ratio} shows the resulting uncertainty 
in the ratio of the number of detected neutrino events at a detector 
at an arbitrary position to that of the near detector, assuming 
uncorrelated uncertainty of 1\%, as a reference, in the power output for each reactor.
For the RENO's far detector position, the uncertainty in the ratio is 
less than 0.4\% for uncorrelated uncertainty of 1.6\% in 
power output in each reactor. 

\subsubsection{Energy Released Per Fission}
Depending on the composition of fissile materials in the nuclear 
fuel, the average energy, the number of neutrinos per fission, and 
the shape of the neutrino spectrum vary. The evolution of the 
fissile material composition can be calculated using {\sc origen} 
code~\cite{Systematics:origen} using the reactor operational 
conditions and initially loaded fuel composition. The effect is
under investigation. However, the effect should be minimal
since this effect is correlated among reactors. 

\subsubsection{Distances}
The distances between the center of the fuel assembly in the reactors 
and the detectors can be determined with a precision better than 10~cm 
The barycenter of the neutrino source can change from the
fuel load-up over the reactor operation due to fuel burnup.
The Burgey experiment was able to determine the barycenter
position at a level of a few centimeters laterally~\cite{Systematics:Decl95}. 
Therefore, uncertainty of 5~cm was used to calculate the effect.
The combined uncertainty on the ratio of the event rate at the far detector 
to that of at the near detector is 0.06\%.

\subsection{Detector Related Uncertainties}
In CHOOZ experiment, the detector related uncertainty is 
1.7\%. However, uncertainties are greatly reduced by using two identical 
detectors at RENO. The relative normalization between the two detectors is 
the main source of the uncertainties.
The liquid scintillator for the near and far detectors will be prepared 
from a single batch. 
Therefore, the compositions of the liquids will be, in principle, same 
for both near and far detectors and there will be no systematic 
uncertainties originating from the free proton density and the 
Gadolinium (Gd) concentration.

\subsubsection{Number of Target Protons}
The number of observed IBD events in the detector is proportional to
the number of free protons in the target.
If the ratio of hydrogen to carbon (H/C) is known, the number of
free protons can be determined from the target mass.
The uncertainty on the target mass can be reduced by the accurate
measurement of the target volume and weight. 

Since the target liquid will be prepared in a single batch, the chemical
compositions of the liquid in the near and far detectors are, in principle, 
the same. Therefore, the uncertainty in the Hydrogen to Carbon ratio, H/C,
will vanish in the normalization. 
However, we assign 0.1\% uncertainty on the H/C measurement which we can 
achieve with Gas Chromatography-Mass Spectrometer (GC-MS) method. 

The target liquid mass will be measured by two methods; mass
flow meters and weight sensors. A Coriolis mass flow meter measures 
the mass and density of the liquid flowing through the device. 
Commercially available Coriolis mass flow meters have the typical 
absolute precision of 0.1\% and repeatability of $\sim 0.05$\%.
The weight sensors will be used to cross check the mass flow meters.
To maintain a constant fluid density, the variation of the target 
temperature will be kept under 1$^\circ$C. 
Our goal is to measure the difference between the near and far target 
masses with an accuracy better than 0.2\%.

\subsubsection{Positron Event Selection}
If an inverse beta decay event occurs close to the target vessel, 
sometimes gammas may escape the scintillator volume (target and 
$\gamma$ catcher) and not fully deposit energy in the scintillator. 
However, the effect of geometrical 
difference of detectors is negligible and the effect of energy 
scale difference between two detectors is dominant. 

We will require the prompt events to have $>1$~MeV energy to remove 
low energy backgrounds. Although the minimum energy from the positron
event is 1.022~MeV, the value of measured energy may go below 1~MeV due to 
energy resolution of the detector. A 2\% uncertainty in the energy scale
leads to a 0.1\% uncertainty in the positron event selection.

\subsubsection{Gadolinium Capture Fraction}
The Gadolinium (Gd) concentration of the liquid scintillator in the 
target affects the neutron capture time as well as the fraction of
neutron captured by Gd. The Gd-doped liquid scintillator from the 
same batch will be used for the near and far detectors. Therefore, 
as with the H/C, the Gd concentration in the target will be, in 
principle, identical for both detectors. 

The CHOOZ experiment showed that the neutron capture time can be 
measured at a level of 0.3\%~\cite{Systematics:Apollonio03}.
Also, we can measure the neutron capture time using a Californium 
(Cf) source. Using the same Cf source will allow us to estimate the 
difference in the fraction of the neutrons captured by Gd between 
the near and far detectors better than 0.4\%.

\subsubsection{Neutron Capture Energy}
As with the positron identification, the effect of geometrical 
difference of detectors is negligible and the effect of energy 
scale difference between two detectors is dominant. 
An energy scale uncertainty of 2\% gives 0.4\% uncertainty in the
neutron event selection. 

\subsubsection{Neutron Capture Time}
The time distribution of neutrons captured by Gd exhibits an 
exponential like behavior. The mean capture time is  $\sim 30~\mu$s. 
Using the acceptance window of $0.3\sim 200~\mu{\rm s}$ and the
assuming the time resolution of electronics of $\sim 10$~ns, we
expect the relative uncertainty between near and far detector to
be less than 0.1\%. 

\subsubsection{Neutron Multiplicity}
Muon induced spallation neutrons can give multiple neutron 
capture event and can mimic a signal event. In CHOOZ experiment, 
such events were removed, resulting in a $2.6\pm 0.5\%$ 
inefficiency. 
However, RENO detectors have veto layer that can identify   
such events with a very high efficiency and we consider
the uncertainty to be negligible. 

\subsubsection{Dead Time Measurement}
Due to small overburden, the near detector is expected to experience a 
large flux of cosmic muons compared to the far detector. This will incur 
a large fraction of dead time in the near detector to veto the cosmic 
muons. The dead time for near and far detectors are expected to be 
$\sim 25$\% and $\sim 4$\%, respectively, assuming 0.5~ms veto after 
each muon event. The electronics dead time is expected to be much 
smaller than the veto caused dead time.

Because the dead time will be significantly different for near and
far detectors, it is crucial to measure the dead time precisely to
reduce the overall systematic uncertainty. The methods of dead time 
measurement and its uncertainties are being investigated but the
uncertainty is expected to be negligible.

\subsection{Background Subtraction Uncertainties}
The studies on backgrounds are in progress and the uncertainties from the
background subtraction will be estimated shortly.

\subsubsection{Spent Fuel}
A third of fuel assembly in each reactor at Yonggwang power plant 
is replaced with a fresh batch about every 18 months. 
All the replaced spent fuel since the start of the operation of the 
power plant is stored in on-site water pools located within each
reactor complex.
Some of fission products in the spent fuel emit neutrinos above 
1.8~MeV of energy, which are in equilibrium with long lived 
predecessors. Therefore, the effect will be present even long
after the spent fuel is placed in the pool. 
The effects of the spent fuel is under investigation.

\section{Sensitivity and Discovery Potentials}
\label{section: sensitivity}
\subsection{Experiment Parameters}
The average total thermal output of the Yonggwang power plant is
16.4~GW. It is assumed that each reactor has equal thermal power 
output. The fuel composition is assumed to be constant at 0.556, 
0.326, 0.071, and 0.047 for $^{235}$U, $^{239}$Pu, $^{238}$U, and 
$^{241}$Pu, respectively.  

Each detector has $1.2\times 10^{30}$ free protons in the target
vessel with a volume of 18.7~m$^3$. 
We expect $4.7\times 10^5$ and $4.2\times 10^4$ inverse beta
decay interactions within the target volumes of the far and
near detectors, respectively. 
We assume 40\% and 70\% event acceptance and selection
efficiencies for the near and far detectors, respectively,
accounting for the dead time incurred by cosmic muon veto. 
Therefore, we expect $5.6\times 10^5$ and $8.7\times 10^4$ 
events for three years at the near and far detectors, respectively.
The reactor--detector distances are shown in 
Table~\ref{sensitivity:reactor} and the detector parameters are 
summarized in Table~\ref{sensitivity:detector}.

\begin{table}
\begin{center}
\begin{tabular}{lcccccc}\hline
        &Reactor 1   &Reactor 2    &Reactor 3   &Reactor 4   &Reactor 5   &Reactor 6    \\\hline
Near &668.9~m  &453.2~m  &306.9~m  &338.0~m  &515.2~m  &740.0~m\\
Far  &1557.0~m &1456.6~m &1396.4~m &1381.8~m &1414.2~m &1490.5~m \\
\hline
\end{tabular}
\end{center}
\caption{
Reactor-detector distances used in the sensitivity calculation.
}
\label{sensitivity:reactor}
\end{table}

\begin{table}
\begin{center}
\begin{tabular}{lcc}\hline
	&Near	&Far	\\\hline
Target Free Protons	&\multicolumn{2}{c}{$1.21\times 10^{30}$}\\
Overall Efficiency	&40\%	&70\% \\
Number of events (3 yrs.)	&$5.6\times 10^5$ &$8.7\times 10^4$\\\hline
\end{tabular}
\end{center}
\caption{Detector parameters used in the sensitivity calculations. 
Overall efficiencies include the dead time.
The thermal power of each reactor is assumed to be 2.73~GW. 
}
\label{sensitivity:detector}
\end{table}

The backgrounds contributions are under investigation and are not 
included in the sensitivity calculations.
  
\subsection{Sensitivity}
The sensitivity of the experiment to $\sin^2(2\theta_{13})$ is calculated
using the pull approach described in Ref.~\cite{Systematics:Fog02}, where the 
correlations in the uncertainties are naturally accounted for. 
The $\chi^2$ function, constructed using positron spectral information, 
is written as 
\begin{eqnarray}
\chi^2 = \min_{\alpha} &&\left\{\sum_{d=N,F}\left\{
\sum_{i=1}^{N_b} 
\left[
{\left(O_i^d-N_i^d \over U_i^d\right)^2}
+\left({c_i\over \sigma_{shape}}\right)^2\right]
+\left({b^d\over \sigma_b}\right)^2
+\left({g^d\over \sigma_{cal}}\right)^2
\right\}\right.\nonumber\\
&&+\left.\left(a\over\sigma_a\right)^2
+\sum_{r=1}^{N_c}\left(f_r\over\sigma_{cfl}\right)^2\right\},
\end{eqnarray}
where $O_i^d$ is the number observed events in the $i$th bin of 
the positron energy spectrum of near ($d=N$) and far ($d=F$) detectors,
and $\alpha =\{a, b^d, c_i, f_r, g^d\}$ are the parameters to minimize 
$\chi^2$ against.
If we define $T_{ir}^d$ to be the expected number of events in the
$i$th bin of the energy spectrum on detector $d$ of positrons 
originating from reactor $r$, then $N_i^d$ is written as 
\begin{equation}
N_i^d = (1+a+b^d+c_i)\sum_{r=1}^{N_c}(1+f_r)T_{ir}^d + g^dM_i^d,
\end{equation}
where $N_c$ is the number of reactors, and $M_i^d$ is the 
first order differential of the $\sum_{r=1}^{N_c}T_{ir}^d$
with respect to $g^d$.
The uncorrelated uncertainty is 
\begin{equation}
U_i^d = \sqrt{O_i^d+B_{ij}^{d}(1+\sigma^d_{bj} B_{ij}^d)},
\end{equation}
where $B_{ij}^d$ and $\sigma^d_{bj}$ are the number of events of the $j$th 
background in the $i$th energy bin in detector $d$ and its uncertainty.
To estimate the sensitivity of the experiment, we use the expected
positron spectrum with oscillation as the observed spectrum.
The energy scale uncertainty is accounted for in $g^d$ to the first 
order by 
\begin{equation}
T_i^d\simeq T_i^d(g^d=0) + g^d M_i^d,
~\hbox{with   }
M_i^d=\left.{dT_i^d\over dg^d}\right|_{g^d=0},
\end{equation}
where $T_i^d$ is the number of events.

The uncertainty on the overall normalization, $\sigma_a$, on 
the event rates of the near and far detectors
accounts for correlated uncertainties and any bias that 
would affect the near and far detectors in the same way and, therefore,
does not degrade the sensitivity of the experiment. 
This uncertainty is taken to be 2.0\%.

The relative uncertainty, $\sigma_b$, accounts for uncorrelated 
uncertainty between the event rates of the near and far detectors. 
We take 0.6\% as the relative uncertainty as shown in 
Sect.~\ref{systematic uncertainties}. 
The uncorrelated reactor power output uncertainty is taken to be 
$\sigma_{cfl}=1.6\%$ as shown in Sect.~\ref{section:reactor related uncertainties},

The uncorrelated bin-to-bin spectrum shape uncertainty is 
taken to be $\sigma_{shape}=2.0$\%~\cite{Systematics:Decl95}.
The energy scale uncertainty is taken to be $\sigma_{cal}=1.0$\%. 
The uncertainties $\sigma_{b1}$, $\sigma_{b2}$,
and $\sigma_{b3}$ are the uncertainties for the fast neutron backgrounds,
accidental backgrounds, and cosmogenic isotope backgrounds, respectively.

\begin{figure}
\begin{center}
\includegraphics[width=4.0in]{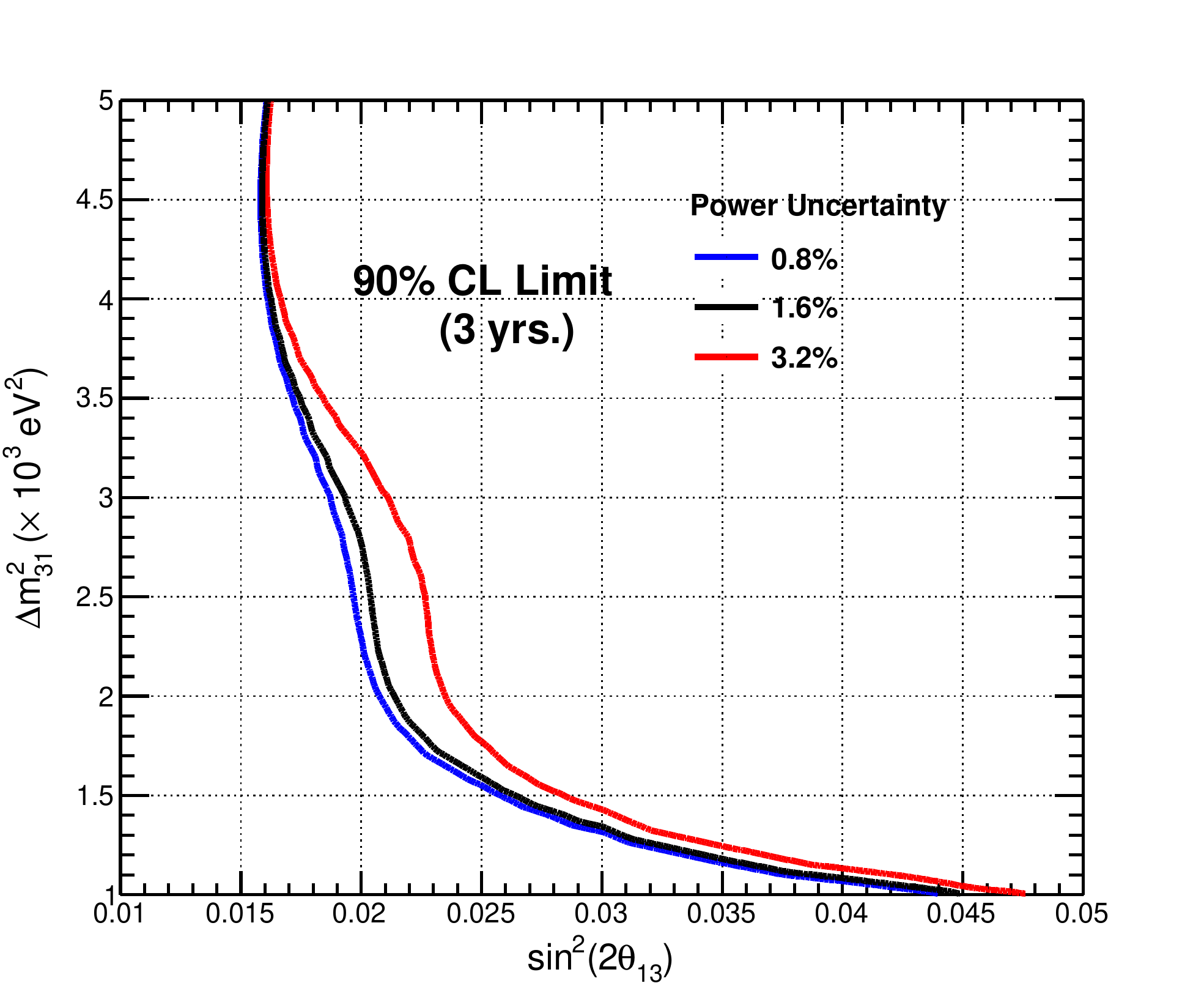}
\includegraphics[width=4.0in]{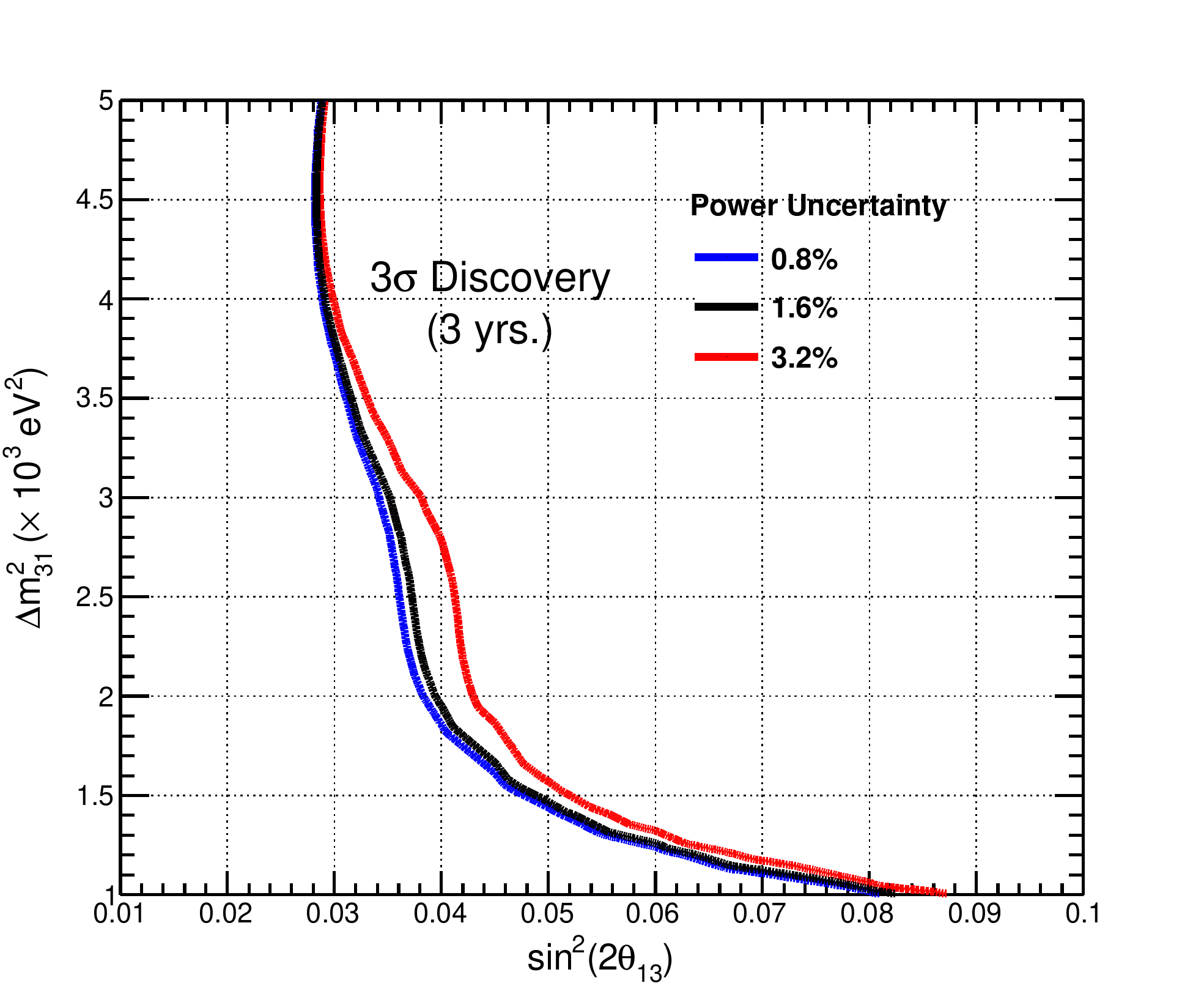}
\end{center}
\caption{Expected 90\% CL limits (top) and discovery potential at
$3\sigma$ (bottom) for three years of data taking
with the systematic uncertainties given in 
Tables~\protect\ref{reactor systematic uncertainty summary}
and power
uncertainties of 0.8\%, 1.6\%, and 3.2\% for each 
reactor. 
The total efficiency of 70 and 40\% for inverse beta 
decay event acceptance for near and far detector are assumed,
respectively. 
Background contributions are not included. 
}
\label{sensitivity}
\end{figure}

The $\chi^2$ is minimized against the parameters $a$, $b^d$, 
$c_i$, $f_r$, and $g^d$ for a given point in the oscillation parameter 
phase space.
We used energy range from 1.0~MeV to 8.0~MeV with a bin size of 0.25~MeV.
Figure~\ref{sensitivity} shows the 90\% CL limits on the
$\Delta m_{31}^2$ {\it vs} $\sin^2(2\theta_{13})$ space for three
years of data taking. 
The discovery potential with a $3\sigma$ significance is also shown
in Fig.~\ref{sensitivity}.
The total efficiencies of 70 and 40\% for inverse beta 
decay events for near and far detectors, respectively, are assumed.
Since the background contributions are not yet fully understood,
background contributions are not included in this calculations.
The limits with the power uncertainty of 0.8\% and 3.2\% ($1/2$ 
and 2 times of the nominal uncertainty of 1.6\%) are also shown.
For the region of interest, $\Delta  m_{31}^2=0.002\sim0.003~{\rm eV}^2$, 
the expected exclusion region for $\sin^2(2\theta_{13})$ lies around 
$\sin^2(2\theta_{13})=0.02$ for the nominal set of uncertainty values.
The limit in this region is somewhat sensitive to the reactor power 
uncertainty worsening by as much as 0.003 for the change in the 
power uncertainty from 1.6\% to 3.2\%.

\newpage

\end{document}